\pdfoutput=1
\documentclass[useAMS,usenatbib]{mnras}
\usepackage[british]{babel}             
\usepackage{newtxtext}               
\usepackage[slantedGreek]{newtxmath}    
\let\la=\lesssim 
\usepackage[T1]{fontenc}   
\usepackage{ae,aecompl}
\usepackage{graphicx}
\usepackage{graphics}
\usepackage{caption}

\graphicspath{./{Figs/}}
\DeclareGraphicsExtensions{.pdf}

\newcommand{\Hii}{H{\sc ii}\ } 
\newcommand{\Te}{T$_\text{e}$}

\newcommand{\mstar}{M$_{\star}$}

\newcommand{\oiii}{[\ion{O}{iii}]$\lambda5007$}
\newcommand{\oii}{[\ion{O}{ii}]$\lambda3727,29$}
\newcommand{\sii}{[\ion{S}{ii}]$\lambda6717,31$}
\newcommand{\nii}{[\ion{N}{ii}]$\lambda6584$}
\newcommand{\siii}{[\ion{S}{iii}]$\lambda9068,9530$}

\def\arcsec{\hbox{$^{\prime\prime}$}}

\title[KLEVER: metallicity gradients]{The KLEVER Survey: Spatially resolved metallicity maps and gradients in a sample of $1.2<\textit{z}<2.5$ lensed galaxies}	

\author[M. Curti et al.]{Mirko Curti,$^{1,2}$ 
\thanks{E-mail: mc2041@cam.ac.uk}
Roberto Maiolino,$^{1,2}$
Michele Cirasuolo,$^{3}$ 
Filippo Mannucci,$^{4}$
\newauthor Rebecca J. Williams,$^{5,1}$
Matt Auger,$^{6}$
Amata Mercurio,$^{7}$
Connor Hayden-Pawson,$^{1,2}$
\newauthor Giovanni Cresci,$^{4}$
Alessandro Marconi,$^{8}$
Francesco Belfiore,$^{3}$
 Michele Cappellari,$^{9}$
\newauthor Claudia Cicone,$^{10,11}$
 Fergus Cullen,$^{12}$
Massimo Meneghetti,$^{13}$
Kazuaki Ota,$^{14}$
\newauthor Yingjie Peng,$^{15}$
Max Pettini,$^{6}$
Mark Swinbank,$^{16,17}$
and Paulina Troncoso$^{18}$
\\
\\
$^{1}$Cavendish Laboratory, University of Cambridge, 19 J. J. Thomson Ave., Cambridge CB3 0HE, UK\\
$^{2}$Kavli Institute for Cosmology, University of Cambridge, Madingley Road, Cambridge CB3 0HA, UK\\
$^{3}$European Southern Observatory, Karl-Schwarzschild-Strasse 2, D-85748 Garching bei Muenchen, Germany\\
$^{4}$INAF - Osservatorio Astrofisico di Arcetri, Largo E. Fermi 5, I-50125, Firenze, Italy\\
$^{5}$Observatory Sciences Ltd, 1 New Road, St Ives, PE27 5BG, Cambridgeshire, UK\\
$^{6}$Institute of Astronomy, University of Cambridge, Cambridge CB3 0HA, UK\\
$^{7}$INAF - Osservatorio Astronomico di Capodimonte, Via Moiariello 16, I-80131 Napoli, Italy\\
$^{8}$Dipartimento di Fisica e Astronomia, Universit\`a di Firenze, Via G. Sansone 1, I-50019, Sesto Fiorentino (Firenze), Italy\\
$^{9}$Sub-department of Astrophysics, Department of Physics, University of Oxford, Denys Wilkinson Building, Keble Road, Oxford OX1 3RH, UK\\
$^{10}$INAF - Osservatorio Astronomico di Brera, via Brera 28, 20121 Milan, Italy \\
$^{11}$Institute of Theoretical Astrophysics, University of Oslo, P.O. Box 1029, Blindern, 0315 Oslo, Norway \\
$^{12}$ Institute for Astronomy, University of Edinburgh, Royal Observatory, Edinburgh EH9 3HJ, UK\\
$^{13}$INAF—Osservatorio Astronomico di Bologna, Via Ranzani 1, I-40127 Bologna, Italy\\
$^{14}$Kyoto University Research Administration Office, Yoshida-Honmachi, Sakyo-ku, Kyoto 606-8501 Japan\\
$^{15}$ Kavli Institute for Astronomy and Astrophysics, Peking University, Beijing 100871, China\\
$^{16}$Institute for Computational Cosmology, Durham University, South Road, Durham DH1 3LE UK\\
$^{17}$Center for Extra-galactic Astronomy, Durham University, South Road, Durham DH1 3LE UK\\
$^{18}$Universidad Aut\'onoma de Chile, Facultad de Ingenier\'ia, N\'ucleo de Astroqu\'imica \& Astrof\'isica, Av. Pedro de Valdivia 425,
Providencia, Santiago, Chile \\
}

\begin{document}

\date{Accepted 2019 November 25. Received 2019 October 28; in original form 2019 June 5}
\pagerange{\pageref{firstpage}--\pageref{fig;append_lastfig}} 
\pubyear{2019}
\maketitle
\label{firstpage}

%%%%%%%%%%%%%%%%%%%%% ABSTRACT %%%%%%%%%%%%%%%%%%%
\begin{abstract}
We present near-infrared observations of $42$ gravitationally lensed galaxies obtained in the framework of the KMOS Lensed Emission Lines and VElocity Review (KLEVER) Survey, a program aimed at investigating the spatially resolved properties of the ionized gas in $1.2<\text{z}<2.5$ galaxies by means of a full coverage of the YJ, H and K near-infrared bands.
Detailed metallicity maps and gradients are derived for a sub-sample of $28$ galaxies from reconstructed source plane emission line maps, exploiting the variety of different emission line diagnostics provided by the broad wavelength coverage of the survey.
About $85 \%$ of these galaxies are characterized by metallicity gradients shallower than $0.05 \rm dex/kpc$ and $89\%$ are consistent with flat slope within $3\sigma$ ($67\%$ within $1\sigma$), suggesting a mild evolution with cosmic time.
In the context of cosmological simulations and chemical evolution models, the presence of efficient feedback mechanisms and/or extended star formation profiles on top of the classical ``inside-out” scenario of mass assembly is generally required to reproduce the observed flatness of the metallicity gradients beyond $z\sim1$ .
Three galaxies with significantly ($>3\sigma$) ``inverted'' gradients are also found, showing an anti-correlation between metallicity and star formation rate density on local scales, possibly suggesting recent episodes of pristine gas accretion or strong radial flows in place.
Nevertheless, the individual metallicity maps are characterised by a variety of different morphologies, with flat radial gradients sometimes hiding non-axisymmetric variations on kpc scales which are washed out by azimuthal averages, especially in interacting systems or in those undergoing local episodes of recent star formation.
\end{abstract}


%%%%%%%%%%%%%%%%%%% INTRODUCTION %%%%%%%%%%%%%%%%%%%
\section{Introduction}
\label{sect:intro}
During the epoch characterized by the peak of the cosmic star formation history (i.e. $1.5<\text{z}<3$) galaxies were experiencing dramatic transformations affecting their morphology and dynamics. % and level of chemical enrichment.
The enhanced star formation activity, regulated by the interplay between cosmic gas accretion, merger events and gas outflows due to the stellar and AGN winds, was responsible for the the bulk of the cosmic evolution of galaxies \citep[e.g.][]{Somerville:2015aa}.
% across the epochs.
All these processes have also left a clear imprint on the content of heavy elements (i.e. metals) in the interstellar medium (ISM) and their spatial distribution across a galaxy \citep{Dave:2011aa}.
Measurements of the gas-phase metallicity (which we simply refer henceforth as metallicity) at these epochs therefore provides unique insights on the history of the baryonic cycling and its influence on the evolution of galaxies (see \citealt{Maiolino:2019aa} for an extensive review).

%Usually, the relative oxygen abundance with respect to hydrogen in the ISM i.e. $12+\text{log(O/H)}$, is chosen as the observational proxy of metallicity.
The existence of global scaling relations involving metallicity and other intrinsic galaxy properties, like stellar mass (i.e. the mass-metallicty relation, MZR, \citealt{Tremonti:2004aa}, \citealt{Andrews:2013aa} and many others), has been assessed through the years thanks to the advent of large astronomical databases such as the Sloan Digital Sky Survey (SDSS, \citealt{Yuan:2011aa}).
These relationships have been further investigated at higher redshifts to search for clues of a possible cosmic evolution. %and set tighter constraints on galaxy evolution models.
Several lines of evidence supporting an overall decrease in metallicity at fixed stellar mass have been found from deep observational campaigns conducted in the near-infrared (e.g. \citealt{Erb:2006ad}, \citealt{Maiolino:2008aa}, \citealt{Mannucci:2009aa},\citealt{Zahid:2011aa}, \citealt{Cullen:2014aa}, \citealt{Steidel:2014aa},\citealt{Sanders:2015ab}, \citealt{Guo:2016aa}).
Moreover, the scatter in the local MZR has been observed to exhibit a clear secondary dependence on the star formation rate \citep{Ellison:2008aa};
%a more tight link between stellar mass, metallicity and SFR;
\cite{Mannucci:2010aa} first proposed the existence of a tight (i.e. $\sim 0.05$ dex dispersion) relation, followed by local galaxies, in the three-dimensional space defined by stellar mass (\mstar), metallicity (Z) and star formation rate (SFR), which is usually referred to as the Fundamental Metallicity Relation (FMR).
This tight relation, later tested and revised by different authors (e.g. \citealt{Yates:2012aa, Salim:2014aa, Curti:2019aa}), has been interpreted as a consequence of a long lasting equilibrium between gas accretion, mass growth, metal production and outflows of enriched material, and does not show any clear sign of evolution up to $z \sim 2.5$ \citep{Mannucci:2010aa, Cresci:2018aa}.
%Galaxies in the local Universe are thus found to lie on a surface in the 3D space defined by these three quantities, 
In this picture, the observed MZR (at any cosmic time) just follows from the two-dimensional projection of the FMR on the M$_{\star}$ vs log(O/H) plane, while its cosmic evolution arises from sampling different regions of the FMR due to the increase in the average SFR density with redshift.
Despite a variety of physically motivated theoretical frameworks supporting this scenario 
(e.g. \citealt{Lilly:2013aa, Dayal:2013aa, Hunt:2016ab, Dave:2017aa}), as well as several confirmations from observations of high-z galaxies   \citep[e.g.][]{Richard:2011aa,Belli:2013aa,Wang:2017aa},
the existence and the possible evolution of the M-Z-SFR relation at high redshift is currently debated (\citealt{Steidel:2014aa,Cullen:2014aa,Wuyts:2014aa,Wuyts:2016aa, Sanders:2018aa}, see the discussion in \citealt{Cresci:2018aa}).
%and, recently, even its own existence has been questioned by some studies based on integral field spectroscopy (\citealt{Sanchez:2017aa}, \citealt{Barrera-Ballesteros:2017aa}).

With the increased availability of multi-object and integral field spectrographs, many studies have also turned to investigating the spatial distribution of metals inside galaxies, assessing the presence of radial variations in the chemical enrichment levels. %, i.e. of metallicity gradients.
Tracing the evolution of these metallicity gradients across the cosmic epochs is a crucial benchmark for theoretical models aimed at describing the relative contributions that star formation, gas flows and feedback processes play in driving galaxy evolution.
Although relatively well characterized in the local Universe, the main features of metallicity gradients still remain so far very poorly constrained at higher redshifts. 
%, particularly at $z\sim1.5$ around the peak of star formation in the Universe \citep{hopkins06} and where galaxies accumulated and produced the majority of their stellar mass. 
Locally, the large majority of spiral galaxies exhibit negative metallicity gradients, with inner regions more chemically enriched with respect to the outskirts of galactic discs. 
%can be properly mapped only in the MW (Deharveng et al., 2000; Esteban et al., 2005; Rudolph et al., 2006; Balser et al., 2011) and in a few nearby galaxies (e.g., Bresolin, 2007a; Bresolin et al., 2009b, 2012; Werk et al., 2011; Berg et al., 2012, 2013, 2015b). 
This follows observationally from the spectroscopic analysis of \Hii regions in the Milky Way \citep[e.g.][]{Magrini:2010aa, Stanghellini:2010ab} and in nearby galaxies
\citep[e.g.][]{Zaritsky:1994aa,Kewley:2010aa, Bresolin:2011aa,Berg:2012aa,Berg:2015aa}, from large campaign based on integral field spectroscopy \citep[e.g.][]{Sanchez:2014aa,Ho:2015aa,Belfiore:2017aa}, as well as from abundance measurements from stellar spectroscopy of individual massive young stars \citep{Kudritzki:2015aa,Gazak:2015aa,Bresolin:2016aa} and spatially resolved analysis of stellar population properties (e.g \citealt{Li:2018aa}, see \citealt{Conroy:2013aa} for a review).
%as well as from Galactic archaeology.  %\cite{vanZee:1998aa}
Negative gradients are generally interpreted as indicative of the so called inside-out growth scenario of galaxy formation \citep{Samland:1997aa,Portinari:1999aa,Prantzos:2000aa,Pilkington:2012aa, Gibson:2013aa}.
Indications of flattening gradients beyond a certain radius may indicate the presence of radial mixing processes or (re)accretion of metal-enriched gas in the outer regions \citep{Bresolin:2012aa}, whilst intense accretion af external pristine gas may also produce inverted gradients (as seen for instance in some dwarfs galaxies, \citealt{Sanchez-Almeida:2018ab}).
Merger events could also play an important role in flattening the metallicity gradients \citep{Kewley:2010aa,Rupke:2010aa,Rupke:2010ab}.

On the contrary, there is no general consensus yet in the literature about the behaviour of metallicity gradients at high redshift and their cosmic evolution.
Planetary nebulae have been extensively used to investigate the time evolution of abundance gradients in the Milky Way and in nearby galaxies, as they probe the enrichment of the gas on different (i.e. older) timescales than \Hii regions (\citealt{Maciel:2003aa,Magrini:2009aa,Henry:2010aa,Stanghellini:2010aa,Stanghellini:2014aa}).
In general, it is found that gradients inferred from tracers of long time-scales enrichment tend to be flatter than those inferred from \Hii regions \citep[see e.g.][]{Stanghellini:2014aa}.
However, it is not trivial to account for the effects of radial stellar migration, which could potentially bias the results of stellar archaeology studies. 
The situation is perhaps even more uncertain for high redshift galaxies, where the investigations of metallicity gradients conducted so far have led to diverse and sometimes conflicting conclusions (e.g. \citealt{Cresci:2010aa,Swinbank:2012aa, Queyrel:2012aa,Stott:2014aa,Jones:2013aa,Wuyts:2016aa,Leethochawalit:2016aa, Wang:2017aa}). 
This could be mainly ascribed to the intrinsic challenges in obtaining reliable measurements of metallicity gradients at high-z, due to the poor angular resolution \citep{Yuan:2013aa} and to uncertainties affecting metallicity diagnostics.
If the latter problem is currently still a source of large systematics, the former could be addressed and partly mitigated by exploiting new adaptive optics systems and/or the magnification provided by gravitational lensing. 

The number of direct measurements of metallicity gradients at high redshift has been constantly increasing in recent years thanks to the advent of multi-IFU instrumentation like KMOS \citep{Sharples:2013aa} on the Very Large Telescope (VLT), with surveys targeting hundreds of galaxies at $z \sim 1-2$ (e.g. \citealt{Stott:2014aa,Wuyts:2016aa}).
However, a typical seeing $\geq0.6\arcsec$, which roughly corresponds to $\sim 5 \text{kpc}$ at $\text{z}\sim 2$, does not allow us to properly resolve the inner structure of high-redshift galaxies, especially for low mass objects.
As previously stated, adaptive optics assisted observations could greatly enhance the spatial resolution down to $\sim 1 \text{kpc}$ \citep{Swinbank:2012aa}, which can be further
improved by targeting gravitationally lensed sources \citep{Jones:2010aa,Jones:2010ab,Jones:2013aa,Leethochawalit:2016aa}.
Space-based grism spectroscopy from \emph{HST} has also been recently used to infer sub-kpc resolution metallicity maps and gradients \citep{Jones:2015ab,Wang:2017aa}, which are however affected by poor spectral resolution. 
Despite these efforts, high angular resolution measurements of metallicity gradients are still scarce in terms of statistical significance.
%\cite{Jones:2010aa} \cite{Jones:2010ab}\cite{Jones:2013aa} \cite{Swinbank:2012aa}\cite{Leethochawalit:2016aa}
%\cite{Yuan:2013aa} \cite{Stott:2014aa}
Moreover, the large majority of the studies conducted so far rely only on a limited number of emission line detections, with metallicity estimates that could therefore suffer from potential biases due to the different physical properties of high redshift galaxies with respect to the local samples used to calibrate the abundance diagnostics.

\begin{figure*}
\centering
\includegraphics[width=0.495\textwidth]{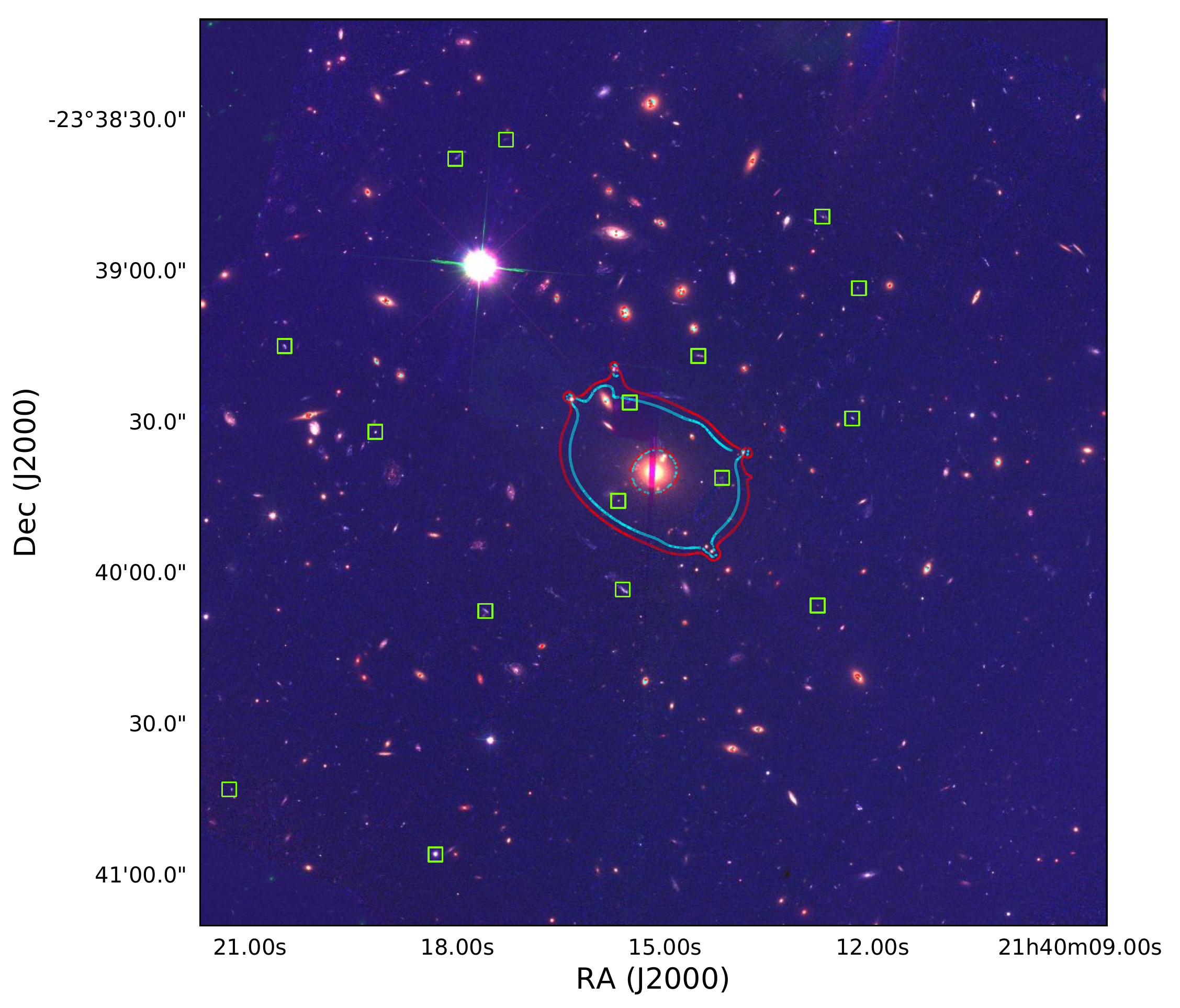}
%\centering
\includegraphics[width=0.495\textwidth]{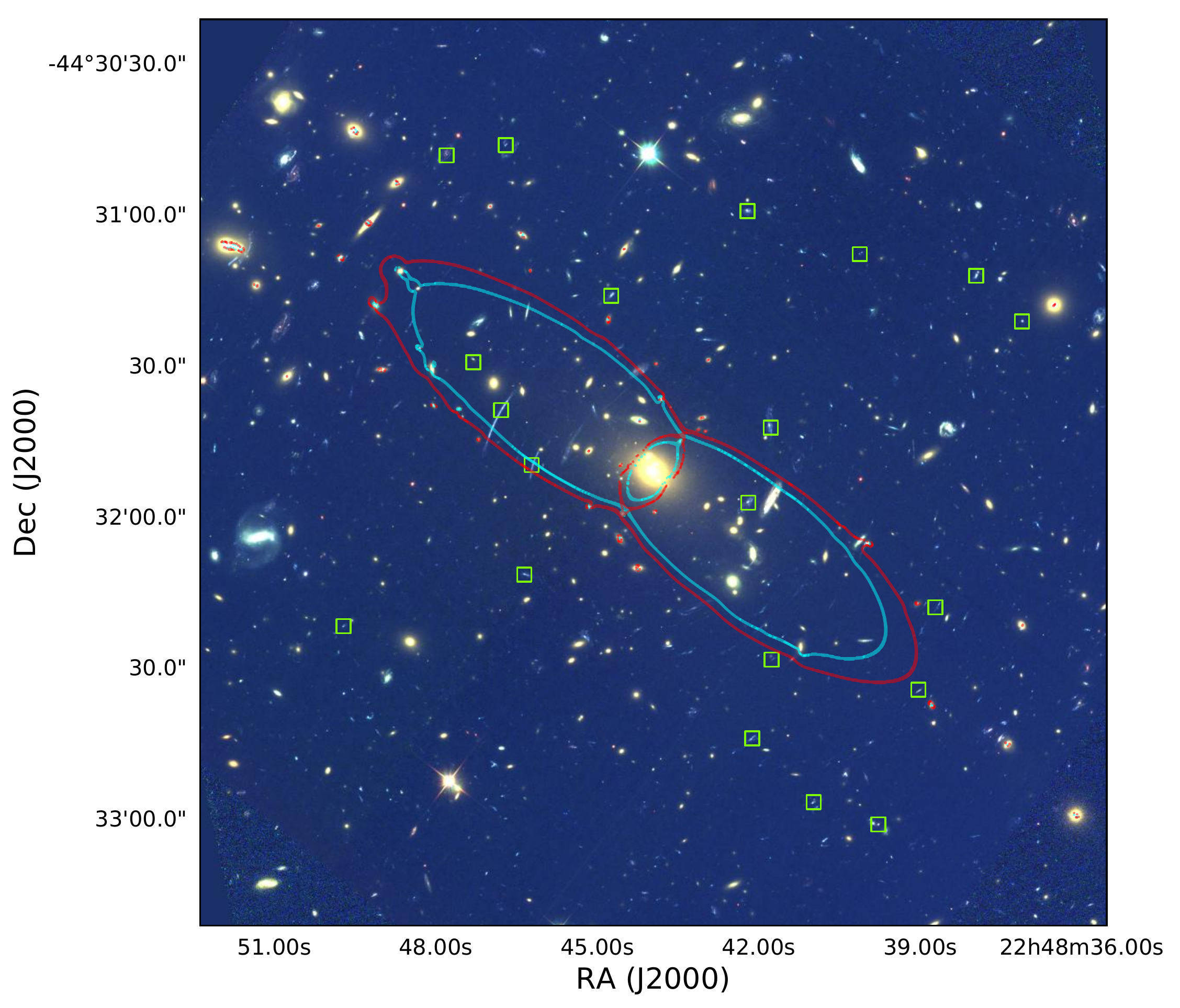} 
\caption{Composite RGB-color image of the MS2137 (\textit{left panel}) and RXJ2248 (\textit{right panel}) clusters. The red, green and blue channels are assigned to broad band F814W, F606W and F435W images obtained in the framework of the CLASH and Frontier Fields program respectively. The green squares superimposed mark the position of deployed KMOS IFUs targeting the high redshift sources analysed in this work, while cyan and red contours represent the critical lines of infinite magnifications predicted, for $\text{z}=1.4$ and $\text{z}=2.2$ source redshift respectively, by the lensing models adopted in this work, i.e. \citet{Zitrin:2013aa}. }
\label{fig:kmos_fov}
\end{figure*}

In this paper we investigate the metallicity properties of a sample of 42 galaxies between $1.2<\text{z}<2.5$, gravitationally lensed by either foreground galaxy clusters (in particular the MS2137 and RXJ2248 clusters) or individual galaxies.
The analysis presented in this work is based on the first observations conducted in the framework of the KMOS LEnsed galaxies Velocity and Emission line Review (KLEVER) Large Programme ($39$ galaxies) and include also $3$ strongly lensed sources observed with SINFONI in the context of different observational programs. 
We take advantage of multi-band observations conducted in the YJ, H and K bands, which allow the simultaneous coverage of many rest-frame optical emission lines, to derive spatially resolved metallicity maps using different strong-line diagnostics and assess the presence and the cosmic evolution of metallicity gradients. 
We manage to obtain at least marginally resolved gradients for 28 out of 42 the galaxies presented in this work: we refer to these galaxies throughout the paper as the \textit{metallicity gradient sample}.

The paper is organized as follows: in Section 2 we give an overview on the KLEVER Programme, describing the target selection, the observing strategy and the data reduction.
Section 3 describes the emission line fitting method, the lens modelling and Source Plane reconstruction for our galaxies and how we measured stellar masses, star formation rates and gas-phase metallicities.
In Section 4 we present our results in terms of global sample properties and spatially resolved metallicity maps and gradients, which are then discussed in Section 5.
Finally, our conclusions are reported in Section 6.

In Table~\ref{tab:line_ratios}, we report the notations used throughout the paper to indicate the line ratios adopted in our analysis.
Throughout this work we assume a standard $\Lambda$CDM cosmology based on the results from \cite{Planck-Collaboration:2016aa}, with H$_{0}$ = $67.8\ \text{km}\  \text{s}^{-1}\  \text{Mpc}^{-1}$, 
$\rm \Omega_{\text{M}}$ = $0.308$, $\rm \Omega_{\rm \Lambda}$ = $0.69$.

\begin{table}
	\centering
	\caption{ Definition of line ratios adopted throughout this paper.}
	\begin{tabular}{|@{} l|c @{}|}\hline
		\text{Notation} & \text{Line Ratio} \\
		\hline
		R$_{2}$  &  \oii/H$\beta$ \\
		R$_{3}$ & \oiii/H$\beta$\\
		N$_{2}$ & \nii/H$\alpha$ \\
		S$_{2}$ & \sii/H$\alpha$\\
%		R$_{23}$ & (\oii + [\ion{O}{iii}]$\lambda4959,5007$)/H$\beta$ \\
		O$_{3}$O$_{2}$ &  \oiii/\oii \\
%		RS$_{32}$ & \oiii/H$\beta$ +  \sii/H$\alpha$ \\
%		O$_{3}$S$_{2}$ & \oiii/H$\beta$ / \sii/H$\alpha$  \\
		O$_{3}$N$_{2}$ & \oiii/H$\beta$ / \nii/H$\alpha$  \\
		\hline
	\end{tabular}
	
	\label{tab:line_ratios}
\end{table}

%%%%%%%%%%%%%%%%%%%%% DATA REDUCT. %%%%%%%%%%%%%%%%%%%
\section{Observations}
\label{sect:sample}
\subsection{The KLEVER Survey: sample selection and observing strategy}

\begin{table*}

\begin{tabular}{@{} cccccccc @{} }
%	\begin{tabularx}{\textwidth}{@{} l *{7}{>{\hsize=0.35\hsize}x} @{} }
	\hline
	Galaxy  & R.A. & Dec. & z [H$\upalpha$]	  &  \textit{J}-Band  & \textit{H}-Band &\textit{K}-Band\\
	\hline
	\textbf{SINFONI Galaxies} && &  &  &&  & \\
	Horseshoe &11:48:32.7  & 19:30:03.5 & 2.383  &  - & [\ion{O}{iii}]+H$\upbeta$ & H$\upalpha$+[\ion{N}{ii}]+[\ion{S}{ii}] \\
	 \ \textit{(Western arc)} && & &   &&  & \\
	Horseshoe &11:48:32.7  & 19:30:03.5 & 2.383  & - &-&H$\upalpha$+[\ion{N}{ii}]+[\ion{S}{ii}] \\
	  \ \textit{(South+western arc)} && & &  &&  & \\
	MACS0451 Arc & 04:51:57.3 &  00:06:19.7 & 2.014   & [\ion{O}{ii}]   &  [\ion{O}{III}]+H$\upbeta$ & H$\upalpha$+[\ion{N}{ii}]+[\ion{S}{ii}] \\
%	MACS0451 & 04:51:57.3 &  00:06:19.7 & 2.014  &-  & [\ion{O}{III}]+H$\upbeta$ &-  \\
	CSWA164 & 02:32:49.8 & -03:23:26.6 & 2.518   & [\ion{O}{ii}] & [\ion{O}{iii}]+H$\upbeta$ & H$\upalpha$+[\ion{N}{ii}]+[\ion{S}{ii}] \\

\textbf{KMOS Galaxies}&& & &  &&  & \\
\textbf{MS2137}&& & &  &&  & \\
sp1 & 21:40:18.031 & -23:38:37.87 & 1.393 & [\ion{O}{iii}]+H$\upbeta$ & H$\upalpha$+[\ion{N}{ii}]+[\ion{S}{ii}] & - \\
sp2 & 21:40:12.730 & -23:38:49.34 & 2.2425 & [\ion{O}{ii}] & [\ion{O}{iii}]+H$\upbeta$ & H$\upalpha$+[\ion{N}{ii}]+[\ion{S}{ii}] \\
sp3 & 21:40:14.522 & -23:39:17.03 & 1.2645 & [\ion{O}{iii}]+H$\upbeta$ & H$\upalpha$+[\ion{N}{ii}]+[\ion{S}{ii}] & [\ion{S}{iii}] \\
sp5 & 21:40:12.298 & -23:39:29.44 & 2.0146 & [\ion{O}{ii}] & [\ion{O}{iii}]+H$\upbeta$ & H$\upalpha$+[\ion{N}{ii}]+[\ion{S}{ii}] \\
sp6 & 21:40:19.188 & -23:39:32.11 & 2.4881 & [\ion{O}{ii}] & [\ion{O}{iii}]+H$\upbeta$ & H$\upalpha$+[\ion{S}{ii}] \\
sp7 & 21:40:17.599 & -23:40:07.70 & 1.6523 & [\ion{O}{iii}]+H$\upbeta$ & H$\upalpha$+[\ion{N}{ii}]+[\ion{S}{ii}] & - \\
sp9$^{*}$ & 21:40:18.319 & -23:40:56.04 & 1.6497 & H$\upbeta$ +[\ion{O}{iii}] & H$\upalpha$+[\ion{N}{ii}] & - \\
sp10 & 21:40:15.679 & -23:39:45.84 & 3.0843 & - & - & [\ion{O}{iii}] \\
sp13 & 21:40:15.510 & -23:39:26.27 & 1.4951 & [\ion{O}{iii}]+H$\upbeta$ & H$\upalpha$+[\ion{S}{ii}] & - \\
sp14 & 21:40:14.178 & -23:39:41.26 & 1.4948 & [\ion{O}{iii}]+H$\upbeta$ & H$\upalpha$+[\ion{S}{ii}] & - \\
sp15 & 21:40:15.614 & -23:40:03.42 & 1.4967 & [\ion{O}{iii}]+H$\upbeta$ & H$\upalpha$+[\ion{N}{ii}]+[\ion{S}{ii}] & - \\
ph6532 & 21:40:21.265 & -23:40:43.11 & 2.071 & [\ion{O}{ii}] & [\ion{O}{iii}]+H$\upbeta$ & H$\upalpha$ \\
ph2594 & 21:40:17.300 & -23:38:34.06 & 1.348 & [\ion{O}{iii}] & H$\upalpha$+[\ion{N}{ii}]+[\ion{S}{ii}] & - \\
ph3729 & 21:40:12.214 & -23:39:3.561 & 1.4291 & [\ion{O}{iii}]+H$\upbeta$ & H$\upalpha$+[\ion{N}{ii}] & - \\
ph3912 & 21:40:20.497 & -23:39:15.06 & 1.3275 & [\ion{O}{iii}]+H$\upbeta$ & H$\upalpha$+[\ion{S}{ii}] & - \\
ph5514 & 21:40:12.795 & -23:40:6.609 & 1.5254 & - & H$\upalpha$+[\ion{N}{ii}]+[\ion{S}{ii}] & - \\
ph7727 & 21:40:12.228 & -23:41:25.74 & 1.5789 & [\ion{O}{iii}] & H$\upalpha$ & - \\
ph8073 & 21:40:10.607 & -23:41:37.73 & 1.2386 & [\ion{O}{iii}]+H$\upbeta$ & H$\upalpha$+[\ion{S}{ii}]& [\ion{S}{iii}] \\

\textbf{RXJ2248 (AS1063)} && & &  &&  & \\
GLASS\_00093-99-99 & 22:48:46.701 & -44:30:46.27 & 1.4317 & [\ion{O}{iii}]+H$\upbeta$ & H$\upalpha$+[\ion{N}{ii}] & - \\
R2248\_LRb\_p1\_M3\_Q4\_58\_\_2 & 22:48:42.213 & -44:30:59.38 & 1.4199 & [\ion{O}{iii}]+H$\upbeta$ & H$\upalpha$+[\ion{N}{ii}]+[\ion{S}{ii}] & - \\
MUSE\_SW\_462-99-99 & 22:48:41.781 & -44:31:42.39 & 1.4286 & [\ion{O}{iii}]+H$\upbeta$ & H$\upalpha$+[\ion{N}{ii}]+[\ion{S}{ii}] & - \\
GLASS\_00333-99-99 & 22:48:40.130 & -44:31:07.90 & 1.4285 & [\ion{O}{iii}]+H$\upbeta$ & H$\upalpha$ & - \\
R2248\_LRb\_p3\_M4\_Q3\_93\_\_1 & 22:48:37.966 & -44:31:12.21 & 1.485 & [\ion{O}{iii}]+H$\upbeta$ & H$\upalpha$+[\ion{N}{ii}]+[\ion{S}{ii}] & [\ion{S}{iii}] \\
R2248\_MR\_p1\_M1\_Q4\_10\_\_1 & 22:48:35.016 & -44:30:30.07 & 1.4312 & [\ion{O}{iii}] & H$\upalpha$ & - \\
R2248\_LRb\_p3\_M4\_Q3\_94\_\_1 & 22:48:37.116 & -44:31:21.29 & 2.0665 & [\ion{O}{ii}] & [\ion{O}{iii}]+H$\upbeta$ & H$\upalpha$+[\ion{N}{ii}]+[\ion{S}{ii}] \\
MUSE\_SW\_48-99-99 & 22:48:38.723 & -44:32:18.05 & 2.5804 & - & - & - \\
MUSE\_SW\_51-99-99 & 22:48:41.764 & -44:32:28.47 & 3.2275 & - & - & [\ion{O}{iii}] \\
R2248\_MR\_p3\_M1\_Q3\_43\_\_1 & 22:48:39.037 & -44:32:34.41 & 3.2421 & - & [\ion{O}{ii}] & [\ion{O}{iii}]+H$\upbeta$ \\
GLASS\_01891-99-99 & 22:48:39.780 & -44:33:01.17 & 1.159 & - & - & - \\
GLASS\_01845-99-99 & 22:48:40.984 & -44:32:56.75 & 2.3014 & [\ion{O}{ii}] & [\ion{O}{iii}]+H$\upbeta$ & H$\upalpha$ \\
MUSE\_SW\_45-99-99 & 22:48:42.125 & -44:32:44.08 & 1.2694 & [\ion{O}{iii}]+H$\upbeta$ & H$\upalpha$+[\ion{N}{ii}]+[\ion{S}{ii}] & [\ion{S}{iii}] \\
MUSE\_SW\_461-99-99 & 22:48:42.199 & -44:31:57.29 & 1.4286 & [\ion{O}{iii}]+H$\upbeta$ & H$\upalpha$+[\ion{S}{ii}] & - \\
MUSE\_NE\_111-99-99 & 22:48:46.358 & -44:32:11.56 & 1.3975 & [\ion{O}{iii}]+H$\upbeta$ & H$\upalpha$+[\ion{S}{ii}] & - \\
GLASS\_01404-99-99 & 22:48:49.712 & -44:32:21.81 & 1.4475 & - & - & [\ion{S}{iii}] \\
R2248\_MR\_p1\_M1\_Q4\_51\_\_1 & 22:48:46.222 & -44:31:49.81 & 1.2593 & [\ion{O}{iii}]+H$\upbeta$ & H$\upalpha$ & [\ion{S}{iii}] \\
GLASS\_00800-99-99 & 22:48:46.787 & -44:31:38.91 & 1.2282 & [\ion{O}{iii}]+H$\upbeta$ & H$\upalpha$+[\ion{N}{ii}]+[\ion{S}{ii}] & [\ion{S}{iii}] \\
MUSE\_NE\_23-99-99 & 22:48:44.742 & -44:31:16.18 & 1.2282 & [\ion{O}{iii}]+H$\upbeta$ & H$\upalpha$+[\ion{S}{ii}] & - \\
R2248\_MR\_p1\_M1\_Q4\_59\_\_1 & 22:48:47.796 & -44:30:48.28 & 1.4281 & - & H$\upalpha$+[\ion{N}{ii}]+[\ion{S}{ii}] & - \\
MUSE\_NE\_117-99-99 & 22:48:47.302 & -44:31:29.37 & 3.4519 & - & - & - \\

	\hline
	\end{tabular}
\caption{The full sample of galaxies analysed in this work. The systemic redshift reported is derived from the H$\upalpha$ detection in the integrated spectra. In case of no H$\upalpha$ detection (e.g. for $z>3$ galaxies), the redshift is computed from \oiii. 
We also report the main emission lines detected above $3 \sigma$ in the integrated spectra in each targeted band. 
A `-' denotes that observations in that band were available, but we did not detect any emission line with at least $3 \sigma$ significance, with the exception of the Horseshoe galaxy for which observations in the J-band were not available. \newline
$^{*}$: type-1 AGN
}
\label{tab:klever_sources} 
\end{table*}

KLEVER is an ESO Large Program (197.A-0717, PI: Michele Cirasuolo) conducted with the multi-object near-IR integral field spectrograph KMOS on the VLT \citep{Sharples:2013aa} and aimed at investigating spatially resolved kinematics, dynamics and properties of the ionised gas in a sample of $\sim 200$ galaxies at $1.2\lesssim \text{z} \lesssim 2.5$.
The survey is designed to provide a full coverage of the near-infrared region of the spectrum by observing each galaxy in the YJ, H and K band (spanning respectively the $1.025-1.344\ \mu$m, $1.456-1.846\ \mu$m and $1.92-2.46\ \mu$m wavelength ranges), hence allowing us to detect and spatially map almost the entire set of the brightest optical rest-frame nebular lines. 
%Compared to the majority of previous programs conducted with KMOS, which focused on the observations of one specific band
%which constitutes a unique and powerful diagnostic tool to characterize the physical conditions of high redshift galaxies.% at these cosmic epochs.
%The targets have been \textbf{primarily} selected to fall within specific redshift ranges, either at z $\in$ [$1.2$ , $1.65$] to have H$\upbeta$+[\ion{O}{iii}]$\lambda 5007$ in the \emph{YJ} band, H$\upalpha$+[\ion{N}{ii}]$\lambda 6584$+[\ion{S}{ii}]$\lambda\lambda 6717,31$ in the \emph{H} band and [\ion{S}{iii}]$\lambda\lambda 9068,9530$ in the \emph{K} band, or at z $\in$ [$2$ , $2.6$] to have [\ion{O}{ii}]$\lambda\lambda 3727,29$ in the \emph{YJ} band, H$\upbeta$+[\ion{O}{iii}]$\lambda 5007$ in the \emph{H} band and H$\upalpha$+[\ion{N}{ii}]$\lambda 6584$+[\ion{S}{ii}]$\lambda\lambda 6717,31$ in the \emph{K} band. 
%as well as to avoid, as much as possible, OH sky line contamination.
The full KLEVER sample comprises both gravitationally lensed galaxies within well studied cluster fields from the CLASH \citep{Postman:2012aa} and FRONTIER FIELDS \citep{Lotz:2017aa} programs, as well as un-lensed galaxies in the southern CANDELS fields UDS, COSMOS and GOODS-S.

%Program ID - MS2137:095.B-0480
The analysis presented in this paper is based on the first available observations in KLEVER, targeting 39 gravitationally lensed galaxies behind the clusters MS2137-2353 (hereafter MS2137) and RXJ2248.7-4431 (also know as AS1063, hereafter RXJ2248) and carried out in Service Mode during Periods 95-97 (from May 2015 to September 2016). 
During the creation of the KMOS mask, the targets within each pointing have been prioritised according to the observability of the emission lines of interest, in order to maximise the number of lines falling within the different NIR bands 
while minimising at the same time the sky contamination from OH lines (as identified from the catalogue provided by \citealt{Rousselot:2000aa}).
The targets have been primarily selected to fall within specific redshift ranges, either at z $\in$ [$1.2$ , $1.65$] to have H$\upbeta$+[\ion{O}{iii}]$\lambda 5007$ in the \emph{YJ} band, H$\upalpha$+[\ion{N}{ii}]$\lambda 6584$+[\ion{S}{ii}]$\lambda\lambda 6717,31$ in the \emph{H} band and [\ion{S}{iii}]$\lambda\lambda 9068,9530$ in the \emph{K} band, or at z $\in$ [$2$ , $2.6$] to have [\ion{O}{ii}]$\lambda\lambda 3727,29$ in the \emph{YJ} band, H$\upbeta$+[\ion{O}{iii}]$\lambda 5007$ in the \emph{H} band and H$\upalpha$+[\ion{N}{ii}]$\lambda 6584$+[\ion{S}{ii}]$\lambda\lambda 6717,31$ in the \emph{K} band. 
Spectroscopic redshifts used for target selection and prioritisation were provided as part of the CLASH-VLT survey \citep{Rosati:2014aa} conducted with VIMOS on the VLT.
Three IFUs have been assigned, for each pointing, to bright continuum sources for alignment purposes (see Sect.~\ref{sect:reduction}), while remaining spare IFUs were assigned to low priority targets or z$>3$ sources. 
No prior screening to identify AGN contamination was performed for the galaxies within the clusters.

Each lensed galaxy belonging to RXJ2248 has been observed for a total exposure time on source of $11$ hours ($3$h in the \emph{YJ} band, $3$h in \emph{H} and $5$h in \emph{K} respectively), whereas galaxies from MS2137-2353 have been observed in total for $13.1$ hours ($4.2$ in YJ, $4.5$ in H and $4.4$ in K band respectively).
For KMOS observations, we adopted an A-B-A nodding (with dithering) strategy for sky sampling and subtraction.
%From the KMOS observations of the two clusters, as mentioned above, we obtained an almost full near-IR coverage in J,H and K band, spanning respectively the $1.025-1.344\ \mu$m, $1.456-1.846\ \mu$m and $1.34-2.46\ \mu$m wavelength windows.
The average seeing of the observations (as inferred from the reference stars observed in three KMOS IFUs) ranged between $0.5\arcsec$ and $0.6\arcsec$.
In Figure \ref{fig:kmos_fov} the position of the KMOS IFUs, deployed on the target galaxies, are shown on top of color-composite RGB images of both cluster fields.
The coloured curves in Fig.~\ref{fig:kmos_fov} denote the critical lines of infinite magnification within the two clusters for source redshift $1.4$ (cyan) and $2.2$ (red) respectively, as predicted by the adopted lensing models (see Section \ref{sect:lensmod}). 

In addition, we have included in the analysis three strongly lensed galaxies at z$>2$ observed with the integral field spectrograph SINFONI and part of galaxy-galaxy lensing systems or lensed by galaxy clusters. 
These galaxies fully match the criteria described above for the KMOS-sample in terms of emission lines detectability.
Two out of the three galaxies were observed as part of some of our previous programs: SDSS J114833+193003 (also known as the \textit{Horseshoe}) and a strongly lensed, arc-like shaped galaxy within the MACS J0451+0006 (MACS0451) cluster.
Seeing-limited mode observations have been performed, with a PSF-FWHM ranging between $0.4\arcsec$ to $0.9\arcsec$, and the $0.25\arcsec$ x $0.25\arcsec$ pixel scale was adopted.
The observations of the third additional galaxy (SDSS J0232-0323, also know as CSWA164 as part of the CASSOWARY Survey \citet{Belokurov:2009aa}) were retrieved instead from the ESO archive. 
This brings the total observed sample to 42 galaxies.
%In particular, we included A68/HLS115 and SDSS J0232-0323 (also know as CSWA64).
We note that, due to scheduling constraints, not all the galaxies observed with SINFONI have the desired full wavelength coverage 
(i.e. \emph{J}, \emph{H} and \emph{K} band observations) as the KMOS galaxies.

The complete list of the targets analysed in this work is given in Table~\ref{tab:klever_sources}, for which we report coordinates, redshift and emission lines detected above the $3\sigma$ level within each band from the integrated spectra.
Unfortunately, low signal-to-noise affected the detection of the faintest nebular lines (i.e H$\upbeta$, \nii, \sii) in some of the KMOS datacubes, reducing the effective number of sources for which we could reliably derive a metallicity gradient to 28 (the \textit{metallicty gradient sample} see Sect.~\ref{sect:grads} for details).

\subsection{Data Reduction}
\label{sect:reduction}
The KMOS data were reduced using the pipeline provided by ESO (v.1.4.3).
Within the pipeline environment, we implemented the advanced sky subtraction technique from \cite{Davies:2007aa} as well as the sky-stretch algorithm, which stretches the sky-cube involving a relatively high degree polynomial in order to align the sky lines with those in the object-cube. 
This slightly improved the residuals from the first correction. 
The final datacubes were then reconstructed onto a $0.1''$ x $0.1''$ pixel scale. 
To properly align and combine the individual exposures, both within a single observing block (OB) 
and between different OBs, three IFUs (i.e. one for each of the KMOS detectors) were devoted to observe bright stars; we then exploited the relative position of their centroids in each exposure to compute the shifts to apply to the scientific sources that were observed on the same detector of the corresponding reference star. 
This method has proven to provide more precise alignments than just relying only on the informations stored in the header, since it mitigates potential offsets introduced by the differential rotation of the instrument's IFUs at different times that are non tracked by the World Coordinate System (WCS). 
The final cubes were then created through a (three) sigma-clipped average.

The SINFONI data were reduced following the latest version of the ESO-SINFONI pipeline to perform the flat fielding, wavelength calibration and reconstruct a non-sky subtracted cube for each observation, after the removal of cosmic rays from raw data using the ``L.A.Cosmic'' procedure by \cite{van-Dokkum:2001aa}.  
The pixels in the datacubes were resampled to a symmetric angular size of $0.125''$ x $0.125''$. 
Then, we implemented the sky subtraction technique from \cite{Davies:2007aa} to perform a better removal of the residual OH airglow emission lines from the data. 
We corrected for the atmospheric absorption and instrumental response using a telluric standard star, which also provides the flux calibration, before finally combining all the single OBs through a sigma-clipped average to produce the final science cubes.

\section{Analysis}
\subsection{Emission line fitting}
\label{sect:fitting}

On spatially resolved basis, we performed the emission line fitting on the fully reduced datacubes, which sample the image plane of each galaxy.
We spatially smoothed the datacubes with a Gaussian kernel of $0.4\arcsec$ FWHM, below the average PSF-FWHM of the observations ($\sim 0.5\arcsec - 0.6\arcsec$, as measured from the three reference stars), to increase the signal-to-noise ratio (SNR) of the weakest emission lines that we aim to detect and map in our galaxies (e.g. [\ion{N}{ii}]$\lambda 6583$, [\ion{S}{ii}]$\lambda 6717,6731$ and H$\upbeta$).

All the emission lines of interest were fitted with single Gaussian components at the location of each spaxel in the datacube and 
we included a linear component to account for continuum emission in each band. 
The H$\upalpha$, [\ion{N}{ii}] and [\ion{S}{ii}] lines were jointly fit by linking their velocity and line width, and the same criteria was applied when fitting [\ion{O}{iii}] and H$\upbeta$. 
When available, the [\ion{O}{ii}]$\lambda\lambda 3726,3729$ and [\ion{S}{iii}]$\lambda\lambda 9068,9532$ line doublets are fitted with two Gaussian components which are linked in velocity and width as well.
Moreover, the [\ion{O}{iii}]$\lambda\lambda 4959,5007$ and [\ion{S}{iii}]$\lambda\lambda 9068,9532$ doublets were fixed in flux ratios to their relative Einstein spontaneous emission coefficients (i.e. $3$ and $2.47$ respectively), and the relative intensity of the two lines of the \sii doublet was constrained to be within the physical range associated to the low- and high-density regimes (i.e. [\ion{S}{ii}]$\lambda 6717$/[\ion{S}{ii}]$\lambda 6731$ $\in [0.4 ,1.45]$). 
We choose to keep the fitting procedure within each band self-consistent: we do not link the line widths between different bands because of the different KMOS resolving powers (namely $3582$, $4045$ and $4227$ in the centre of YJ, H and K band respectively). 
Velocities were not linked as well to avoid any systematic introduced by possible offsets in the wavelength calibration between the three bands.
Each fit on spaxel-by-spaxel basis was inverse weighted by the corresponding noise spectrum extracted from the noise datacube provided by the pipeline, and the spectra were totally masked at the position of the brightest sky lines. 
%The signal-to-noise ratio was estimated for each emission line on the integrated line flux. 

For each galaxy, the emission line maps from different bands were finally re-aligned exploiting the detection of the underlying stellar continuum or, in case of no continuum detection, aligning the peaks of the brightest emission lines detected in each band (e.g. \oiii\ in YJ, H$\upalpha$ in H and [\ion{S}{iii}]$\lambda 9530$ in K for a typical $z\sim1.4$ galaxy).
After the fitting procedure is completed, all the emission line maps are mapped back into the Source Plane of the galaxy after converting them to surface brightness units, which is the physical quantity conserved by gravitational lensing (we refer to the following sub-section for more details).

\begin{figure*}
	\centering
	\includegraphics[width=0.95\columnwidth]{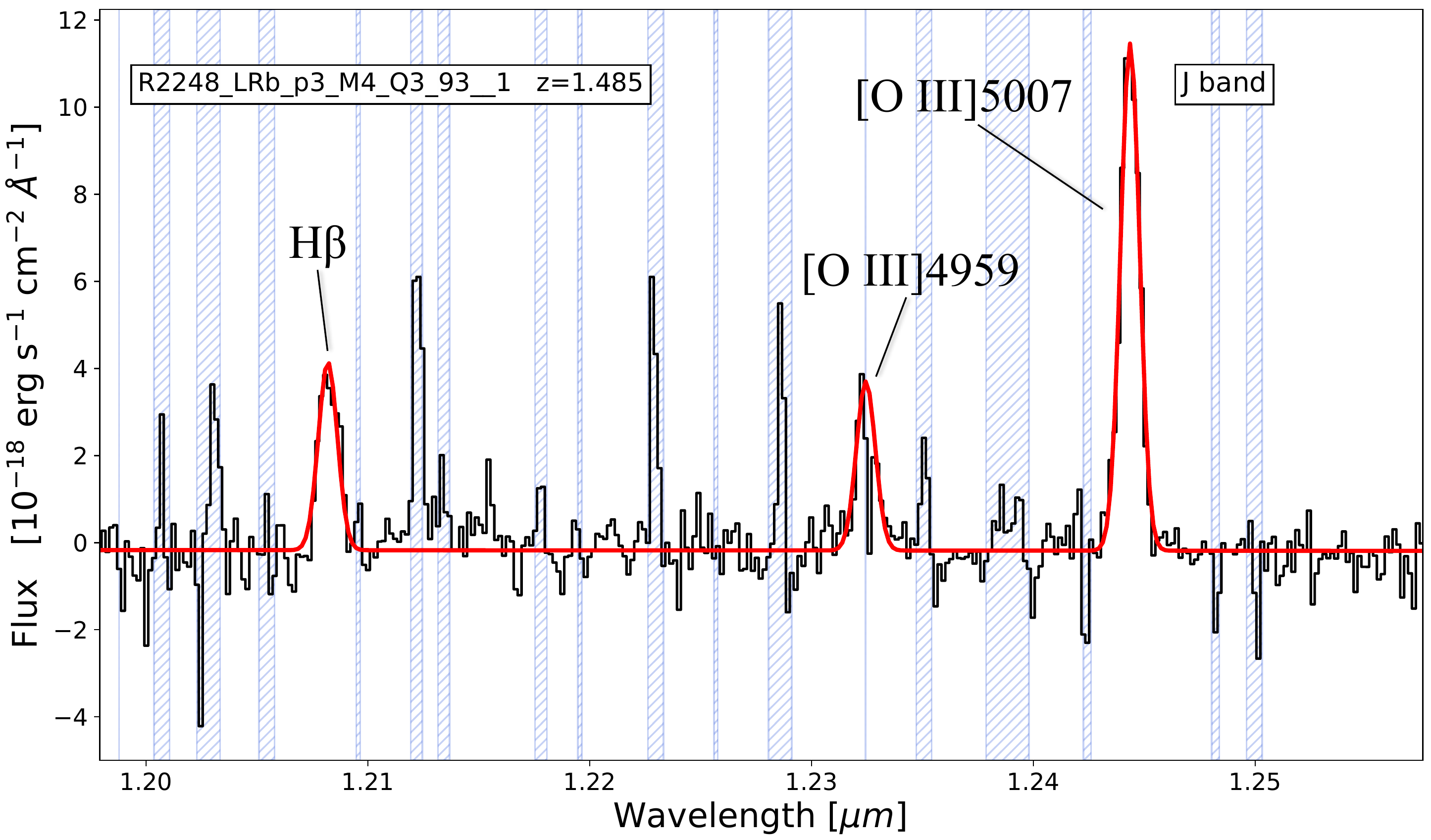} 
	\includegraphics[width=0.95\columnwidth]{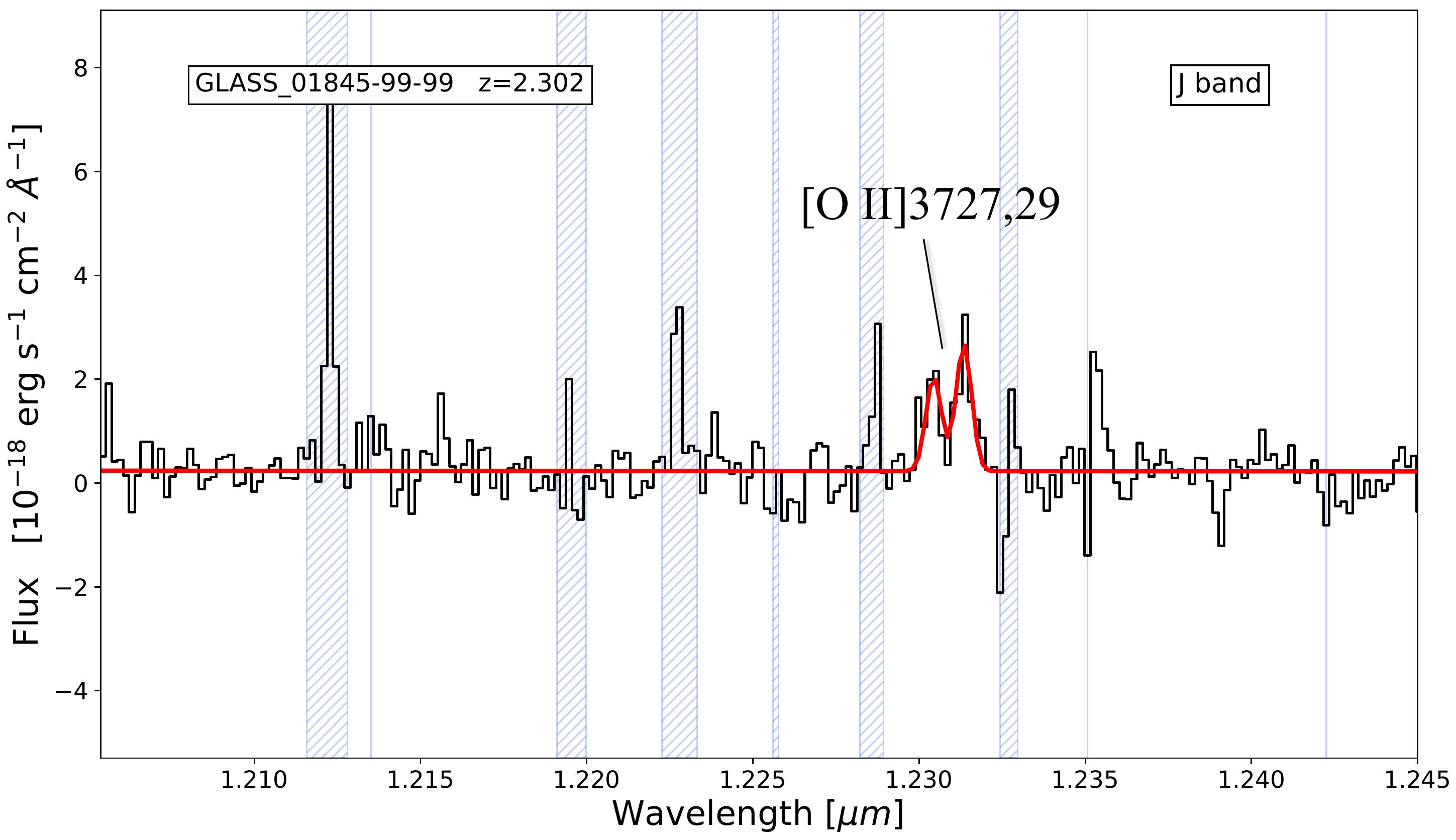} \\

	\includegraphics[width=0.95\columnwidth]{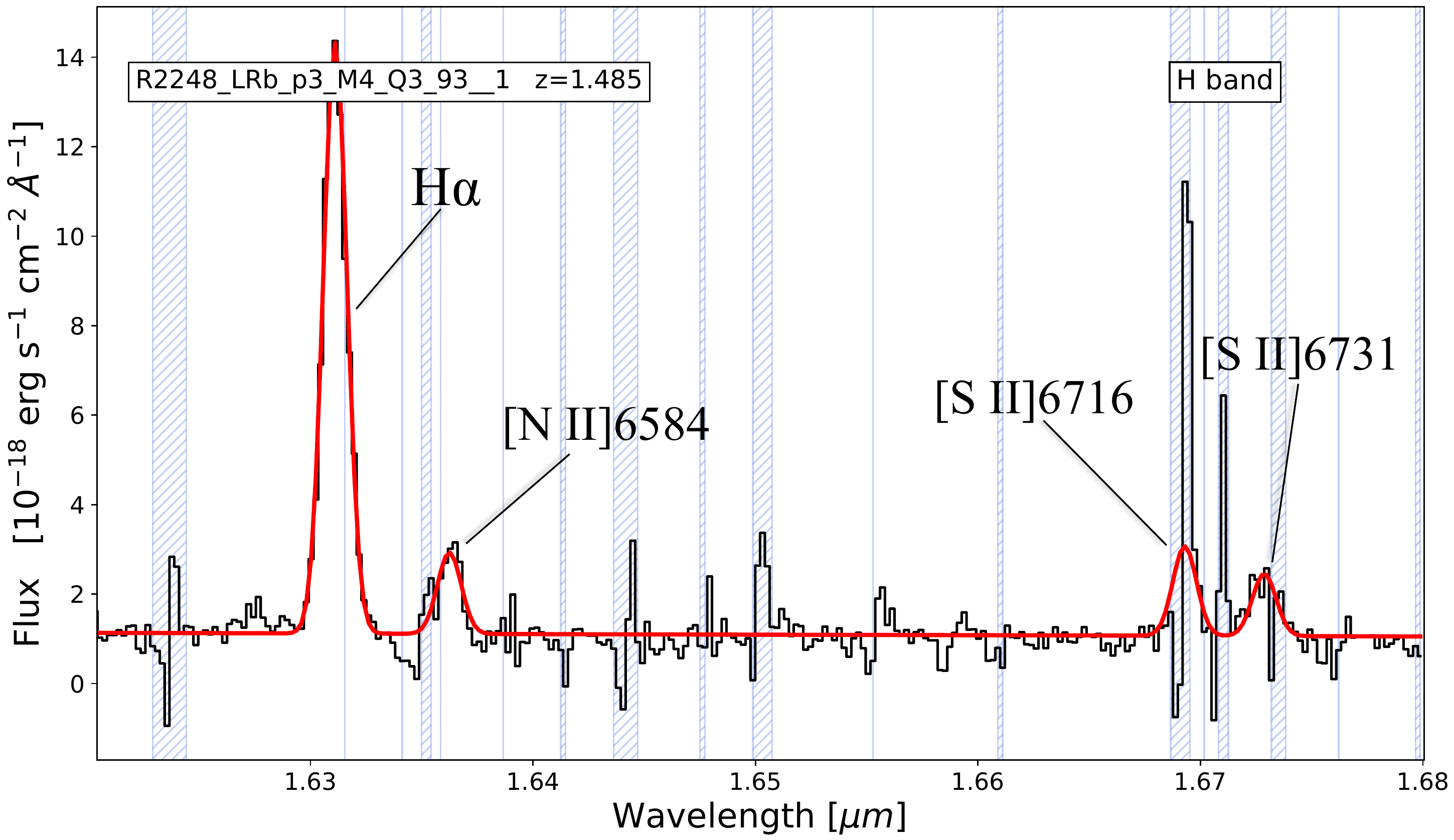} 	
	\includegraphics[width=0.95\columnwidth]{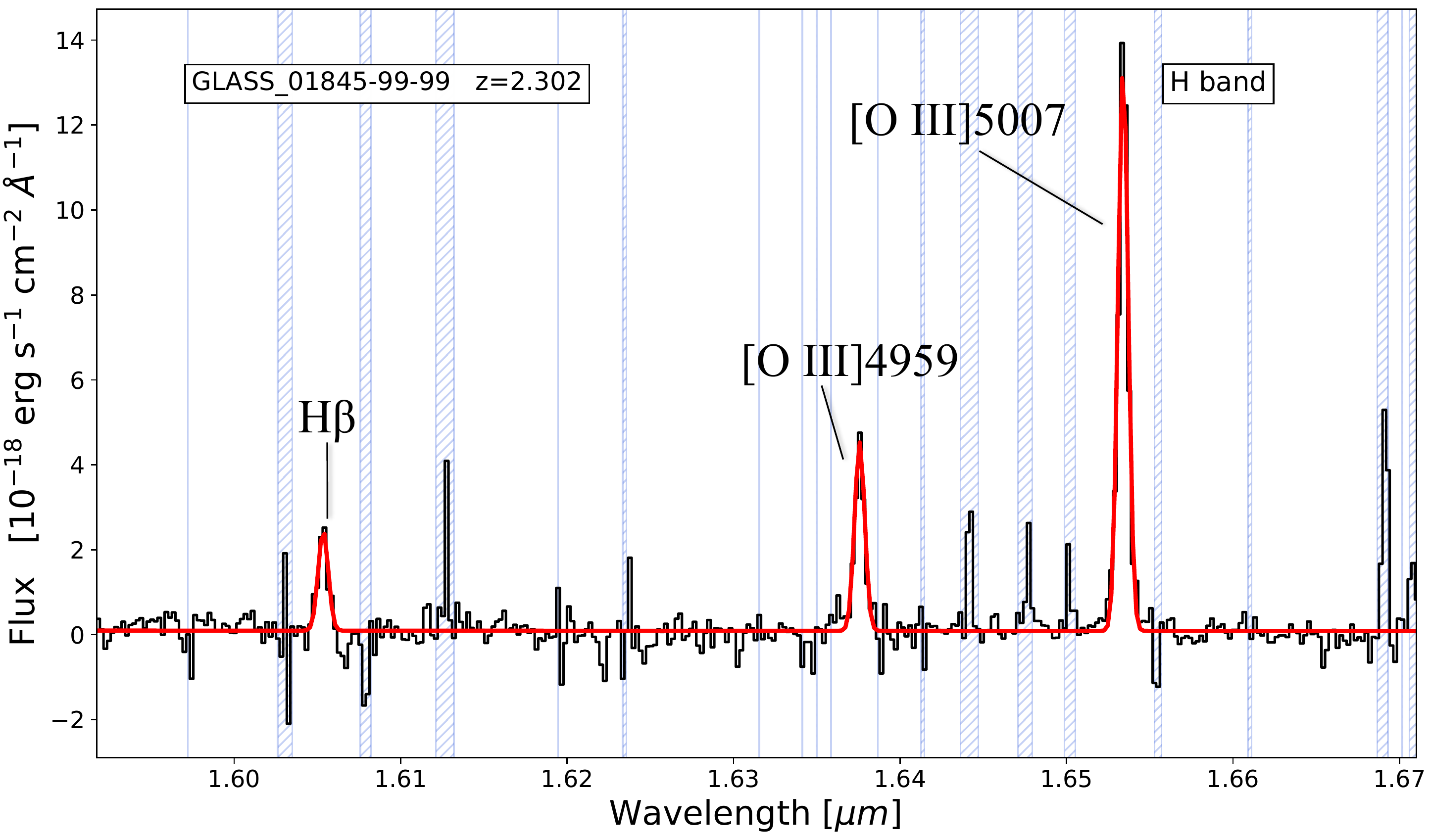}\\ 

	\includegraphics[width=0.95\columnwidth]{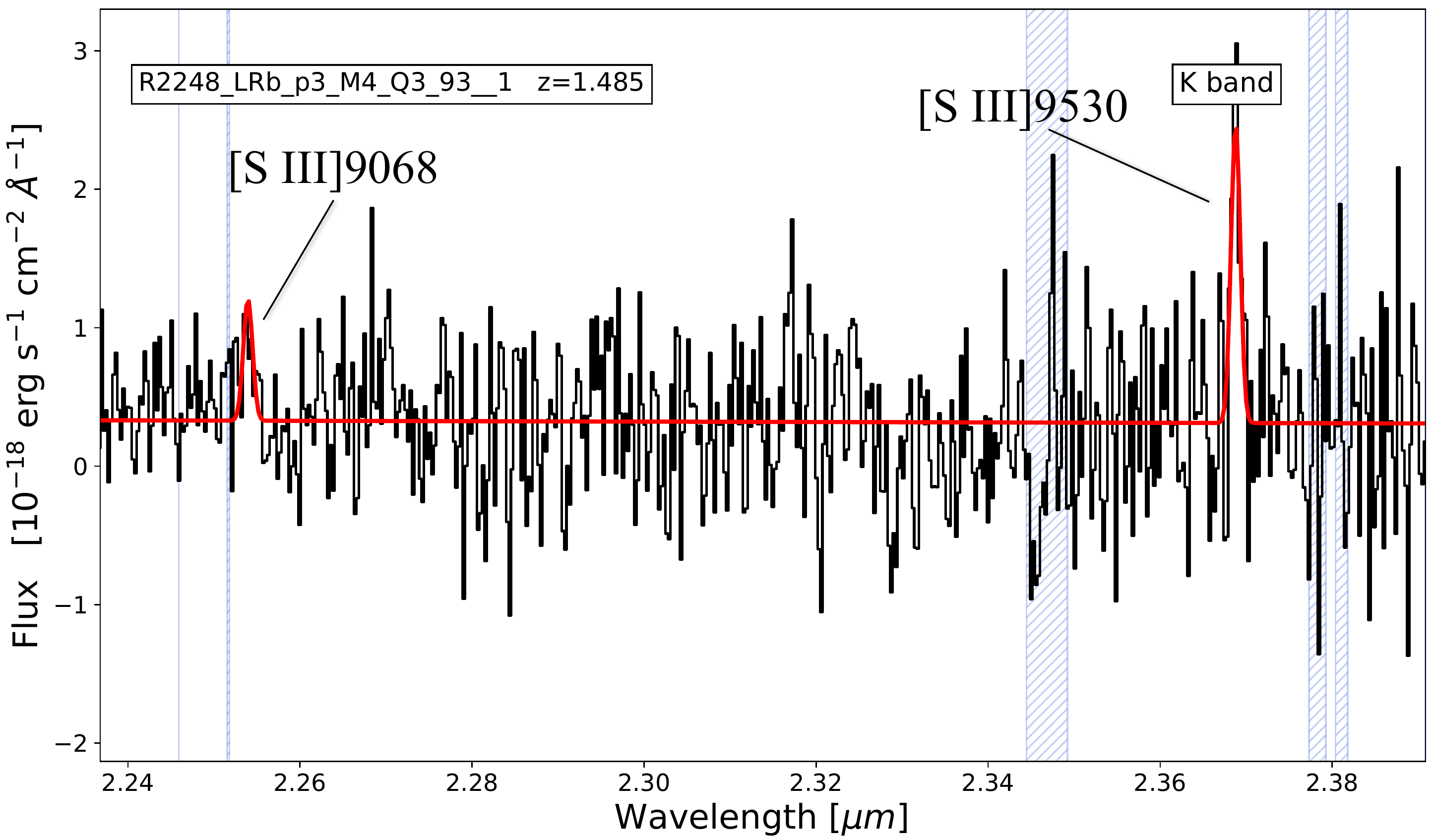}
	\includegraphics[width=0.95\columnwidth]{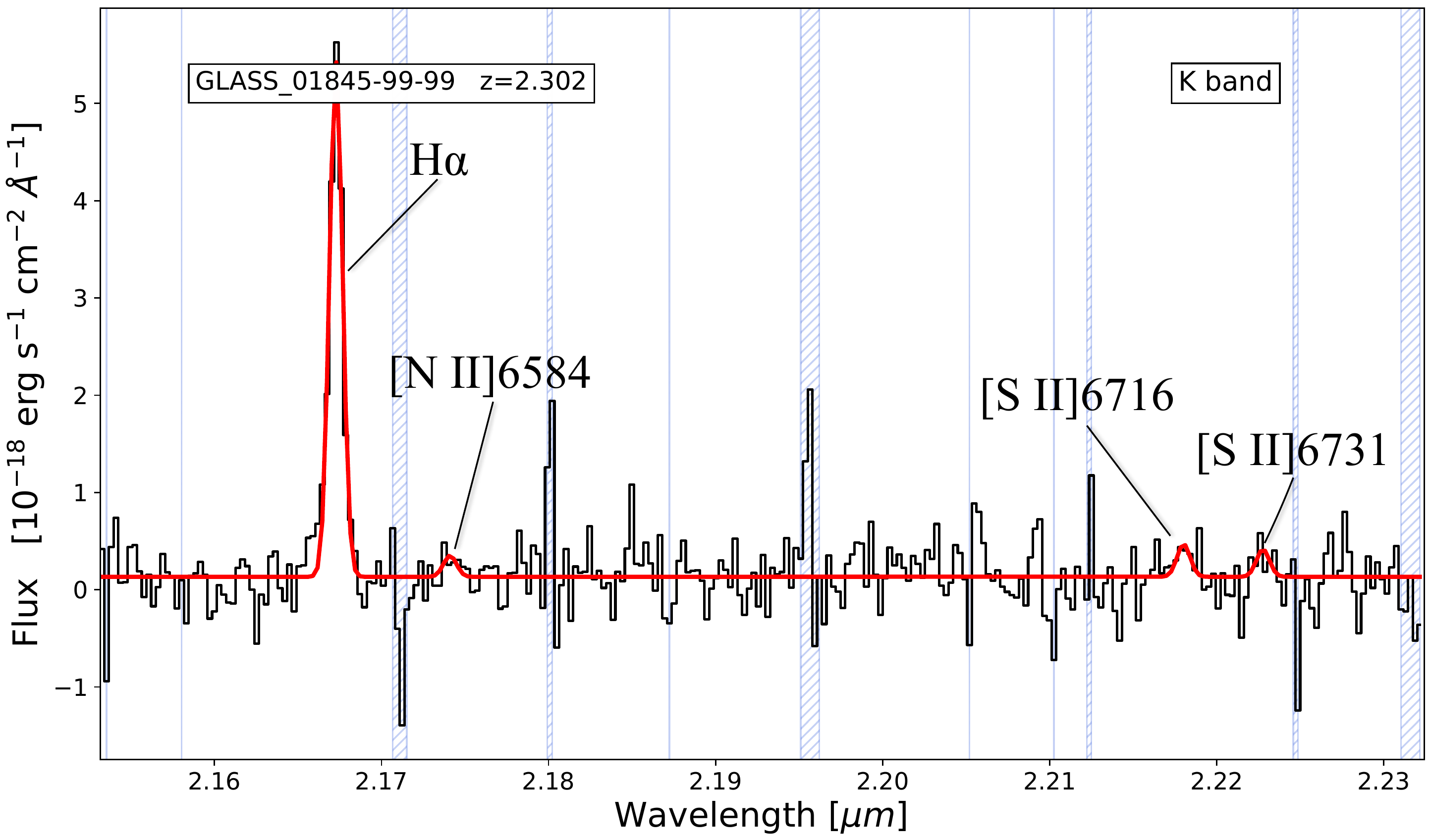} 	

	\caption{Integrated spectra for a $z=1.485$ galaxy (R2248\_LRb\_p3\_M4\_Q3\_93\_\_1, left panels) and for a
	$z = 2.302$ galaxy (GLASS\_01845-99-99, right panels) in the YJ, H and K band. The best-fit to the main emission lines targeted in this study (i.e. \oii, \oiii, H$\beta$, H$\alpha$, \nii, \sii, \siii) are shown by the red component, while the blue shaded areas mark the regions masked out during the fitting procedure. }
	\label{fig:integrated_spec_1} 
\end{figure*}

Representative integrated spectra for each source have also been extracted from circular pseudo-fibres of $0.6\arcsec$-radius (equivalent to the average FWHM of the seeing-limited PSF) centred on the position of the peak of the emission in each band, and we fit the emission lines following the same prescriptions described above.
The choice of the aperture width is arbitrary, but has proven to be an effective compromise to be representative of the bulk of the emission while encompassing at the same time a significant fraction of the total flux.
An example of integrated YJ, H and K band spectra for a $z\sim 1.4$ and a z$\sim 2.2$ galaxy are shown in Fig.\ref{fig:integrated_spec_1}; the best-fit to continuum and nebular line emission is shown in red, while the blue shaded areas mark the regions masked-out in the fitting procedure.

The list of emission lines detected in each galaxy is reported in Table~\ref{tab:klever_sources}.
We report a detection if the signal-to-noise ratio (SNR) on the integrated emission line flux is higher than $3$.
We achieved a $92\%$ detection rate for integrated H$\upalpha$ and $83\%$ for [\ion{O}{iii}]$\lambda 5007$.
%are not only detected in the integrated spectrum, but even spatially resolved in more than $90\%$ of the sample.
The detection rate drops to $71\%$ for H$\upbeta$, $56\%$ for [\ion{N}{ii}]$\lambda 6584$ and $53\%$ for [\ion{S}{ii}]$\lambda 6717,31$ (i.e. where at least one of the lines in the doublet is detected with SNR$>3$.).
In $7$ of the z$\sim 1.4$ sources, the [\ion{S}{iii}]$\lambda 9530$ line (the brightest of the [\ion{S}{iii}] doublet) is also detected in the integrated spectra: these represent some of the very first observations of this emission line at such redshifts, and provide key insights regarding the excitation state of the gas. 
Finally, $64\%$ of the z$>2$ galaxies present a detection of the [\ion{O}{ii}]$\lambda 3727,29$ doublet.

\subsection{Lens modelling and deprojection}
\label{sect:lensmod}
In order to reconstruct our galaxies in the source plane, we have to build lens models that describe how the foreground mass (either a single galaxy or a cluster) has re-distributed the emission from the background lensed galaxy into the image plane, i.e. the image that we see on the sky. 
In the following we discuss in more detail the lens modelling by distinguishing between the cases of lensing by individual galaxies and lensing by galaxy clusters. 
After the lens models have been constrained by the broadband imaging (retrieved from both \textit{HST} and Keck archives), we then use deflection maps generated from these models to map the source properties (line intensities, velocities, etc.) back to the un-distorted source plane.

\subsubsection{Galaxy lenses}
\label{sect:gallens}
Two of our targets are part of a galaxy-galaxy lensing system (i.e. the foreground mass is a single galaxy). For the Horseshoe galaxy, we use \textit{F606W} \textit{HST} observations to perform the modelling where we assume an elliptical power law with external shear lensing mass distribution. We use the adaptive-source-plane technique of \citet{Vegetti:2009aa} to create a galaxy model in the source plane, which we then lens into the image plane using a trial lens model before comparing the results with the \textit{HST} image. We then vary the mass-model parameters until an optimal match to the \textit{HST} data is obtained. This can be done particularly well for the Horseshoe as it forms an almost complete Einstein ring. The same procedure is employed to model the CSWA164 lensed galaxy (which also consists of a nearly complete Einstein ring) using data from the ESI imager on KECK II.

To de-lens the galaxies and reconstruct them back in the source plane we use these lens mass models to compute deflection maps. We first have to align the SINFONI datacubes with the broadband images, and we do so by determining the centre of the foreground galaxy (i.e. the lens) in both datasets and applying a scaling factor to the Einstein radius equal to the ratio of the different pixel scales of SINFONI and the broadband data. The mass model then defines a mapping for every pixel in the SINFONI cube back to its un-distorted location in the source plane.

\subsubsection{Cluster lenses}
\label{sect:clustlens}
\begin{figure}
	\centering
	\includegraphics[width=0.48\columnwidth]{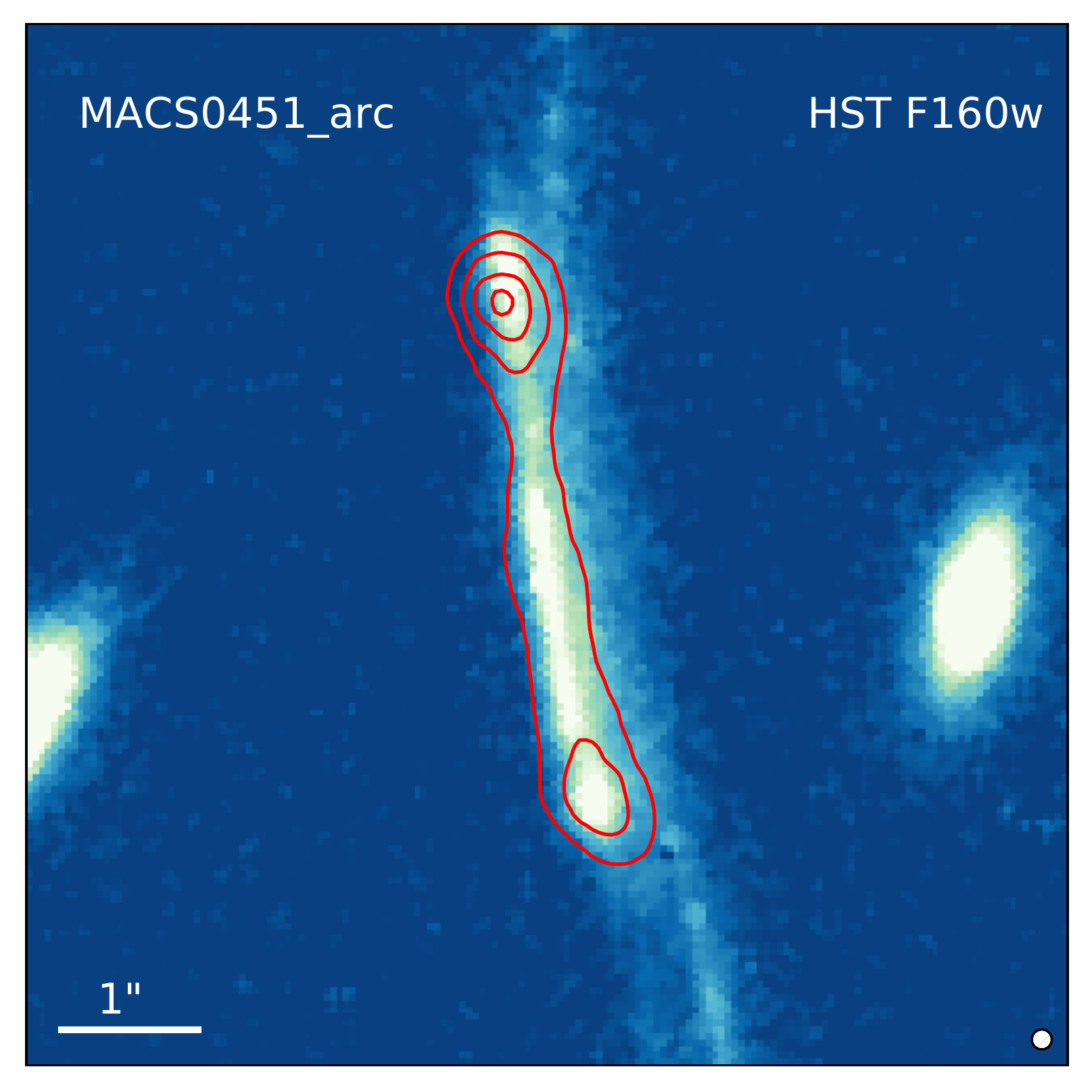}
	\includegraphics[width=0.48\columnwidth]{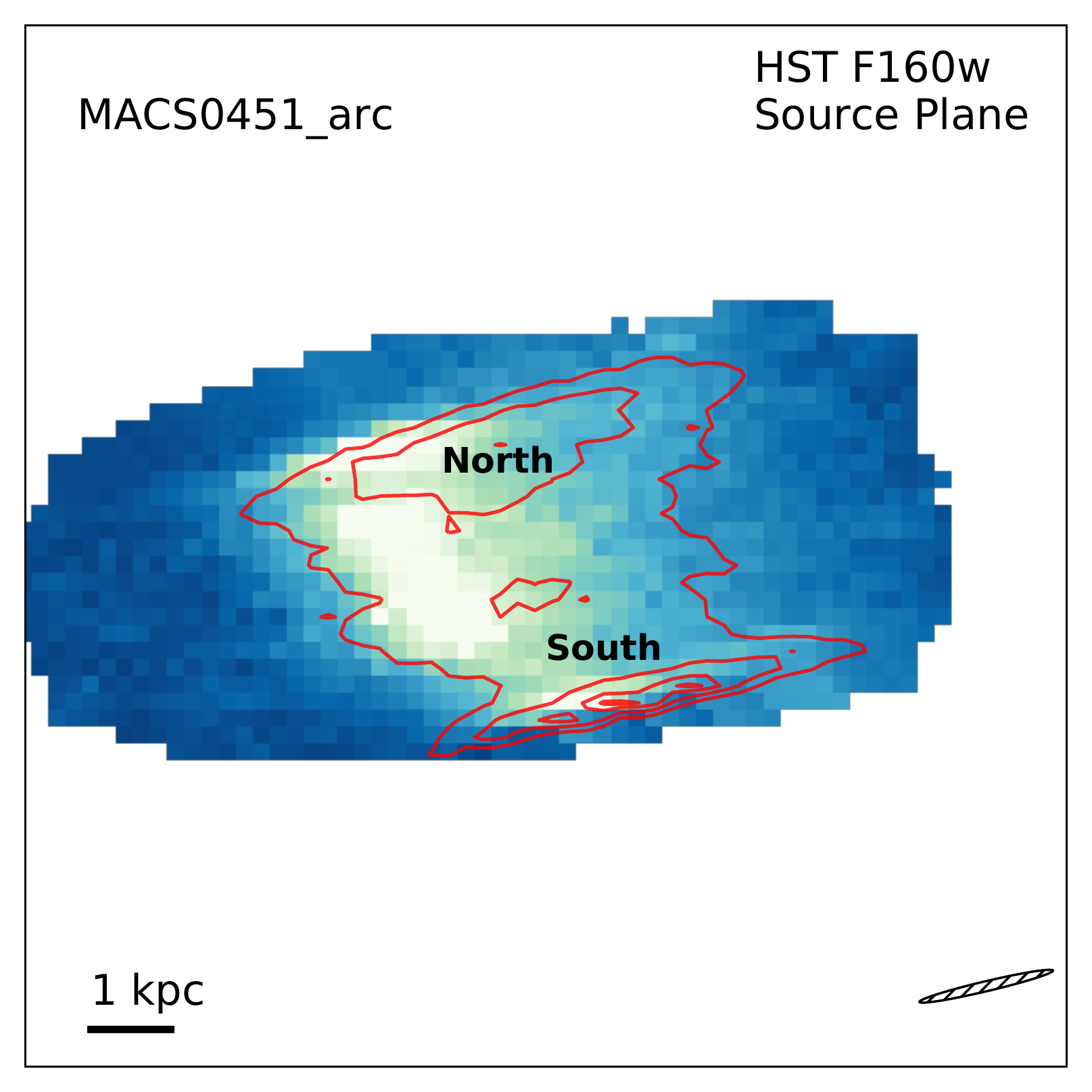}\\
	\includegraphics[width=0.48\columnwidth]{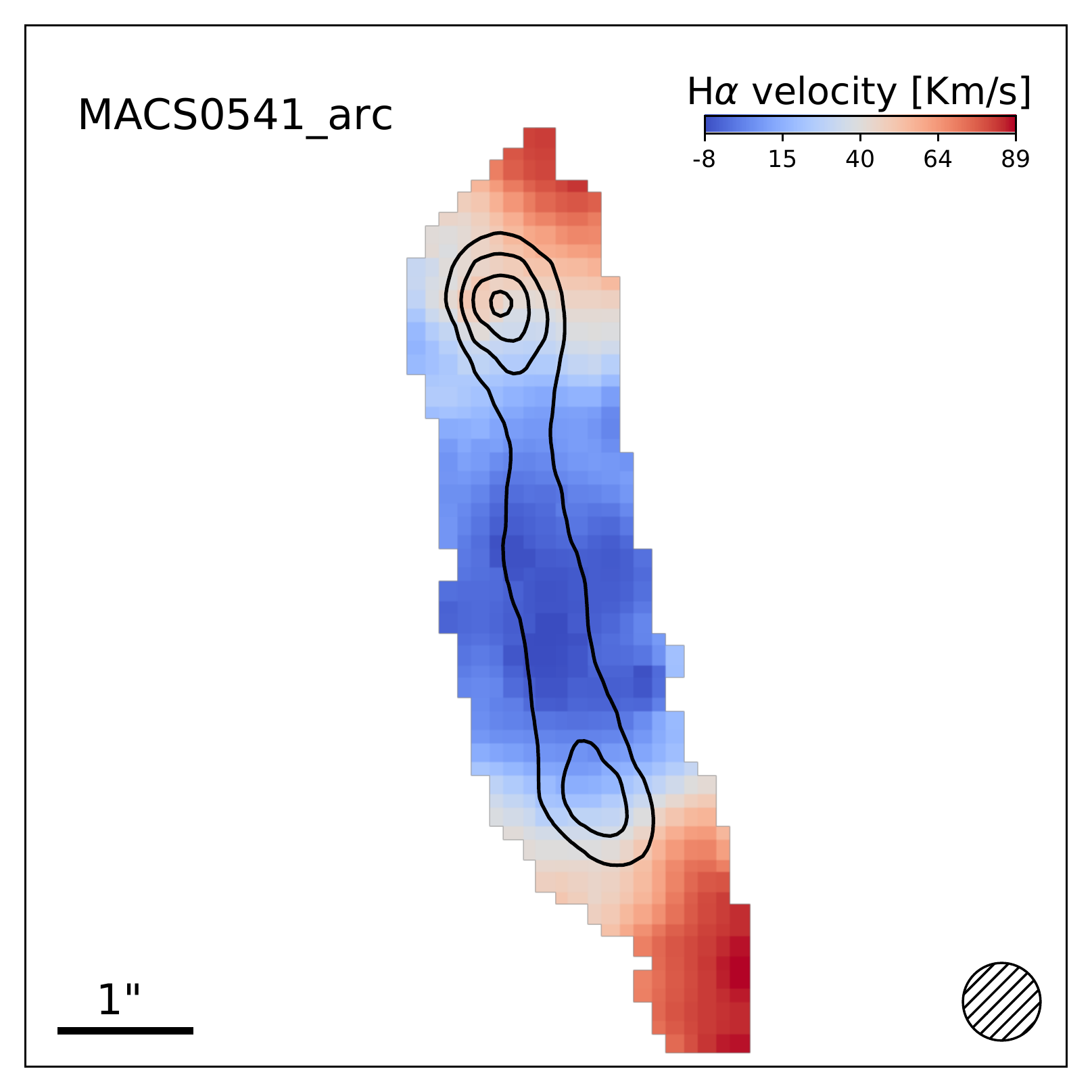}
	\includegraphics[width=0.48\columnwidth]{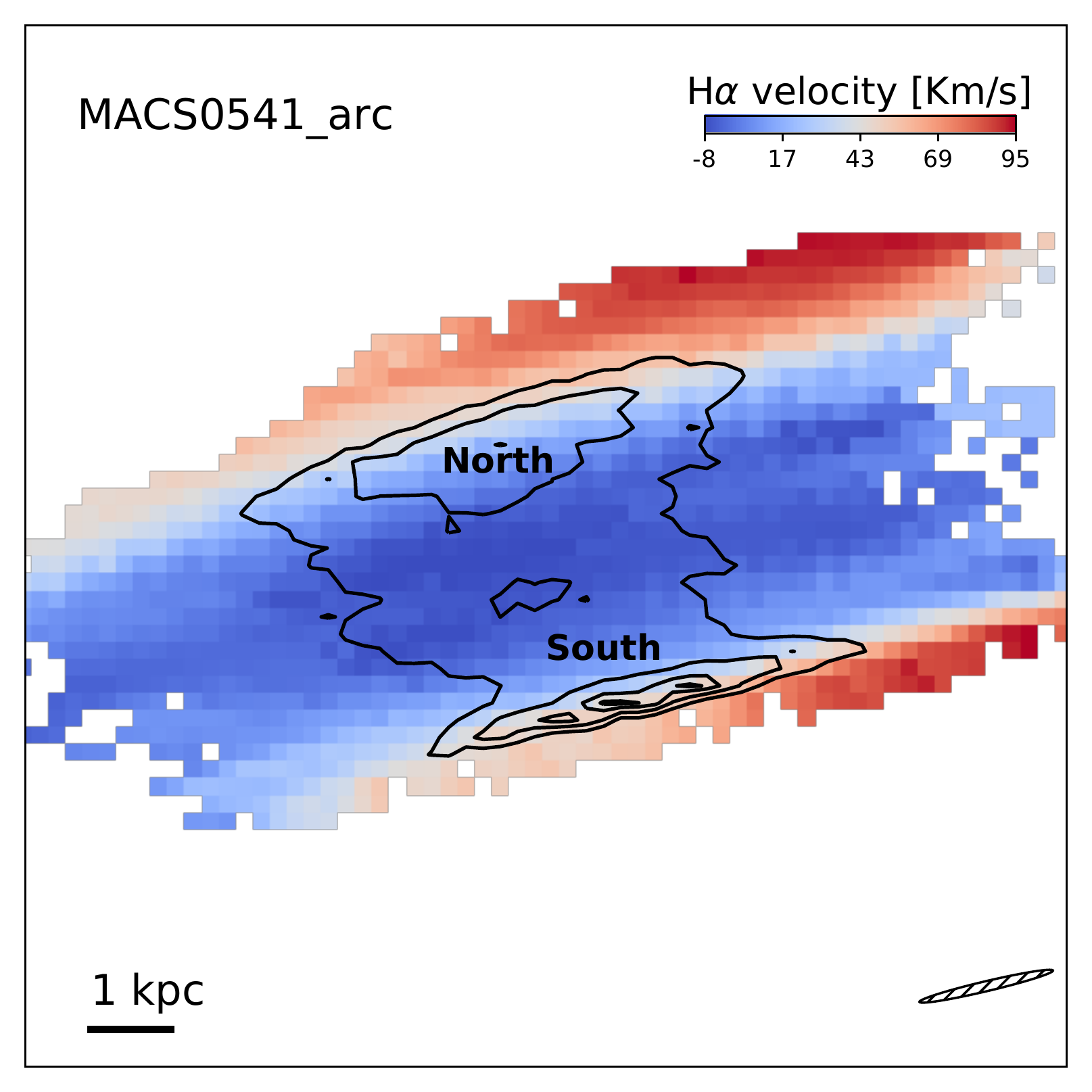}\\
	
	\caption{\textit{Upper left panel:} HST F160W broad filter image of the galaxy lensed by the MACS0451 cluster ($\mu=11.7$).
		\textit{Upper right panel:}  Source plane reconstructed HST image of the source. 
		The contours of the H$\alpha$ emission from SINFONI observations are overlaid in red in both panels.
		\textit{Lower panels:} Velocity field from H$\alpha$ emission in the image plane (left) and reconstructed velocity field in the source plane (right); the source plane maps and velocity field reconstruction of this system suggest the presence of two distinct sources, which we indicate as the \textit{northern} and \textit{southern} galaxy.  H$\alpha$ contours are overlaid in black. 
		The PSF in the image and source plane is shown in the bottom right corner of each panel.}
	\label{fig:MACS_vel}
\end{figure}
In the case of the galaxy strongly lensed by the MACS0451 cluster (and part of the SINFONI sub-sample), we use a different technique. We exploit the multiple imaged system to constrain the model by focussing the images back to the source plane so that the multiple images all map back to the same position. We assume an elliptical power law with external shear for the cluster halo and also use singular isothermal sphere mass profiles to model the mass in each of the individual cluster galaxies. As explained in the Sect. \ref{sect:gallens}, we then use this mass model to create deflection maps that allow us to deproject the SINFONI data and reconstruct the source properties in the un-distorted source plane. 
Interestingly, both the de-lensed HST image and H$\upalpha$ map of the strongly lensed galaxy in MACS0451 (Fig.~\ref{fig:MACS_vel}) highlight what appears to be two separate peaks approximately $\sim 2$ kpc apart. Further investigation conducted on the velocity field confirms the possibility that this is a merging system, as the two H$\upalpha$ peaks are apparently associated with different rotation patterns. 
Therefore, for the purpose of this work, this system is considered as constituted by a pair of galaxies and we thus compute two separate metallicity gradients, one for the \textit{northern} and one for the \textit{southern} region of each map.

The galaxies observed with KMOS are background galaxies of lensing clusters with publicly available mass models\footnote{downloadable from the \href{https://archive.stsci.edu/pub/hlsp/frontier/abells1063/models/}{Frontier Fields} and \href{https://archive.stsci.edu/pub/hlsp/clash/ms2137/models/}{CLASH} models repository} provided by the CLASH and FRONTIER FIELDS collaborations.
In particular, we exploited the mass models described in \citet{Zitrin:2013aa,Zitrin:2015aa}, which are based on assuming a Pseudo Isothermal Elliptical Mass Distributions for the galaxies and an analytical elliptical-NFW dark matter halo, primarily centred on the BCG(s).
We use the deflection maps generated from these models to map the lensed source properties back to the de-lensed source plane; these deflection maps are defined from \textit{HST} images and we therefore re-bin our IFU data to the same pixel scaling as the \textit{HST} observations (i.e. $0.065\arcsec$/pixel) and then align our surface brightness maps with the \textit{HST} images for each emission line. In the cases where we do not have a strong detection of the continuum from our IFU data, we are forced to align the peak of the \textit{H}-band emission line (H$\upalpha$ or [\ion{O}{iii}] depending on the source redshift), with the peak in the \textit{HST} \textit{H}-band image, hence assuming that the latter is dominated by emission line flux, or that continuum and line emission are co-spatial.

\subsubsection{Source plane reconstruction and the PSF}
We reconstruct the de-lensed surface brightness maps by defining a regular grid in the source plane for each system and using the deflection maps described in the previous two sections to define on which source-plane pixel a given observed pixel will fall. The source-plane pixel is then evaluated as the average of all of the image-plane pixels that were mapped to it (the average is used because lensing conserves surface brightness).
The pixel size of the grid was chosen to allow a proper sampling of the source plane PSF (see below).
For KMOS data, maintaining the original HST pixel scale is enough for this purpose, while different resampling factors were applied to the reconstructed SINFONI data. 
The final pixel size of our reconstructed images on the source plane is therefore $0.065\arcsec$/pixel for galaxies observed with KMOS, $0.062\arcsec$/pixel for CSWA164 and $0.025\arcsec$/pixel for both Horseshoe and the galaxy lensed by the MACS0451 cluster.

It is worth mentioning here what effect the de-lensing has on the PSF. Because we do not perform forward modelling of the moment maps, the reconstructed sources are significantly affected by the observational PSF. For example, what may be a circularly symmetric PSF in the image plane will be significantly skewed in the source plane as a result of removing the (preferentially tangential) lensing distortion. To investigate this effect, we take a very small (i.e. point-like) synthetic source in the source plane and use the lens models to produce mock image-plane observations. We simulate what would be observed by SINFONI or KMOS by convolving this with the PSF from each observation, as measured from the alignment stars observed in three dedicated IFUs. 
The PSF FWHM of our seeing-limited observations ranges between $0.5\arcsec-0.6\arcsec$.
%and give the resulting full width half maximum (FWHM) in Table~\ref{tab:lensmod}. 
We then de-lens this image back into the source plane to visualise how the observational PSF affects the morphology of the reconstructed source (i.e. to see the distortions imposed on a source-plane point-like object as a result of our reconstruction technique). We fit an ellipse to model the source plane PSF and give the major and minor FWHM in physical distance scales for each galaxy in Table~\ref{tab:grads}.
On average, the typical source plane resolution obtained is then of the order of $\sim 3$ kpc.
It is also worth recalling that, despite the apparent high spatial sampling obtained in the source plane provided by our procedure, the information encoded in individual pixels in the final reconstructed and interpolated images is not fully independent from that of the neighbouring spatial elements. However, taking into account the shape of the source plane PSF allows us to properly interpret the reconstructed emission line maps.

We estimate the total magnification factor $\mu$ for our sources in two different ways.
In the first case, we compute $\mu_{\text{H}\alpha}$ as the ratio of the total H$\alpha$ flux in the image plane to the total H$\alpha$ flux in the source plane, obtained by co-adding the flux of all the spaxels in the H$\alpha$ map with signal-to-noise ratio $>3$ before and after the reconstruction procedure.
The uncertainty on $\mu_{\text{H}\alpha}$ is estimated by propagating the relative uncertainties on the source plane and image plane H$\alpha$ fluxes.
In the second case, we exploit the magnification maps generated from the mass and shear maps provided by the lens models, once rescaled to the proper source redshift for each galaxy, to compute the magnification ($\mu_{\text{model}}$) at the position of each galaxy.
The statistical uncertainties in this latter case are estimated from the ranges of magnification values 
in $100$ different realizations of each model, i.e. taking the $1\sigma$ interval of the resulting magnification distribution.
Both values are reported in Table~\ref{tab:info_glob}, with the two different estimates consistent within 
$1\sigma$ for almost all sources.

\begin{figure}
	\centering
	\includegraphics[width=0.99\linewidth]{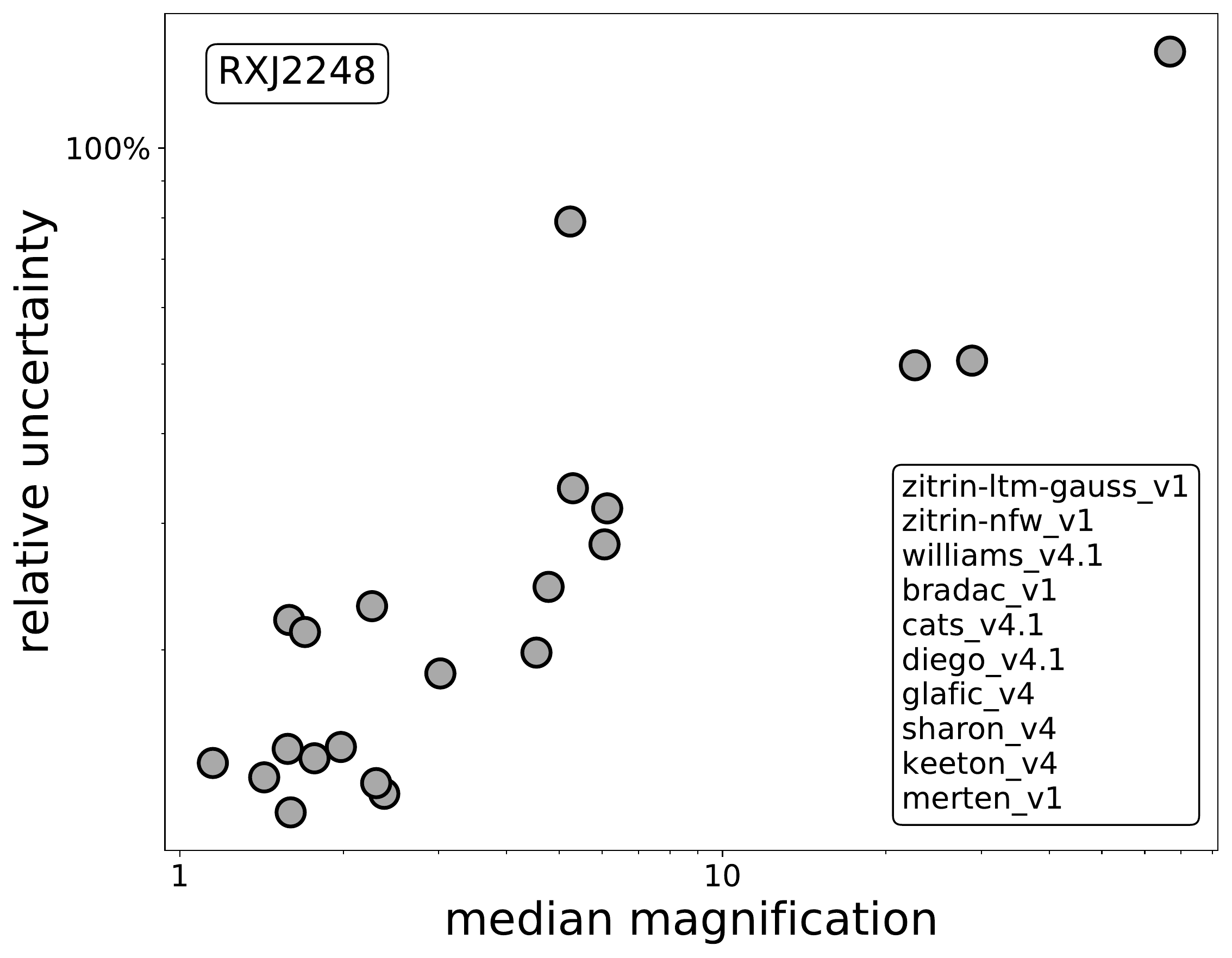}
	\caption{The (relative) systematic uncertainties on the magnification are plotted as a function of the magnifications themselves for the KLEVER galaxies within the RXJ2248 cluster. 
	Adopting $10$ different realisations of the lens model (listed in the bottom-right corner) we estimate the median and $1\sigma$ interval of the magnification distribution for each source, assuming then $\sigma$ as the typical systematic uncertainty on the magnification factor associated to the choice of a particular lens model.}
	\label{fig:systuncmodels}
\end{figure}

It is finally worth recalling that different approaches in the mass reconstruction algorithms of various lensing models could produce significant discrepancies in the estimated magnifications, especially in the proximity of the critical lines where they can reach factors of $\sim 30\%$ \citep{Meneghetti:2017aa}.
For this reason, systematics uncertainties on the lens modelling are likely to dominate the error budget over the statistical uncertainties computed for each individual model.
To test the impact of such effect on our study, we created magnification maps at each source redshift adopting the prescriptions of $10$ different models available for the Frontier Field cluster RXJ2248 and computed the magnifications for each galaxy in our sample.
We can then assume the $1\sigma$ interval of the resulting distributions as an estimate of the systematic uncertainty on the magnification associated with the choice of a particular lens model. 
In Fig.~\ref{fig:systuncmodels} we plot the relative uncertainties, i.e. $1\sigma$ divided by median of the magnification distribution, as a function of the median magnification itself for the KLEVER galaxies within RXJ2248. 
%The typical systematic uncertainties to be considered for our galaxies, whose average magnifications are of the order of $\sim 2$, are hence of the order of $10\% - 20\%$.
The typical systematic uncertainties for the majority of our galaxies are comprised between $10\% - 25\%$, and increase with increasing magnification.
The list of lens models included in this test is reported in the bottom-right corner of the plot and 
each of them can be downloaded from the web page of the Frontier Fields project.

\subsection{Stellar Mass and Star Formation Rate}
\label{sec:mass_sfr}
Stellar masses and star formation rates (SFR) for our sample are listed in columns 2 and 3 of Table~\ref{tab:info_glob} and were derived as follows.
%As previously mentioned in Section \ref{sect:sample}, the galaxies observed with KMOS are distributed around the two clusters MS2137-2353 and RXJ2248, which are among the clusters observed as part of the HST-CLASH program. 
For KMOS galaxies lensed by clusters an SED-fitting was performed on the photometric measurements extracted (with SExtractor) from broadband HST images provided by the CLASH and FRONTIER FIELDS (FF) programs. 
In particular, for galaxies observed in MS2137 we implemented the photometry from the $7$ optical and near-infrared filters covered by CLASH (namely $F105W, F125W, F140W, F160W, F435W, F606W, F814W$), while for RXJ2248 we could exploit the deepest exposures in the same filters provided by the FF images.
Stellar masses were obtained using the high-$z$ extension of the \textsc{magphys} program \citep{da-Cunha:2015aa}, assuming a \cite{Chabrier:2003aa} IMF and exploting the \cite{Bruzual:2003aa} stellar population synthesis models; in particular, \textsc{magphys} adopts the two-component model of \cite{Charlot:2000aa} to describe the attenuation of stellar emission at ultraviolet, optical, and near-infrared wavelengths. 
Stellar masses for a few objects within the clusters are not available because these galaxies are not covered by the HST imaging.
Similarly, for the Horseshoe, we used the available HST images in multiple bands ($F110W, F160W, F475W, F660W, F814W$) to perform the SED-fitting and derive the stellar mass.
The uncertainties on stellar masses are derived from the $1\sigma$ interval of the resulting likelihood distribution and include also the contribution from statistical uncertainties on the magnification, but does not account for systematic uncertainties on the lensing model.

For the remaining two SINFONI galaxies we adopt the stellar masses provided in the literature, after a proper scaling (when needed) to the same \cite{Chabrier:2003aa} IMF. 
In detail, the stellar mass for MACS0451 is taken from \cite{Richard:2011aa} and has been derived from a SED fitting including HST and Spitzer/IRAC data in the $3.6\mu m$ and $4.5\mu m$ filters with a \cite{Calzetti:2000aa} extinction law.
As pointed out in Section~\ref{sect:clustlens}, we are here considering this system as constituted by a pair of galaxies for which we can compute separate metallicity maps and gradients. Therefore, in addition to the global M$_{\star}$ value, we used the continuum flux from the de-lensed HST H-band images as a proxy for the stellar mass distribution in order to split the global value and provide an M$_{\star}$ estimate for both the \textit{northern} and \textit{southern} region of this system.
The M$_{\star}$ for CSWA164 is taken instead from \cite{Kostrzewa-Rutkowska:2014aa} and has been derived from SED fitting to the SDSS photometry.
%, SGAS from \cite{Wuyts:2012aa}, A68 from \cite{Sklias:2014aa} 
All values reported in Table~\ref{tab:info_glob} are corrected for the $\mu_{\text{H}\alpha}$ magnification factor (or alternatively $\mu_{\text{maps}}$ if the former is not available.)
 
The global star formation rate for our sources is calculated from the extinction corrected H$\upalpha$ luminosity, which is converted to SFR assuming the \cite{Kennicutt:2012aa} relation and applying a scaling factor of $1.06$ to convert from \cite{Kroupa:1993aa} to \cite{Chabrier:2003aa} IMF. 
The amount of nebular reddening has been deduced from the Balmer decrement and corrected assuming an intrinsic ratio of H$\upalpha / $H$\upbeta = 2.87$, adopting the \cite{Cardelli:1989aa} extinction law. 
We finally corrected for the magnification factor derived from the de-lensing procedure (see Section \ref{sect:lensmod}) and reported in Table~\ref{tab:info_glob} as $\mu_{\text{H}\alpha}$. Both individual measurements errors on E(B-V)$_{neb}$ and the magnification factors are included in the total uncertainties on SFR.

The total H$\upalpha$ flux required to compute the SFR was estimated by co-adding the flux from all individual spaxels with robust H$\upalpha$ detection (i.e. above $5 \sigma$). 
Indeed, for extensive quantities like the SFR, different choices of the aperture width adopted to extract the integrated spectra would provide different measurements of the H$\upalpha$ flux and thus change the final inferred SFR, whereas
for physical quantities derived from line ratios, like the gas-phase metallicity, changing the aperture width has negligible impact on the results.
On average, the fraction of the total H$\upalpha$ flux collected within the $0.6\arcsec$-wide aperture is $\sim 75\%$.
The lowest fractions (i.e around $50\%$) occur in composite or interacting systems, whose spatial emission line profiles can be significantly smeared or even double-peaked (for instance MUSE\_SW\_461-99-99).
%The SFRs computed from the \textit{total}, co-added H$\upalpha$ flux are reported in Table~\ref{tab:info_glob} too.
%are slightly higher, but always within $1 \sigma$, than those inferred from the integrated spectra. 

%For some of our galaxies H$\upbeta$ is weakly detected on resolved spatial scales, therefore we cannot produce a reliable E(B-V) map to correct the H$\upalpha$ on a spaxel by spaxel basis. In such cases, we inferred a global E(B-V) value from the total H$\upalpha$ and H$\upbeta$ fluxes and applied it to all spaxels, assuming that the extinction does not change dramatically across the galaxy.
%The resulting, de-lensed SFRs are reported in Table \ref{tab:info_glob} in units of $\rm M_{\odot}~yr^{-1}$. 

Additional care has to be taken when computing the SFR and magnification for the Horseshoe galaxy. Our IFU observations of this object only encompass the western arc, hence do not allow us to sample the entire galaxy when re-constructed in the source plane.
%(this is clearly visible if compared to the total de-lensed \textit{HST} image). 
%This is shown in Fig.~\ref{fig:Hshoe} where we take the \textit{HST} \textit{V}-band image (left) and de-lens different regions into the source plane (right) by masking different components of the image. The top panel shows the results for the whole arc and therefore maps the whole of the galaxy in the source plane. The bottom panel instead shows the western arc where we have our SINFONI observation. This shows how our observation has missed both of the high surface brightness peaks to the north and south east and so misses this component of the galaxy in the source plane. 
%Thus, in order to compute the intrinsic SFR of the Horseshoe, we need to cover the complete galaxy in the source plane. 
Therefore, we use additional SINFONI observations of the Horseshoe, conducted in 2009, which sample both the western arc and the southern high surface brightness peak. 
%as demonstrated in the middle panel of Fig.~\ref{fig:Hshoe} showing this indeed has enough coverage of the arc to reconstruct the entire galaxy in the source plane. 
%Then, we perform the lensing analysis on the H$\upalpha$ emission line map from these SINFONI observations to compute the total SFR. 
However, the observations from 2009, covering the full arc, only provide K-band data; therefore, for the rest of the current analysis, which depends on detecting multiple emission lines in different bands, we will use only the latest SINFONI multi-band observations (from 2013-2014) of the western arc of this system.

\subsection{Metallicity determination}
\label{sect:metcal}
We derive the gas-phase metallicity using different diagnostics, exploiting the coverage of multiple emission lines.
%\textcolor{red}{discutere applicabilità dei diagnostici ad alto z}
%Since for the large majority of the sources in our sample we target, detect and spatially resolve at least H$\upalpha$, [\ion{N}{ii}]$\lambda6584$, [\ion{O}{iii}]$\lambda5007$ and H$\upbeta$, a combination of such emission lines is used as the standard diagnostic to create the metallicity maps.
In this work we exploit the strong-line calibrations presented in \cite{Curti:2017aa,Curti:2019aa} (hereinafter C17 and C19).
These works redefined the diagnostics from \cite{Maiolino:2008aa} to fully anchor them to the \Te -abundance scale defined in the local Universe by the SDSS galaxies.
A combination of the R$_{3}$ (log([\ion{O}{iii}]$\lambda5007$/H$\upbeta$)), N$_{2}$ 
(log([\ion{N}{ii}]$\lambda6484$/H$\upalpha$)), S$_{2}$ (log([\ion{S}{ii}]$\lambda6717,31$/H$\upalpha$)) and O$_{3}$O$_{2}$ (log([\ion{O}{iii}]$\lambda5007$/[\ion{O}{ii}]$\lambda3727,29$)) diagnostics, depending on source redshift and detectability of each emission line at $\geq 3 \sigma$, are jointly used to tightly constrain the metallicity. 
We ran a  Monte Carlo Markov chain algorithm (MCMC, implemented through the \textit{emcee} package in python) to sample the log(O/H) posterior distribution, minimizing at each step the chi-square defined as:
\begin{equation}
\chi^{2} = \sum_{i} \frac{(R_{i}^{\text{obs}}  - R_{i}^{\text{exp}})^{2}}{\sigma_{\text{obs}}^{2} + \sigma_{R_{i}}^{2}} \ ,
\label{eq:met}
\end{equation}
where R$_{i}^{\text{obs}}$ is the observed line ratio while R$_{i}^{\text{exp}}$ is the expected value, according to each calibration, 
at fixed metallicity. 
The median of the resulting distribution is then assumed as the \textit{true} metallicity and its $1 \sigma$ interval defines the associated uncertainties.
Both the uncertainty on the observed line ratio $\sigma_{\text{obs}}$ and the intrinsic dispersion of each calibrated indicator $\sigma_{R_{i}}$ are taken into account in the procedure.
For most of the targets, the metallicity determination is based solely on the R$_{3}$ and N$_{2}$ (and, in some cases, S$_{2}$, see Sect.\ref{sect:grads}) indicators, hence we do not require any reddening correction since all these diagnostics are constituted by ratios of very nearby emission lines.
For z $>2$ galaxies, where [\ion{O}{ii}]$\lambda3727,29$ is detected and included in our routine, line fluxes are extinction corrected exploiting the Balmer decrement by assuming an intrinsic H$\upalpha$/H$\upbeta$ ratio equal to $2.87$ and the \cite{Cardelli:1989aa} extinction law.
We note that in these cases the extinction correction often represents the largest contribution to the uncertainty in the metallicity determination, given the relatively low SNR of the H$\beta$ line.
In fact, for some of our galaxies we could not produce a reliable E(B-V) map to correct the line fluxes on a spaxel by spaxel basis. In such cases, we inferred the global E(B-V) value from the total H$\upalpha$ and H$\upbeta$ fluxes and applied it to all spaxels, assuming that the extinction does not change dramatically across the galaxy.

\subsubsection{Metallicity diagnostics at high redshift}

We briefly discuss here the robustness of the metallicity measurements performed on our galaxies by means of strong-line diagnostics.
At the present time, any result involving gas-phase metallicity measurements from strong-line calibrations at high redshift should always be interpreted with full awareness of these potential caveats.
Nonetheless we also stress that, since the main results presented in this paper are based on relative metallicity measurements, they can be considered more robust against biases in the metallicity calibrations than those relying on absolute measurements.

As a general remark, whether the locally calibrated metallicity diagnostics are applicable to high-redshift galaxies is still a matter of great debate.
Diagnostics that are expected to be little affected by the ionization conditions of the gas (see e.g. \citealt{Dopita:2016aa}) have been suggested to be valuable for high-redshift galaxies, where strong variations in ionization parameter and excitation conditions compared to local galaxies have been invoked to explain the observed evolution in the emission line ratios (as seen for example from the offset of high-z sources in the classical BPT diagrams with respect to the local sequence \citep{Kewley:2013aa,Nakajima:2013aa,Steidel:2014aa,Kashino:2017aa,Strom:2017aa}.
However, since such diagnostics usually involve the [\ion{N}{ii}]/[\ion{O}{ii}] or the [\ion{N}{ii}]/[\ion{S}{ii}] line ratios, they are strongly dependent on the assumed relation between the N/O ratio as a function of the oxygen abundance O/H, which is affected by a large scatter and whose evolution with cosmic time and/or dependence on galaxy mass is also indicated as a possible origin of the observed evolution of the emission line properties in high-z galaxies \citep{Masters:2014aa,Masters:2016ab,Shapley:2015aa}. 
%Within this context, we show in a companion paper (Williams et al., in prep.) that an increase in N/O at fixed O/H could indeed be responsible for the observed offset in the N2-BPT diagram for some of the KLEVER galaxies.
Therefore, strong-line indicators based only on alpha-elements (like, e.g., oxygen) have also been suggested as appropriate to high redshift studies, since galaxies at z$\sim 1.5-2.5$ seem to show no appreciable offset from local trends in oxygen based diagnostic diagrams (e.g. R$_{23}$ vs O$_{32}$, \citealt{Shapley:2015aa}). 
However, the location on the abovementioned diagram could even be sensitive to a variation in the hardening of the radiation field at fixed metallicity rather than a variation in abundances \citep{Steidel:2016aa,Strom:2017aa}.
In any case, at redshifts $\sim1.5$ (where the majority of KLEVER galaxies considered in this work lie), the lack of the [\ion{O}{ii}] doublet in the NIR bands observable from KMOS prevents us from using purely oxygen diagnostics, thus forcing us to exploit the nitrogen-based ones.
When the survey will be complete, we will investigate the spatially resolved behaviour of z$\sim 2$ galaxies on the R$_{23}$ vs O$_{32}$ diagram in a more statistically robust manner.
%With high-z surveys providing more and more larger statistics and 
%Nevertheless, only a fully Te-based calibration of metallicity indicators at these redshifts would represent the keystone to overcome all these potential discrepancies; unfortunately, just a handful of robust auroral line detections have been reported so far in high-z sources (e.g. \citealt{Jones:2015aa,Sanders:2016aa}, Patricio et al and references therein), 
%due to the intrinsic observational challenges in detecting these faint emissions with current instrumentation.
Recently, \cite{Patricio:2018aa} have shown that oxygen-based diagnostics z$\sim 2$ provide metallicities comparable to those inferred from the electron temperature method; unfortunately, just a handful of robust auroral line detections have been reported so far in high-z sources (e.g. \citealt{Jones:2015aa,Sanders:2016aa}, see also \citealt{Patricio:2018aa} and references therein), due to the intrinsic observational challenges in detecting the faint auroral lines with current instrumentation.
Only the advent of new facilities like JWST or the MOONS spectrograph on the VLT will ultimately allow us to tackle this issue in the next few years, allowing to properly calibrate the metallicity diagnostics against fully Te-based abundances determination at high redshifts and providing the key to overcome all these potential discrepancies.

We finally recall here that the strong-line calibrations adopted throughout this work are valid only if the gas is photoionised by stellar continuum from young massive stars, with no contribution to ionisation due to AGN or shocks.
The sp9 galaxy presents clear signatures of a Seyfert 1 galaxy (with emission lines as broad as $\gtrsim1000\ \text{km}\ \text{s}^{-1}$) and has been thus removed form the analysis.
We further checked the possible contamination from AGN or shocks in the other galaxies of our sample in two different ways, which will be described more in detail in a forthcoming paper (Curti et al., in prep.).
% which are described in more detail in Williams et al (in prep.) and that we briefly summarize here.
First, we found no clear correlation between spaxels lying above the theoretical dividing line of \cite{Kewley:2001aa} and their distance from the central regions of the galaxy, as it would have reasonably been if the ionisation was driven by an AGN.
%In other words, if an AGN were the primary source of ionisation, it would be reasonable to expect that the pixels showing the most AGN-like line ratios should be concentrated near the centre of the galaxy, but this is not what is observed in our galaxies.
Moreover, we find no clear trend between the observed BPT-offsets and velocity dispersion, as one would expect in case of shock-driven line ratios.
Finally, we stacked the spectra of these spaxels from both the [\ion{N}{ii}] and [\ion{S}{ii}]-based BPT diagrams, to look for clear spectral signatures of AGN or Wolf-Rayet contamination (like prominent \ion{He}{ii}$\lambda4686$ emission), but we did not find any evidence for those either.
Therefore, we conclude that the contribution from AGN or shocks is negligible in the majority of cases and that our metallicity determination based on emission line ratios is reliable for the purposes that we pursue in this work.

\begin{figure*}
\centering
\includegraphics[width=0.98\textwidth]{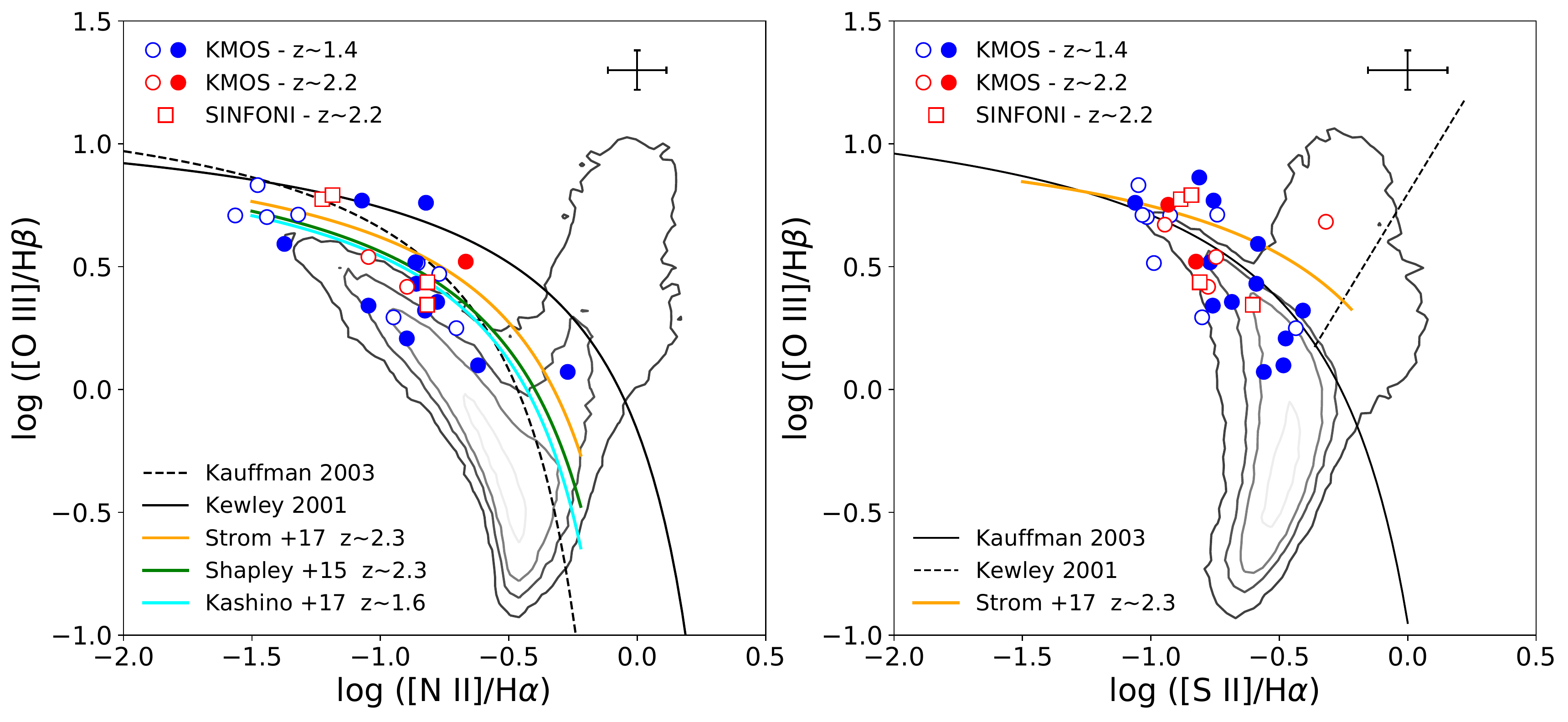}

\caption{Global BPT diagrams for the KLEVER sample analysed in this work. In both panels the positions of the $\text{z} \in [1.2,1.6]$ sources are marked in blue, while that of the $\text{z} \in [2,2.4]$ ones are marked in red. Empty and filled circles denote galaxies belonging to the MS2137 and RXJ2248 cluster respectively, while empty squares represent the sample of strongly lensed galaxies observed with SINFONI.
The theoretical demarcation line between the star forming locus and AGN/LINER galaxies from \citealt{Kewley:2001aa} and \citealt{Kauffmann:2003aa} are shown with solid and dashed black lines respectively, while the underlying grey contours encompass the $25$, $75$, $90$ and $97.5 \%$ of the distribution of SDSS galaxies. 
%a fit to the local tight sequence of SDSS star-forming galaxies from\citealt{Kewley:2013aa} is also shown in purple. 
On the N2-BPT in the \textit{left panel} a fit to the location of high redshift galaxies from the FMOS survey (\citealt{Kashino:2017aa}, $\text{z} \sim 1.6$), MOSDEF survey (\citealt{Shapley:2015aa}, $\text{z} \sim 2.3$) and KBSS survey (\citealt{Strom:2017aa}, $\text{z} \sim 2.3$) is shown with cyan, green and orange lines respectively. The typical error on the emission line ratios is indicated in the upper right corner of each panel. Consistently with findings reported by the abovementioned surveys, many of our galaxies appear shifted from the position occupied by the bulk of the local galaxy distribution, especially in the N2-BPT. }
\label{fig:global_bpt}
\end{figure*}
\begin{figure*}
\centering

\includegraphics[width=0.485\textwidth]{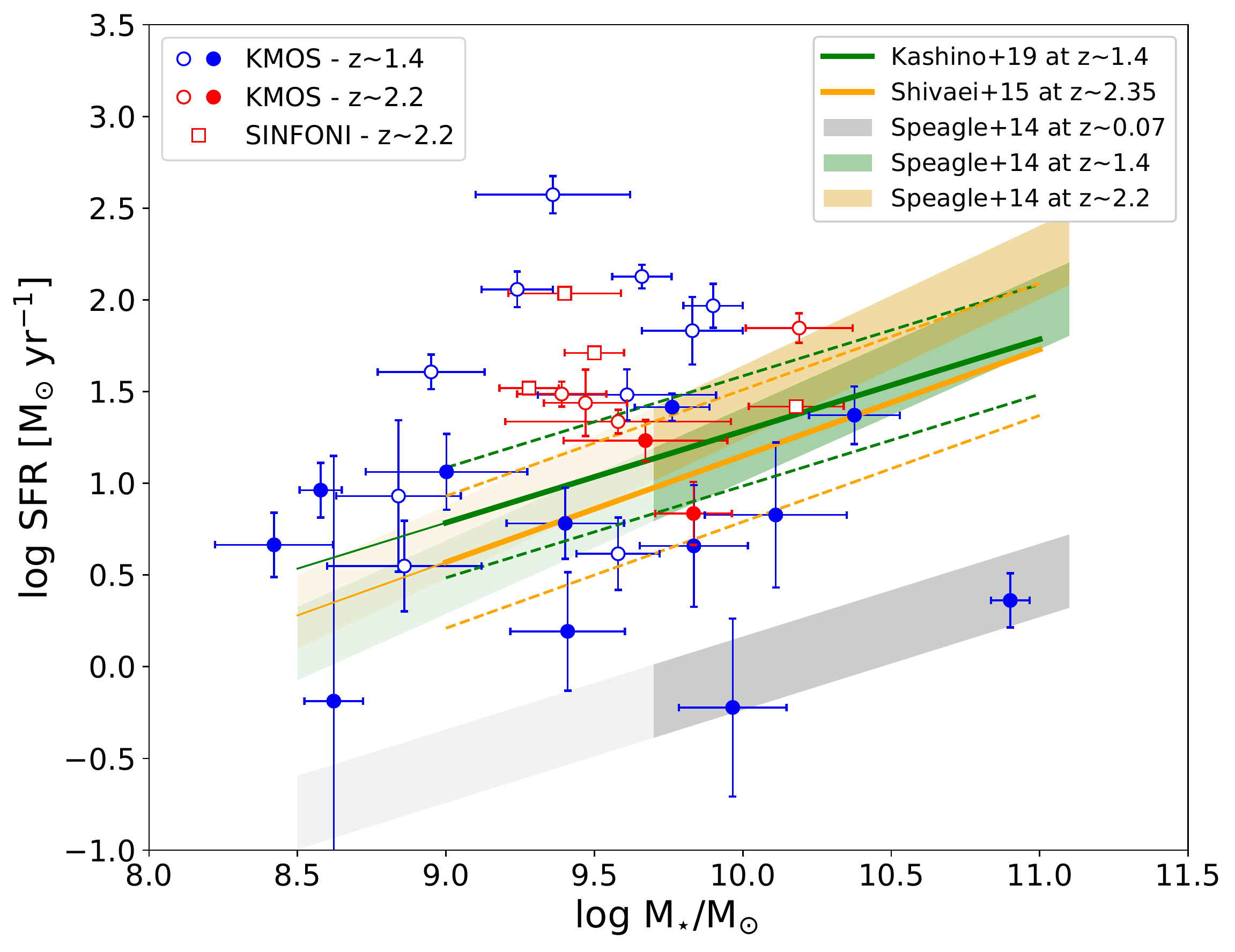}
\includegraphics[width=0.47\textwidth]{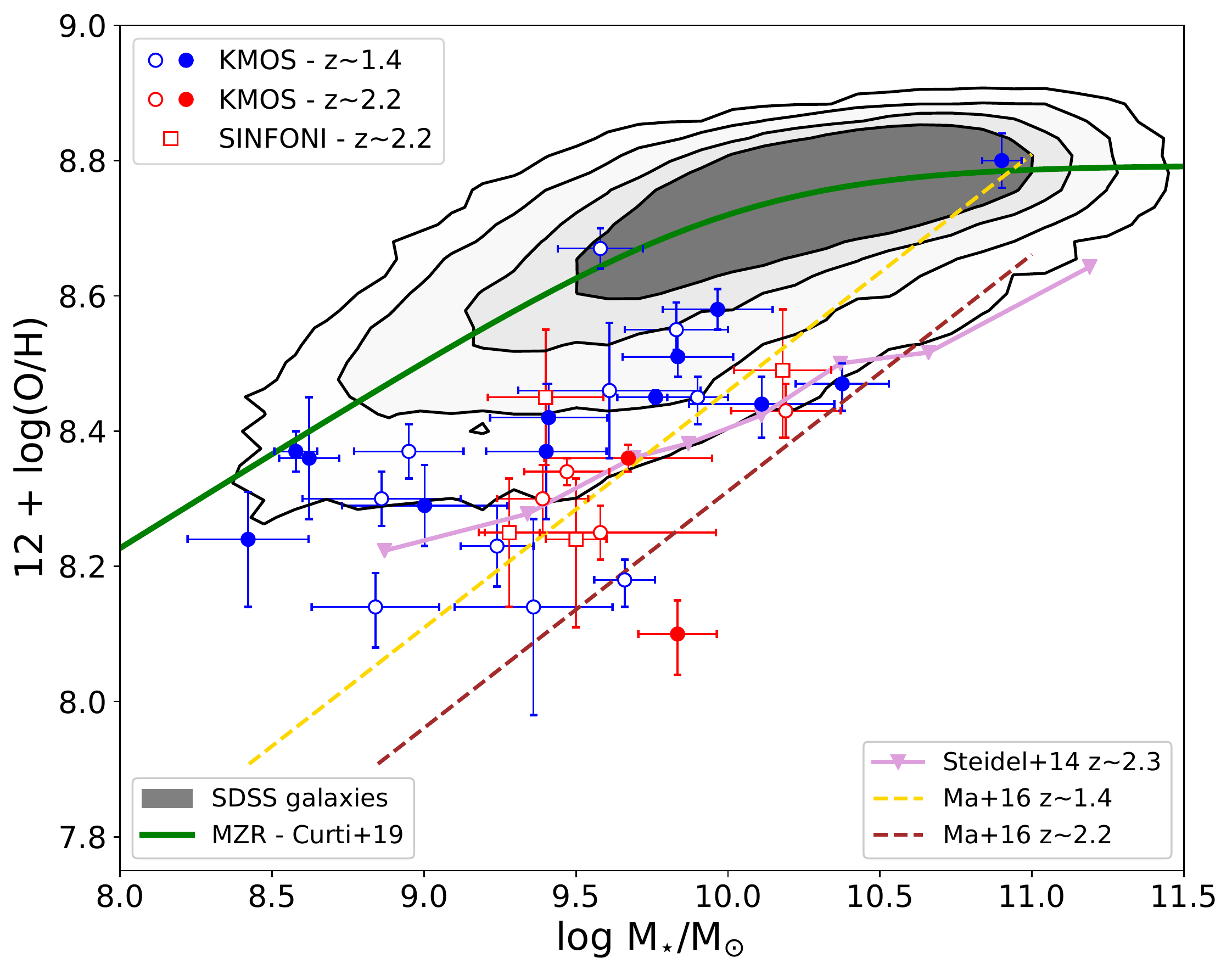}

\caption{\textit{Left Panel}: Distribution of our galaxy sample in the  SFR vs \mstar plane. 
Symbols and colours are the same as in Fig.~\ref{fig:global_bpt}.
%Galaxies visually identified as interacting systems/merger candidates are marked with a green X.
Both SFR and \mstar measurements are corrected for lensing magnification and the error bars account for the uncertainties associated to the reddening correction and the statistical uncertainty on the magnification itself.
The representative star-forming main sequence (SFMS) at redshift $0.07$, $1.4$ and $2.2$ from \citet{Speagle:2014aa} is represented by grey, green and yellow shaded regions respectively, with a nominal intrinsic scatter of $0.2$ \rm{dex}. 
The best-fit to SFMS (based on dust corrected H$\alpha$-based SFR) from \citet{Kashino:2019aa} (FMOS Survey, z$\sim 1.4$) and \citet{Shivaei:2015ab} (MOSDEF Survey, z$\sim2.3$) are also plotted with green and golden solid lines, with an observational scatter of $0.3$ and $0.36$ \rm{dex} respectively.
%The fit to the SFMS at $\text{z}=1.4$ and $\text{z}=2.2$ from \citet{Whitaker:2014aa} are also shown as purple and orange dashed lines. 
All SFRs are reported to a common \citet{Chabrier:2003aa} IMF.
\textit{Right Panel}: Relationship between stellar mass and gas-phase metallicity for our sample, derived with the C17 and C19 calibrations as described in the text. Colors and symbols are the same as in the left panel.
The region within grey contours encloses the $68\%$,$84\%$ $95\%$ and $99\%$ of the distribution of local SDSS galaxies in the M$_{\star}$ vs O/H plane, with abundances calculated from the same C17 and C19 calibrations, and the relative best-fit MZR to median points in bins of stellar mass from \citet{Curti:2019aa} in shown in green.
%The local mass-metallicity relation (MZR) from \citet{Andrews:2013aa} derived with the T$_{e}$ method from stacked spectra in bins of stellar mass is shown in green, 
We also plot the MZR at z$\sim 2.3$ from \citet{Steidel:2014aa} in bins of stellar mass, shown in violet, as obtained from the O3N2 diagnostic rescaled to the C17 calibrations.
All these metallicity measurements are consistent with the abundance scale defined by the T$_{e}$ method.
For comparison, we also show,with yellow and brown dashed curves respectively, the predicted MZR at z$=1.4$ and z$=2.2$ from high-resolution FIRE simulations, as presented in \citet{Ma:2016aa}. In agreement with the prescriptions of an evolving MZR, our galaxies are characterised by lower metallicities, at fixed \mstar, compared to local galaxies ($0.24$ and $0.36$ dex lower on average than the local MZR for z$\sim1.4$ and z$\sim2.2$ sources respectively). However, C17-C19 calibrations provide higher metallicities compared to the abundance scale defined by both \citet{Ma:2016aa} simulations at the reference redshift, possibly due to the uncertainties affecting the stellar yields adopted in the simulations.
}
\label{fig:global_prop}
\end{figure*}

%%%%%%%%%% RESULTS %%%%%%%
\section{Results and discussion}
\label{sect:results}

\subsection{Global properties}
\label{sec:global_prop}

\begin{table*}
\centering

\begin{tabular}{@{} cccccc @{}}
	\hline
Galaxy 		 				& SFR(H$\alpha$) [M$_{\odot} \text{yr}^{-1}$]  & log(M$_{\star}$/M$_{\odot}$) 	& 12 + log(O/H)	 & $\mu^{(1)}_{H\alpha}$ &		$\mu^{(2)}_{maps}$	\\
\hline
\textbf{SINFONI Galaxies} &&&& \\
Horseshoe & 54.60$^{\star}$$\pm$12.02 & 10.18$\pm$0.16 & 8.49$\substack{+0.04 \\ -0.05}$ & 12.11$\pm$1.56 &-- \\
MACS0451 Arc (Total) & 93.04$\pm$11.08 & 9.71$\pm$0.10 & 8.24$\substack{+0.05 \\ -0.06}$ & 11.85 $\pm$0.59 		& --\\
MACS0451 Arc (North) & 56.08$\pm$6.74 & 9.50$\pm$0.12 & 8.24$\substack{+0.05 \\ -0.07}$ & 9.18$\pm$1.01 		& --\\
MACS0451 Arc (South) & 36.04$\pm$4.20 & 9.28$\pm$0.10 & 8.25$\substack{+0.04 \\ -0.06}$ & 14.51$\pm$1.45 		& -- \\
CSWA164 & 108.50$\pm$20.30 & 9.40$\pm$0.19 & 8.45$\substack{+0.06 \\ -0.05}$ & 12.62$\pm$2.25 				&-- \\

\textbf{KMOS Galaxies} &&&& \\
\textbf{MS 2137} &&&& \\
sp1 & 40.48$\pm$8.81 & 8.95$\pm$0.18 & 8.37$\substack{+0.04 \\ -0.04}$ & 1.12$\pm$0.16 & 1.22$\substack{+0.01 \\ -0.01}$ \\
sp2 & 70.24$\pm$13.02 & 10.19$\pm$0.18 & 8.43$\substack{+0.04 \\ -0.04}$ & 1.23$\pm$0.18 & 1.27$\substack{+0.01 \\ -0.02}$ \\
sp3 & 92.89$\pm$25.71 & 9.90$\pm$0.10 & 8.45$\substack{+0.03 \\ -0.04}$ & 2.26$\pm$0.19 & 2.57$\substack{+0.12 \\ -0.12}$ \\
sp5 & 30.67$\pm$4.81 & 9.39$\pm$0.15 & 8.30$\substack{+0.05 \\ -0.05}$ & 1.73$\pm$0.15 & 1.76$\substack{+0.04 \\ -0.04}$ \\
sp6 & 27.45$\pm$11.43 & 9.47$\pm$0.14 & 8.34$\substack{+0.02 \\ -0.02}$ & 1.43$\pm$0.4 & 1.55$\substack{+0.02 \\ -0.02}$ \\
sp7 & 67.9$\pm$28.69 & 9.83$\pm$0.17 & 8.55$\substack{+0.04 \\ -0.03}$ & 1.37$\pm$0.23 & 1.4$\substack{+0.02 \\ -0.02}$ \\
sp9 & -- & -- & -- & -- & 1.1$\substack{+0.01 \\ -0.01}$ \\
sp10 & -- & -- & -- & -- & 3.62$\substack{+0.22 \\ -0.19}$ \\
sp13 & 8.53$\pm$8.1 & 8.84$\pm$0.21 & 8.14$\substack{+0.05 \\ -0.06}$ & 13.48$\pm$0.71 & 25.51$\substack{+15.84 \\ -7.13}$ \\
sp14 & 3.54$\pm$2.01 & 8.86$\pm$0.26 & 8.3$\substack{+0.04 \\ -0.04}$ & 4.93$\pm$0.54 & 5.69$\substack{+0.6 \\ -0.53}$ \\
sp15 & 134.14$\pm$19.86 & 9.66$\pm$0.10 & 8.18$\substack{+0.03 \\ -0.04}$ & 1.97$\pm$0.1 & 2.28$\substack{+0.09 \\ -0.1}$ \\
ph6532 & 21.70$\pm$3.21 & 9.58$\pm$0.38 & 8.25$\substack{+0.04 \\ -0.04}$ & 1.07$\pm$0.14 & 1.08$\substack{+0.0 \\ -0.0}$ \\
ph2594 & 4.13$^{\star\star}$$\pm$1.88 & 9.58$\pm$0.14 & 8.67$\substack{+0.03 \\ -0.03}$ & 1.28$\pm$0.44 & 1.28$\substack{+0.01 \\ -0.01}$ \\
ph3729 & 30.37$\pm$9.67 & 9.61$\pm$0.30 & 8.46$\substack{+0.1 \\ -0.1}$ & 1.33$\pm$0.21 & 1.34$\substack{+0.02 \\ -0.02}$ \\
ph3912 & 114.12$\pm$25.57 & 9.24$\pm$0.12 & 8.23$\substack{+0.06 \\ -0.06}$ & 1.22$\pm$0.13 & 1.27$\substack{+0.01 \\ -0.01}$ \\
ph5514 & 4.01$^{\star\star}$$\pm$0.07 & -- & 8.79$\substack{+0.04 \\ -0.04}$ & 1.40$\pm$0.38 & 1.71$\substack{+0.02 \\ -0.03}$ \\
ph7727 & -- & -- & -- & -- & 1.07$\substack{+0.0 \\ -0.0}$ \\
ph8073 & 199.55$\pm$46.72 & 9.36$\pm$0.26 & 8.14$\substack{+0.13 \\ -0.16}$ & 1.0$\pm$0.19 & 1.05$\substack{+0.0 \\ -0.0}$ \\
\textbf{RXJ2248} &&&& \\
GLASS\_00093-99-99 & 6.05$\pm$1.13 & 9.40$\pm$0.13 & 8.37$\substack{+0.12 \\ -0.12}$ & 1.62$\pm$0.02 & 1.93$\substack{+0.17 \\ -0.17}$ \\
R2248\_LRb\_p1\_M3\_Q4\_58\_\_2 & 23.51$\pm$2.62 & 10.38$\pm$0.07 & 8.47$\substack{+0.04 \\ -0.04}$ & 1.36$\pm$0.01 & 1.55$\substack{+0.13 \\ -0.07}$ \\
MUSE\_SW\_462-99-99 & 4.55$\pm$1.51 & 9.83$\pm$0.12 & 8.51$\substack{+0.06 \\ -0.06}$ & 3.21$\pm$0.03 & 4.35$\substack{+1.15 \\ -0.4}$ \\
GLASS\_00333-99-99 & 4.54$\pm$0.76 & 8.41$\pm$0.11 & 8.25$\substack{+0.10 \\ -0.12}$ & 1.24$\pm$0.01 & 1.44$\substack{+0.13 \\ -0.05}$ \\
R2248\_LRb\_p3\_M4\_Q3\_93\_\_1 & 26.02$\pm$1.38 & 9.76$\pm$0.08 & 8.45$\substack{+0.01 \\ -0.02}$ & 1.23$\pm$0.01 & 1.35$\substack{+0.1 \\ -0.03}$ \\
R2248\_MR\_p1\_M1\_Q4\_10\_\_1 & 1.35$^{\star\star}$$\pm$0.07 & -- & -- & 1.01$\pm$0.02 & 1.14$\substack{+0.05 \\ -0.02}$ \\
R2248\_LRb\_p3\_M4\_Q3\_94\_\_1 & 17.24$\pm$0.66 & 9.67$\pm$0.19 & 8.36$\substack{+0.04 \\ -0.04}$ & 1.22$\pm$0.01 & 1.53$\substack{+0.11 \\ -0.09}$ \\
MUSE\_SW\_48-99-99 & -- & 8.30$\pm$0.21 & -- & -- & 8.28$\substack{+2.78 \\ -2.58}$ \\
MUSE\_SW\_51-99-99 & -- & 7.30$\pm$0.22 & -- & -- & 26.15$\substack{+50.92 \\ -13.78}$ \\
R2248\_MR\_p3\_M1\_Q3\_43\_\_1 & -- & 9.58$\pm$0.07 & -- & -- & 28.61$\substack{+57.4 \\ -16.91}$ \\
GLASS\_01891-99-99 & -- & 9.82$\pm$0.06 & -- & -- & 1.88$\substack{+0.15 \\ -0.16}$ \\
GLASS\_01845-99-99 & 6.85$\pm$0.24 & 9.83$\pm$0.04 & 8.09$\substack{+0.06 \\ -0.07}$ & 1.76$\pm$0.02 & 2.19$\substack{+0.25 \\ -0.25}$ \\
MUSE\_SW\_45-99-99 & 11.61$\pm$2.03 & 9.01$\pm$0.20 & 8.32$\substack{+0.06 \\ -0.05}$ & 1.69$\pm$0.02 & 2.15$\substack{+0.21 \\ -0.22}$ \\
MUSE\_SW\_461-99-99 & 6.72$\pm$2.02 & 10.11$\pm$0.15 & 8.52$\substack{+0.05 \\ -0.04}$ & 2.81$\pm$0.03 & 3.66$\substack{+0.4 \\ -0.55}$ \\
MUSE\_NE\_111-99-99 & 9.18$\pm$1.13 & 8.58$\pm$0.02 & 8.35$\substack{+0.04 \\ -0.04}$ & 1.9$\pm$0.01 & 2.28$\substack{+0.33 \\ -0.16}$ \\
GLASS\_01404-99-99 & -- & 9.45$\pm$0.13 & -- & -- & 1.4$\substack{+0.12 \\ -0.04}$ \\
R2248\_MR\_p1\_M1\_Q4\_51\_\_1 & 0.65$\pm$2.0 & 8.62$\pm$0.10 & 8.36$\substack{+0.12 \\ -0.10}$ & -- & 20.65$\substack{+16.55 \\ -8.34}$ \\
GLASS\_00800-99-99 & 0.6$\pm$0.07 & 9.97$\pm$0.09 & 8.58$\substack{+0.04 \\ -0.04}$ & 4.83$\pm$0.08 & 10.52$\substack{+7.45 \\ -2.87}$ \\
MUSE\_NE\_23-99-99 & 1.40$\pm$0.07 & 9.37$\pm$0.07 & 8.42$\substack{+0.06 \\ -0.06}$ & 2.64$\pm$0.03 & 3.64$\substack{+0.41 \\ -0.55}$ \\
R2248\_MR\_p1\_M1\_Q4\_59\_\_1 & 2.89$^{\star\star}$$\pm$0.78 & 10.90$\pm$0.07 & 8.81$\substack{+0.06 \\ -0.06}$ & 1.86$\pm$0.02 & 2.34$\substack{+0.2 \\ -0.23}$ \\
MUSE\_NE\_117-99-99 & -- & 7.75$\pm$0.25 & -- & -- & 3.05$\substack{+0.73 \\ -0.66}$ \\

	\hline
\end{tabular}
\caption{Global properties of the analysed sample. The star formation rate, the stellar mass and the oxygen abundance are reported for each source. 
SFRs are derived from the total H$_{\alpha}$ flux in the image plane and corrected for magnification.
Stellar masses are derived from SED fitting, while global metallicities are computed from the integrated spectra adopting the \citet{Curti:2017aa,Curti:2019aa} calibrations. Galaxies with no measured M$_{\star}$ were not covered by multi-band HST images (but sometimes they are still covered by the lensing maps), while galaxies with no measured SFR have no H$\alpha$ detection.\newline
In the last columns we report two different estimates of the magnification factor: $\rm \mu_{\text{H}\alpha}$
is derived from the ratio between the H$\alpha$ flux in the image plane and that in the source plane, while
$\rm \mu_{\text{maps}}$ exploits the magnification maps computed, for each source redshift, from the output files of the lens models adopted in this work \citep{Zitrin:2013aa}. The uncertainties on the latter value are purely statistical, as computed from the standard deviation of $100$ different realisations of the magnification maps, and do not take into account the systematic uncertainties on the lens models, which are of the order of $10\%-20\%$ (see Section~\ref{sect:lensmod} for more details).\newline
\textit{Notes} - $^{\star}$ Total SFR derived from SINFONI observations of both the western and southern arc. The SFR computed for the western arc only, which is the only region covered also by observations in the J and H band, is $26.2 \pm 4.3$ 
M$_{\odot} \text{yr}^{-1}$. \newline
 $^{\star\star}$: H$\alpha$ not corrected for reddening due to absence of H$\beta$ detection.
}
\label{tab:info_glob} 
\end{table*}

We briefly discuss here the global properties of our sample by examining line ratios and physical quantities inferred from integrated spectra. 
%Emission line fitting was performed on spectra extracted from $0.6\arcsec$-radius circular pseudo apertures, equivalent to the average seeing of our observations and large enough to encompass the majority of the light from our sources. 
%These integrated fluxes are fully consistent with those obtained by summing the individual pixel fluxes within the the emission line maps.

In Fig~ \ref{fig:global_bpt} we show the position of our galaxies on the BPT diagrams, named after Baldwin, Phillips \& Telervich \citep{Baldwin:1981aa} and defined as [\ion{N}{ii}]$\lambda6584$/H$\upalpha$ vs [\ion{O}{iii}]$\lambda5007$/H$\upbeta$ (the N2-BPT)  and [\ion{S}{ii}]$\lambda6716,31$/H$\upalpha$ vs [\ion{O}{iii}]$\lambda5007$/H$\upbeta$ (the S2-BPT).
We indicate in blue galaxies at $\text{z}\sim1.4$, while in red those at $\text{z}\sim2.2$. 
The sources observed within the RXJ2248 cluster are represented by filled circles, while empty circles mark those observed in MS2137. 
Finally, empty squares are assigned to the sample of strongly lensed galaxies observed with SINFONI, where both the northern and southern component of the MACS041 Arc are shown.

Consistently with typical findings at these redshifts, a systematic offset from the tight sequence occupied by galaxies in the local Universe (encompassed by the grey contours represented by SDSS galaxies) is seen on the N2-BPT diagram, observed towards higher [\ion{O}{iii}]/H$\upbeta$ and/or [\ion{N}{ii}]/H$\upalpha$ ratios.
Nonetheless, the position of our galaxies is still consistent, within the uncertainties, with the star forming region predicted by theoretical classification schemes like those proposed by \cite{Kewley:2001aa} and \cite{Kauffmann:2003aa} (the solid and dashed black lines respectively). 
On the S2-BPT diagram the offset is less prominent than in the N2-BPT diagram, and the points are more scattered.
For comparison, we also show in Fig.~\ref{fig:global_bpt} the fits to the average position of high-z galaxies on BPT diagrams as inferred from previous large surveys in the near-infrared (i.e. \citealt{Shapley:2015aa} for MOSDEF, \citealt{Kashino:2017aa} for FMOS and \citealt{Strom:2017aa} for KBSS).
Compared to these studies, KLEVER provides spatially resolved information which could be used to investigate radial trends and to what extent the observed evolution in line ratios is driven locally (and by which physical effect) or whether it is a global property of the galaxy as a whole.
For a more in-depth, spatially resolved analysis and discussion on the physical drivers of the BPT-offset for the KLEVER sample, we refer to a forthcoming paper by the collaboration (Curti et al., in prep.).

In the left panel of Figure~\ref{fig:global_prop} we show instead the distribution of our sample on the \mstar vs log(SFR) plane;
	colours and symbols are the same as in Fig.~\ref{fig:global_bpt}.	
	%indicating in blue galaxies at $\text{z}\sim1.4$  and in red those at $\text{z}\sim2.2$. 
	%In particular, we adopt here the SFR computed from summing the H$\upalpha$ flux in all spaxels with SNR$>5$ (see Sect.~\ref{sec:mass_sfr}).
	%The sources observed within the RXJ2248 cluster are represented by filled circles, while empty circles mark those observed in MS2137.
	%Finally, empty squares are assigned to the sample of three strongly lensed galaxies observed with SINFONI; again, both the northern and southern component of the MACS041 Arc are shown here.
	The plotted error bars include the statistical errors on the magnification, but do not take into account the systematic uncertainties on the lensing models (see Sect.~\ref{sect:lensmod}).
	%global SFR (as measured from H$\upalpha$) as a function of stellar mass for the sample analysed in this work.
	For comparison, the Star-Forming Main Sequence (SFMS) at $\text{z}=0.07$, $\text{z}=1.4$ and $\text{z}=2.2$ (i.e. the local Universe and the two average redshift intervals probed by the KLEVER sample) from \cite{Speagle:2014aa} are shown as shaded regions, assuming a nominal intrinsic scatter of $0.2$ \rm{dex}.
This is based on a compilation of many different works in the literature, which are predominantly based on UV/IR SFR indicators. 
	Although \cite{Speagle:2014aa} assume a \cite{Kroupa:1993aa} IMF, converting to a \cite{Chabrier:2003aa} IMF introduce negligible effects. 
	%The grey shaded region represents the $1\sigma$ dispersion of the local SFMS from \cite{Speagle:2014aa}.
	In addition, the best-fit to the SFMS derived in the context of the FMOS Survey \citep[][their power-law equation at z$\sim 1.4$]{Kashino:2019aa} and the MOSDEF Survey \citep[][z$\sim 2.3$]{Shivaei:2015ab} are also shown with green and orange solid lines respectively, with an observational scatter of $0.3$ and $0.36$ \rm{dex} as marked by the dashed lines.
	The latter relations are based on dust-corrected H$\upalpha$-SFR measurements, hence they are more consistent with the methodology followed in this work.

The sample of lensed galaxies analysed in this work allows us to extend the probe of ionised gas properties in the SFR vs \mstar\  plane towards lower masses (in particular below $10^{9.5}$M$_{\odot}$) compared to typical magnitude-limited surveys of field galaxies, populating both the region below and above the representative SFMS at z$\sim 1.5-2.5$. 
	In particular, $8$ galaxies at z$\sim1.4$ lie within $1\sigma$ of the \cite{Kashino:2019aa} relation, while $7$ galaxies fall more than $1\sigma$ below and $7$ galaxies more than $1\sigma$ above the best-fit SFMS. 
	However, we note that assuming a different parametrisation of the SFMS, like the broken power-law proposed in \cite{Kashino:2019aa} (or by \cite{Whitaker:2014aa}), or even considering the extrapolation of the \cite{Speagle:2014aa} relation, would result in a steepening of the slope of the low-mass end of the relation, making the bias towards high specific star formation rates (sSFR) of the KLEVER sample in that regime more prominent.
	At z$\sim2.2$, our sample comprises only $1$ galaxy below the \cite{Shivaei:2015ab} SFMS parametrisation, while the remaining 
	$9$ lie above (although $3$ of them are still consistent within the observational scatter).
	Remarkably, a few galaxies in the sample are approximately ten times more star forming than the average population at those epochs.
	The impact that the distribution in sSFR of our sample has in the interpretation of the observed metallicity gradients is discussed in Sect.~\ref{sect:grads}.
	The full KLEVER sample will comprise also a large number of more massive, unlensed sources from the CANDELS fields, which are complementary to this sample for a full exploration of the \mstar-SFR plane.

The right panel of Figure~\ref{fig:global_prop} shows instead the relation between stellar mass and oxygen abundance (the mass-metallicity relation, MZR) for our sample. 
The gas-phase metallicity is derived exploiting emission line diagnostics measured from the integrated spectra, following the scheme presented in Sect.~\ref{sect:metcal} and the C17, C19 calibrations.
%For comparison, the MZR derived by \citet{Andrews:2013aa} with the T$_{e}$ method from stacked spectra in bins of stellar mass is shown with the green solid line. 
%while in magenta we report the local MZR computed exploiting the \citet{Pettini:2004aa} calibration of the O3N2 indicator.
The grey shaded area encloses the $68\%, 84\%, 95\%$ and $99\%$ of local SDSS galaxies and the relative best-fit MZR from \cite{Curti:2019aa} is shown in green.
%For consistency, all the mass-metallicity relations shown here are based on the Te abundance scale.
In agreement with typical findings at these redshifts, our sample is characterized by lower metallicities, at fixed stellar mass, compared to the values assumed by local galaxies, indicative of the cosmic evolution of the MZR (see \citealt{Maiolino:2019aa} and references therein). 
%As an example, a linear fit to the MZR at z$\sim 2.3$, derived by \cite{Steidel:2014aa} on individual KBSS galaxies, is shown in violet.
The mean offsets in metallicity, at fixed \mstar, from the local MZR are $-0.25$ and $-0.36~\rm{dex}$ for $\text{z}\sim 1.4$ and $\text{z}\sim 2.2$ galaxies respectively.
In violet we also show the MZR at z$\sim 2.3$ from bins in stellar mass of individual KBSS galaxies \citep{Steidel:2014aa}, derived exploiting the C17 calibration of the O3N2 indicator for consistency with the abundance scale adopted in this work; $8$ out of $10$ KLEVER galaxies at z$\sim2.2$ with robust metallicity determination are consistent within the uncertainties with the \cite{Steidel:2014aa} curve.
%However, our galaxies present systematically higher metallicities compared to the \cite{Steidel:2014aa} curve. 
%This is most likely due to the different sample adopted to derive the C17-C19 calibrations and those adopted by \cite{Steidel:2014aa}, whose metallicity measurements are based on a slightly revised version of the \cite{Pettini:2004aa} O3N2 indicator.
%A detailed study on the MZR within KLEVER will follow when the observations of the program are completed.
Finally, we also plot for comparison the predicted mass-metallicity relations from the FIRE cosmological simulations at redshift $1.4$ and $2.2$, which are shown by the yellow and brown curves respectively, as presented in \citet{Ma:2016aa}.
Despite the very different methodologies considered (\Te-based strong line methods versus prescriptions from zoom-in cosmological simulations), the predicted MZR  are in reasonable agreement with  our measurements for both redshift intervals considered, although showing a small systematic offset towards lower metallicities, possibly due to the large uncertainties affecting the stellar yields used in simulations \citep[e.g.][]{Wiersma:2009aa}.

\begin{table*}
	%\tiny
\centering
%\resizebox{\columnwidth}{!}{
%	\setlength{\extrarowheight}{.3em}
%\scalebox{0.75}{
\begin{tabular}{@{\extracolsep{\fill}} ccccc @{}}
\hline
Galaxy & Metallicity Gradient & Resolved & PSF-FWHM (SP) & Diagnostics \\
 		   &  [dex kpc$^{-1}$]     &               &[kpc $\times$ kpc] \\ 

\hline
\textbf{SINFONI Galaxies} &&& \\
Horseshoe & 0.012$\pm$0.014 & Yes & 1.82$\times$0.37 & R$_{3}$,N$_{2}$ \\
MACS Arc (North)& -0.011$\pm$0.012 & Yes & 1.74$\times$1.33 & R$_{3}$,N$_{2}$,O$_{3}$O$_{2}$ \\
MACS Arc (South) & -0.011$\pm$0.016 & Yes & 1.64$\times$0.76 & R$_{3}$,N$_{2}$,O$_{3}$O$_{2}$ \\
CSWA164 & -0.006$\pm$0.013 & Yes & 2.58$\times$0.36 & R$_{3}$,N$_{2}$,O$_{3}$O$_{2}$ \\

\textbf{KMOS Galaxies} &&& \\
\textbf{MS2137} &&&& \\
sp1 & 0.019$\pm$0.012 & Yes & 3.57$\times$2.8 & R$_{3}$,S$_{2}$ \\
sp2 & -0.007$\pm$0.015 & Yes & 3.48$\times$2.65 & R$_{3}$,O$_{3}$O$_{2}$ \\
sp3 & 0.025$\pm$0.013 & Yes & 3.28$\times$1.52 & R$_{3}$,N$_{2}$ \\
sp5 & 0.001$\pm$0.020 & Yes & 3.6$\times$1.96 & R$_{3}$,O$_{3}$O$_{2}$ \\
sp6 & 0.063$\pm$0.012 & Yes & 3.39$\times$2.18 & R$_{3}$,O$_{3}$O$_{2}$ \\
sp7 & 0.024$\pm$0.014 & Yes & 3.67$\times$2.44 & R$_{3}$,N$_{2}$ \\
sp13 & 0.060$\pm$0.084 & Yes & 2.61$\times$0.49 & R$_{3}$,N$_{2}$ \\
sp15 & 0.005$\pm$0.010 & Yes & 3.63$\times$1.56 & R$_{3}$,S$_{2}$ \\
ph6532 & 0.036$\pm$0.012 & Yes & 3.65$\times$3.01 & R$_{3}$,O$_{3}$O$_{2}$\\
ph3729 & 0.002$\pm$0.048 & Yes & 3.62$\times$2.56 & R$_{3}$,N$_{2}$ \\
ph3912 & 0.051$\pm$0.012 & Yes & 3.58$\times$2.67 & R$_{3}$,S$_{2}$ \\
ph8073 & 0.006$\pm$0.105 & Marg & 3.59$\times$3.15 &R$_{3}$,S$_{2}$ \\
\textbf{RXJ2248} &&&& \\
GLASS\_00093-99-99 & 0.011$\pm$0.039 & Yes & 4.14$\times$2.61 & R$_{3}$,N$_{2}$ \\
R2248\_LRb\_p1\_M3\_Q4\_58\_\_2 & 0.002$\pm$0.010 & Yes & 4.46$\times$3.01 & R$_{3}$,N$_{2}$ \\
MUSE\_SW\_462-99-99 & -0.005$\pm$0.012 & Yes & 3.14$\times$1.62 & R$_{3}$,N$_{2}$ \\
GLASS\_00333-99-99 & 0.031$\pm$0.033 & Marg & 4.58$\times$3.13 & R$_{3}$,S$_{2}$ \\
R2248\_LRb\_p3\_M4\_Q3\_93\_\_1 & 0.008$\pm$0.006 & Yes & 4.66$\times$3.28 & R$_{3}$,N$_{2}$ \\
R2248\_LRb\_p3\_M4\_Q3\_94\_\_1 & 0.051$\pm$0.018 & Yes & 4.62$\times$2.97 & R$_{3}$,O$_{3}$O$_{2}$ \\
GLASS\_01845-99-99 & 0.029$\pm$0.023 & Yes & 4.1$\times$2.27 & R$_{3}$,O$_{3}$O$_{2}$ \\
MUSE\_SW\_45-99-99 & 0.009$\pm$0.015 & Yes & 4.1$\times$2.38 & R$_{3}$,S$_{2}$ \\
MUSE\_SW\_461-99-99 & -0.019$\pm$0.018 & Yes & 2.56$\times$2.52 & R$_{3}$,N$_{2}$ \\
MUSE\_NE\_111-99-99 & 0.012$\pm$0.038 & Marg & 3.8$\times$2.42 & R$_{3}$,N$_{2}$ \\
GLASS\_00800-99-99 & -0.014$\pm$0.019 & Yes & 3.36$\times$0.75 & R$_{3}$,N$_{2}$ \\
MUSE\_NE\_23-99-99 & 0.011$\pm$0.021 & Yes & 3.6$\times$1.68 & R$_{3}$,S$_{2}$ \\

\hline
\end{tabular}
%}

\caption{Radial gradients for galaxies for which they could be at least marginally resolved (the \textit{metallicity gradients sample}).
We indicate to what degree each gradient has been resolved following the criteria described in Section~\ref{sect:grads}. ``Resolved'' indicates galaxies where the gradient is extracted out to a radius larger than $2$ times the linear size of the PSF-HWMH ($\sqrt{\text{HWHM}_{x} \times \text{HWHM}_{y}}$), while ``Marginal'' denotes that the gradient is extracted within a radial distance between $1$ and $2$ times this value. The third column report the major- and minor-axis FWHM (in kpc) of the ellipse used to model the
PSF in the source plane, while in the last one we indicate the combination of strong-line diagnostics adopted to derive the metallicity map in the source plane for each galaxy.}
\label{tab:grads} 
\end{table*}

\subsection{Spatially resolved metallicity maps and gradients}
\label{sect:grads}
We have derived the metallicity maps for our galaxy sample from the de-lensed source plane emission line maps, using the scheme presented in Sec.~\ref{sect:metcal} and a combination of the diagnostics and calibrations presented in \cite{Curti:2017aa, Curti:2019aa}.
As general criteria we impose, on pixel-by-pixel basis, a SNR threshold of $5$ on H$\upalpha$ and \oiii and of $2.5$ on the other lines in order to include them in the metallicity calculation. 
To maintain a sufficient level of self-consistency, the same combination of diagnostics was used for all spatial elements across a given galaxy, in order to avoid possible systematics introduced by the differential dependence of each of the line ratios considered on the gas properties of that particular source.
However, the set of diagnostics involved in the metallicity calculation can vary from galaxy to galaxy, 
according to the availability and signal-to-noise of the different emission lines for each source. % as illustrated in Table~\ref{tab:klever_sources}.
In general, the majority of our metallicity maps involves the simultaneous use of the R$_{3}$ and N$_{2}$ diagnostics. The S$_{2}$ line ratio is adopted instead of N$_{2}$ when a resolved gradient can not be obtained with N$_{2}$ or the latter is heavily contaminated by residuals of badly subtracted sky-lines.
Moreover, for galaxies at $\text{z}>2$ we could include the information from the O$_{3}$O$_{2}$ line ratio (once properly corrected for reddening) in the metallicity determination.
The list of diagnostics adopted for each galaxy to derive the relative metallicity map and gradient is reported in the last column of Table~\ref{tab:grads}.

To derive the metallicity gradients, we radially bin our data defining a series of elliptical apertures according to the shape of the source plane PSF. Increasing radii are taken in steps of a fraction of the linear PSF size, defined as the geometric mean of the model ellipse, i.e. $r = \sqrt{x\times y}$, where $x$ and $y$ represent the FWHM along the major and minor axis respectively (which are reported for each source in Table~\ref{tab:grads}). 
This is aimed at partially overcoming the smearing effect introduced by the distorted source plane PSF, which would be particularly intense, especially at large distances from the centre, if we adopted simple circular apertures of increasing radius.
The potential systematics associated to this choice are discussed in Appendix A. %~\ref{sect:appenA}.
We then estimate the weighted average of the metallicity within each annulus; each value is weighted on the fraction of the relative pixel falling into the considered aperture.
We note that the result does not change if we compute instead the average emission line ratios and re-compute the metallicity accordingly.
The uncertainty associated with the average metallicity within each annulus takes into account both the $1\sigma$ dispersion of values of individual pixels inside the aperture and the error on each individual metallicity measurement (introduced by adopting our set of strong-line calibrations).  
%To avoid that the most external apertures are dominated by a few isolated pixels in the outskirts of the galaxy, we consider only apertures including a number of pixels covering at least the $10\%$ of the total aperture area.
The centre of our apertures is assumed to be the peak of the continuum emission in the H band or, alternatively, the peak of the H$\upalpha$ emission (hence the peak of the SFR surface density) for galaxies with no continuum detection.
This choice does not affect the inferred gradients for the majority of the cases, with the exception of a few interacting systems with double-peaked (or strongly smeared) distribution of the H$\upalpha$ emission (see discussion in Sect.~\ref{sec:mergers}).
Finally, the metallicity gradients are derived by performing a linear fit to the extracted metallicity values in each annulus and their uncertainties are evaluated by sampling the posterior distribution of the parameters through a Markov Chain Monte Carlo.

\begin{figure*}
	\centering
	\textbf{Resolved} \\
  	\includegraphics[width=0.95\textwidth]{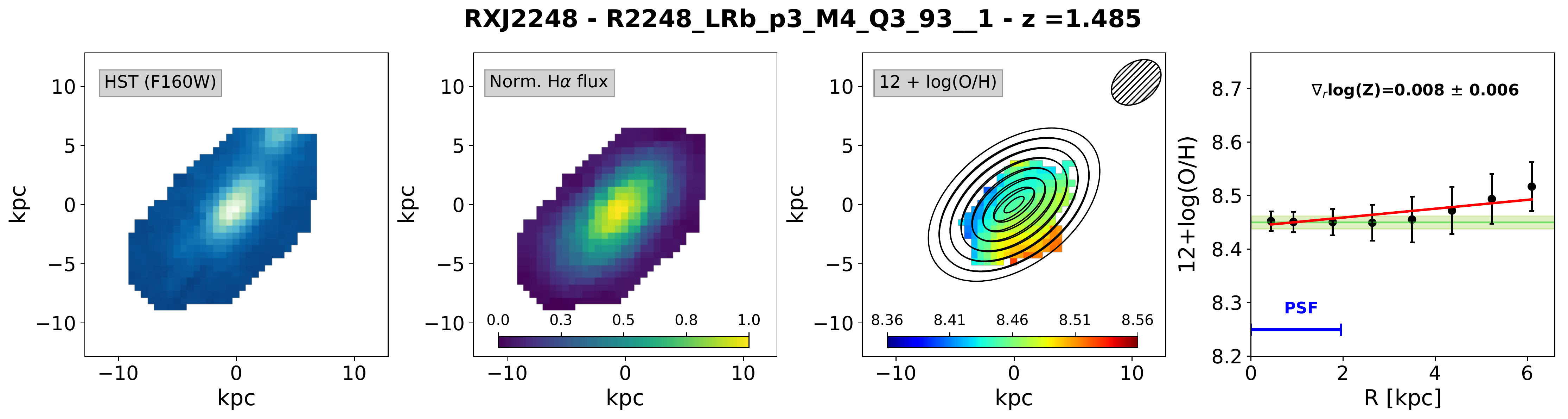}\\ 
    	\textbf{Marginally Resolved} \\
 	\includegraphics[width=0.95\textwidth]{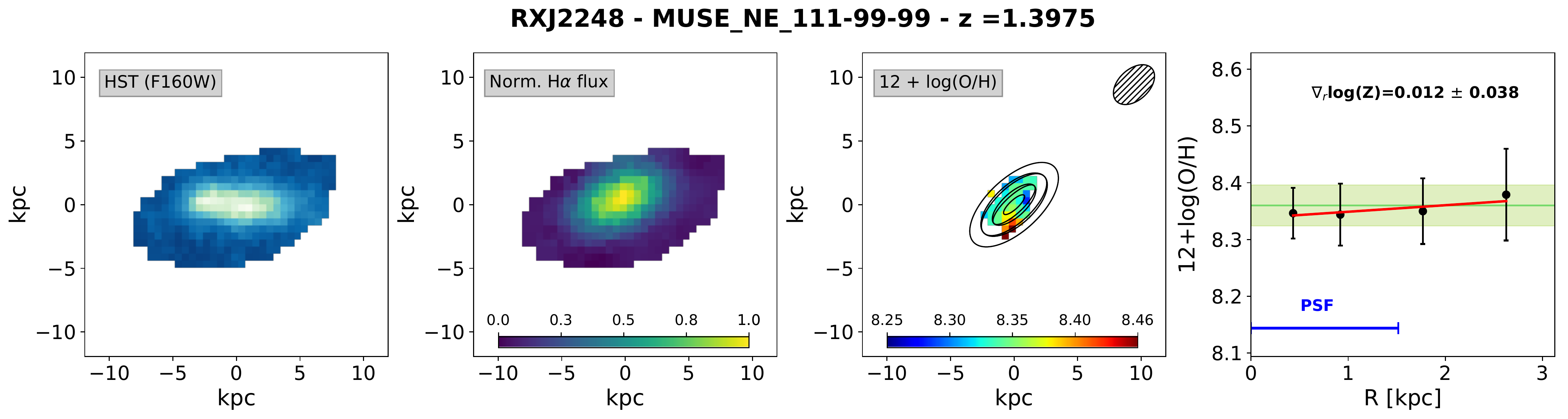}\\
	
\caption{Two examples of metallicity maps and gradients for KLEVER galaxies, illustrating a resolved (\textit{top} panel) and a marginal resolved (\textit{bottom} panel) case. We show, from left to right respectively, the source plane HST reconstructed image in HST-F160w filter, the normalized  source plane H$\upalpha$ flux maps, the full metallicity map and the extracted metallicities at increasing radii across each galaxy.
The shaded black region in the \textit{third} panel reproduce the shape and size of the PSF when mapped back into the source plane.
The elliptical apertures used to derive the radial gradient overlay the 2D metallicity map. 
In the \textit{fourth} panel, the linear fit to the metallicity gradient (in red) and the linear size of the PSF-HWHM ($\sqrt{\text{HWHM}_{x} \times \text{HWHM}_{y}}$, in blue) are also shown, while the green region encompass the uncertainty associated with the global metallicity of the galaxy (inferred from the integrated spectrum and marked in lime green by the horizontal line).
}
\label{fig:EGgrad}
\end{figure*}

%
%\begin{figure*}
%	\centering
%	\includegraphics[width=0.46\linewidth]{Figs/ell_vs_circ_grad}
%	\includegraphics[width=0.45\linewidth]{Figs/grads/circ_vs_ell_grad_fig}
%	
%	\caption{\textit{Left Panel:} Metallicity gradients computed by averaging within elliptical apertures in the source plane are plotted against those derived assuming purely circular apertures. The points are colour coded by the corresponding value of the axis ratio of the PSF in the source plane. The two estimates are consistent for more than $80\%$ of the sources, with only two with significant slope inversion.
%	\textit{Right Panels:} The difference between elliptically-averaged (top panel) and circularly-averaged (bottom panel) metallicity gradients is shown for one of the most extreme cases in our sample (i.e. the CSWA164 galaxy, $\mu \approx 13$) .
%	%observed with SINFONI ($\mu \approx 13$), where the sign of the gradient in the two configurations is inverted.
%	Here, the significant elongation of the PSF causes the information from very different radii to be combined within the same annulus, potentially biasing the determination of the gradient.
%	Indeed, the circularly-averaged gradient seems to be more representative of the metallicity distribution of individual spaxels across the map.
%}
%	\label{fig:ellvscircgrad}
%\end{figure*}

From the analysis of the source plane PSF we determine whether we are truly resolving a metallicity gradient or we are affected by poor angular resolution. 
We claim our gradients to be fully resolved if we can extract metallicities up to a radial distance at least twice the linear size of the PSF-HFHM (i.e. $\sqrt{\text{HWHM}_{x} \times \text{HWHM}_{y}}$), while we consider them marginally resolved if the gradient is evaluated between $1$ and $2$ times the linear PSF-HFHM; we finally classify them as unresolved otherwise.
We show an example of a resolved and marginally resolved metallicity gradient in Fig.~\ref{fig:EGgrad}.
From left to right we show respectively the source plane HST F160W image, the normalised source plane H$\upalpha$ map,
the metallicity map and the extracted metallicity at increasing radii, with the linear fit to the points represented by the red line.
The elliptical apertures adopted in the extraction of the radial gradient overlay with black contours the 2D metallicity map.
The shape of the source plane PSF (third panel), as well as the linear size of the PSF-HWHM (in blue, fourth panel), are reported for each source to aid visualization.

Following the procedure described above, we manage to construct reliable metallicity maps and compute resolved (or, at least, marginally resolved) metallicity gradients for a total of $28$ galaxies out of the $42$ presented in Table~\ref{tab:klever_sources}. 
We refer to this subsample as the \textit{metallicity gradients sample}.
The derived values of the metallicity gradients for these sources are reported in Table ~\ref{tab:grads}, while the associated metallicity and line ratios maps are shown in Appendix B. %~\ref{sect:appenB}.
Overall, we report $6$ negative gradients and $22$ positive gradients respectively, with gradients resolved for $25$ out of $28$ sources while only ``marginally resolved'' for the remaining $3$.
The $86\%$ of the galaxies present gradients shallower than $0.05 \rm dex/kpc$ (and $72\%$ shallower than $0.025 \rm dex/kpc$).
Moreover, we note that $25$ out of $28$ gradients (i.e. $89\%$ of the sample) are totally consistent with a flat slope within their $3 \sigma$ uncertainty ($67\%$ within $1\sigma$), with only three galaxies significantly diverging from zero in showing inverted gradients.

\begin{figure*}
	\centering
%	\centerline{\hbox to \textwidth{\hspace{5.8cm}(i) Global\hfill}}
	\includegraphics[width=0.95\textwidth]{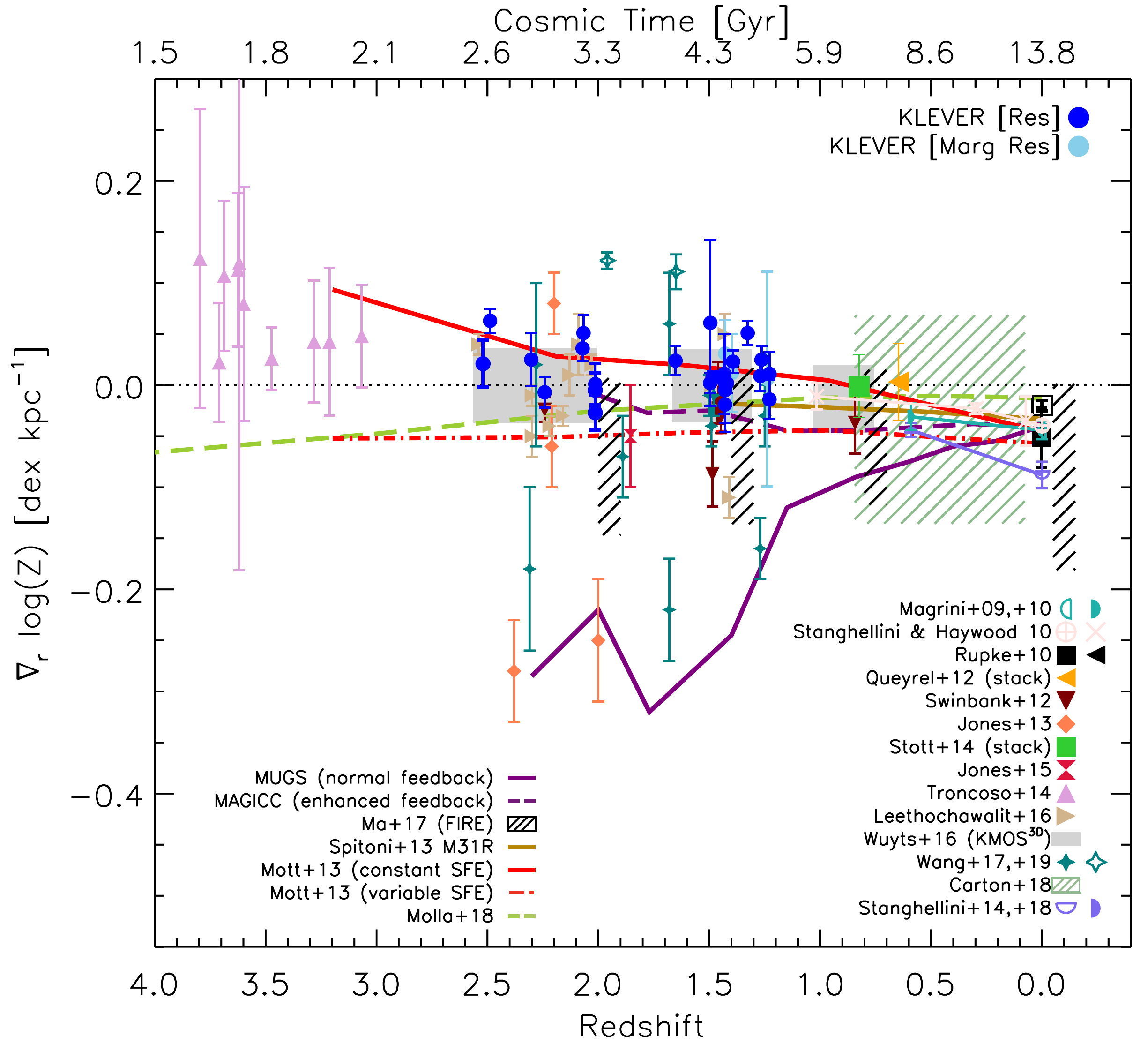}

\caption{
Compilation of metallicity gradients plotted as a function of redshift (or equivalently the age of the Universe). 
Blue points mark the results derived in this work, where we distinguish between resolved (blue circles) or marginally resolved (light blue circles) gradients. 
Additional measurements of metallicity gradients from previous works, derived with different techniques on both local and high-$z$ galaxies, are included with different symbols/colors as explained in the legend. 
%\citep[][see text for more details]{Magrini:2009aa,Magrini:2010aa,Rupke:2010aa,Swinbank:2012aa,Queyrel:2012aa,Jones:2013aa,Stanghellini:2014aa,Stott:2014aa,Troncoso:2014aa,Jones:2015ab,Leethochawalit:2016aa,Wuyts:2016aa, Wang:2017aa,Wang:2018aa,Carton:2018aa,Stanghellini:2018aa}.
The metallicity gradients by \citet{Carton:2018aa} are based on MUSE observations of $0.1<z<0.8$ galaxies, for which we report the spread in their measurements both in redshift and $\nabla Z$.
The sample from \citet{Queyrel:2012aa} reports SINFONI observations of z$\sim 1.2$ galaxies from the MASSIV survey, while \citet{Stott:2014aa} reports KMOS observations of z$\sim 0.8$ galaxies from the HiZELS survey.
For these works, we report the average and $1\sigma$ dispersion for their metallicity gradients measurements at the mean redshift of each sample. 
The grey shaded areas, instead, represent the spread in the metallicity gradients measured in the framework of the KMOS$^{3\text{D}}$ Survey within three different redshift intervals and presented by \citet{Wuyts:2016aa}.
\citet{Jones:2013aa} and \citet{Leethochawalit:2016aa} report observations conducted with AO-assisted spectroscopy with OSIRIS at Keck on gravitationally lensed systems, while \citet{Jones:2015ab}, \citet{Wang:2017aa} and \citet{Wang:2019aa} are based on HST-grism slitless spetroscopy of lensed galaxies behind the Frontier Fields clusters as part of the GLASS program (\citealt{Treu:2015aa}).
The metallicity gradients a z$>3$ from \citet{Troncoso:2014aa} are based on SINFONI observations in the framework of the AMAZE and LSD programs (\citealt{Maiolino:2008aa,Mannucci:2009aa}).
For what concern the metallicity gradients in the local Universe, we show the average of local gradients for isolated (filled square) and interacting (open square) systems from \citet{Rupke:2010aa}.
We also report the time evolution of metallicity gradients for the Milky Way, M33 and M81, 
as inferred from abundances computed for \Hii regions (open half-circles) and Planetary Nebulae (PNe, filled half-circles, associated to older stellar population progenitors) and presented in \citet{Magrini:2009aa,Magrini:2010aa,Stanghellini:2010ab,Stanghellini:2014aa,Stanghellini:2018aa}.
The predictions from MUGS and MaGICC cosmological simulations reported in \citet{Gibson:2013aa}, which tracks the evolution of a typical disc galaxy implementing two different feedback modes, are shown with a dashed purple line for the enhanced feedback scenario and with a straight purple line for the normal feedback scenario respectively. 
The black hatched regions cover instead the predictions of metallicity gradients at four different epochs (i.e. z$=0,0.8,1.5,2$) from the FIRE simulations, as presented by \citet{Ma:2017aa}.
The brown curve shows the predicted cosmic evolution of the metallicity gradient from the \citet{Spitoni:2013aa} modelling of M31, while in light-green those from the chemical evolution models by \citet{Molla:2019aa}.
Finally, the predictions of chemical evolution models by \citet{Mott:2013aa} assuming constant (solid line) and variable (dot-dashed line) star formation efficiencies are plotted in red.
Our results are consistent with little or no evolution of the metallicity gradients with redshift, in better agreement with predictions by simulations and models including strong feedback mechanisms and/or gas radial flows in place.
}
	\label{fig:redshift}
\end{figure*}

\subsection{Cosmic evolution of metallicity gradients}
\label{sect:redshift}

We discuss here the cosmic evolution of metallicity gradients, including the sample derived in this work in the context of the current observational framework and comparing the trends with those predicted by chemical evolution models and simulations.
In Fig.~\ref{fig:redshift} we plot the \textit{metallicity gradients sample} as a function of redshift, together with a collection of radial abundance gradients (in units of dex/kpc) from previous works, both at high redshift \citep[namely][]{Swinbank:2012aa,Queyrel:2012aa,Jones:2013aa,Jones:2015ab,Stott:2014aa,Troncoso:2014aa,Leethochawalit:2016aa,Wuyts:2016aa,Wang:2017aa,Wang:2019aa, Carton:2018aa} and based on local galaxies \citep[namely][]{Rupke:2010aa,Magrini:2009aa,Magrini:2010aa,Stanghellini:2010ab,Stanghellini:2014aa,Stanghellini:2018aa}.
A brief description of the samples involved and the observational techniques adopted in the determination of the metallicity gradients collected from the literature is given in the caption of Fig.~\ref{fig:redshift}.
The gradients derived in this work for the KLEVER galaxies are shown with dark blue filled circles (for resolved gradients) and light blue circles (for marginally resolved gradients, see Sect.~\ref{sect:grads}).
The results presented in this paper are generally consistent with previous investigations of metallicity gradients at high redshift: 
despite the variety of observational set up, resolution, sample selection, metallicity calibration etc.. the bulk of observed star forming galaxies does not exhibit, on average, strong radial trends (i.e. above $0.05 \rm dex/kpc$) in the metal distributions between $1<$z$<2.5$, with a large part of them characterised by gradients consistent with being flat within their $1\sigma$ uncertainty. 

We can compare these results with the predictions from cosmological hydrodynamical simulations which trace the gas phase metallicity of disc galaxies from $z\sim2$ to the present day under different feedback modes in the framework of the MUGS \citep{Stinson:2010aa} and MaGICC \citep{Gibson:2013ab} schemes.
We show the evolutionary track for the simulated galaxy g15784 under both \textit{normal} and \textit{enhanced} feedback conditions. The ``normal" feedback scenario, in which $10$ to $40 \%$  of the energy from each supernova is used to heat the surrounding ISM, predicts steeper metallicity gradients at earlier redshifts (i.e. $\sim -0.2 \text{dex/kpc}$) which become flatter over cosmic time.
The totality of our measurements and even the largest part of those reported in the literature are inconsistent with this scenario,
with just a few example of observed gradients that match its predictions at redhifts higher than $1.5$ \citep{Jones:2013aa, Wang:2017aa}.
The enhanced feedback scheme instead, where outflows re-distribute energy and re-cycle ISM material over larger galactic scales, with preferential re-accretion into the outer regions (i.e. metal re-cycling), is characterised by a milder evolution with a progressive 
(minimal) flattening of the metallicity gradients with increasing redshift. 
Our results, together with the majority of previous indications from the literature, seem to point towards this latter regime, which suggests that at high-$z$, stronger star formation feedback is in place causing flatter gradients due to rapid and efficient recycling of enriched gas and re-distribution of the ISM, which occurs on timescales shorter that those related to star formation and metal enrichment.
Within the context of the high resolution FIRE simulations, \cite{Ma:2017aa} presented gas-phase metallicity maps and measured radial gradients within $0.25-1$R$_{90}$\footnote{the radius enclosing $90\%$  of the star formation within $10$ kpc} for a sample of simulated galaxies up to z$\sim2$.
They are capable of predicting a variety of metallicity distributions and gradients at each cosmic epoch (represented by the black hatched regions in Fig.~\ref{fig:redshift}), spanning from strongly negative ones in ordered discs to flatter gradients in more perturbed systems affected by mergers and significant gas flows, which mix the ISM on large galactic scales ($\lesssim 10$kpc) on relatively short timescales ($\sim 10-50$Myr), nicely reproducing the scatter in the observational results.

In Fig~\ref{fig:redshift} we also show the prediction for the metallicity gradient of M31 at different redshifts according to the \cite{Spitoni:2013aa} chemical evolution model (solid brown line) which assumes no gas threshold on star formation, an inside-out formation of the disk, constant star formation efficiency along the disk and radial gas flows; this model is consistent with enhanced-feedback simulation schemes in prescribing a mild steepening of the gradient with time. 
%and reproduces the trend followed by the majority of the currently available observations.
Interestingly, no or little evolution in the predicted metallicity gradients can be achieved also by purely- or semi-analytical chemical evolution models which do not include any prescription about feedback processes mixing up the gas. 
The models from \cite{Molla:2005aa} (then revisited by \citealt{Molla:2017aa,Molla:2019aa}, green dashed line), for example, assuming an inside-out disc growth, an ISM structured as a multiphase mix of diffuse gas and cold clouds and prescribing star formation efficiencies, stellar yields, gas infall rates and atomic to molecular hydrogen conversion factors, predict flat metallicity gradients at all redshifts within $2.5$ effective radii in disc galaxies, with no assumption on radial gas mixing.
%Tidal and gravitational interactions can also strongly contribute to the flattening of the observed gradients.

Among our galaxies, we  report several examples of positive, 'inverted' metallicity gradients, despite only three determined  
with $>3\sigma$ significance (i.e. sp6, ph3912 and ph6532, with R2248\_LRb\_p3\_M4\_Q3\_94\_\_1 at $2.8\sigma$).
Previous claims of such gradients have been reported both at the same redshifts \citep{Wuyts:2016aa, Wang:2019aa} and at earlier epochs (z $> 3$), mostly in seeing-limited data of unlensed galaxies \citep{Cresci:2010aa,Troncoso:2014aa}; similarly to these findings, the central metal poor regions in our galaxies are co-spatial with the regions of highest star formation rate density (and hence higher gas fractions).
These observations have been interpreted as evidence of cold flows of pristine gas into the central regions of such galaxies, which both fuels new star formation episodes and dilutes the metallicity \citep{Dekel:2009aa, Cresci:2010aa}. 
An alternative explanation consist in considering these galaxies affected by strong feedback processes, in the form of outflows that rain material from the inner regions back down in the outskirts of the galaxy (the so called `galactic fountain' effect, \citealt{Fraternali:2008aa,Werk:2011aa}). 
This process, whose primary effect results into an average flattening of the abundance gradients, can contribute in some cases to invert the gradient itself, as the highest mass loss rate is observed in the central regions \citep{Wang:2019aa}.
However, the impact of these outflows does not seem to be powerful enough to represent the dominant contribution to such a flattening for the ensemble population, since the measured mass outflow rates are often too small \citep{Cresci:2010aa, Troncoso:2014aa}.
In the framework of chemical evolution models, \cite{Mott:2013aa} managed to reproduce inverted gradients at z$>1$ assuming the inside-out formation of the disc, a threshold in the gas density for star formation, a variable star formation efficiency along the disc and the presence of radial flows with varying speed. In particular, they find the velocity pattern for radial gas flows to be a crucial parameter in shaping the chemical evolution of the disc and that a constant star formation efficiency along the disc is needed to reproduce the observed inverse gradients (solid red line in Fig.~\ref{fig:redshift}).

Finally, It is worth noting here that the large majority of currently available results rely on a variety of different indicators for metallicity determination: many of them are derived only from nitrogen based diagnostics ([\ion{N}{ii}]/H$\upalpha$ in particular, being one of the most easily accessible at high-z), some from purely oxygen based ones.
Moreover, beside the fact that the metallicity gradients shown in Fig.~\ref{fig:redshift} are obtained with different diagnostics, the latter are sometimes even based on different calibration methods (e.g. \Te\ versus photoionisation modelling).
This means not only that the absolute metallicity estimates could disagree between various methods, but also that the dynamic range of possible values spanned by the calibration itself could be different, which might translate in a change of the slope of the calibrated relation. 
In general, theoretical calibrations provide a wider range of possible metallicities with respect to \Te -based methods, even when considering the same set of strong-line ratios, as they often provide higher normalisations in the high-metallicity regime.
Therefore, the use of different diagnostics/calibrations could introduce systematics in the 
slope of the inferred gradients.
However, given the generally small range of abundances spanned by individual regions across a galaxy (very strong gradients are rare), and considering that relative metallicity measurements are always much more robust than absolute ones, this effect is likely to be negligible in the majority of cases, making the comparison of samples derived with different techniques more fair. 

\subsection{Correlation with stellar mass and SFR}
\label{c:ssfr}
%

%GRAD vs sSFR : $\alpha$ = 0.013379 +- 0.004147
%GRAD vs Delta SFMS : $\alpha$ = 0.011650 +- 0.005610
%GRAD vs Mass : $\alpha$ = -0.013415 +- 0.004958

\begin{figure}
	\centering
	\includegraphics[width=0.97\linewidth]{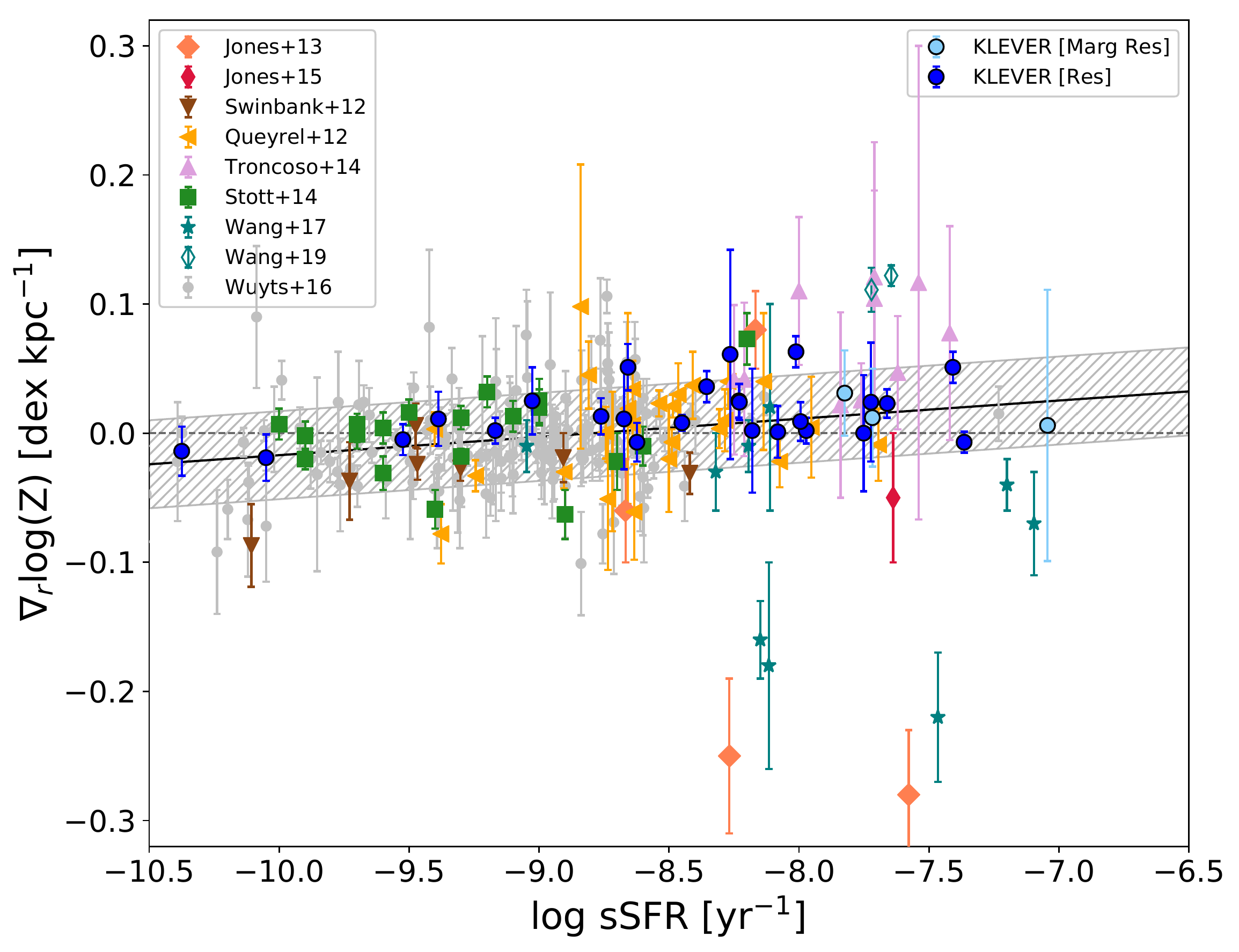}
	\includegraphics[width=0.97\linewidth]{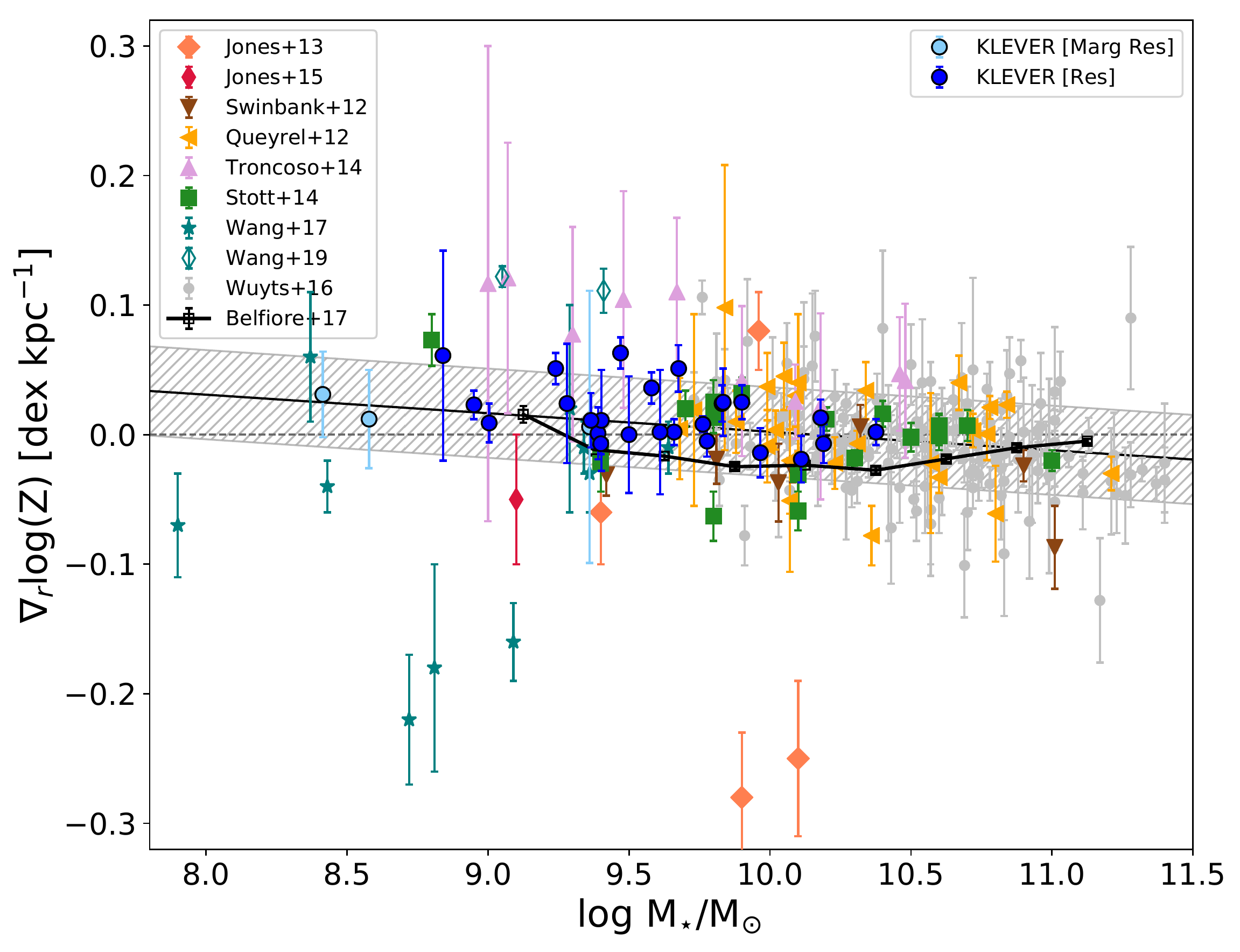}

	\caption{Metallicity gradients as a function of global specific star formation rate (sSFR, \textit{Upper Panel}) and stellar mass (\textit{Bottom Panel}).
	The sample includes both objects from KLEVER and additional gradients from previous works on z$\gtrsim 0.8$ sources; symbols and colors are the same as in Fig.~\ref{fig:redshift}.	 
	A robust linear regression fit to the data is shown in black in both panels, with the RMS region marked in grey. 
	A mild correlation is present between metallicity gradients and both quantities, in the sense of a positive trend with increasing sSFR and a negative trend with increasing M$_{\star}$, significant at $3.2 \sigma$ and $2.7 \sigma$ and with a Spearman correlation coefficient $\rho=0.29$ and $\rho=-0.25$ respectively.
	In the bottom panel, the trend between gradients and stellar mass in the local Universe (inferred from a sample of galaxies in MANGA, \citet{Belfiore:2017aa}) is traced by the black line.
}
\label{fig:ssfr}
\end{figure}

\begin{figure}
	\centering
	\includegraphics[width=0.47\linewidth]{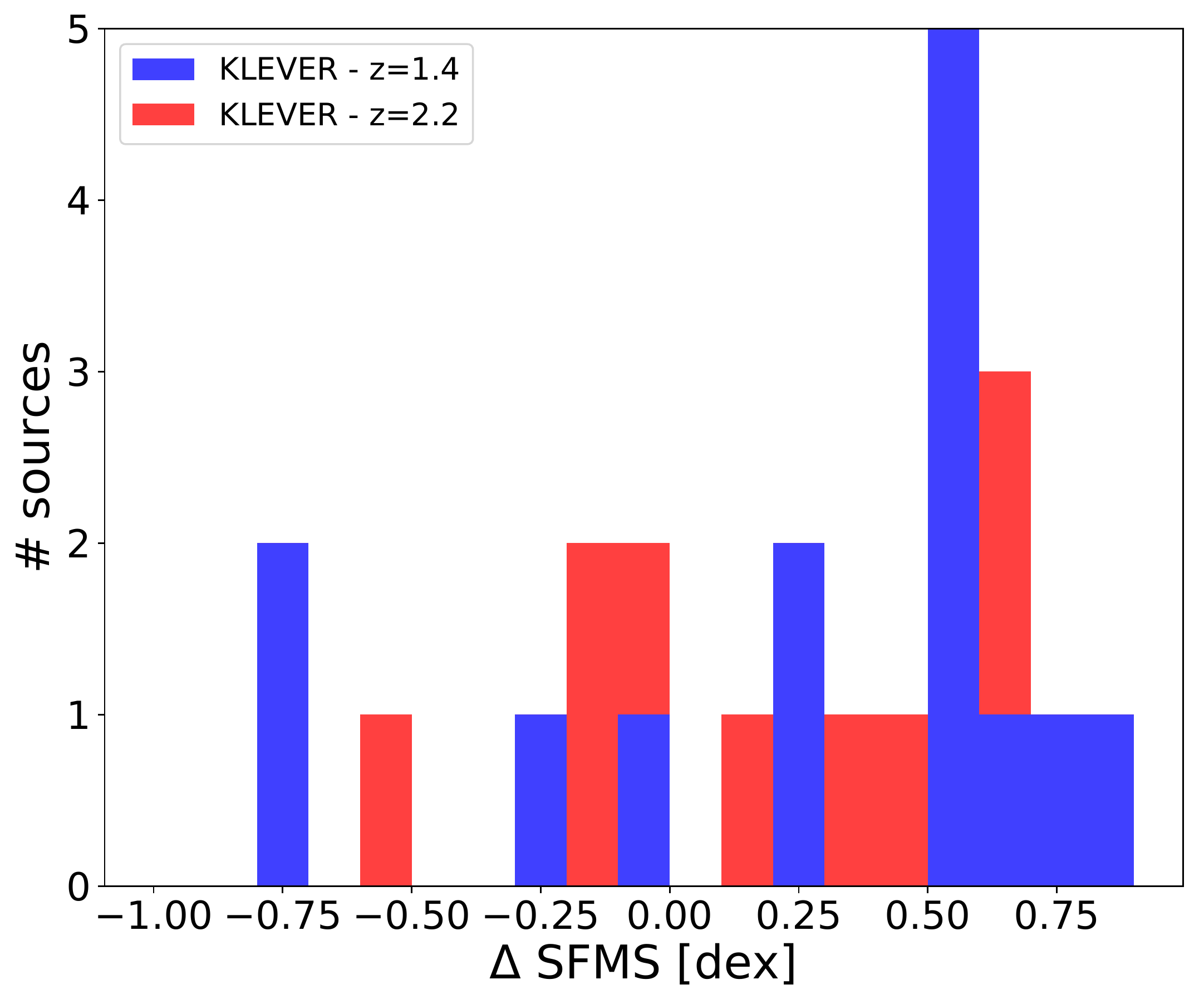}
	\includegraphics[width=0.47\linewidth]{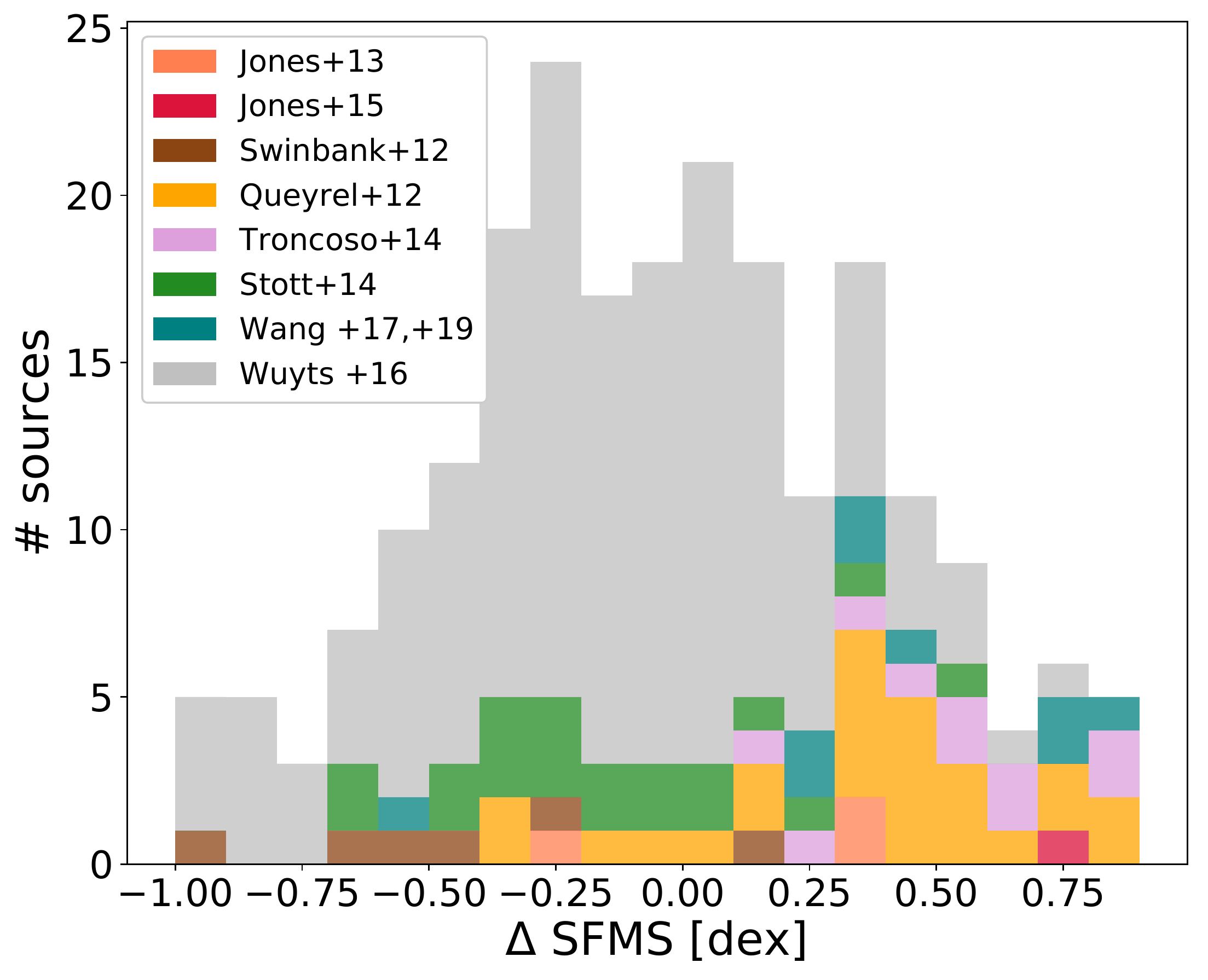} \\
	\includegraphics[width=0.99\linewidth]{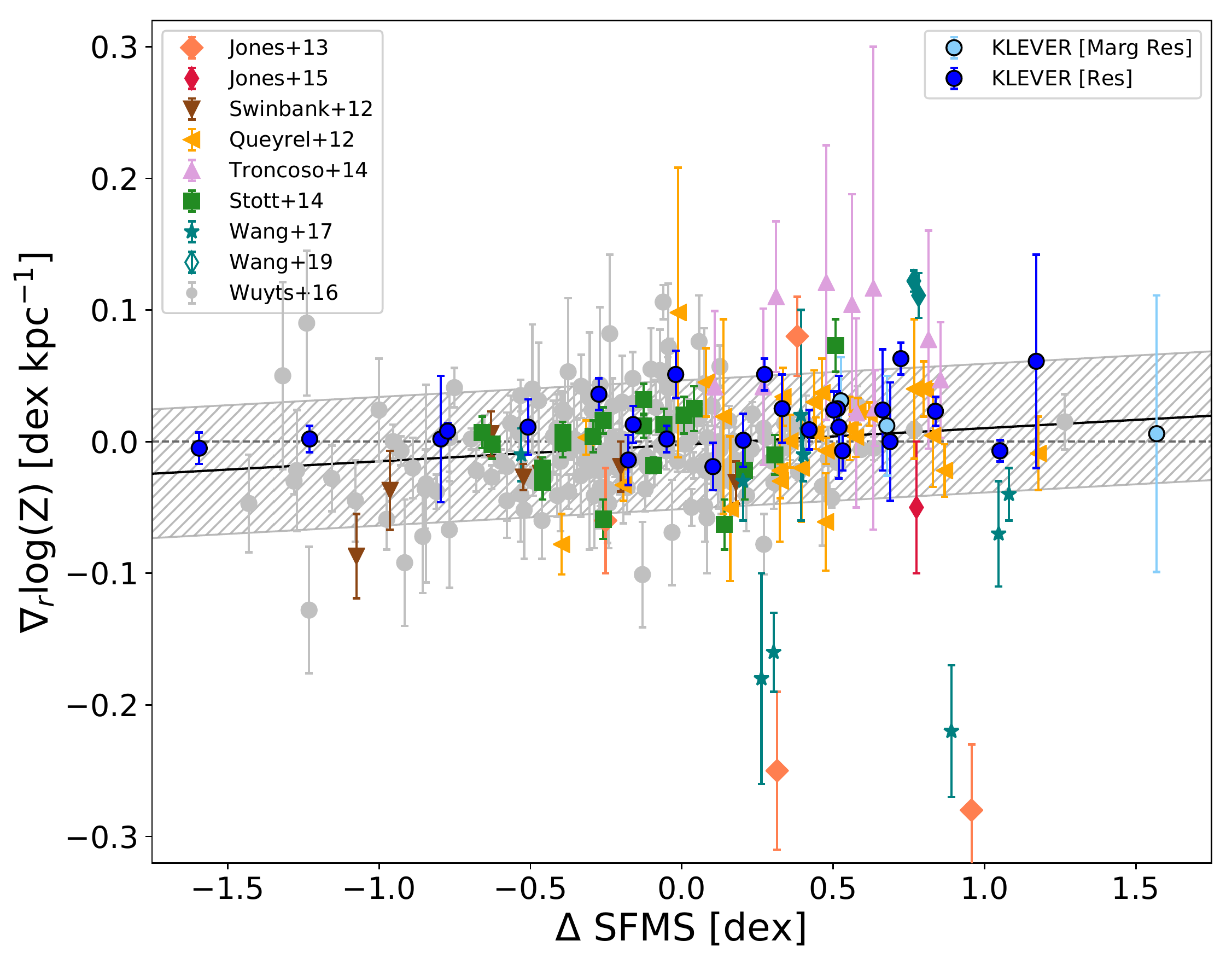}
	
	\caption{\textit{Upper Panels:} Stacked histograms of the distribution in $\rm \Delta$ SFMS, defined as the distance from the star-forming main sequence of \citet{Speagle:2014aa} at the mean redshift of each reference sample, for the KLEVER galaxies hereby analysed (left panel) and the compiled sample from the literature at z$\gtrsim 0.8$ (right panel).
	\textit{Bottom Panel:} Metallicity gradients as a function of $\rm \Delta$ SFMS. Compared to the trends shown in Fig.~\ref{fig:ssfr}, the significance is reduced to $\sim2\sigma$, with Spearman coefficient $\rho=0.20$.
}
	\label{fig:deltasfmshist}
\end{figure}

%Despite the lack of a clear and unambiguous trend in the behaviour of metallicity gradients with redshift, the sample is clearly affected by scatter in its distribution.
We discuss here whether the metallicity gradients correlate with global galaxy properties.
As previously mentioned, simulations of the inside-out growth of galaxies predicts initial steep negative gradients which flatten at later times, whereas observed positive or flat gradients are predicted by simulations and models as due to efficient re-accretion and gas mixing processes \citep[e.g.][]{Gibson:2013ab,Mott:2013aa}. 
In particular, the specific star formation rate (sSFR=SFR/\mstar) is a measured quantity that holds information about galaxy star formation history: galaxies exhibiting high sSFR are experiencing a starbursting phase and might be in an early stage of their cosmic evolution, possibly fuelled by inflowing gas into their central regions and affected by strong feedback episodes.
One can therefore expect that signatures of this process might also be observed in the metallicity gradients, in the sense that galaxies with higher sSFR exhibit on average flatter (or more positive) gradients than those characterised by low sSFR.

In the upper panel of Fig.~\ref{fig:ssfr} we plot the \textit{metallicity gradients sample} as a function of sSFR, along with gradients from previous works at $z\gtrsim 0.8$ \citep[i.e.][]{Swinbank:2012aa,Queyrel:2012aa,Jones:2013aa,Troncoso:2014aa,Stott:2014aa, Jones:2015ab,Wuyts:2016aa,Wang:2017aa,Wang:2019aa}).
We exploited the \textit{sklearn} python package to perform a robust linear regression and quantify the level of correlation: the result is shown with the black line in Fig.~\ref{fig:ssfr}, with the shaded region indicating the r.m.s. of the fit.
The slope of the relation is equal to $0.013 \pm 0.004$ (i.e. $3.2 \sigma$ from being flat), with an r.m.s. of $0.05$\rm{dex}, suggesting that a correlation between metallicity gradients and sSFR might be present in the sense of most active galaxies (i.e. higher sSFR) being characterized by flatter or positive gradients. 
This is consistent with the trend originally found by \cite{Stott:2014aa} and the predictions of the FIRE simulations in \cite{Ma:2017aa}.
Nevertheless, the level of correlation is very mild, as a Spearman test gives only $\rho=0.29$ with a $99.9\%$ level of confidence.

We investigate also whether our metallicity gradients correlate with total stellar mass in the bottom panel of Fig.~\ref{fig:ssfr}.
%0.013415 +- 0.004958
We perform the same linear regression and find a slightly negative slope ($-0.013 \pm 0.005$), significant at the $2.6 \sigma$ level. A Spearman correlation test gives $\rho=-0.25$ with the $99.9\%$ of confidence.
Again, this is consistent with previous observational assessments \citep{Stott:2014aa,Wuyts:2016aa} and simulations \citep{Ma:2017aa}.
In the local Universe, a clear trend of metallicity gradients with stellar mass is observed, as recently shown for example by \cite{Belfiore:2017aa} for a large sample of galaxies within the MANGA survey: less massive systems are characterized by flatter metallicity gradients, steepening (i.e. becoming negative) with increasing stellar mass up to $10^{10.5}$ M$_{\odot}$, before flattening again at higher masses (especially in their central regions). 
In the framework of the classical inside-out growth scenario, and assuming a radially decreasing star formation efficiency, feedback and/or gas recycling mechanisms are required to reproduce the observed flattening of the gradients at low masses \citep[see also][]{Belfiore:2019aa}.

{As also seen in Sect.~\ref{sec:mass_sfr}, the sample analysed in the present paper is distributed across a wide region of the \mstar-SFR plane, although being preferentially constituted by higher sSFR galaxies than representative samples at the corresponding redshifts, especially at $\text{z}>2$.
Fig.~\ref{fig:deltasfmshist} shows how the \textit{metallicity gradient sample} analysed in this paper and the compiled sample from the literature are distributed as a function of $\rm \Delta$ SFMS, defined as the distance from the star forming main sequence computed at each source redshift, adopting the redshift-dependent parametrisation of \cite{Speagle:2014aa}.
In the bottom panel of Fig.~\ref{fig:deltasfmshist} we investigate the correlation between metallicity gradients and $\rm \Delta$ SFMS. 
When normalised to the main sequence at each redshift, the correlation is found to be weaker compared to the trends previously discussed, with $\rho=0.20$ and a slope significantly different from being flat only at the $2\sigma$ level.

In light of the trends shown and discussed above, we note that it might be therefore more likely for KLEVER galaxies to show flatter gradients compared to the rest of the complied sample, although flat gradients are found at all \mstar and SFR within our sample.
Nonetheless, the full high-z sample shown in Fig.~\ref{fig:redshift} is almost uniformly distributed in 
$\rm \Delta$ SFMS, hence the overall bias introduced in the interpretation of the cosmic evolution of the gradients is supposed to be minimal.
The only exception is the sample of z$>3$ galaxies from \cite{Troncoso:2014aa}, which appears to be constituted by galaxies lying systematically above the SFMS at that redshift. Therefore, it is possible that the incidence of inverted gradients in the population, as inferred from that particular sample, is overestimated, modulo the uncertainties associated with the parametrisation of the SFMS at high redshift which should always taken into account when performing this type of analysis.

Interestingly, \cite{Wang:2017aa} report a few examples of steep, negative gradients ($< -0.1$ {\rm dex}/kpc) at low masses (i.e. $\text{M}_{\star}<10^{9}\text{M}_{\odot}$) from their HST-grism spectroscopic survey of lensed galaxies, which deviate both from the average trend followed by the rest of the high-z sample and from the picture outlined in the local Universe.
%These represent the only claim to date of so negative gradients measured for such low mass objects and, 
According to the authors, they could be interpreted as systems with a rapid and highly centrally concentrated star formation history and limited feedback mechanisms and are, to date, the only reported systems that fit the ``purely inside-out growth' scenario of mass assembly.
However, a larger number of robust measurements of metallicity gradients in low mass galaxies is required in order to carefully estimate the occurrence rate of such systems and whether they might represent a different galaxy population.
In this sense, the full KLEVER sample will strongly contribute to increase the statistics in the low-mass regime.

\subsection{Are radial gradients always meaningful ?}
A first look to the derived metallicity gradients reveals how the vast majority of the sample is characterised by radial gradients consistent with being flat within their uncertainties.
However, we note that some of the metallicity maps are characterised by the presence of non-axisymmetric patterns, revealing a more complex distribution in the level of chemical enrichment.
This means that a formally flat radial metallicity gradient may be either the consequence, on one hand, of a fairly homogeneous metallicity map, resulting from the mixing of the ISM on large scales or from a flat, extended star formation profile, while on the other hand can hide the presence of non-radial variations (even on $\sim$ kpc scales) which are washed out (producing a flat gradient) when azimuthally averaged.
On average, the maximum metallicity variation between different regions in most of our galaxies is around $\sim 0.1-0.2 $ {\rm dex}; however, when this regions are not symmetrically distributed around the  galaxy centre, they might be hidden by the azimuthal averages and the resulting gradient would result artificially flatter. 
Interestingly, some of those galaxies showing irregularities in their emission line maps and/or velocity fields, present clear signatures of disturbed morphology in the HST images, which could reveal ongoing interactions or suggest clumpy structures in their disks.
A few examples are discussed in the following and shown in Fig.~\ref{fig:disturbed}, where both image plane and source plane HST images, velocity fields and metallicity maps are reported for four different systems.

\label{sec:mergers}
\begin{figure*}
	\includegraphics[width=0.195\textwidth]{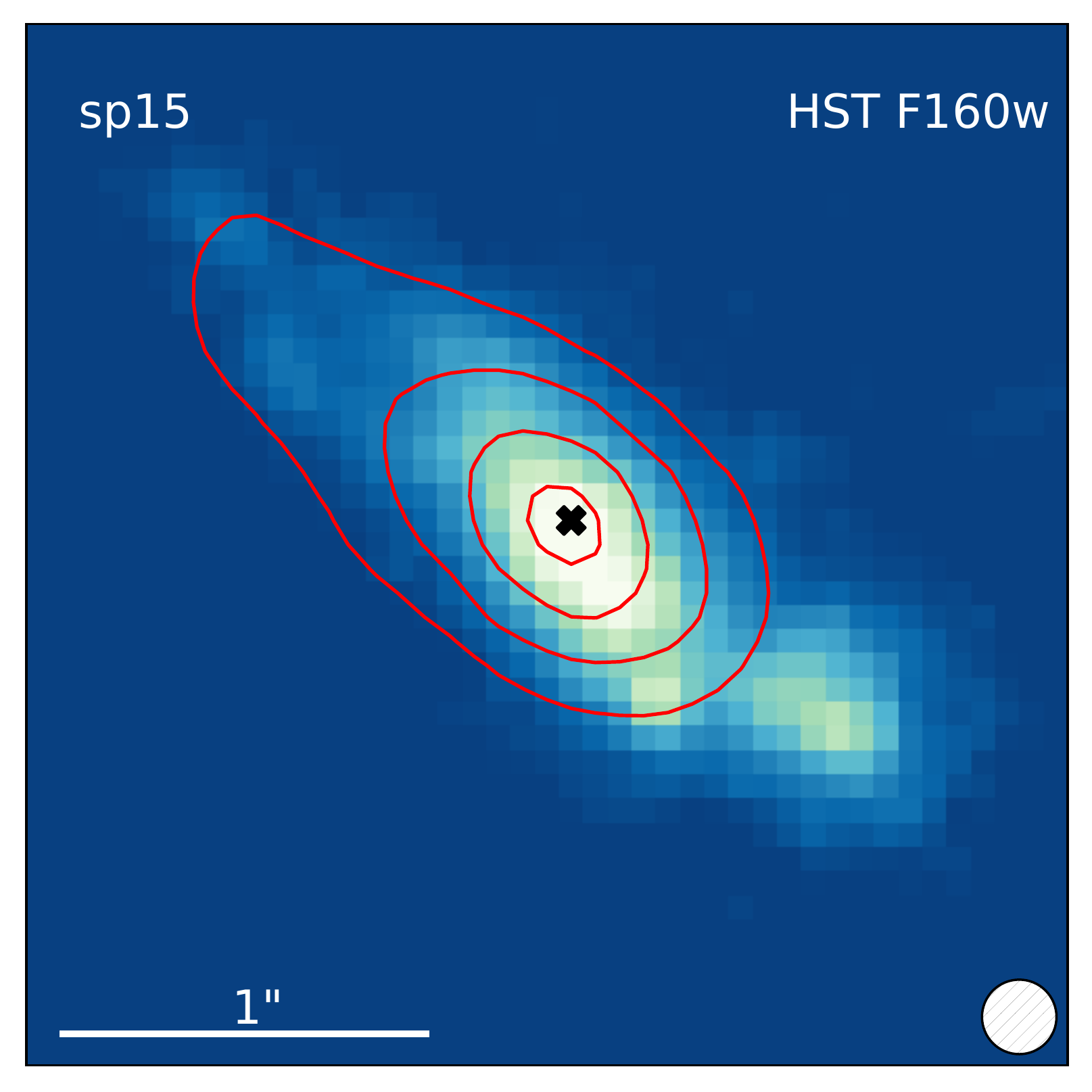}
	\includegraphics[width=0.195\textwidth]{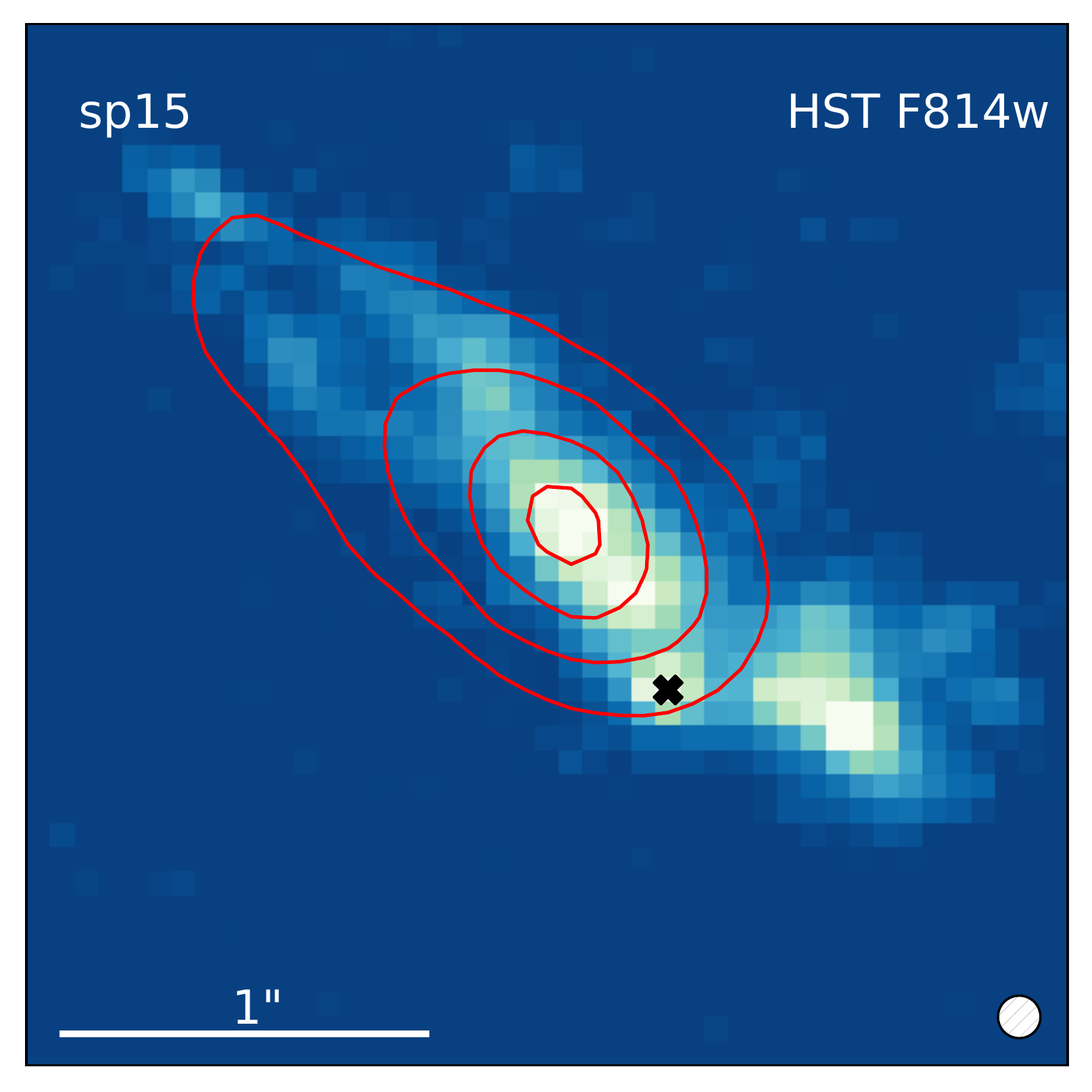}
	\includegraphics[width=0.195\textwidth]{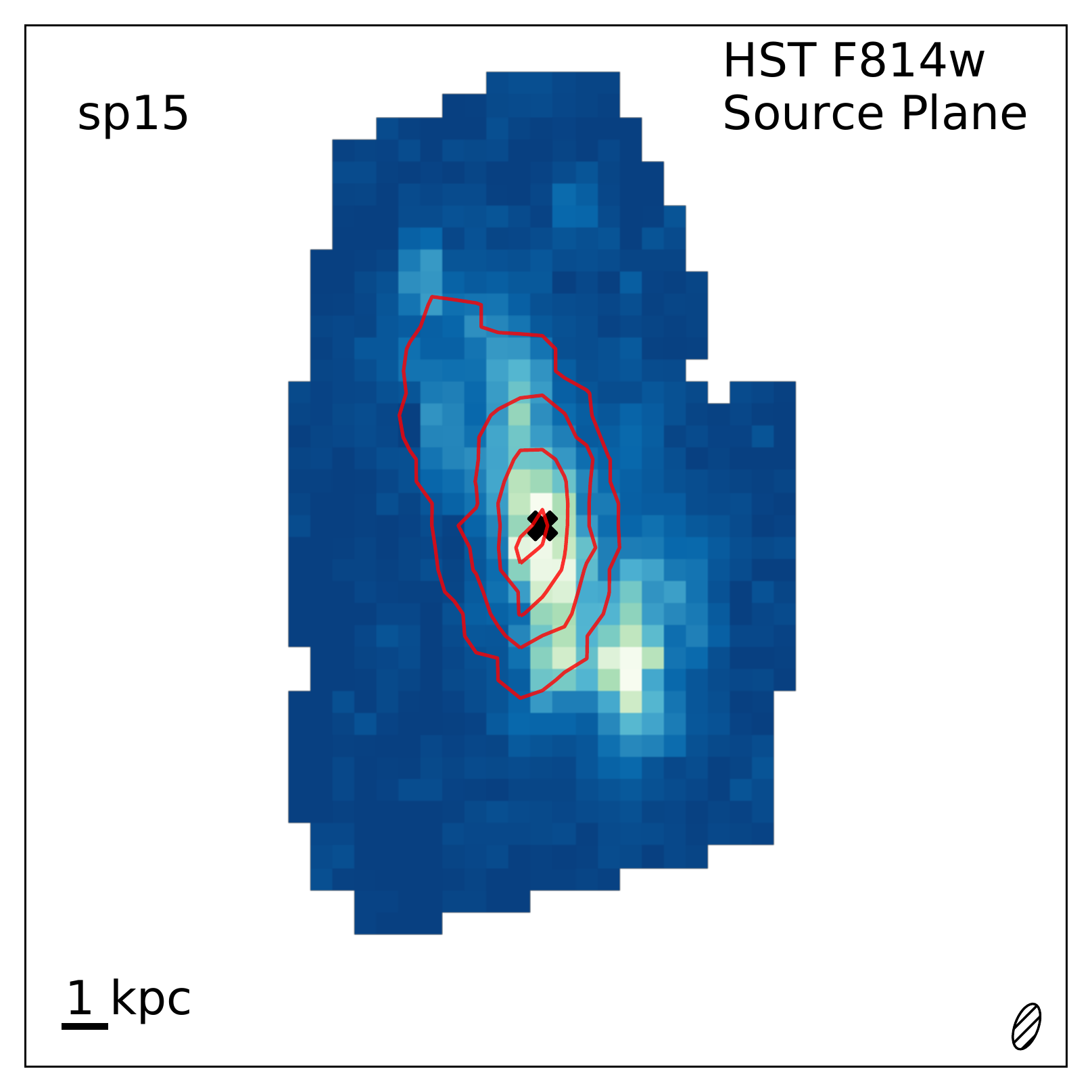}
	\includegraphics[width=0.195\textwidth]{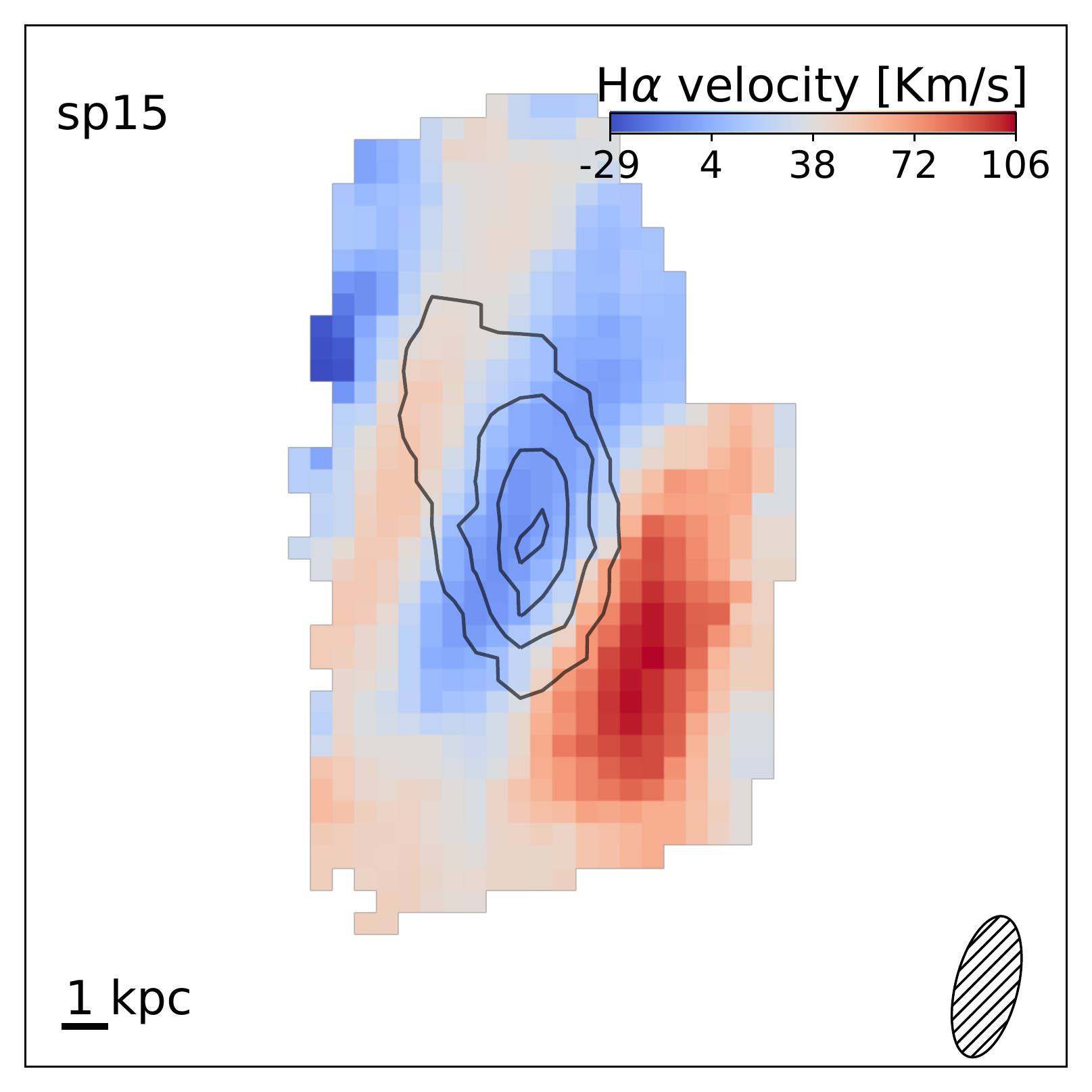}
	\includegraphics[width=0.195\textwidth]{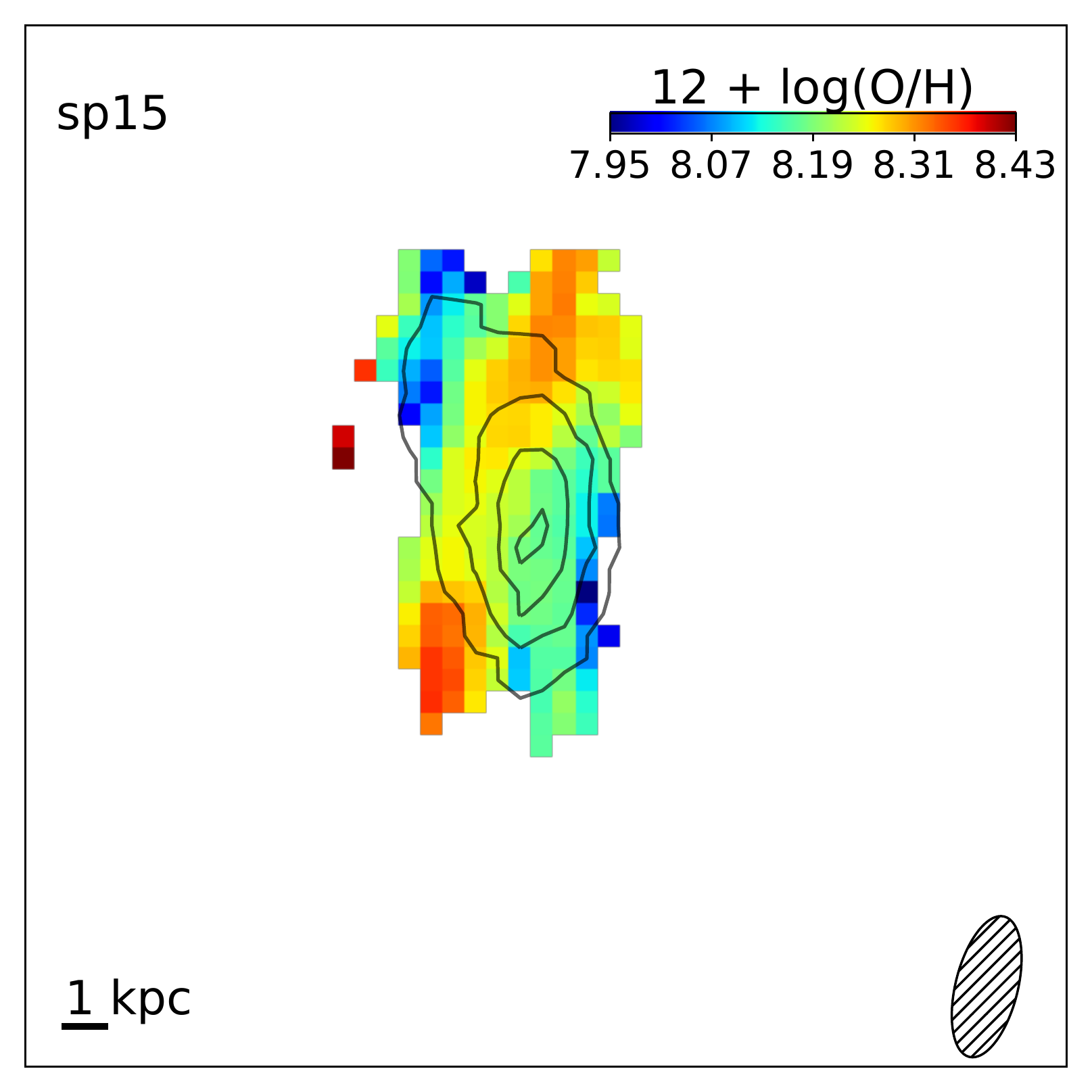}\\
	
	\includegraphics[width=0.195\textwidth]{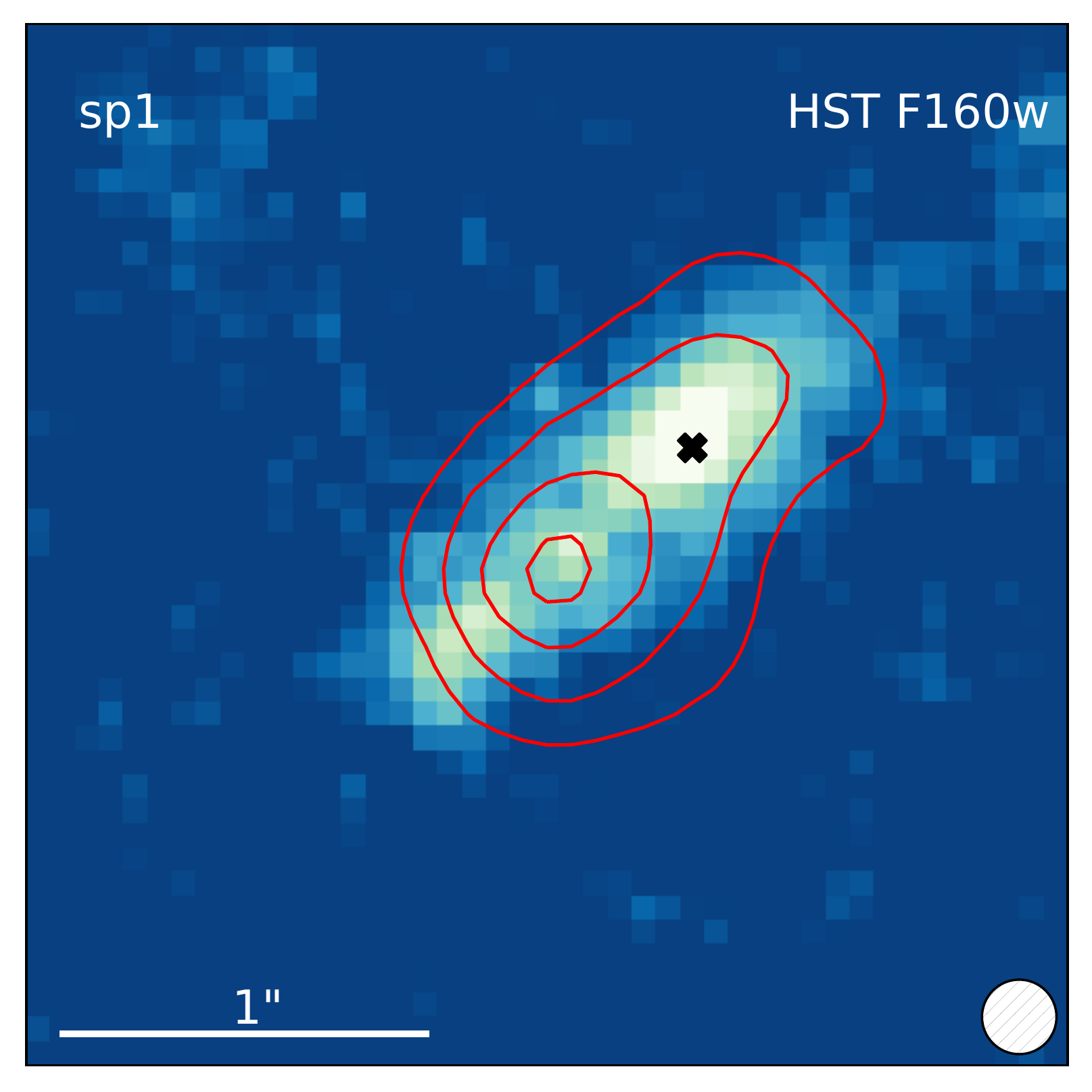}
	\includegraphics[width=0.195\textwidth]{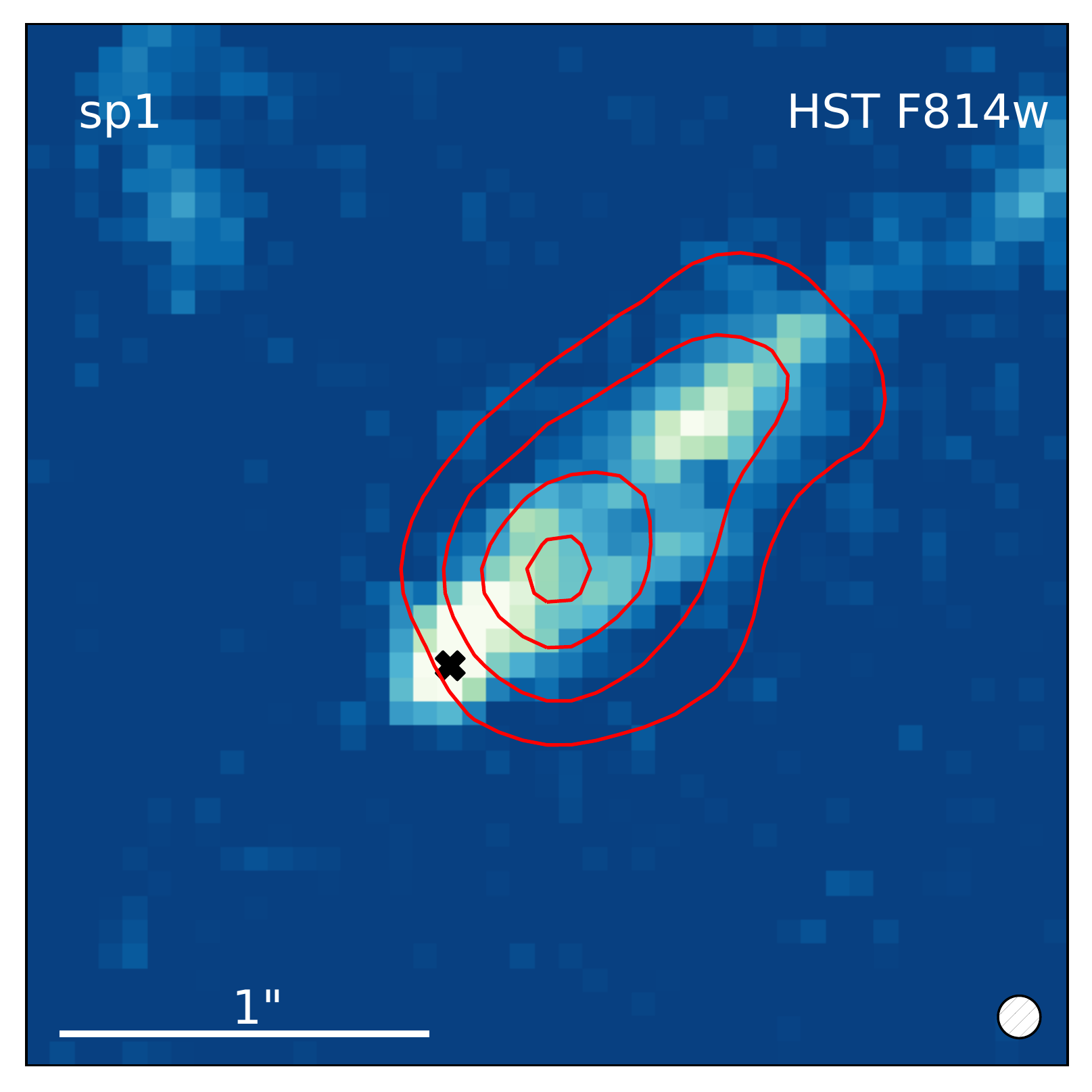}
	\includegraphics[width=0.195\textwidth]{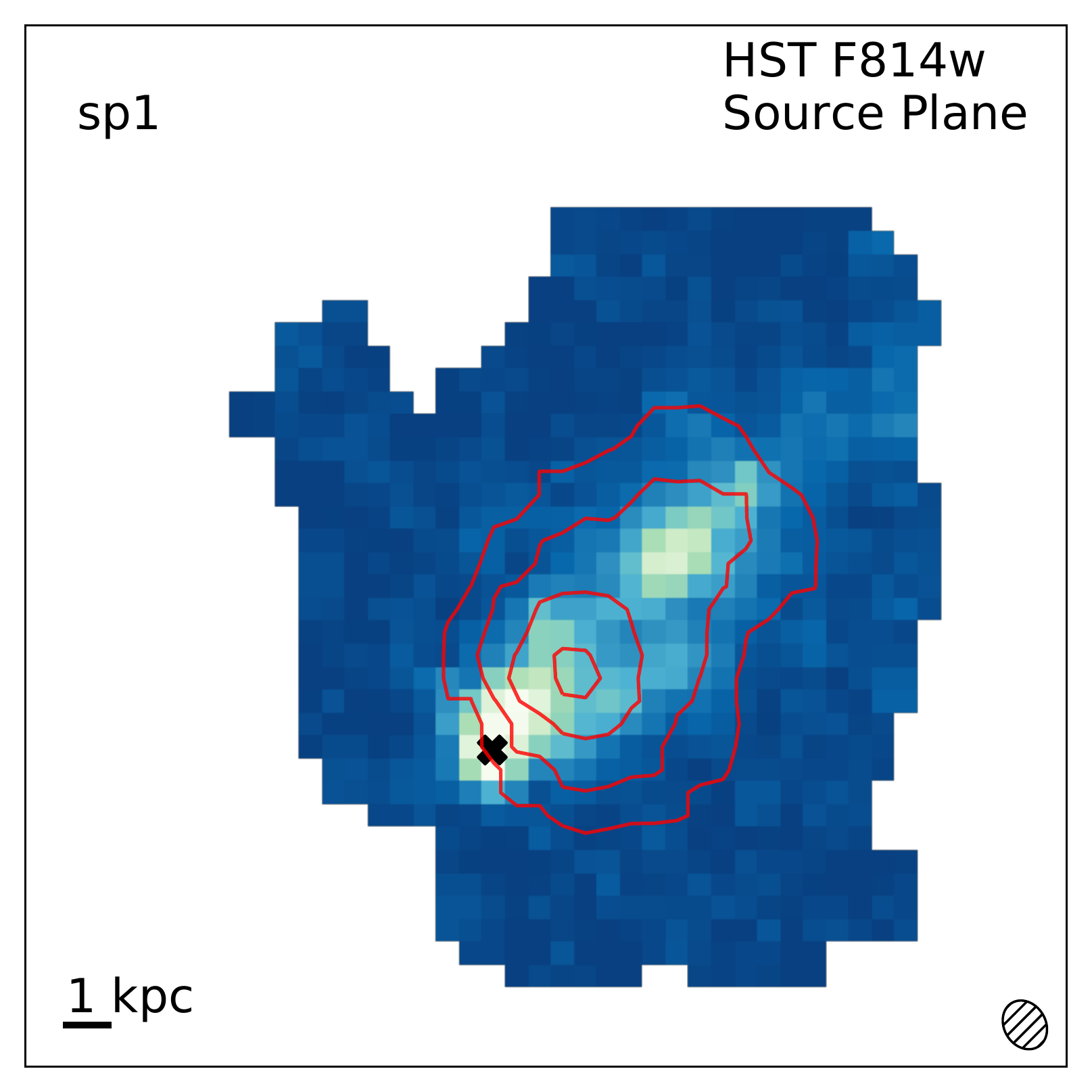}
	\includegraphics[width=0.195\textwidth]{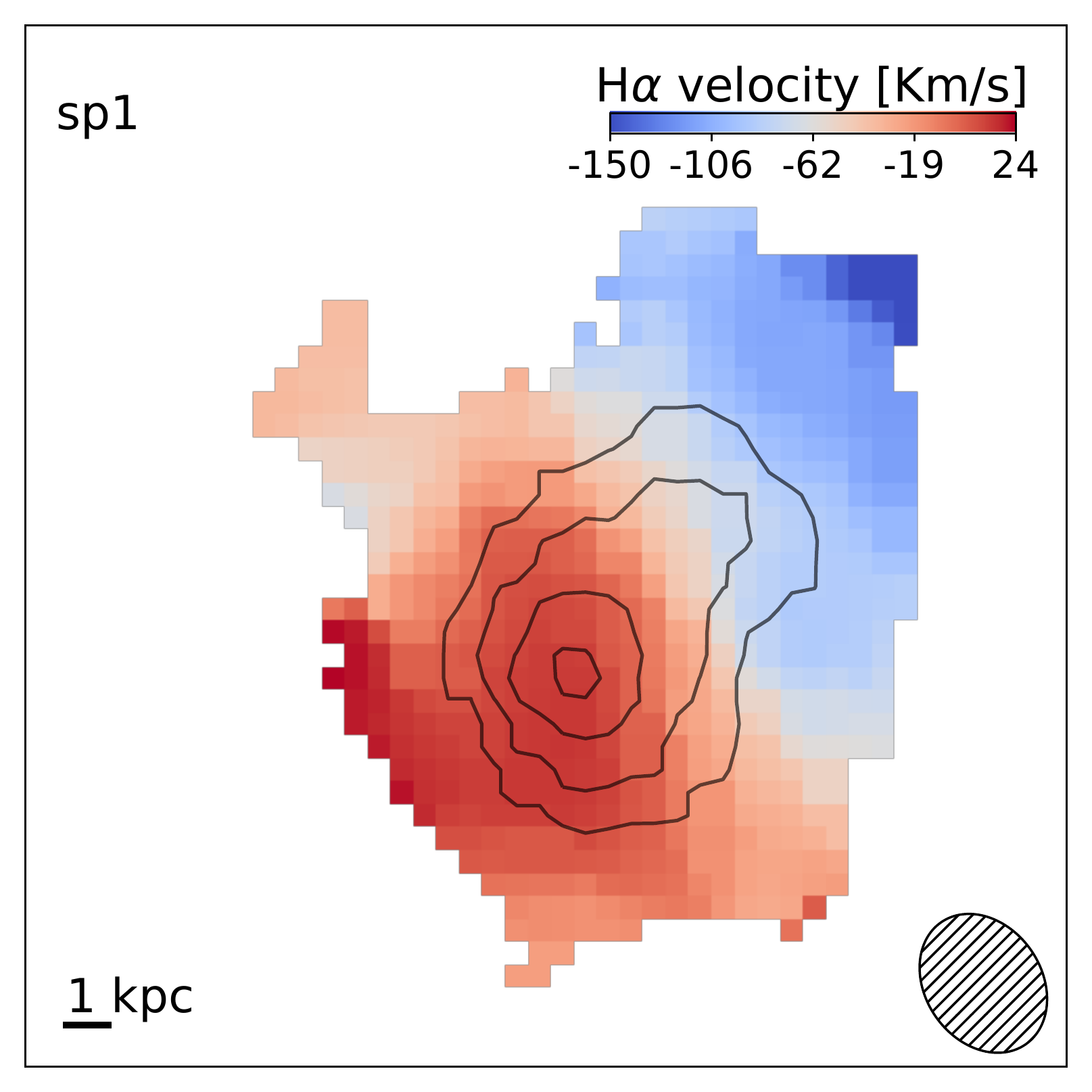}
	\includegraphics[width=0.195\textwidth]{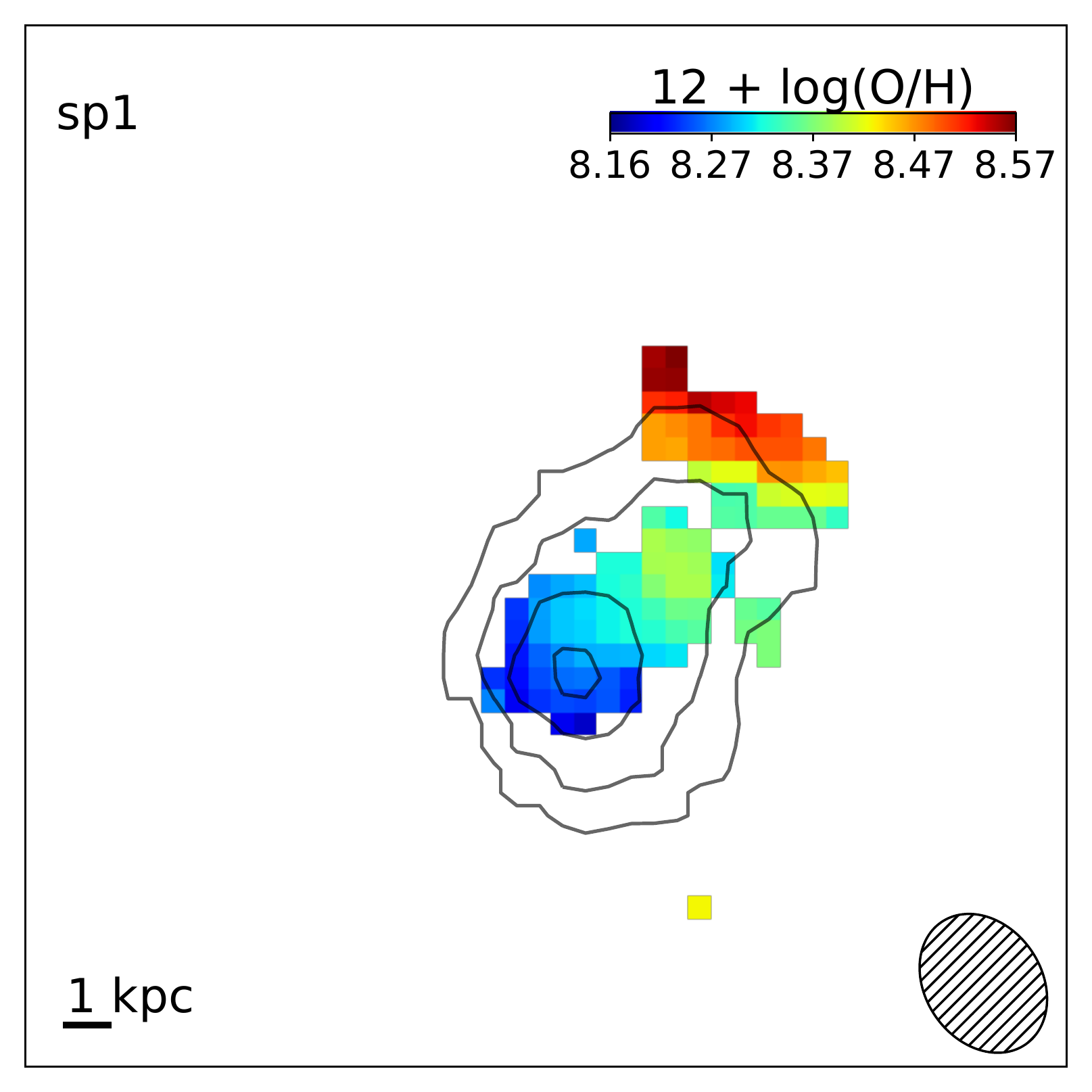}\\

	\includegraphics[width=0.195\textwidth]{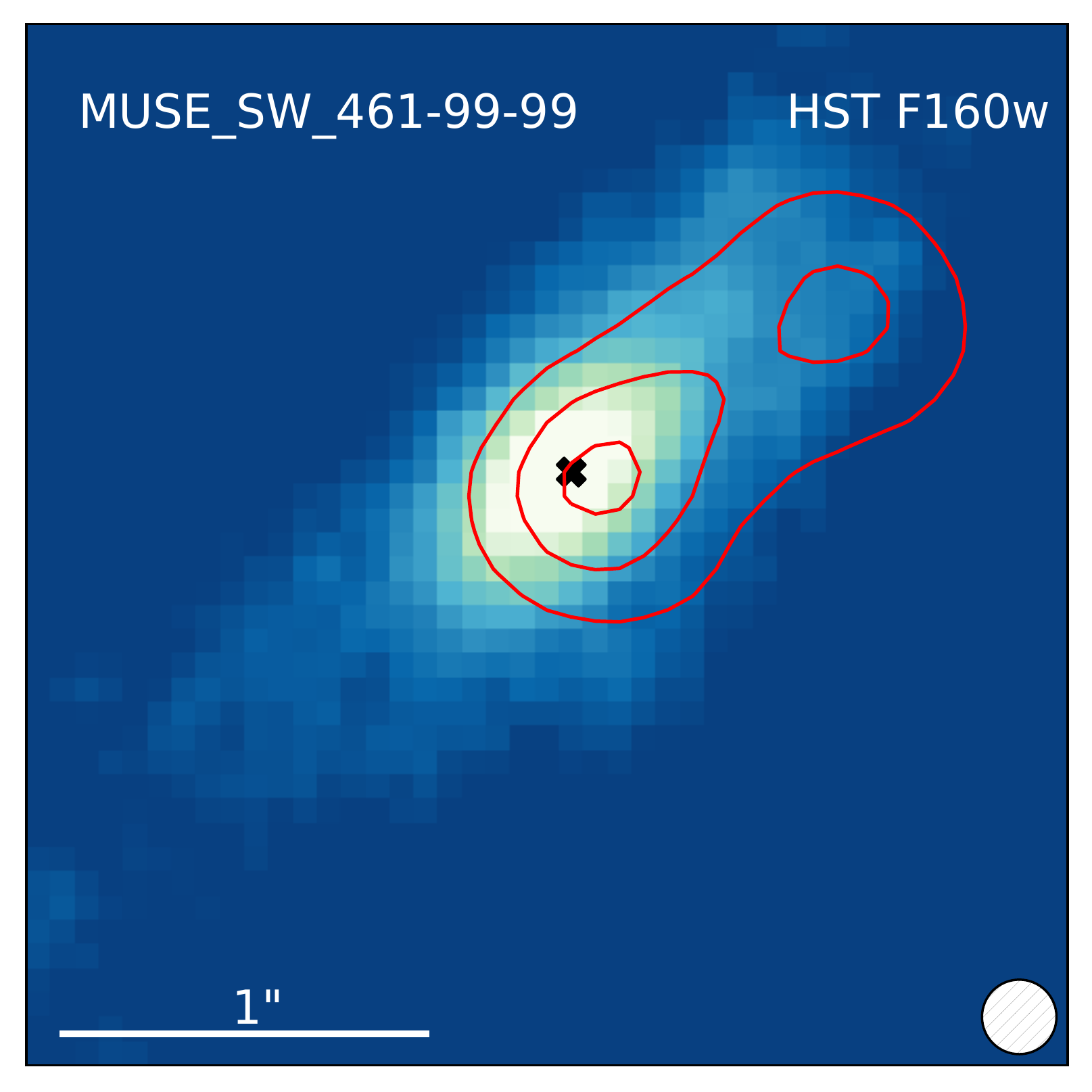}
	\includegraphics[width=0.195\textwidth]{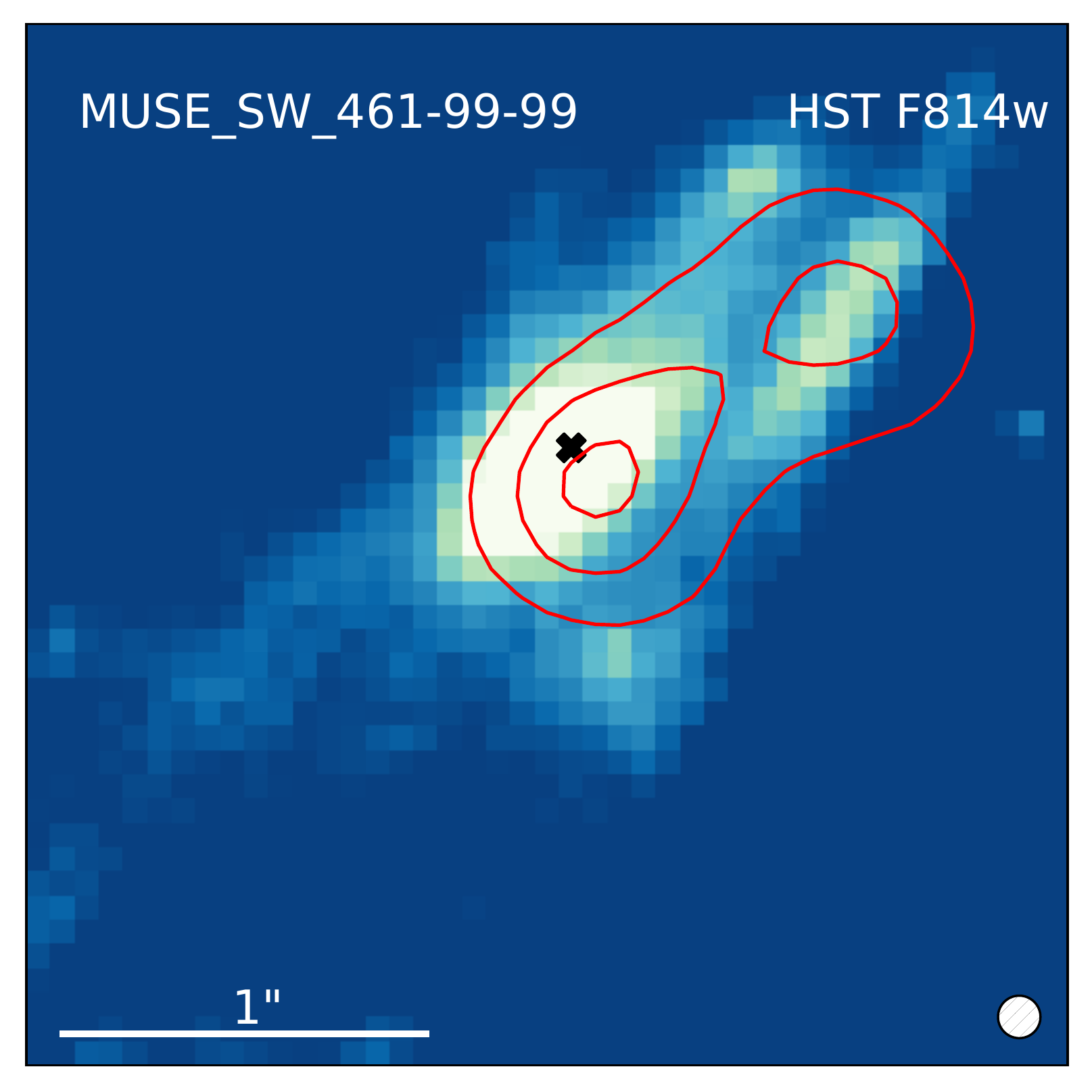}
	\includegraphics[width=0.195\textwidth]{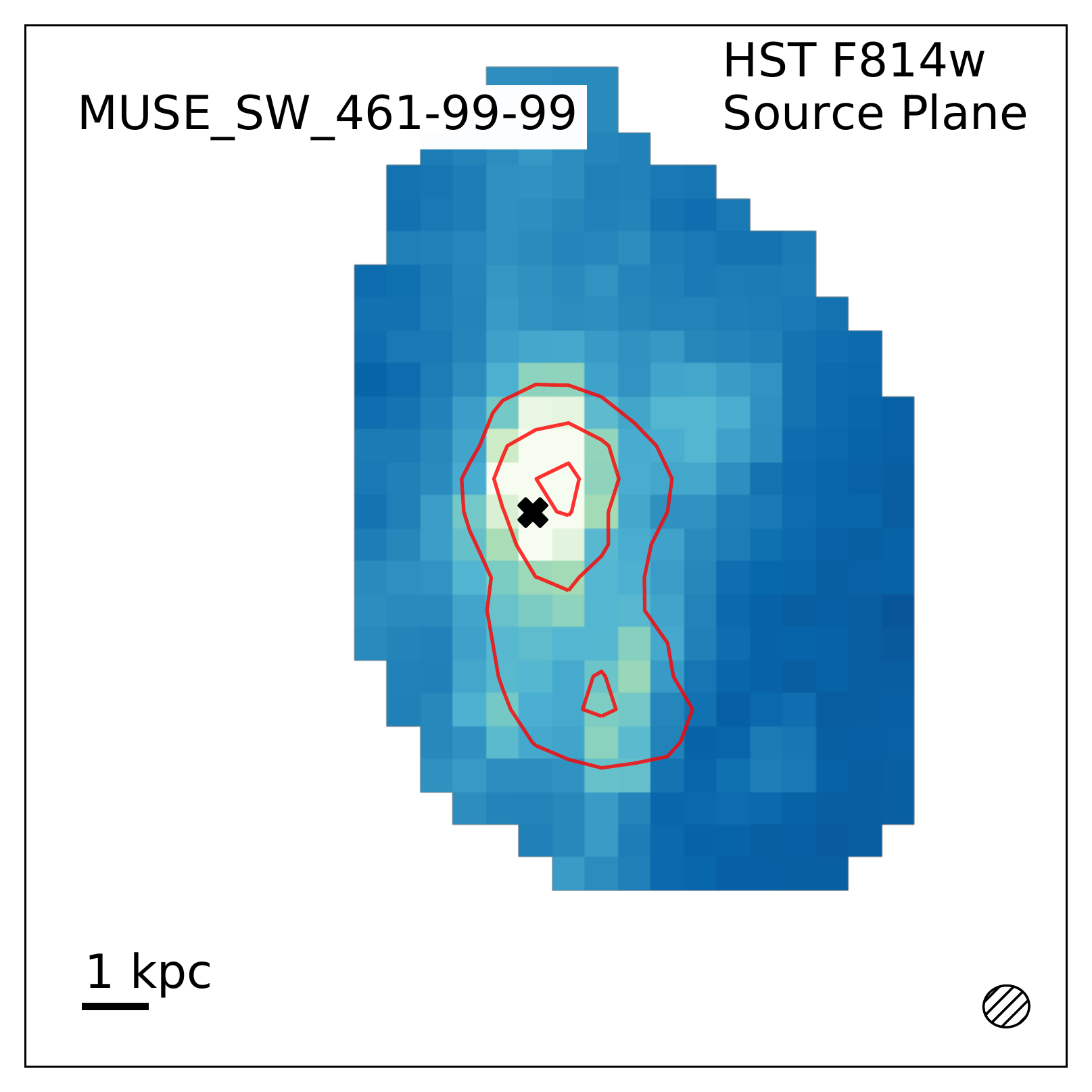}
	\includegraphics[width=0.195\textwidth]{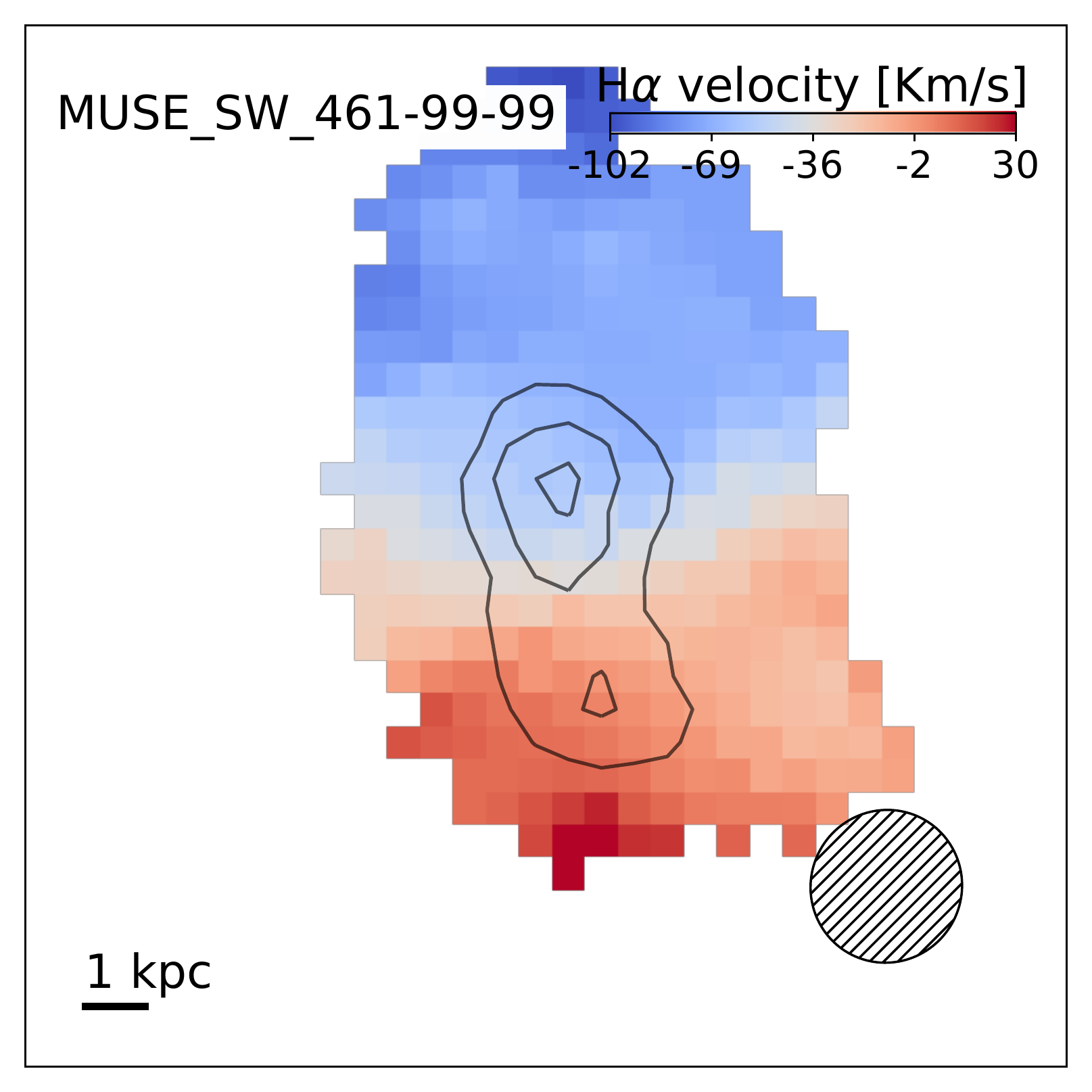}
	\includegraphics[width=0.195\textwidth]{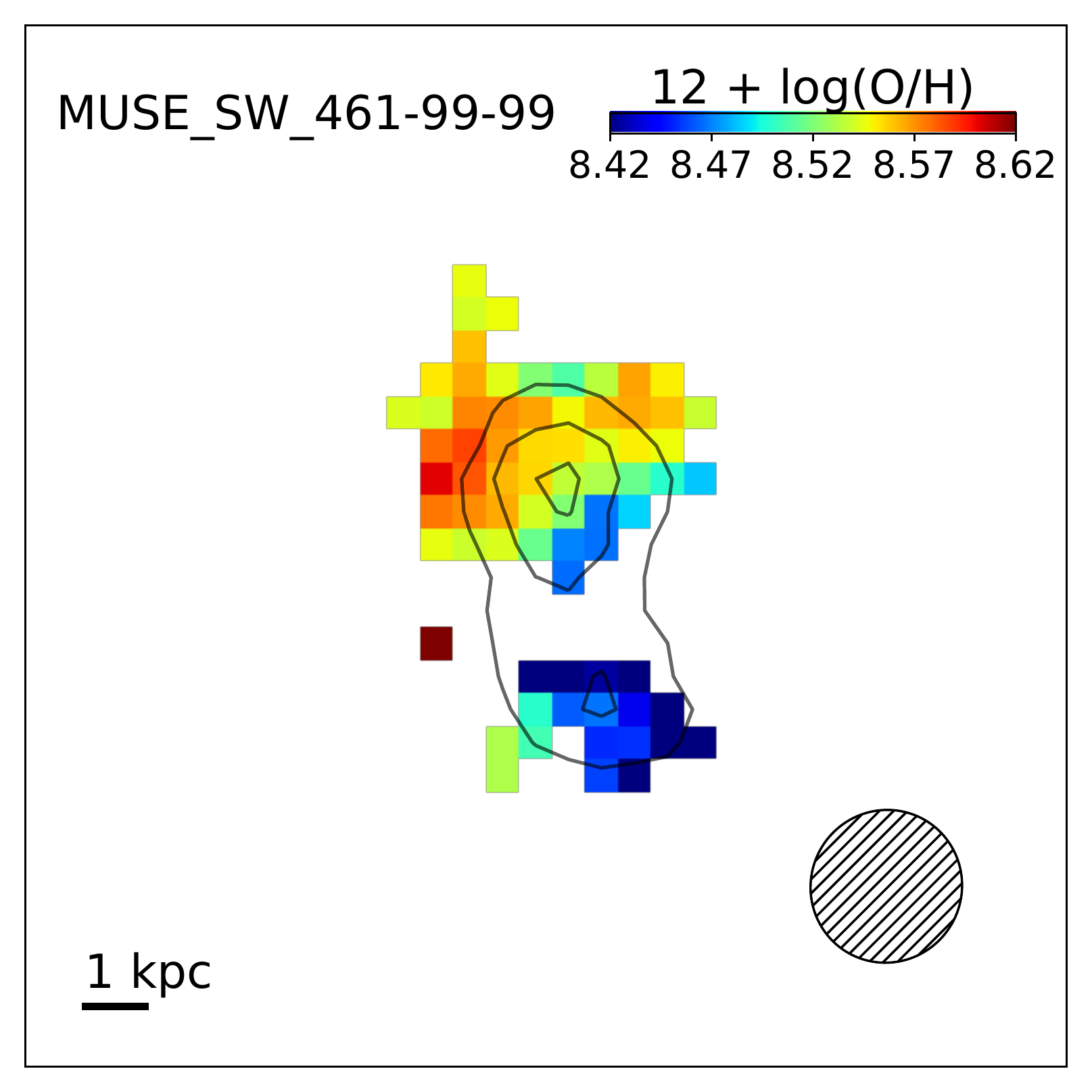}\\

	\includegraphics[width=0.195\textwidth]{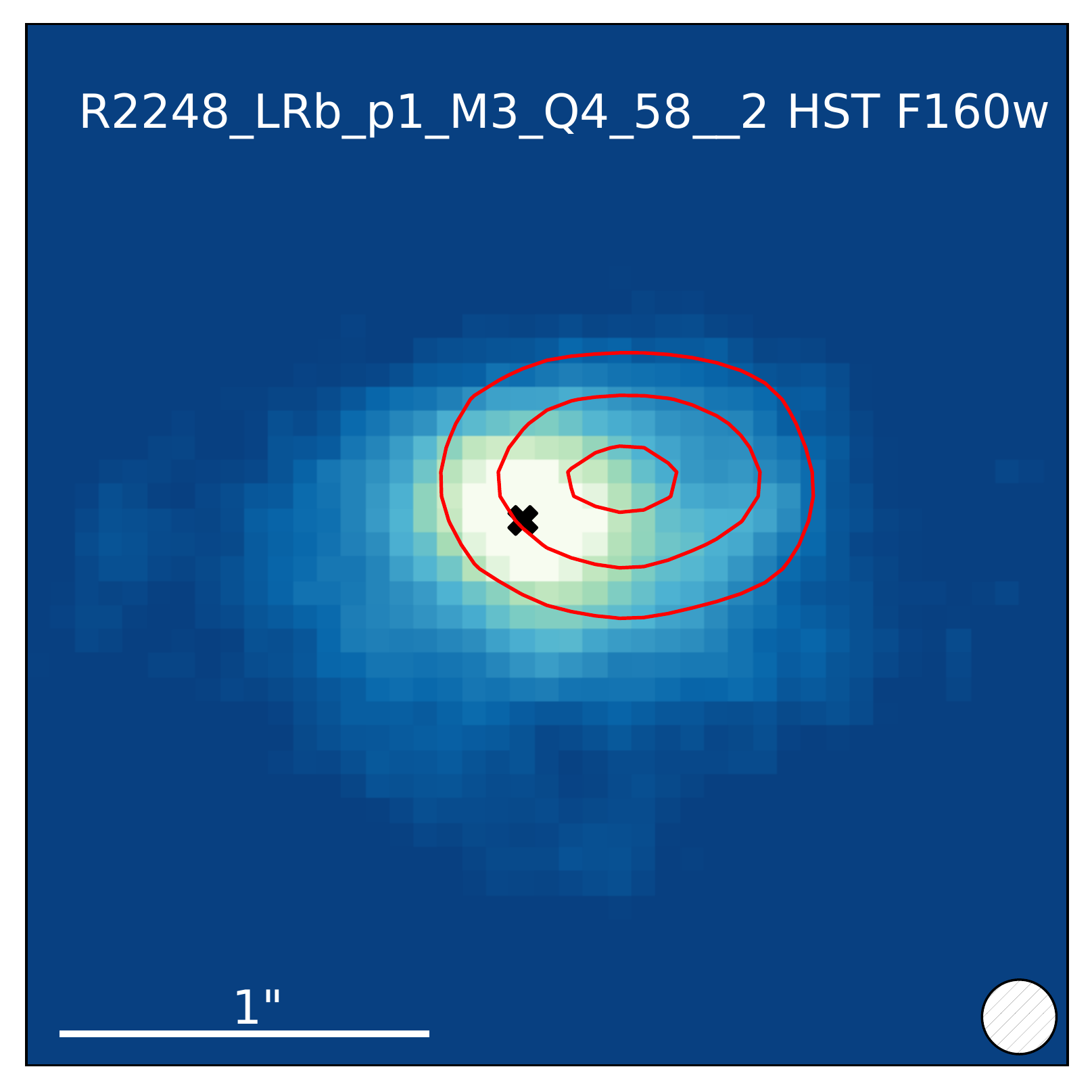}
	\includegraphics[width=0.195\textwidth]{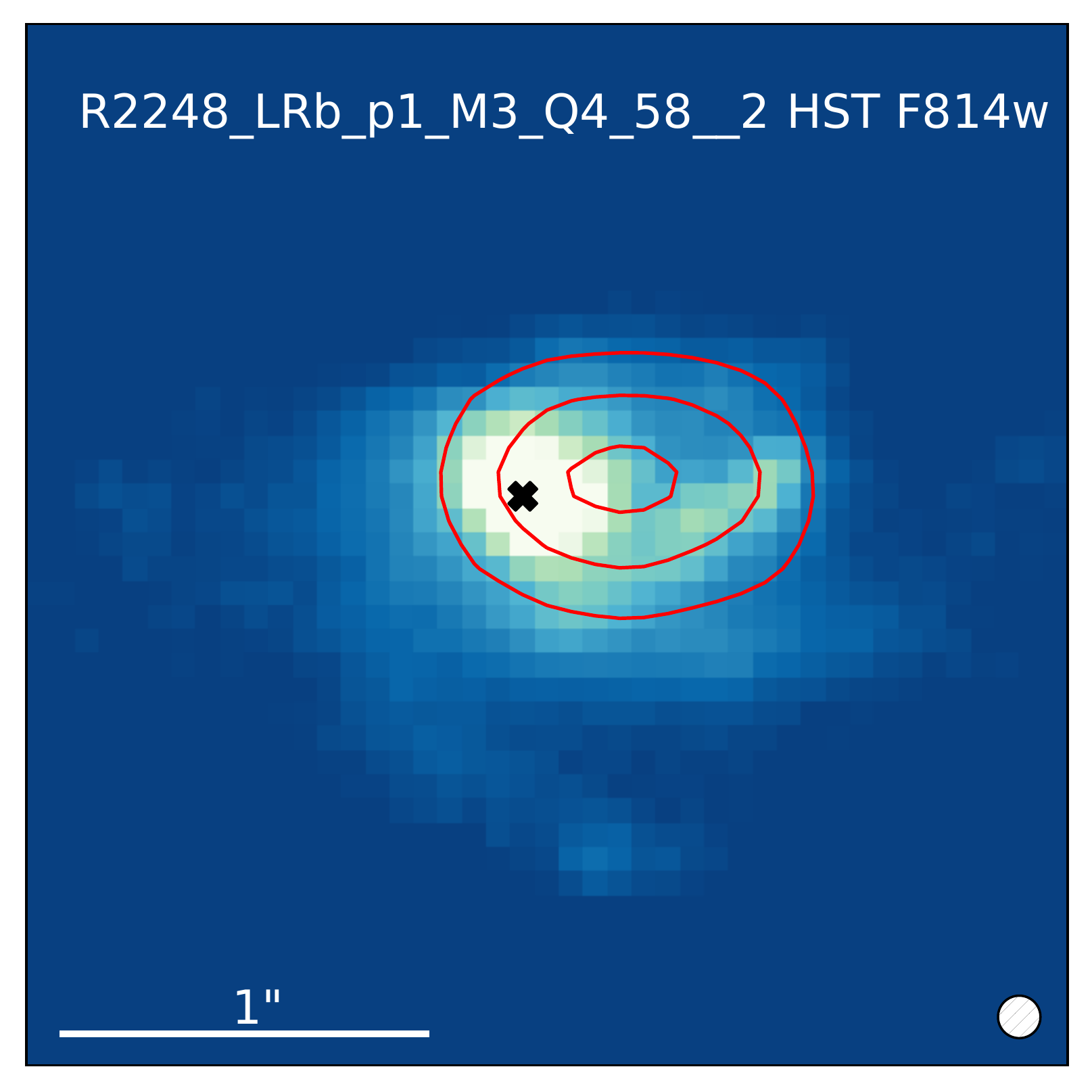}
	\includegraphics[width=0.195\textwidth]{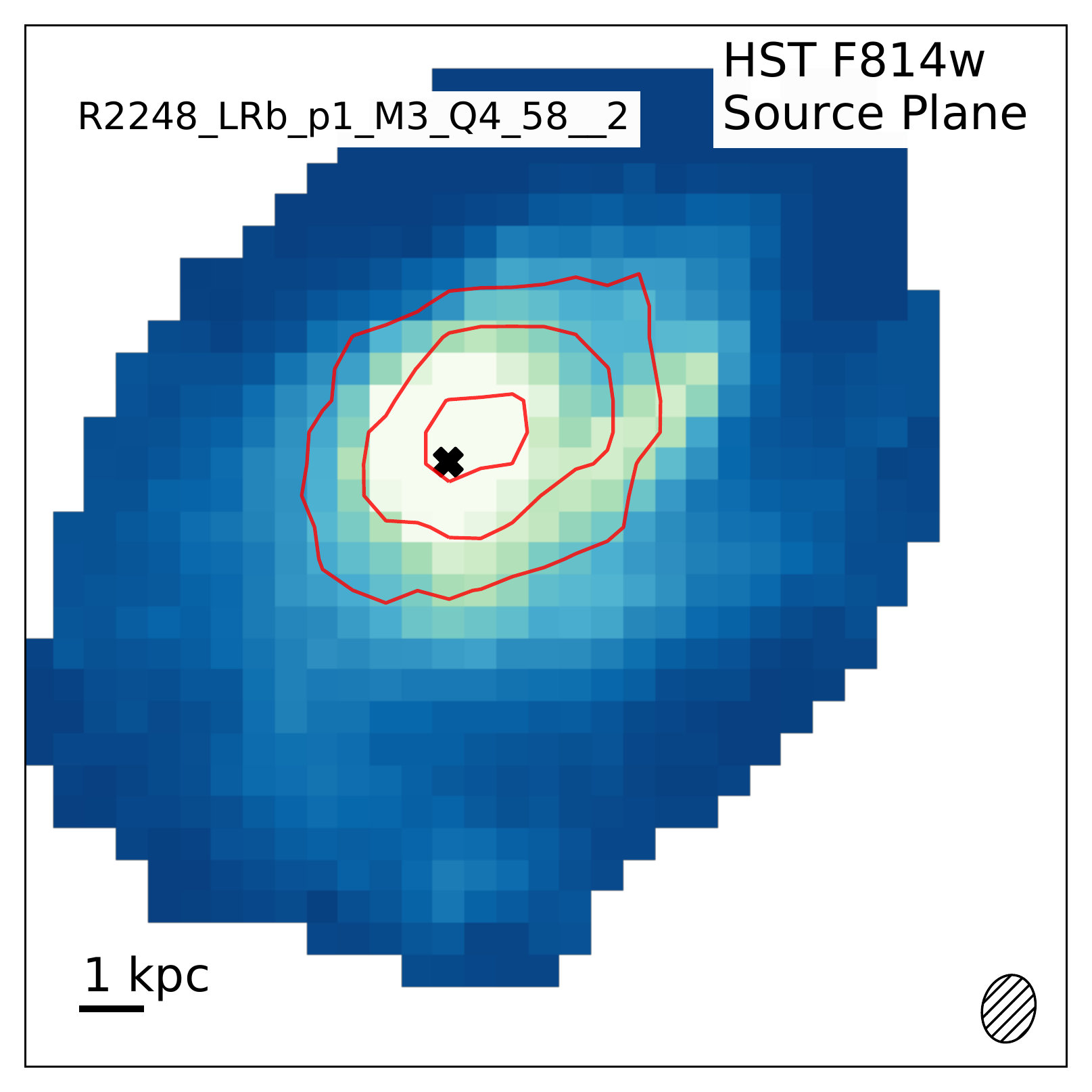}
	\includegraphics[width=0.195\textwidth]{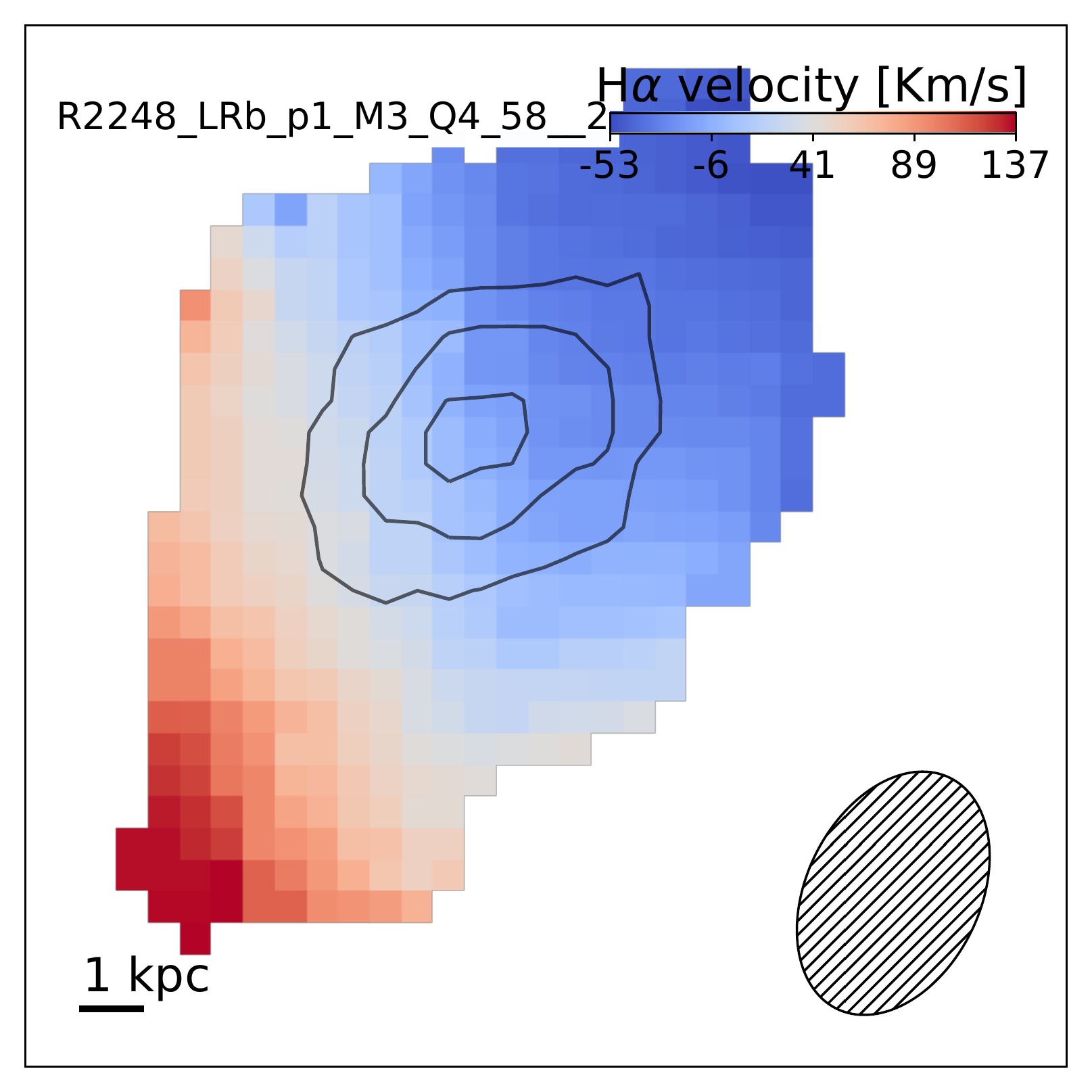}
	\includegraphics[width=0.195\textwidth]{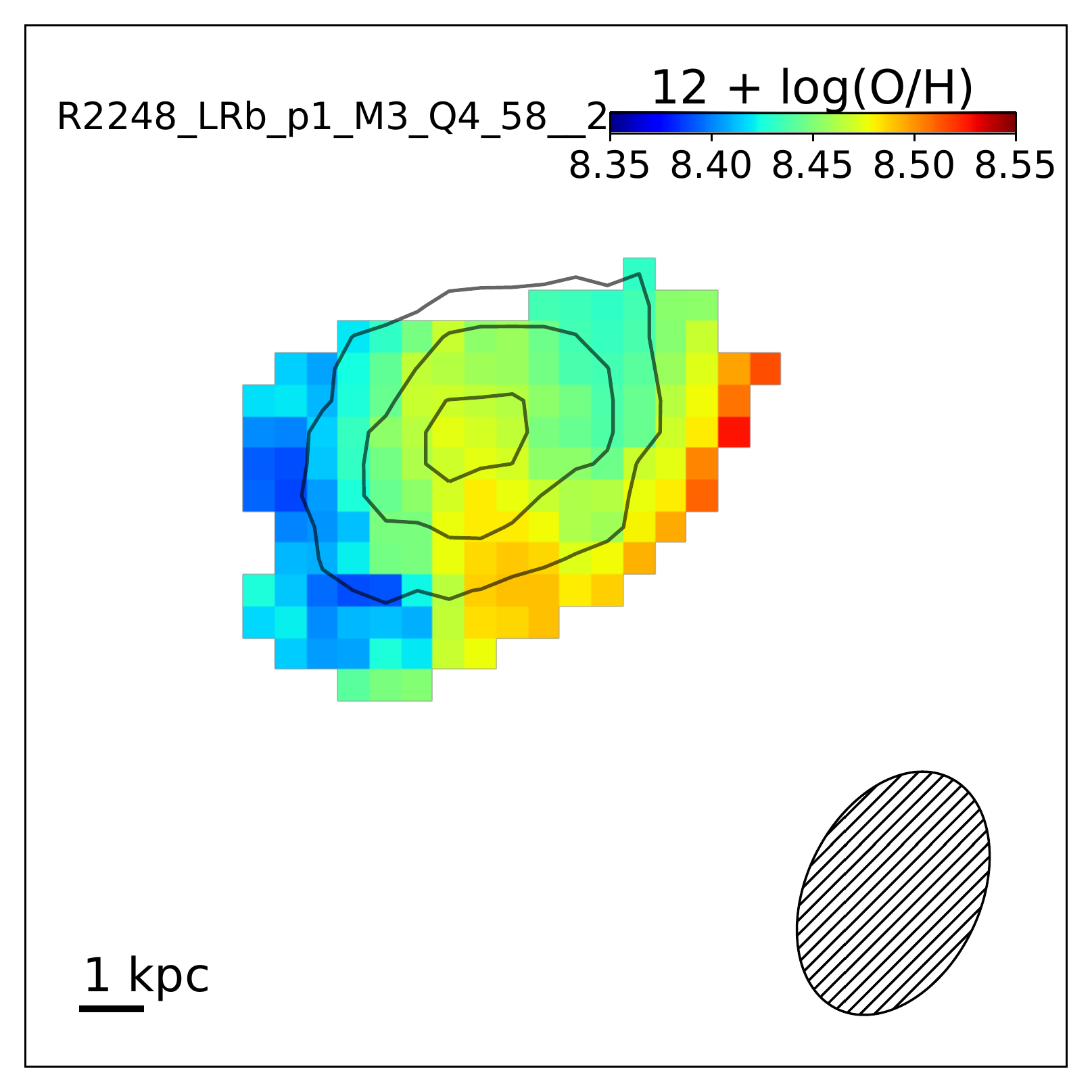}
	
	\caption{Examples of systems showing disturbed morphologies, kinematics and metallicity maps.
		From \textit{left} to \textit{right}: HST F160W image, HST F814W image,  reconstructed HST F814W source plane image,  H$\alpha$  velocity field and metallicity map are shown for four galaxies analysed in this work.
		These galaxies present extended H$\alpha$ emission or secondary peaks misaligned from the bulk of the underlying H-band continuum. These regions of intense off-nuclear H$\alpha$ emission may be either associated with episodes of recent star formation occurring in compact clumps which form in the turbulent gas-rich disk or the results of merger events which, however, can not always be obviously confirmed from the velocity fields. These regions are often characterised by a different level of chemical enrichment compared to the bulk of the system.
		In each panel, the overlaid contours encompass the $50\%,68\%, 85\%$ and $97\%$ of the H$\alpha$ emission, while black crosses mark the position of the peak of the underlying HST continuum image.
	}
	\label{fig:disturbed}
\end{figure*}

The complex kinematics seen in the velocity fields of sp15 suggests a primary galaxy (associated with the H$\upalpha$ peak and with a clear velocity gradient) interacting with a fainter companion responsible for the redshifted H$\upalpha$ emission towards the bottom right part of the image. 
The H$\upalpha$ emission is elongated towards the north-est, co-spatial with the ``tail"  visible in the HST images.
The metallicity map is strongly irregular, with an apparent spatial anti-correlation between the intensity of the H$\upalpha$ and the  metallicity. However, the azimuthal asymmetry is partially washed out when averaging, with the final gradient equal to $-0.005 \pm 0.010$.
A similar behaviour can be seen for the galaxy sp1, which shows the presence of multiple components, clearly visible in the F814W image, with the centroid of the extended H$\upalpha$ emission misaligned from the peak of the F160W continuum, more representative of the stellar mass density distribution. 
The velocity map is nevertheless quite regular and does not show any clear deviation from an ordered disk-like rotation pattern (making its interpretation often non trivial in seeing-limited data, see e.g. \citealt{Simons:2019aa}).
A clear linear  gradient in the metallicity map can be seen from the most active, metal poor region in the south-est area to the more metal rich region towards the north-west direction, which is associated with the main disk component (as marked by the peak of H-band continuum image); however, the gradient is lost if radially averaging around the centroid of the emission in the system.
MUSE\_SW\_461-99-99 is characterized by the presence of a distinct H$\upalpha$ blob in the north-west region of the image plane map (south-west in the source plane), which is not detected in the KMOS H-band continuum map nor can be clearly associated with any visible structure in the broad band HST-F160w image (nor in the de-lensed velocity field).
However, the HST-F814w image reveals a more complex morphology, with an offset component (likely a companion galaxy) aligned with the secondary H$\upalpha$ peak.
It is worth noting how in this case the metallicity of the region co-spatial with the secondary H$\upalpha$ blob is 
significantly lower than that of the central galaxy.
Finally, R2248\_LRb\_p1\_M3\_Q4\_58\_\_2 presents a bright, elongated knot westward ($\sim 1.5$ kpc far) of the centre in the F814W image which is not detected in H band continuum. 
However, the centroid of the H$\upalpha$ emission is shifted in the same direction, and the region is characterised by 
a slightly enhanced level of enrichment.

As we have seen, these configurations may be either related to the presence of close interacting companions or even suggest that different processes, occurring on shorter timescales than secular processes and mainly associated to gas flows, could contribute to the complexity seen in some of the 2D metallicity maps.
High resolution simulations of disc galaxies (Milky Way-like at redshift zero) have shown in fact that strong fluctuations in their star formation history and gas outflow rate occurring on relatively short timescales (i.e. $\lesssim 1 $Gyr) can induce significant fluctuations in the gas-phase metallicity maps and hence in the derived radial gradients \citep{Ma:2017aa}.
%Episodes of local accretion of metal poor gas could favor the formation of low metallicity regions, contributing to explain some of the complex morphologies seen in the metallicity maps.
In general, at the peak of the cosmic star formation history (z$\sim 1.5-2.5$, the epoch probed by KLEVER), galaxies are often characterized by irregular morphologies and their emission lines luminosity might be dominated by the presence of young, massive star forming clumps \citep{Elmegreen:2009aa,Forster-Schreiber:2011aa,Genzel:2011aa}.
Recent cosmological simulations by \cite{Ceverino:2016aa} also show that in $50 \%$ of cases at redshifts lower than $4$, pristine gas accretion from the cosmic web can give rise to the formation of local star-forming clumps and subsequent drop in metallicity ($\sim 0.3 \rm~dex$) compared to the surrounding ISM. They further assert that the accretion should be rapid enough to sustain such a metallicity drop, as the turbulence mixing would dissolve these features within a few disc dynamical timescales. 
Another possibility is that these clumps may form within their gas-rich disks due to gravitational instabilities \citep{Bournaud:2007aa,Elmegreen:2008aa,Genzel:2008aa}; in this case, they might leave a different imprint on the metallicity maps, showing similar level of chemical enrichment to those of the material from which they have formed.
The presence of such regions of different star formation and chemical enrichment histories within the same galaxy may reflect itself in the azimuthal asymmetries seen in some of the metallicity maps, despite the impossibility to resolve them spatially given the intrinsic limitations of seeing-limited observations in reaching the required resolution \citep[$\sim 100 $pc, ][]{Tamburello:2017aa}.
Nonetheless, recent observations conducted with SINFONI AO-assisted spectroscopy have also reported similar asymmetries in the distribution of the \nii/H$\upalpha$ line ratio in $\text{z}\sim2$ galaxies \citep{Forster-Schreiber:2018ab}.
Therefore, although azimuthally averaged gradients are still useful and commonly adopted as benchmark for theoretical predictions, this suggests that only the full characterization of the spatial distribution of the heavy elements within a galaxy (i.e. the 2D metallicity maps) can ultimately provide the key to resolve the interplay between physical processes occurring on local scales (gas accretion, gas flows, turbulent mixing, etc...) and global, secular processes, constraining their role in shaping the formation of galactic disks at these epochs.

\section{Summary and Conclusions}
We have analysed a sample of gravitationally lensed galaxies between redshift $1.2 < z <2.5$, observed with KMOS and SINFONI as a first part of the KLEVER Large Programme. 
The observations conducted in the YJ, H and K bands provided spatially resolved mapping of many rest-frame optical nebular emission lines, allowing us to assess different ISM properties in these objects on both global and local scales.
In particular, in this work we have constrained gas-phase metallicities by simultaneously adopting different strong-line diagnostics and calibrations tied to the \Te\ abundance scale. 
The sample covers a range of $10^{8.5} - 10^{10.5}$ M$_{\odot}$ in stellar mass and $\sim 1-200$ M$_{\odot}$yr$^{-1}$ in SFR, sampling galaxies on, above and below the star-forming main sequence at these redshifts, with a slight bias towards high specific star formation rates especially at $\text{z}>2$.
On global scales (i.e. from the analysis of their integrated spectra) the galaxies analysed are consistent with the observed evolution in the BPT diagrams and in the mass-metallicity relation reported in previous studies (Fig.~\ref{fig:global_bpt} and ~\ref{fig:global_prop}).

In this paper, we mainly focused on the analysis of metallicity gradients.
We have exploited robust lens modelling of foreground sources (mainly galaxy clusters) to obtain source plane emission line maps with a typical resolution of $\sim3\ \text{kpc}$, from which we have derived metallicity maps and extracted radial metallicity gradients by means of azimuthal averages within elliptical apertures defined according to the shape of the source plane PSF (e.g. Fig.~\ref{fig:EGgrad}).
The main conclusions drawn in this paper are summarized as follows. 
\begin{itemize}
	\item The $86\%$ of our galaxy sample present radial metallicity gradients shallower than $0.05 \rm dex/kpc$ ($72\%$ within $0.025 \rm dex/kpc$), with $89\%$ of the gradients consistent with being flat within $3\sigma$ uncertainty ($67\%$ within $1\sigma$). 
	This is in agreement with the majority of previous results reported in the literature which investigated metallicity gradients at z $\sim 1-2.5$, despite the diversity of observational techniques and diagnostics adopted (Fig.~\ref{fig:redshift}).
	Comparison with cosmological simulations that explore the effect of different feedback modes suggest a scenario where efficient mixing processes, that redistribute a significant amount of gas over large scales, are in place at these epochs \citep{Gibson:2013ab}.	
	However, predictions from analytical models which assumes fairly constant star formation profiles and no prescriptions about radial gas mixing are also broadly consistent with the observed distribution of metallicity gradients observed in the galaxy population at $1\lesssim \text{z} \lesssim 2.5$ \citep{Molla:2019aa}.
	Galaxies showing a relatively homogeneous metal distribution across large spatial scales might also be consistent with a scenario of uniform disk mass assembly. 

	\item Three sources in the sample are characterised by significantly (i.e. above $3 \sigma$) positive (``inverted") metallicity gradients. Similarly with previous findings in the literature \citep{Cresci:2010aa, Troncoso:2014aa}, in these galaxies the central metal poor regions are associated with the highest level of H$\upalpha$ emission, hence with the highest star formation rate surface density, suggesting recent accretion of pristine gas.
	Alternatively, chemical evolution models assuming a constant SFE along the disc and allowing for large variations in the velocity of radial gas flows have also shown to be able to predict an ``inversion" of the metallicity gradients \citep{Mott:2013aa}.

	\item We explored the correlation between the sSFR and metallicity gradients at $\text{z}>0.8$, extending the relation towards the highest sSFRs probed by our sample (Fig.~\ref{fig:ssfr}, upper panel). We find a mild trend in which galaxies with a higher sSFR exhibit flatter or 'more positive' gradients, with a significance of $3.2 \sigma$, in agreement with previous findings \citep[][]{Stott:2014aa, Wuyts:2016aa}. 
	When normalising the SFR to that expected from the star forming main sequence at each redshift, the significance of the trend reduces to $2 \sigma$ (Fig.~\ref{fig:deltasfmshist}).
	We have found a milder, negative correlation between metallicity gradients and stellar mass (significant at the $2.7 \sigma$ level, lower panel of Fig.~\ref{fig:ssfr}), consistent though with the trend followed in the local Universe (i.e. gradients steepening with increasing stellar mass, \citealt{Belfiore:2017aa}). 

	\item Despite the apparent radial invariance, some of our galaxies exhibit complex patterns in the metallicity maps, with variations in log(O/H) of the order of $0.1-0.2$ {\rm dex} on kpc scales which are, however, not symmetrically distributed around the galaxy centre (Fig.~\ref{fig:disturbed}).
	In these cases, flat gradients might artificially arise from azimuthal averages, washing out and therefore hiding the presence of regions with different level of chemical enrichment.
	Strong fluctuations in the star formation history and outflow rates or local episodes of gas accretion occurring on short timescales could explain the observed irregular morphologies.
	This warns against using radial gradients as the only constrains for galaxy evolution models. Extracting more information from the full 2D metallicity maps is key to discriminate between the contribution that different local processes (like gas recycling via stellar feedback, pristine gas accretion, turbulent mixing and merger events) have in shaping the observed morphologies.

\end{itemize}

In conclusion, mutual interactions between star formation and different processes involving efficient radial mixing driven by intense galactic feedback or merger events could strongly affect the complex morphology of the maps and the average radial flattening of the gradients that we see at these epochs.
A more comprehensive and self-consistent view of spatially resolved metallicity properties in high-z galaxies will be possible as soon as the KLEVER Programme is completed, comprising a larger sample spanning a wider range in stellar masses and star formation rates.
The complexity revealed by current generation IFU instruments is also paving the way for observations with next generation facilities like NIRSPEC on JWST and ERIS (the AO-assisted Enhanced Resolution Imager and Spectrograph, aimed at substituting SINFONI on the VLT from 2020, \citealt{Davies:2018ab}), that will offer in the near future the opportunity to probe the metal distribution in high redshfit galaxies with unprecedented detail, helping to discriminate between different theoretical predictions and setting new and powerful constraint for cosmological simulations and chemical evolution models.

\section*{Acknowledgements}
Based on observations made with ESO Telescopes at the La Silla Paranal Observatory under programme ID 083.A-0918(A), 083.B-0108(A), 090.B-0313(A), 092.A-0426(A), 092.B-0677(A), 094.A-0279(A) , 095.B-0480(A) and 197.A-0717(A). 
The mass models for the lensed galaxies in galaxy clusters were developed by \cite{Zitrin:2009aa,Zitrin:2013aa}, and obtained through the Hubble Space Telescope Archive, as a high-end science product of the CLASH \citep{Postman:2012aa} and Frontier Fields \citep{Lotz:2017aa} programs.

MC and RM acknowledge support by the Science and Technology Facilities Council (STFC) and from ERC Advanced Grant 695671 ``QUENCH".
GC acknowledges the support by INAF/Frontiera through the "Progetti Premiali" funding scheme of the Italian Ministry of Education, University, and Research; GC has been also supported by the INAF PRIN-SKA 2017 program 1.05.01.88.04.
MC acknowledges support from a Royal Society University Research Fellowship.
CC acknowledges funding from the European Union’s Horizon 2020 research and innovation programme under the Marie Sklodowska-Curie grant agreement no. 664931.
Y. P. acknowledges support from the National Key Program for Science and Technology Research and Development under grant number 2016YFA0400702, and the NSFC grant no. 11773001, 11721303.

This work utilizes gravitational lensing models produced by PIs Bradac, Natarajan \& Kneib (CATS), Merten \& Zitrin, Sharon, Williams, Keeton, Bernstein and Diego, and the GLAFIC group. This lens modeling was partially funded by the HST Frontier Fields program conducted by STScI. STScI is operated by the Association of Universities for Research in Astronomy, Inc. under NASA contract NAS 5-26555. The lens models were obtained from the Mikulski Archive for Space Telescopes (MAST).

%%%%%%%%%%%%%%%%%%%%% REFERENCES %%%%%%%%%%%%%%%%%%%
\bibliographystyle{mnras}
\setlength{\labelwidth}{0pt}
\bibliography{/Users/mirkocurti/Google_Drive_pro/astro_biblio.bib}

\begin{thebibliography}{}
\makeatletter
\relax
\def\mn@urlcharsother{\let\do\@makeother \do\$\do\&\do\#\do\^\do\_\do\%\do\~}
\def\mn@doi{\begingroup\mn@urlcharsother \@ifnextchar [ {\mn@doi@}
  {\mn@doi@[]}}
\def\mn@doi@[#1]#2{\def\@tempa{#1}\ifx\@tempa\@empty \href
  {http://dx.doi.org/#2} {doi:#2}\else \href {http://dx.doi.org/#2} {#1}\fi
  \endgroup}
\def\mn@eprint#1#2{\mn@eprint@#1:#2::\@nil}
\def\mn@eprint@arXiv#1{\href {http://arxiv.org/abs/#1} {{\tt arXiv:#1}}}
\def\mn@eprint@dblp#1{\href {http://dblp.uni-trier.de/rec/bibtex/#1.xml}
  {dblp:#1}}
\def\mn@eprint@#1:#2:#3:#4\@nil{\def\@tempa {#1}\def\@tempb {#2}\def\@tempc
  {#3}\ifx \@tempc \@empty \let \@tempc \@tempb \let \@tempb \@tempa \fi \ifx
  \@tempb \@empty \def\@tempb {arXiv}\fi \@ifundefined
  {mn@eprint@\@tempb}{\@tempb:\@tempc}{\expandafter \expandafter \csname
  mn@eprint@\@tempb\endcsname \expandafter{\@tempc}}}

\bibitem[\protect\citeauthoryear{{Andrews} \& {Martini}}{{Andrews} \&
  {Martini}}{2013}]{Andrews:2013aa}
{Andrews} B.~H.,  {Martini} P.,  2013, \mn@doi [\apj]
  {10.1088/0004-637X/765/2/140}, \href
  {http://adsabs.harvard.edu/abs/2013ApJ...765..140A} {765, 140}

\bibitem[\protect\citeauthoryear{{Baldwin}, {Phillips}  \&
  {Terlevich}}{{Baldwin} et~al.}{1981}]{Baldwin:1981aa}
{Baldwin} J.~A.,  {Phillips} M.~M.,   {Terlevich} R.,  1981, \mn@doi [\pasp]
  {10.1086/130766}, \href {http://adsabs.harvard.edu/abs/1981PASP...93....5B}
  {93, 5}

\bibitem[\protect\citeauthoryear{{Belfiore} et~al.,}{{Belfiore}
  et~al.}{2017}]{Belfiore:2017aa}
{Belfiore} F.,  et~al., 2017, \mn@doi [\mnras] {10.1093/mnras/stx789}, \href
  {http://adsabs.harvard.edu/abs/2017MNRAS.469..151B} {469, 151}

\bibitem[\protect\citeauthoryear{{Belfiore}, {Vincenzo}, {Maiolino}  \&
  {Matteucci}}{{Belfiore} et~al.}{2019}]{Belfiore:2019aa}
{Belfiore} F.,  {Vincenzo} F.,  {Maiolino} R.,   {Matteucci} F.,  2019, \mn@doi
  [\mnras] {10.1093/mnras/stz1165}, \href
  {https://ui.adsabs.harvard.edu/abs/2019MNRAS.487..456B} {487, 456}

\bibitem[\protect\citeauthoryear{{Belli}, {Jones}, {Ellis}  \&
  {Richard}}{{Belli} et~al.}{2013}]{Belli:2013aa}
{Belli} S.,  {Jones} T.,  {Ellis} R.~S.,   {Richard} J.,  2013, \mn@doi [\apj]
  {10.1088/0004-637X/772/2/141}, \href
  {http://adsabs.harvard.edu/abs/2013ApJ...772..141B} {772, 141}

\bibitem[\protect\citeauthoryear{{Belokurov}, {Evans}, {Hewett}, {Moiseev},
  {McMahon}, {Sanchez}  \& {King}}{{Belokurov} et~al.}{2009}]{Belokurov:2009aa}
{Belokurov} V.,  {Evans} N.~W.,  {Hewett} P.~C.,  {Moiseev} A.,  {McMahon}
  R.~G.,  {Sanchez} S.~F.,   {King} L.~J.,  2009, \mn@doi [\mnras]
  {10.1111/j.1365-2966.2008.14075.x}, \href
  {http://adsabs.harvard.edu/abs/2009MNRAS.392..104B} {392, 104}

\bibitem[\protect\citeauthoryear{{Berg} et~al.,}{{Berg}
  et~al.}{2012}]{Berg:2012aa}
{Berg} D.~A.,  et~al., 2012, \mn@doi [\apj] {10.1088/0004-637X/754/2/98}, \href
  {https://ui.adsabs.harvard.edu/abs/2012ApJ...754...98B} {754, 98}

\bibitem[\protect\citeauthoryear{{Berg}, {Skillman}, {Croxall}, {Pogge},
  {Moustakas}  \& {Johnson-Groh}}{{Berg} et~al.}{2015}]{Berg:2015aa}
{Berg} D.~A.,  {Skillman} E.~D.,  {Croxall} K.~V.,  {Pogge} R.~W.,  {Moustakas}
  J.,   {Johnson-Groh} M.,  2015, \mn@doi [\apj] {10.1088/0004-637X/806/1/16},
  \href {http://adsabs.harvard.edu/abs/2015ApJ...806...16B} {806, 16}

\bibitem[\protect\citeauthoryear{{Bournaud}, {Elmegreen}  \&
  {Elmegreen}}{{Bournaud} et~al.}{2007}]{Bournaud:2007aa}
{Bournaud} F.,  {Elmegreen} B.~G.,   {Elmegreen} D.~M.,  2007, \mn@doi [\apj]
  {10.1086/522077}, \href {http://adsabs.harvard.edu/abs/2007ApJ...670..237B}
  {670, 237}

\bibitem[\protect\citeauthoryear{{Bresolin}}{{Bresolin}}{2011}]{Bresolin:2011aa}
{Bresolin} F.,  2011, \mn@doi [\apj] {10.1088/0004-637X/730/2/129}, \href
  {https://ui.adsabs.harvard.edu/abs/2011ApJ...730..129B} {730, 129}

\bibitem[\protect\citeauthoryear{{Bresolin}, {Kennicutt}  \&
  {Ryan-Weber}}{{Bresolin} et~al.}{2012}]{Bresolin:2012aa}
{Bresolin} F.,  {Kennicutt} R.~C.,   {Ryan-Weber} E.,  2012, \mn@doi [\apj]
  {10.1088/0004-637X/750/2/122}, \href
  {https://ui.adsabs.harvard.edu/abs/2012ApJ...750..122B} {750, 122}

\bibitem[\protect\citeauthoryear{{Bresolin}, {Kudritzki}, {Urbaneja}, {Gieren},
  {Ho}  \& {Pietrzy{\'n}ski}}{{Bresolin} et~al.}{2016}]{Bresolin:2016aa}
{Bresolin} F.,  {Kudritzki} R.-P.,  {Urbaneja} M.~A.,  {Gieren} W.,  {Ho}
  I.-T.,   {Pietrzy{\'n}ski} G.,  2016, \mn@doi [\apj]
  {10.3847/0004-637X/830/2/64}, \href
  {http://adsabs.harvard.edu/abs/2016ApJ...830...64B} {830, 64}

\bibitem[\protect\citeauthoryear{{Bruzual} \& {Charlot}}{{Bruzual} \&
  {Charlot}}{2003}]{Bruzual:2003aa}
{Bruzual} G.,  {Charlot} S.,  2003, \mn@doi [\mnras]
  {10.1046/j.1365-8711.2003.06897.x}, \href
  {http://adsabs.harvard.edu/abs/2003MNRAS.344.1000B} {344, 1000}

\bibitem[\protect\citeauthoryear{{Calzetti}, {Armus}, {Bohlin}, {Kinney},
  {Koornneef}  \& {Storchi-Bergmann}}{{Calzetti}
  et~al.}{2000}]{Calzetti:2000aa}
{Calzetti} D.,  {Armus} L.,  {Bohlin} R.~C.,  {Kinney} A.~L.,  {Koornneef} J.,
   {Storchi-Bergmann} T.,  2000, \mn@doi [\apj] {10.1086/308692}, \href
  {http://adsabs.harvard.edu/abs/2000ApJ...533..682C} {533, 682}

\bibitem[\protect\citeauthoryear{{Cardelli}, {Clayton}  \& {Mathis}}{{Cardelli}
  et~al.}{1989}]{Cardelli:1989aa}
{Cardelli} J.~A.,  {Clayton} G.~C.,   {Mathis} J.~S.,  1989, \mn@doi [\apj]
  {10.1086/167900}, \href {http://adsabs.harvard.edu/abs/1989ApJ...345..245C}
  {345, 245}

\bibitem[\protect\citeauthoryear{{Carton} et~al.,}{{Carton}
  et~al.}{2018}]{Carton:2018aa}
{Carton} D.,  et~al., 2018, \mn@doi [\mnras] {10.1093/mnras/sty1343}, \href
  {http://adsabs.harvard.edu/abs/2018MNRAS.478.4293C} {478, 4293}

\bibitem[\protect\citeauthoryear{{Ceverino}, {S{\'a}nchez Almeida}, {Mu{\~n}oz
  Tu{\~n}{\'o}n}, {Dekel}, {Elmegreen}, {Elmegreen}  \& {Primack}}{{Ceverino}
  et~al.}{2016}]{Ceverino:2016aa}
{Ceverino} D.,  {S{\'a}nchez Almeida} J.,  {Mu{\~n}oz Tu{\~n}{\'o}n} C.,
  {Dekel} A.,  {Elmegreen} B.~G.,  {Elmegreen} D.~M.,   {Primack} J.,  2016,
  \mn@doi [\mnras] {10.1093/mnras/stw064}, \href
  {http://adsabs.harvard.edu/abs/2016MNRAS.457.2605C} {457, 2605}

\bibitem[\protect\citeauthoryear{{Chabrier}}{{Chabrier}}{2003}]{Chabrier:2003aa}
{Chabrier} G.,  2003, \mn@doi [\pasp] {10.1086/376392}, \href
  {http://adsabs.harvard.edu/abs/2003PASP..115..763C} {115, 763}

\bibitem[\protect\citeauthoryear{{Charlot} \& {Fall}}{{Charlot} \&
  {Fall}}{2000}]{Charlot:2000aa}
{Charlot} S.,  {Fall} S.~M.,  2000, \mn@doi [\apj] {10.1086/309250}, \href
  {https://ui.adsabs.harvard.edu/abs/2000ApJ...539..718C} {539, 718}

\bibitem[\protect\citeauthoryear{{Conroy}}{{Conroy}}{2013}]{Conroy:2013aa}
{Conroy} C.,  2013, \mn@doi [\araa] {10.1146/annurev-astro-082812-141017},
  \href {http://adsabs.harvard.edu/abs/2013ARA%26A..51..393C} {51, 393}

\bibitem[\protect\citeauthoryear{{Cresci}, {Mannucci}, {Maiolino}, {Marconi},
  {Gnerucci}  \& {Magrini}}{{Cresci} et~al.}{2010}]{Cresci:2010aa}
{Cresci} G.,  {Mannucci} F.,  {Maiolino} R.,  {Marconi} A.,  {Gnerucci} A.,
  {Magrini} L.,  2010, \mn@doi [\nat] {10.1038/nature09451}, \href
  {http://adsabs.harvard.edu/abs/2010Natur.467..811C} {467, 811}

\bibitem[\protect\citeauthoryear{{Cresci}, {Mannucci}  \& {Curti}}{{Cresci}
  et~al.}{2018}]{Cresci:2018aa}
{Cresci} G.,  {Mannucci} F.,   {Curti} M.,  2018, arXiv e-prints, \href
  {http://adsabs.harvard.edu/abs/2018arXiv181106015C} {}

\bibitem[\protect\citeauthoryear{{Cullen}, {Cirasuolo}, {McLure}, {Dunlop}  \&
  {Bowler}}{{Cullen} et~al.}{2014}]{Cullen:2014aa}
{Cullen} F.,  {Cirasuolo} M.,  {McLure} R.~J.,  {Dunlop} J.~S.,   {Bowler}
  R.~A.~A.,  2014, \mn@doi [\mnras] {10.1093/mnras/stu443}, \href
  {http://adsabs.harvard.edu/abs/2014MNRAS.440.2300C} {440, 2300}

\bibitem[\protect\citeauthoryear{{Curti}, {Cresci}, {Mannucci}, {Marconi},
  {Maiolino}  \& {Esposito}}{{Curti} et~al.}{2017}]{Curti:2017aa}
{Curti} M.,  {Cresci} G.,  {Mannucci} F.,  {Marconi} A.,  {Maiolino} R.,
  {Esposito} S.,  2017, \mn@doi [\mnras] {10.1093/mnras/stw2766}, \href
  {http://adsabs.harvard.edu/abs/2017MNRAS.465.1384C} {465, 1384}

\bibitem[\protect\citeauthoryear{{Curti}, {Mannucci}, {Cresci}  \&
  {Maiolino}}{{Curti} et~al.}{2019}]{Curti:2019aa}
{Curti} M.,  {Mannucci} F.,  {Cresci} G.,   {Maiolino} R.,  2019, arXiv
  e-prints, \href {https://ui.adsabs.harvard.edu/abs/2019arXiv191000597C} {}

\bibitem[\protect\citeauthoryear{{Dav{\'e}}, {Finlator}  \&
  {Oppenheimer}}{{Dav{\'e}} et~al.}{2011}]{Dave:2011aa}
{Dav{\'e}} R.,  {Finlator} K.,   {Oppenheimer} B.~D.,  2011, \mn@doi [\mnras]
  {10.1111/j.1365-2966.2011.19132.x}, \href
  {http://adsabs.harvard.edu/abs/2011MNRAS.416.1354D} {416, 1354}

\bibitem[\protect\citeauthoryear{{Dav{\'e}}, {Rafieferantsoa}, {Thompson}  \&
  {Hopkins}}{{Dav{\'e}} et~al.}{2017}]{Dave:2017aa}
{Dav{\'e}} R.,  {Rafieferantsoa} M.~H.,  {Thompson} R.~J.,   {Hopkins} P.~F.,
  2017, \mn@doi [\mnras] {10.1093/mnras/stx108}, \href
  {http://adsabs.harvard.edu/abs/2017MNRAS.467..115D} {467, 115}

\bibitem[\protect\citeauthoryear{{Davies}}{{Davies}}{2007}]{Davies:2007aa}
{Davies} R.~I.,  2007, \mn@doi [\mnras] {10.1111/j.1365-2966.2006.11383.x},
  \href {http://adsabs.harvard.edu/abs/2007MNRAS.375.1099D} {375, 1099}

\bibitem[\protect\citeauthoryear{{Davies} et~al.,}{{Davies}
  et~al.}{2018}]{Davies:2018ab}
{Davies} R.,  et~al., 2018, in Ground-based and Airborne Instrumentation for
  Astronomy VII. p. 1070209 (\mn@eprint {arXiv} {1807.05089}),
  \mn@doi{10.1117/12.2311480}

\bibitem[\protect\citeauthoryear{{Dayal}, {Ferrara}  \& {Dunlop}}{{Dayal}
  et~al.}{2013}]{Dayal:2013aa}
{Dayal} P.,  {Ferrara} A.,   {Dunlop} J.~S.,  2013, \mn@doi [\mnras]
  {10.1093/mnras/stt083}, \href
  {http://adsabs.harvard.edu/abs/2013MNRAS.430.2891D} {430, 2891}

\bibitem[\protect\citeauthoryear{{Dekel} et~al.,}{{Dekel}
  et~al.}{2009}]{Dekel:2009aa}
{Dekel} A.,  et~al., 2009, \mn@doi [\nat] {10.1038/nature07648}, \href
  {http://adsabs.harvard.edu/abs/2009Natur.457..451D} {457, 451}

\bibitem[\protect\citeauthoryear{{Dopita}, {Kewley}, {Sutherland}  \&
  {Nicholls}}{{Dopita} et~al.}{2016}]{Dopita:2016aa}
{Dopita} M.~A.,  {Kewley} L.~J.,  {Sutherland} R.~S.,   {Nicholls} D.~C.,
  2016, \mn@doi [\apss] {10.1007/s10509-016-2657-8}, \href
  {http://adsabs.harvard.edu/abs/2016Ap%26SS.361...61D} {361, 61}

\bibitem[\protect\citeauthoryear{{Ellison}, {Patton}, {Simard}  \&
  {McConnachie}}{{Ellison} et~al.}{2008}]{Ellison:2008aa}
{Ellison} S.~L.,  {Patton} D.~R.,  {Simard} L.,   {McConnachie} A.~W.,  2008,
  \mn@doi [\apjl] {10.1086/527296}, \href
  {http://adsabs.harvard.edu/abs/2008ApJ...672L.107E} {672, L107}

\bibitem[\protect\citeauthoryear{{Elmegreen}}{{Elmegreen}}{2009}]{Elmegreen:2009aa}
{Elmegreen} B.~G.,  2009, in {Jogee} S.,  {Marinova} I.,  {Hao} L.,   {Blanc}
  G.~A.,  eds,  Astronomical Society of the Pacific Conference Series Vol. 419,
  Galaxy Evolution: Emerging Insights and Future Challenges. p.~23 (\mn@eprint
  {arXiv} {0903.1937})

\bibitem[\protect\citeauthoryear{{Elmegreen}, {Bournaud}  \&
  {Elmegreen}}{{Elmegreen} et~al.}{2008}]{Elmegreen:2008aa}
{Elmegreen} B.~G.,  {Bournaud} F.,   {Elmegreen} D.~M.,  2008, \mn@doi [\apj]
  {10.1086/592190}, \href {http://adsabs.harvard.edu/abs/2008ApJ...688...67E}
  {688, 67}

\bibitem[\protect\citeauthoryear{{Erb}, {Shapley}, {Pettini}, {Steidel},
  {Reddy}  \& {Adelberger}}{{Erb} et~al.}{2006}]{Erb:2006ad}
{Erb} D.~K.,  {Shapley} A.~E.,  {Pettini} M.,  {Steidel} C.~C.,  {Reddy} N.~A.,
    {Adelberger} K.~L.,  2006, \mn@doi [\apj] {10.1086/503623}, \href
  {http://adsabs.harvard.edu/abs/2006ApJ...644..813E} {644, 813}

\bibitem[\protect\citeauthoryear{{F{\"o}rster Schreiber} et~al.,}{{F{\"o}rster
  Schreiber} et~al.}{2011}]{Forster-Schreiber:2011aa}
{F{\"o}rster Schreiber} N.~M.,  et~al., 2011, \mn@doi [\apj]
  {10.1088/0004-637X/739/1/45}, \href
  {http://adsabs.harvard.edu/abs/2011ApJ...739...45F} {739, 45}

\bibitem[\protect\citeauthoryear{{F{\"o}rster Schreiber} et~al.,}{{F{\"o}rster
  Schreiber} et~al.}{2018}]{Forster-Schreiber:2018ab}
{F{\"o}rster Schreiber} N.~M.,  et~al., 2018, \mn@doi [\apjs]
  {10.3847/1538-4365/aadd49}, \href
  {https://ui.adsabs.harvard.edu/abs/2018ApJS..238...21F} {238, 21}

\bibitem[\protect\citeauthoryear{{Fraternali} \& {Binney}}{{Fraternali} \&
  {Binney}}{2008}]{Fraternali:2008aa}
{Fraternali} F.,  {Binney} J.~J.,  2008, \mn@doi [\mnras]
  {10.1111/j.1365-2966.2008.13071.x}, \href
  {http://adsabs.harvard.edu/abs/2008MNRAS.386..935F} {386, 935}

\bibitem[\protect\citeauthoryear{{Gazak} et~al.,}{{Gazak}
  et~al.}{2015}]{Gazak:2015aa}
{Gazak} J.~Z.,  et~al., 2015, \mn@doi [\apj] {10.1088/0004-637X/805/2/182},
  \href {https://ui.adsabs.harvard.edu/abs/2015ApJ...805..182G} {805, 182}

\bibitem[\protect\citeauthoryear{{Genzel} et~al.,}{{Genzel}
  et~al.}{2008}]{Genzel:2008aa}
{Genzel} R.,  et~al., 2008, \mn@doi [\apj] {10.1086/591840}, \href
  {http://adsabs.harvard.edu/abs/2008ApJ...687...59G} {687, 59}

\bibitem[\protect\citeauthoryear{{Genzel} et~al.,}{{Genzel}
  et~al.}{2011}]{Genzel:2011aa}
{Genzel} R.,  et~al., 2011, \mn@doi [\apj] {10.1088/0004-637X/733/2/101}, \href
  {http://adsabs.harvard.edu/abs/2011ApJ...733..101G} {733, 101}

\bibitem[\protect\citeauthoryear{{Gibson}, {Courty}, {Cunnama}  \&
  {Moll{\'a}}}{{Gibson} et~al.}{2013a}]{Gibson:2013aa}
{Gibson} B.~K.,  {Courty} S.,  {Cunnama} D.,   {Moll{\'a}} M.,  2013a,
  Asociacion Argentina de Astronomia La Plata Argentina Book Series, \href
  {http://adsabs.harvard.edu/abs/2013AAABS...4...57G} {4, 57}

\bibitem[\protect\citeauthoryear{{Gibson}, {Pilkington}, {Brook}, {Stinson}  \&
  {Bailin}}{{Gibson} et~al.}{2013b}]{Gibson:2013ab}
{Gibson} B.~K.,  {Pilkington} K.,  {Brook} C.~B.,  {Stinson} G.~S.,   {Bailin}
  J.,  2013b, \mn@doi [\aap] {10.1051/0004-6361/201321239}, \href
  {http://adsabs.harvard.edu/abs/2013A%26A...554A..47G} {554, A47}

\bibitem[\protect\citeauthoryear{{Guo} et~al.,}{{Guo}
  et~al.}{2016}]{Guo:2016aa}
{Guo} Y.,  et~al., 2016, \mn@doi [\apj] {10.3847/0004-637X/822/2/103}, \href
  {http://adsabs.harvard.edu/abs/2016ApJ...822..103G} {822, 103}

\bibitem[\protect\citeauthoryear{{Henry}, {Kwitter}, {Jaskot}, {Balick},
  {Morrison}  \& {Milingo}}{{Henry} et~al.}{2010}]{Henry:2010aa}
{Henry} R.~B.~C.,  {Kwitter} K.~B.,  {Jaskot} A.~E.,  {Balick} B.,  {Morrison}
  M.~A.,   {Milingo} J.~B.,  2010, \mn@doi [\apj]
  {10.1088/0004-637X/724/1/748}, \href
  {http://adsabs.harvard.edu/abs/2010ApJ...724..748H} {724, 748}

\bibitem[\protect\citeauthoryear{{Ho}, {Kudritzki}, {Kewley}, {Zahid},
  {Dopita}, {Bresolin}  \& {Rupke}}{{Ho} et~al.}{2015}]{Ho:2015aa}
{Ho} I.-T.,  {Kudritzki} R.-P.,  {Kewley} L.~J.,  {Zahid} H.~J.,  {Dopita}
  M.~A.,  {Bresolin} F.,   {Rupke} D.~S.~N.,  2015, \mn@doi [\mnras]
  {10.1093/mnras/stv067}, \href
  {http://adsabs.harvard.edu/abs/2015MNRAS.448.2030H} {448, 2030}

\bibitem[\protect\citeauthoryear{{Hunt}, {Dayal}, {Magrini}  \&
  {Ferrara}}{{Hunt} et~al.}{2016}]{Hunt:2016ab}
{Hunt} L.,  {Dayal} P.,  {Magrini} L.,   {Ferrara} A.,  2016, \mn@doi [\mnras]
  {10.1093/mnras/stw2091}, \href
  {http://adsabs.harvard.edu/abs/2016MNRAS.463.2020H} {463, 2020}

\bibitem[\protect\citeauthoryear{{Jones}, {Swinbank}, {Ellis}, {Richard}  \&
  {Stark}}{{Jones} et~al.}{2010a}]{Jones:2010aa}
{Jones} T.~A.,  {Swinbank} A.~M.,  {Ellis} R.~S.,  {Richard} J.,   {Stark}
  D.~P.,  2010a, \mn@doi [\mnras] {10.1111/j.1365-2966.2010.16378.x}, \href
  {http://adsabs.harvard.edu/abs/2010MNRAS.404.1247J} {404, 1247}

\bibitem[\protect\citeauthoryear{{Jones}, {Ellis}, {Jullo}  \&
  {Richard}}{{Jones} et~al.}{2010b}]{Jones:2010ab}
{Jones} T.,  {Ellis} R.,  {Jullo} E.,   {Richard} J.,  2010b, \mn@doi [\apjl]
  {10.1088/2041-8205/725/2/L176}, \href
  {http://adsabs.harvard.edu/abs/2010ApJ...725L.176J} {725, L176}

\bibitem[\protect\citeauthoryear{{Jones}, {Ellis}, {Richard}  \&
  {Jullo}}{{Jones} et~al.}{2013}]{Jones:2013aa}
{Jones} T.,  {Ellis} R.~S.,  {Richard} J.,   {Jullo} E.,  2013, \mn@doi [\apj]
  {10.1088/0004-637X/765/1/48}, \href
  {https://ui.adsabs.harvard.edu/abs/2013ApJ...765...48J} {765, 48}

\bibitem[\protect\citeauthoryear{{Jones} et~al.,}{{Jones}
  et~al.}{2015a}]{Jones:2015ab}
{Jones} T.,  et~al., 2015a, \mn@doi [\aj] {10.1088/0004-6256/149/3/107}, \href
  {http://adsabs.harvard.edu/abs/2015AJ....149..107J} {149, 107}

\bibitem[\protect\citeauthoryear{{Jones}, {Martin}  \& {Cooper}}{{Jones}
  et~al.}{2015b}]{Jones:2015aa}
{Jones} T.,  {Martin} C.,   {Cooper} M.~C.,  2015b, \mn@doi [\apj]
  {10.1088/0004-637X/813/2/126}, \href
  {http://adsabs.harvard.edu/abs/2015ApJ...813..126J} {813, 126}

\bibitem[\protect\citeauthoryear{{Kashino} et~al.,}{{Kashino}
  et~al.}{2017}]{Kashino:2017aa}
{Kashino} D.,  et~al., 2017, \mn@doi [\apj] {10.3847/1538-4357/835/1/88}, \href
  {https://ui.adsabs.harvard.edu/abs/2017ApJ...835...88K} {835, 88}

\bibitem[\protect\citeauthoryear{{Kashino} et~al.,}{{Kashino}
  et~al.}{2019}]{Kashino:2019aa}
{Kashino} D.,  et~al., 2019, \mn@doi [\apjs] {10.3847/1538-4365/ab06c4}, \href
  {https://ui.adsabs.harvard.edu/abs/2019ApJS..241...10K} {241, 10}

\bibitem[\protect\citeauthoryear{{Kauffmann} et~al.,}{{Kauffmann}
  et~al.}{2003}]{Kauffmann:2003aa}
{Kauffmann} G.,  et~al., 2003, \mn@doi [\mnras]
  {10.1111/j.1365-2966.2003.07154.x}, \href
  {http://adsabs.harvard.edu/abs/2003MNRAS.346.1055K} {346, 1055}

\bibitem[\protect\citeauthoryear{{Kennicutt} \& {Evans}}{{Kennicutt} \&
  {Evans}}{2012}]{Kennicutt:2012aa}
{Kennicutt} R.~C.,  {Evans} N.~J.,  2012, \mn@doi [\araa]
  {10.1146/annurev-astro-081811-125610}, \href
  {http://adsabs.harvard.edu/abs/2012ARA%26A..50..531K} {50, 531}

\bibitem[\protect\citeauthoryear{{Kewley}, {Dopita}, {Sutherland}, {Heisler}
  \& {Trevena}}{{Kewley} et~al.}{2001}]{Kewley:2001aa}
{Kewley} L.~J.,  {Dopita} M.~A.,  {Sutherland} R.~S.,  {Heisler} C.~A.,
  {Trevena} J.,  2001, \mn@doi [\apj] {10.1086/321545}, \href
  {http://adsabs.harvard.edu/abs/2001ApJ...556..121K} {556, 121}

\bibitem[\protect\citeauthoryear{{Kewley}, {Rupke}, {Zahid}, {Geller}  \&
  {Barton}}{{Kewley} et~al.}{2010}]{Kewley:2010aa}
{Kewley} L.~J.,  {Rupke} D.,  {Zahid} H.~J.,  {Geller} M.~J.,   {Barton} E.~J.,
   2010, \mn@doi [\apjl] {10.1088/2041-8205/721/1/L48}, \href
  {https://ui.adsabs.harvard.edu/abs/2010ApJ...721L..48K} {721, L48}

\bibitem[\protect\citeauthoryear{{Kewley}, {Maier}, {Yabe}, {Ohta}, {Akiyama},
  {Dopita}  \& {Yuan}}{{Kewley} et~al.}{2013}]{Kewley:2013aa}
{Kewley} L.~J.,  {Maier} C.,  {Yabe} K.,  {Ohta} K.,  {Akiyama} M.,  {Dopita}
  M.~A.,   {Yuan} T.,  2013, \mn@doi [\apjl] {10.1088/2041-8205/774/1/L10},
  \href {https://ui.adsabs.harvard.edu/abs/2013ApJ...774L..10K} {774, L10}

\bibitem[\protect\citeauthoryear{{Kostrzewa-Rutkowska}, {Wyrzykowski}, {Auger},
  {Collett}  \& {Belokurov}}{{Kostrzewa-Rutkowska}
  et~al.}{2014}]{Kostrzewa-Rutkowska:2014aa}
{Kostrzewa-Rutkowska} Z.,  {Wyrzykowski} {\L}.,  {Auger} M.~W.,  {Collett}
  T.~E.,   {Belokurov} V.,  2014, \mn@doi [\mnras] {10.1093/mnras/stu783},
  \href {http://adsabs.harvard.edu/abs/2014MNRAS.441.3238K} {441, 3238}

\bibitem[\protect\citeauthoryear{{Kroupa}, {Tout}  \& {Gilmore}}{{Kroupa}
  et~al.}{1993}]{Kroupa:1993aa}
{Kroupa} P.,  {Tout} C.~A.,   {Gilmore} G.,  1993, \mn@doi [\mnras]
  {10.1093/mnras/262.3.545}, \href
  {http://adsabs.harvard.edu/abs/1993MNRAS.262..545K} {262, 545}

\bibitem[\protect\citeauthoryear{{Kudritzki}, {Ho}, {Schruba}, {Burkert},
  {Zahid}, {Bresolin}  \& {Dima}}{{Kudritzki} et~al.}{2015}]{Kudritzki:2015aa}
{Kudritzki} R.-P.,  {Ho} I.-T.,  {Schruba} A.,  {Burkert} A.,  {Zahid} H.~J.,
  {Bresolin} F.,   {Dima} G.~I.,  2015, \mn@doi [\mnras]
  {10.1093/mnras/stv522}, \href
  {http://adsabs.harvard.edu/abs/2015MNRAS.450..342K} {450, 342}

\bibitem[\protect\citeauthoryear{{Leethochawalit}, {Jones}, {Ellis}, {Stark},
  {Richard}, {Zitrin}  \& {Auger}}{{Leethochawalit}
  et~al.}{2016}]{Leethochawalit:2016aa}
{Leethochawalit} N.,  {Jones} T.~A.,  {Ellis} R.~S.,  {Stark} D.~P.,  {Richard}
  J.,  {Zitrin} A.,   {Auger} M.,  2016, \mn@doi [\apj]
  {10.3847/0004-637X/820/2/84}, \href
  {http://adsabs.harvard.edu/abs/2016ApJ...820...84L} {820, 84}

\bibitem[\protect\citeauthoryear{{Li} et~al.,}{{Li} et~al.}{2018}]{Li:2018aa}
{Li} H.,  et~al., 2018, \mn@doi [\mnras] {10.1093/mnras/sty334}, \href
  {http://adsabs.harvard.edu/abs/2018MNRAS.476.1765L} {476, 1765}

\bibitem[\protect\citeauthoryear{{Lilly}, {Carollo}, {Pipino}, {Renzini}  \&
  {Peng}}{{Lilly} et~al.}{2013}]{Lilly:2013aa}
{Lilly} S.~J.,  {Carollo} C.~M.,  {Pipino} A.,  {Renzini} A.,   {Peng} Y.,
  2013, \mn@doi [\apj] {10.1088/0004-637X/772/2/119}, \href
  {http://adsabs.harvard.edu/abs/2013ApJ...772..119L} {772, 119}

\bibitem[\protect\citeauthoryear{{Lotz} et~al.,}{{Lotz}
  et~al.}{2017}]{Lotz:2017aa}
{Lotz} J.~M.,  et~al., 2017, \mn@doi [\apj] {10.3847/1538-4357/837/1/97}, \href
  {http://adsabs.harvard.edu/abs/2017ApJ...837...97L} {837, 97}

\bibitem[\protect\citeauthoryear{{Ma}, {Hopkins}, {Faucher-Gigu{\`e}re},
  {Zolman}, {Muratov}, {Kere{\v s}}  \& {Quataert}}{{Ma}
  et~al.}{2016}]{Ma:2016aa}
{Ma} X.,  {Hopkins} P.~F.,  {Faucher-Gigu{\`e}re} C.-A.,  {Zolman} N.,
  {Muratov} A.~L.,  {Kere{\v s}} D.,   {Quataert} E.,  2016, \mn@doi [\mnras]
  {10.1093/mnras/stv2659}, \href
  {http://adsabs.harvard.edu/abs/2016MNRAS.456.2140M} {456, 2140}

\bibitem[\protect\citeauthoryear{{Ma}, {Hopkins}, {Feldmann}, {Torrey},
  {Faucher-Gigu{\`e}re}  \& {Kere{\v s}}}{{Ma} et~al.}{2017}]{Ma:2017aa}
{Ma} X.,  {Hopkins} P.~F.,  {Feldmann} R.,  {Torrey} P.,  {Faucher-Gigu{\`e}re}
  C.-A.,   {Kere{\v s}} D.,  2017, \mn@doi [\mnras] {10.1093/mnras/stx034},
  \href {http://adsabs.harvard.edu/abs/2017MNRAS.466.4780M} {466, 4780}

\bibitem[\protect\citeauthoryear{{Maciel}, {Costa}  \& {Uchida}}{{Maciel}
  et~al.}{2003}]{Maciel:2003aa}
{Maciel} W.~J.,  {Costa} R.~D.~D.,   {Uchida} M.~M.~M.,  2003, \mn@doi [\aap]
  {10.1051/0004-6361:20021530}, \href
  {http://adsabs.harvard.edu/abs/2003A%26A...397..667M} {397, 667}

\bibitem[\protect\citeauthoryear{{Magrini}, {Stanghellini}  \&
  {Villaver}}{{Magrini} et~al.}{2009}]{Magrini:2009aa}
{Magrini} L.,  {Stanghellini} L.,   {Villaver} E.,  2009, \mn@doi [\apj]
  {10.1088/0004-637X/696/1/729}, \href
  {http://adsabs.harvard.edu/abs/2009ApJ...696..729M} {696, 729}

\bibitem[\protect\citeauthoryear{{Magrini}, {Stanghellini}, {Corbelli}, {Galli}
   \& {Villaver}}{{Magrini} et~al.}{2010}]{Magrini:2010aa}
{Magrini} L.,  {Stanghellini} L.,  {Corbelli} E.,  {Galli} D.,   {Villaver} E.,
   2010, \mn@doi [\aap] {10.1051/0004-6361/200913564}, \href
  {http://adsabs.harvard.edu/abs/2010A%26A...512A..63M} {512, A63}

\bibitem[\protect\citeauthoryear{{Maiolino} \& {Mannucci}}{{Maiolino} \&
  {Mannucci}}{2019}]{Maiolino:2019aa}
{Maiolino} R.,  {Mannucci} F.,  2019, \mn@doi [\aapr]
  {10.1007/s00159-018-0112-2}, \href
  {http://adsabs.harvard.edu/abs/2019A%26ARv..27....3M} {27, 3}

\bibitem[\protect\citeauthoryear{{Maiolino} et~al.,}{{Maiolino}
  et~al.}{2008}]{Maiolino:2008aa}
{Maiolino} R.,  et~al., 2008, \mn@doi [\aap] {10.1051/0004-6361:200809678},
  \href {http://adsabs.harvard.edu/abs/2008A%26A...488..463M} {488, 463}

\bibitem[\protect\citeauthoryear{{Mannucci} et~al.,}{{Mannucci}
  et~al.}{2009}]{Mannucci:2009aa}
{Mannucci} F.,  et~al., 2009, \mn@doi [\mnras]
  {10.1111/j.1365-2966.2009.15185.x}, \href
  {http://adsabs.harvard.edu/abs/2009MNRAS.398.1915M} {398, 1915}

\bibitem[\protect\citeauthoryear{{Mannucci}, {Cresci}, {Maiolino}, {Marconi}
  \& {Gnerucci}}{{Mannucci} et~al.}{2010}]{Mannucci:2010aa}
{Mannucci} F.,  {Cresci} G.,  {Maiolino} R.,  {Marconi} A.,   {Gnerucci} A.,
  2010, \mn@doi [\mnras] {10.1111/j.1365-2966.2010.17291.x}, \href
  {http://adsabs.harvard.edu/abs/2010MNRAS.408.2115M} {408, 2115}

\bibitem[\protect\citeauthoryear{{Masters} et~al.,}{{Masters}
  et~al.}{2014}]{Masters:2014aa}
{Masters} D.,  et~al., 2014, \mn@doi [\apj] {10.1088/0004-637X/785/2/153},
  \href {http://adsabs.harvard.edu/abs/2014ApJ...785..153M} {785, 153}

\bibitem[\protect\citeauthoryear{{Masters}, {Faisst}  \& {Capak}}{{Masters}
  et~al.}{2016}]{Masters:2016ab}
{Masters} D.,  {Faisst} A.,   {Capak} P.,  2016, \mn@doi [\apj]
  {10.3847/0004-637X/828/1/18}, \href
  {http://adsabs.harvard.edu/abs/2016ApJ...828...18M} {828, 18}

\bibitem[\protect\citeauthoryear{{Meneghetti} et~al.,}{{Meneghetti}
  et~al.}{2017}]{Meneghetti:2017aa}
{Meneghetti} M.,  et~al., 2017, \mn@doi [\mnras] {10.1093/mnras/stx2064}, \href
  {https://ui.adsabs.harvard.edu/abs/2017MNRAS.472.3177M} {472, 3177}

\bibitem[\protect\citeauthoryear{{Moll{\'a}} \& {D{\'{\i}}az}}{{Moll{\'a}} \&
  {D{\'{\i}}az}}{2005}]{Molla:2005aa}
{Moll{\'a}} M.,  {D{\'{\i}}az} A.~I.,  2005, \mn@doi [\mnras]
  {10.1111/j.1365-2966.2005.08782.x}, \href
  {http://adsabs.harvard.edu/abs/2005MNRAS.358..521M} {358, 521}

\bibitem[\protect\citeauthoryear{{Moll{\'a}}, {D{\'{\i}}az}, {Ascasibar}  \&
  {Gibson}}{{Moll{\'a}} et~al.}{2017}]{Molla:2017aa}
{Moll{\'a}} M.,  {D{\'{\i}}az} {\'A}.~I.,  {Ascasibar} Y.,   {Gibson} B.~K.,
  2017, \mn@doi [\mnras] {10.1093/mnras/stx419}, \href
  {http://adsabs.harvard.edu/abs/2017MNRAS.468..305M} {468, 305}

\bibitem[\protect\citeauthoryear{{Moll{\'a}}, {D{\'{\i}}az}, {Cavichia},
  {Gibson}, {Maciel}, {Costa}, {Ascasibar}  \& {Few}}{{Moll{\'a}}
  et~al.}{2018}]{Molla:2018aa}
{Moll{\'a}} M.,  {D{\'{\i}}az} {\'A}.~I.,  {Cavichia} O.,  {Gibson} B.~K.,
  {Maciel} W.~J.,  {Costa} R.~D.~D.,  {Ascasibar} Y.,   {Few} C.~G.,  2018,
  preprint, \href {http://adsabs.harvard.edu/abs/2018arXiv181009182M} {}
  (\mn@eprint {arXiv} {1810.09182})

\bibitem[\protect\citeauthoryear{{Moll{\'a}}, {D{\'\i}az}, {Cavichia},
  {Gibson}, {Maciel}, {Costa}, {Ascasibar}  \& {Few}}{{Moll{\'a}}
  et~al.}{2019}]{Molla:2019aa}
{Moll{\'a}} M.,  {D{\'\i}az} {\'A}.~I.,  {Cavichia} O.,  {Gibson} B.~K.,
  {Maciel} W.~J.,  {Costa} R.~D.~D.,  {Ascasibar} Y.,   {Few} C.~G.,  2019,
  \mn@doi [\mnras] {10.1093/mnras/sty2877}, \href
  {https://ui.adsabs.harvard.edu/abs/2019MNRAS.482.3071M} {482, 3071}

\bibitem[\protect\citeauthoryear{{Mott}, {Spitoni}  \& {Matteucci}}{{Mott}
  et~al.}{2013}]{Mott:2013aa}
{Mott} A.,  {Spitoni} E.,   {Matteucci} F.,  2013, \mn@doi [\mnras]
  {10.1093/mnras/stt1495}, \href
  {http://adsabs.harvard.edu/abs/2013MNRAS.435.2918M} {435, 2918}

\bibitem[\protect\citeauthoryear{{Nakajima}, {Ouchi}, {Shimasaku}, {Hashimoto},
  {Ono}  \& {Lee}}{{Nakajima} et~al.}{2013}]{Nakajima:2013aa}
{Nakajima} K.,  {Ouchi} M.,  {Shimasaku} K.,  {Hashimoto} T.,  {Ono} Y.,
  {Lee} J.~C.,  2013, \mn@doi [\apj] {10.1088/0004-637X/769/1/3}, \href
  {http://adsabs.harvard.edu/abs/2013ApJ...769....3N} {769, 3}

\bibitem[\protect\citeauthoryear{{Patr{\'{\i}}cio}, {Christensen}, {Rhodin},
  {Ca{\~n}ameras}  \& {Lara-L{\'o}pez}}{{Patr{\'{\i}}cio}
  et~al.}{2018}]{Patricio:2018aa}
{Patr{\'{\i}}cio} V.,  {Christensen} L.,  {Rhodin} H.,  {Ca{\~n}ameras} R.,
  {Lara-L{\'o}pez} M.~A.,  2018, \mn@doi [\mnras] {10.1093/mnras/sty2508},
  \href {http://adsabs.harvard.edu/abs/2018MNRAS.481.3520P} {481, 3520}

\bibitem[\protect\citeauthoryear{{Pilkington} et~al.,}{{Pilkington}
  et~al.}{2012}]{Pilkington:2012aa}
{Pilkington} K.,  et~al., 2012, \mn@doi [\mnras]
  {10.1111/j.1365-2966.2012.21353.x}, \href
  {http://adsabs.harvard.edu/abs/2012MNRAS.425..969P} {425, 969}

\bibitem[\protect\citeauthoryear{{Planck Collaboration} et~al.,}{{Planck
  Collaboration} et~al.}{2016}]{Planck-Collaboration:2016aa}
{Planck Collaboration} et~al., 2016, \mn@doi [\aap]
  {10.1051/0004-6361/201525830}, \href
  {http://adsabs.harvard.edu/abs/2016A%26A...594A..13P} {594, A13}

\bibitem[\protect\citeauthoryear{{Portinari} \& {Chiosi}}{{Portinari} \&
  {Chiosi}}{1999}]{Portinari:1999aa}
{Portinari} L.,  {Chiosi} C.,  1999, \aap, \href
  {http://adsabs.harvard.edu/abs/1999A%26A...350..827P} {350, 827}

\bibitem[\protect\citeauthoryear{{Postman} et~al.,}{{Postman}
  et~al.}{2012}]{Postman:2012aa}
{Postman} M.,  et~al., 2012, \mn@doi [\apjs] {10.1088/0067-0049/199/2/25},
  \href {http://adsabs.harvard.edu/abs/2012ApJS..199...25P} {199, 25}

\bibitem[\protect\citeauthoryear{{Prantzos} \& {Boissier}}{{Prantzos} \&
  {Boissier}}{2000}]{Prantzos:2000aa}
{Prantzos} N.,  {Boissier} S.,  2000, \mn@doi [\mnras]
  {10.1046/j.1365-8711.2000.03228.x}, \href
  {http://adsabs.harvard.edu/abs/2000MNRAS.313..338P} {313, 338}

\bibitem[\protect\citeauthoryear{{Queyrel} et~al.,}{{Queyrel}
  et~al.}{2012}]{Queyrel:2012aa}
{Queyrel} J.,  et~al., 2012, \mn@doi [\aap] {10.1051/0004-6361/201117718},
  \href {http://adsabs.harvard.edu/abs/2012A%26A...539A..93Q} {539, A93}

\bibitem[\protect\citeauthoryear{{Richard}, {Jones}, {Ellis}, {Stark},
  {Livermore}  \& {Swinbank}}{{Richard} et~al.}{2011}]{Richard:2011aa}
{Richard} J.,  {Jones} T.,  {Ellis} R.,  {Stark} D.~P.,  {Livermore} R.,
  {Swinbank} M.,  2011, \mn@doi [\mnras] {10.1111/j.1365-2966.2010.18161.x},
  \href {http://adsabs.harvard.edu/abs/2011MNRAS.413..643R} {413, 643}

\bibitem[\protect\citeauthoryear{{Rosati} et~al.,}{{Rosati}
  et~al.}{2014}]{Rosati:2014aa}
{Rosati} P.,  et~al., 2014, The Messenger, \href
  {http://adsabs.harvard.edu/abs/2014Msngr.158...48R} {158, 48}

\bibitem[\protect\citeauthoryear{{Rousselot}, {Lidman}, {Cuby}, {Moreels}  \&
  {Monnet}}{{Rousselot} et~al.}{2000}]{Rousselot:2000aa}
{Rousselot} P.,  {Lidman} C.,  {Cuby} J.-G.,  {Moreels} G.,   {Monnet} G.,
  2000, \aap, \href {http://adsabs.harvard.edu/abs/2000A%26A...354.1134R} {354,
  1134}

\bibitem[\protect\citeauthoryear{{Rupke}, {Kewley}  \& {Barnes}}{{Rupke}
  et~al.}{2010a}]{Rupke:2010ab}
{Rupke} D.,  {Kewley} L.,   {Barnes} J.,  2010a, in {Smith} B.,  {Higdon} J.,
  {Higdon} S.,   {Bastian} N.,  eds,  Astronomical Society of the Pacific
  Conference Series Vol. 423, Galaxy Wars: Stellar Populations and Star
  Formation in Interacting Galaxies. p.~355

\bibitem[\protect\citeauthoryear{{Rupke}, {Kewley}  \& {Chien}}{{Rupke}
  et~al.}{2010b}]{Rupke:2010aa}
{Rupke} D.~S.~N.,  {Kewley} L.~J.,   {Chien} L.-H.,  2010b, \mn@doi [\apj]
  {10.1088/0004-637X/723/2/1255}, \href
  {https://ui.adsabs.harvard.edu/abs/2010ApJ...723.1255R} {723, 1255}

\bibitem[\protect\citeauthoryear{{Salim}, {Lee}, {Ly}, {Brinchmann},
  {Dav{\'e}}, {Dickinson}, {Salzer}  \& {Charlot}}{{Salim}
  et~al.}{2014}]{Salim:2014aa}
{Salim} S.,  {Lee} J.~C.,  {Ly} C.,  {Brinchmann} J.,  {Dav{\'e}} R.,
  {Dickinson} M.,  {Salzer} J.~J.,   {Charlot} S.,  2014, \mn@doi [\apj]
  {10.1088/0004-637X/797/2/126}, \href
  {http://adsabs.harvard.edu/abs/2014ApJ...797..126S} {797, 126}

\bibitem[\protect\citeauthoryear{{Samland}, {Hensler}  \& {Theis}}{{Samland}
  et~al.}{1997}]{Samland:1997aa}
{Samland} M.,  {Hensler} G.,   {Theis} C.,  1997, \mn@doi [\apj]
  {10.1086/303627}, \href {http://adsabs.harvard.edu/abs/1997ApJ...476..544S}
  {476, 544}

\bibitem[\protect\citeauthoryear{{S{\'a}nchez Almeida}, {Caon},
  {Mu{\~n}oz-Tu{\~n}{\'o}n}, {Filho}  \& {Cervi{\~n}o}}{{S{\'a}nchez Almeida}
  et~al.}{2018}]{Sanchez-Almeida:2018ab}
{S{\'a}nchez Almeida} J.,  {Caon} N.,  {Mu{\~n}oz-Tu{\~n}{\'o}n} C.,  {Filho}
  M.,   {Cervi{\~n}o} M.,  2018, \mn@doi [\mnras] {10.1093/mnras/sty510}, \href
  {http://adsabs.harvard.edu/abs/2018MNRAS.476.4765S} {476, 4765}

\bibitem[\protect\citeauthoryear{{S{\'a}nchez} et~al.,}{{S{\'a}nchez}
  et~al.}{2014}]{Sanchez:2014aa}
{S{\'a}nchez} S.~F.,  et~al., 2014, \mn@doi [\aap]
  {10.1051/0004-6361/201322343}, \href
  {http://adsabs.harvard.edu/abs/2014A%26A...563A..49S} {563, A49}

\bibitem[\protect\citeauthoryear{{Sanders} et~al.,}{{Sanders}
  et~al.}{2015}]{Sanders:2015ab}
{Sanders} R.~L.,  et~al., 2015, \mn@doi [\apj] {10.1088/0004-637X/799/2/138},
  \href {http://adsabs.harvard.edu/abs/2015ApJ...799..138S} {799, 138}

\bibitem[\protect\citeauthoryear{{Sanders} et~al.,}{{Sanders}
  et~al.}{2016}]{Sanders:2016aa}
{Sanders} R.~L.,  et~al., 2016, \mn@doi [\apjl] {10.3847/2041-8205/825/2/L23},
  \href {http://adsabs.harvard.edu/abs/2016ApJ...825L..23S} {825, L23}

\bibitem[\protect\citeauthoryear{{Sanders} et~al.,}{{Sanders}
  et~al.}{2018}]{Sanders:2018aa}
{Sanders} R.~L.,  et~al., 2018, \mn@doi [\apj] {10.3847/1538-4357/aabcbd},
  \href {http://adsabs.harvard.edu/abs/2018ApJ...858...99S} {858, 99}

\bibitem[\protect\citeauthoryear{{Shapley} et~al.,}{{Shapley}
  et~al.}{2015}]{Shapley:2015aa}
{Shapley} A.~E.,  et~al., 2015, \mn@doi [\apj] {10.1088/0004-637X/801/2/88},
  \href {http://adsabs.harvard.edu/abs/2015ApJ...801...88S} {801, 88}

\bibitem[\protect\citeauthoryear{{Sharples} et~al.,}{{Sharples}
  et~al.}{2013}]{Sharples:2013aa}
{Sharples} R.,  et~al., 2013, The Messenger, \href
  {http://adsabs.harvard.edu/abs/2013Msngr.151...21S} {151, 21}

\bibitem[\protect\citeauthoryear{{Shivaei} et~al.,}{{Shivaei}
  et~al.}{2015}]{Shivaei:2015ab}
{Shivaei} I.,  et~al., 2015, \mn@doi [\apj] {10.1088/0004-637X/815/2/98}, \href
  {https://ui.adsabs.harvard.edu/abs/2015ApJ...815...98S} {815, 98}

\bibitem[\protect\citeauthoryear{{Simons} et~al.,}{{Simons}
  et~al.}{2019}]{Simons:2019aa}
{Simons} R.~C.,  et~al., 2019, \mn@doi [\apj] {10.3847/1538-4357/ab07c9}, \href
  {http://adsabs.harvard.edu/abs/2019ApJ...874...59S} {874, 59}

\bibitem[\protect\citeauthoryear{{Somerville} \& {Dav{\'e}}}{{Somerville} \&
  {Dav{\'e}}}{2015}]{Somerville:2015aa}
{Somerville} R.~S.,  {Dav{\'e}} R.,  2015, \mn@doi [\araa]
  {10.1146/annurev-astro-082812-140951}, \href
  {http://adsabs.harvard.edu/abs/2015ARA%26A..53...51S} {53, 51}

\bibitem[\protect\citeauthoryear{{Speagle}, {Steinhardt}, {Capak}  \&
  {Silverman}}{{Speagle} et~al.}{2014}]{Speagle:2014aa}
{Speagle} J.~S.,  {Steinhardt} C.~L.,  {Capak} P.~L.,   {Silverman} J.~D.,
  2014, \mn@doi [\apjs] {10.1088/0067-0049/214/2/15}, \href
  {http://adsabs.harvard.edu/abs/2014ApJS..214...15S} {214, 15}

\bibitem[\protect\citeauthoryear{{Spitoni}, {Matteucci}  \&
  {Marcon-Uchida}}{{Spitoni} et~al.}{2013}]{Spitoni:2013aa}
{Spitoni} E.,  {Matteucci} F.,   {Marcon-Uchida} M.~M.,  2013, \mn@doi [\aap]
  {10.1051/0004-6361/201220401}, \href
  {http://adsabs.harvard.edu/abs/2013A%26A...551A.123S} {551, A123}

\bibitem[\protect\citeauthoryear{{Stanghellini} \& {Haywood}}{{Stanghellini} \&
  {Haywood}}{2010}]{Stanghellini:2010ab}
{Stanghellini} L.,  {Haywood} M.,  2010, \mn@doi [\apj]
  {10.1088/0004-637X/714/2/1096}, \href
  {http://adsabs.harvard.edu/abs/2010ApJ...714.1096S} {714, 1096}

\bibitem[\protect\citeauthoryear{{Stanghellini} \& {Haywood}}{{Stanghellini} \&
  {Haywood}}{2018}]{Stanghellini:2018aa}
{Stanghellini} L.,  {Haywood} M.,  2018, preprint, \href
  {http://adsabs.harvard.edu/abs/2018arXiv180602276S} {} (\mn@eprint {arXiv}
  {1806.02276})

\bibitem[\protect\citeauthoryear{{Stanghellini}, {Magrini}, {Villaver}  \&
  {Galli}}{{Stanghellini} et~al.}{2010}]{Stanghellini:2010aa}
{Stanghellini} L.,  {Magrini} L.,  {Villaver} E.,   {Galli} D.,  2010, \mn@doi
  [\aap] {10.1051/0004-6361/201014911}, \href
  {http://adsabs.harvard.edu/abs/2010A%26A...521A...3S} {521, A3}

\bibitem[\protect\citeauthoryear{{Stanghellini}, {Magrini}, {Casasola}  \&
  {Villaver}}{{Stanghellini} et~al.}{2014}]{Stanghellini:2014aa}
{Stanghellini} L.,  {Magrini} L.,  {Casasola} V.,   {Villaver} E.,  2014,
  \mn@doi [\aap] {10.1051/0004-6361/201423423}, \href
  {http://adsabs.harvard.edu/abs/2014A%26A...567A..88S} {567, A88}

\bibitem[\protect\citeauthoryear{{Steidel} et~al.,}{{Steidel}
  et~al.}{2014}]{Steidel:2014aa}
{Steidel} C.~C.,  et~al., 2014, \mn@doi [\apj] {10.1088/0004-637X/795/2/165},
  \href {http://adsabs.harvard.edu/abs/2014ApJ...795..165S} {795, 165}

\bibitem[\protect\citeauthoryear{{Steidel}, {Strom}, {Pettini}, {Rudie},
  {Reddy}  \& {Trainor}}{{Steidel} et~al.}{2016}]{Steidel:2016aa}
{Steidel} C.~C.,  {Strom} A.~L.,  {Pettini} M.,  {Rudie} G.~C.,  {Reddy} N.~A.,
    {Trainor} R.~F.,  2016, \mn@doi [\apj] {10.3847/0004-637X/826/2/159}, \href
  {http://adsabs.harvard.edu/abs/2016ApJ...826..159S} {826, 159}

\bibitem[\protect\citeauthoryear{{Stinson}, {Bailin}, {Couchman}, {Wadsley},
  {Shen}, {Nickerson}, {Brook}  \& {Quinn}}{{Stinson}
  et~al.}{2010}]{Stinson:2010aa}
{Stinson} G.~S.,  {Bailin} J.,  {Couchman} H.,  {Wadsley} J.,  {Shen} S.,
  {Nickerson} S.,  {Brook} C.,   {Quinn} T.,  2010, \mn@doi [\mnras]
  {10.1111/j.1365-2966.2010.17187.x}, \href
  {http://adsabs.harvard.edu/abs/2010MNRAS.408..812S} {408, 812}

\bibitem[\protect\citeauthoryear{{Stott} et~al.,}{{Stott}
  et~al.}{2014}]{Stott:2014aa}
{Stott} J.~P.,  et~al., 2014, \mn@doi [\mnras] {10.1093/mnras/stu1343}, \href
  {http://adsabs.harvard.edu/abs/2014MNRAS.443.2695S} {443, 2695}

\bibitem[\protect\citeauthoryear{{Strom}, {Steidel}, {Rudie}, {Trainor},
  {Pettini}  \& {Reddy}}{{Strom} et~al.}{2017}]{Strom:2017aa}
{Strom} A.~L.,  {Steidel} C.~C.,  {Rudie} G.~C.,  {Trainor} R.~F.,  {Pettini}
  M.,   {Reddy} N.~A.,  2017, \mn@doi [\apj] {10.3847/1538-4357/836/2/164},
  \href {http://adsabs.harvard.edu/abs/2017ApJ...836..164S} {836, 164}

\bibitem[\protect\citeauthoryear{{Swinbank}, {Sobral}, {Smail}, {Geach},
  {Best}, {McCarthy}, {Crain}  \& {Theuns}}{{Swinbank}
  et~al.}{2012}]{Swinbank:2012aa}
{Swinbank} A.~M.,  {Sobral} D.,  {Smail} I.,  {Geach} J.~E.,  {Best} P.~N.,
  {McCarthy} I.~G.,  {Crain} R.~A.,   {Theuns} T.,  2012, \mn@doi [\mnras]
  {10.1111/j.1365-2966.2012.21774.x}, \href
  {http://adsabs.harvard.edu/abs/2012MNRAS.426..935S} {426, 935}

\bibitem[\protect\citeauthoryear{{Tamburello}, {Rahmati}, {Mayer}, {Cava},
  {Dessauges-Zavadsky}  \& {Schaerer}}{{Tamburello}
  et~al.}{2017}]{Tamburello:2017aa}
{Tamburello} V.,  {Rahmati} A.,  {Mayer} L.,  {Cava} A.,  {Dessauges-Zavadsky}
  M.,   {Schaerer} D.,  2017, \mn@doi [\mnras] {10.1093/mnras/stx784}, \href
  {http://adsabs.harvard.edu/abs/2017MNRAS.468.4792T} {468, 4792}

\bibitem[\protect\citeauthoryear{{Tremonti} et~al.,}{{Tremonti}
  et~al.}{2004}]{Tremonti:2004aa}
{Tremonti} C.~A.,  et~al., 2004, \mn@doi [\apj] {10.1086/423264}, \href
  {http://adsabs.harvard.edu/abs/2004ApJ...613..898T} {613, 898}

\bibitem[\protect\citeauthoryear{{Treu} et~al.,}{{Treu}
  et~al.}{2015}]{Treu:2015aa}
{Treu} T.,  et~al., 2015, \mn@doi [\apj] {10.1088/0004-637X/812/2/114}, \href
  {https://ui.adsabs.harvard.edu/\#abs/2015ApJ...812..114T} {812, 114}

\bibitem[\protect\citeauthoryear{{Troncoso} et~al.,}{{Troncoso}
  et~al.}{2014}]{Troncoso:2014aa}
{Troncoso} P.,  et~al., 2014, \mn@doi [\aap] {10.1051/0004-6361/201322099},
  \href {http://adsabs.harvard.edu/abs/2014A%26A...563A..58T} {563, A58}

\bibitem[\protect\citeauthoryear{{Vegetti} \& {Koopmans}}{{Vegetti} \&
  {Koopmans}}{2009}]{Vegetti:2009aa}
{Vegetti} S.,  {Koopmans} L.~V.~E.,  2009, \mn@doi [\mnras]
  {10.1111/j.1365-2966.2008.14005.x}, \href
  {http://adsabs.harvard.edu/abs/2009MNRAS.392..945V} {392, 945}

\bibitem[\protect\citeauthoryear{{Wang} et~al.,}{{Wang}
  et~al.}{2017}]{Wang:2017aa}
{Wang} X.,  et~al., 2017, \mn@doi [\apj] {10.3847/1538-4357/aa603c}, \href
  {http://adsabs.harvard.edu/abs/2017ApJ...837...89W} {837, 89}

\bibitem[\protect\citeauthoryear{{Wang} et~al.,}{{Wang}
  et~al.}{2019}]{Wang:2019aa}
{Wang} X.,  et~al., 2019, \mn@doi [\apj] {10.3847/1538-4357/ab3861}, \href
  {https://ui.adsabs.harvard.edu/abs/2019ApJ...882...94W} {882, 94}

\bibitem[\protect\citeauthoryear{{Werk}, {Putman}, {Meurer}  \&
  {Santiago-Figueroa}}{{Werk} et~al.}{2011}]{Werk:2011aa}
{Werk} J.~K.,  {Putman} M.~E.,  {Meurer} G.~R.,   {Santiago-Figueroa} N.,
  2011, \mn@doi [\apj] {10.1088/0004-637X/735/2/71}, \href
  {http://adsabs.harvard.edu/abs/2011ApJ...735...71W} {735, 71}

\bibitem[\protect\citeauthoryear{{Whitaker} et~al.,}{{Whitaker}
  et~al.}{2014}]{Whitaker:2014aa}
{Whitaker} K.~E.,  et~al., 2014, \mn@doi [\apj] {10.1088/0004-637X/795/2/104},
  \href {https://ui.adsabs.harvard.edu/abs/2014ApJ...795..104W} {795, 104}

\bibitem[\protect\citeauthoryear{{Wiersma}, {Schaye}, {Theuns}, {Dalla Vecchia}
   \& {Tornatore}}{{Wiersma} et~al.}{2009}]{Wiersma:2009aa}
{Wiersma} R.~P.~C.,  {Schaye} J.,  {Theuns} T.,  {Dalla Vecchia} C.,
  {Tornatore} L.,  2009, \mn@doi [\mnras] {10.1111/j.1365-2966.2009.15331.x},
  \href {https://ui.adsabs.harvard.edu/abs/2009MNRAS.399..574W} {399, 574}

\bibitem[\protect\citeauthoryear{{Wuyts} et~al.,}{{Wuyts}
  et~al.}{2014}]{Wuyts:2014aa}
{Wuyts} E.,  et~al., 2014, \mn@doi [\apjl] {10.1088/2041-8205/789/2/L40}, \href
  {http://adsabs.harvard.edu/abs/2014ApJ...789L..40W} {789, L40}

\bibitem[\protect\citeauthoryear{{Wuyts} et~al.,}{{Wuyts}
  et~al.}{2016}]{Wuyts:2016aa}
{Wuyts} E.,  et~al., 2016, \mn@doi [\apj] {10.3847/0004-637X/827/1/74}, \href
  {http://adsabs.harvard.edu/abs/2016ApJ...827...74W} {827, 74}

\bibitem[\protect\citeauthoryear{{Yates}, {Kauffmann}  \& {Guo}}{{Yates}
  et~al.}{2012}]{Yates:2012aa}
{Yates} R.~M.,  {Kauffmann} G.,   {Guo} Q.,  2012, \mn@doi [\mnras]
  {10.1111/j.1365-2966.2012.20595.x}, \href
  {http://adsabs.harvard.edu/abs/2012MNRAS.422..215Y} {422, 215}

\bibitem[\protect\citeauthoryear{{Yuan}, {Kewley}, {Swinbank}, {Richard}  \&
  {Livermore}}{{Yuan} et~al.}{2011}]{Yuan:2011aa}
{Yuan} T.-T.,  {Kewley} L.~J.,  {Swinbank} A.~M.,  {Richard} J.,   {Livermore}
  R.~C.,  2011, \mn@doi [\apjl] {10.1088/2041-8205/732/1/L14}, \href
  {http://adsabs.harvard.edu/abs/2011ApJ...732L..14Y} {732, L14}

\bibitem[\protect\citeauthoryear{{Yuan}, {Kewley}  \& {Rich}}{{Yuan}
  et~al.}{2013}]{Yuan:2013aa}
{Yuan} T.-T.,  {Kewley} L.~J.,   {Rich} J.,  2013, \mn@doi [\apj]
  {10.1088/0004-637X/767/2/106}, \href
  {https://ui.adsabs.harvard.edu/abs/2013ApJ...767..106Y} {767, 106}

\bibitem[\protect\citeauthoryear{{Zahid}, {Kewley}  \& {Bresolin}}{{Zahid}
  et~al.}{2011}]{Zahid:2011aa}
{Zahid} H.~J.,  {Kewley} L.~J.,   {Bresolin} F.,  2011, \mn@doi [\apj]
  {10.1088/0004-637X/730/2/137}, \href
  {http://adsabs.harvard.edu/abs/2011ApJ...730..137Z} {730, 137}

\bibitem[\protect\citeauthoryear{{Zaritsky}, {Kennicutt}  \&
  {Huchra}}{{Zaritsky} et~al.}{1994}]{Zaritsky:1994aa}
{Zaritsky} D.,  {Kennicutt} Jr. R.~C.,   {Huchra} J.~P.,  1994, \mn@doi [\apj]
  {10.1086/173544}, \href {http://adsabs.harvard.edu/abs/1994ApJ...420...87Z}
  {420, 87}

\bibitem[\protect\citeauthoryear{{Zitrin}, {Broadhurst}, {Rephaeli}  \&
  {Sadeh}}{{Zitrin} et~al.}{2009}]{Zitrin:2009aa}
{Zitrin} A.,  {Broadhurst} T.,  {Rephaeli} Y.,   {Sadeh} S.,  2009, \mn@doi
  [\apjl] {10.1088/0004-637X/707/1/L102}, \href
  {http://adsabs.harvard.edu/abs/2009ApJ...707L.102Z} {707, L102}

\bibitem[\protect\citeauthoryear{{Zitrin} et~al.,}{{Zitrin}
  et~al.}{2013}]{Zitrin:2013aa}
{Zitrin} A.,  et~al., 2013, \mn@doi [\apjl] {10.1088/2041-8205/762/2/L30},
  \href {http://adsabs.harvard.edu/abs/2013ApJ...762L..30Z} {762, L30}

\bibitem[\protect\citeauthoryear{{Zitrin} et~al.,}{{Zitrin}
  et~al.}{2015}]{Zitrin:2015aa}
{Zitrin} A.,  et~al., 2015, \mn@doi [\apj] {10.1088/0004-637X/801/1/44}, \href
  {http://adsabs.harvard.edu/abs/2015ApJ...801...44Z} {801, 44}

\bibitem[\protect\citeauthoryear{{da Cunha} et~al.,}{{da Cunha}
  et~al.}{2015}]{da-Cunha:2015aa}
{da Cunha} E.,  et~al., 2015, \mn@doi [\apj] {10.1088/0004-637X/806/1/110},
  \href {http://adsabs.harvard.edu/abs/2015ApJ...806..110D} {806, 110}

\bibitem[\protect\citeauthoryear{{van Dokkum}}{{van
  Dokkum}}{2001}]{van-Dokkum:2001aa}
{van Dokkum} P.~G.,  2001, \mn@doi [\pasp] {10.1086/323894}, \href
  {http://adsabs.harvard.edu/abs/2001PASP..113.1420V} {113, 1420}

\makeatother
\end{thebibliography}

%%%%%%%%%%%%%%%%%%%% APPENDIX %%%%%%%%%%%%%%%

%\section*{Supporting Information}
%The supplementary material presented in the Appendices is available in the online version of the journal.
%In particular, in Appendix A we address the potential systematics associated to the choice of the radial binning scheme adopted in the source plane to derive the metallicity gradients.
%In Appendix B instead we present the emission line maps, metallicity maps and metallicity gradients for the full 
%\textit{metallicity gradient sample}.

%\newpage
\clearpage

\appendix

\renewcommand{\thefigure}{A\arabic{figure}}
\setcounter{figure}{0}
\renewcommand{\thetable}{A\arabic{table}}  
\setcounter{table}{0}
\renewcommand{\thesection}{A\arabic{figure}}
\setcounter{section}{0}

\section*{Appendix A: Systematics in the choice of the averaging apertures}
\label{sect:appenA}
We discuss here the potential biases associated to the radial binning scheme.
As already discussed in Sect.~\ref{sect:grads}, radial averaging within elliptical apertures is required to account
for the smearing effect introduced by the distorted source plane PSF.
However, this choice is not without its issues.
In fact, the information from different radii is combined within the same aperture when considering the directions perpendicular and parallel to the PSF major axis, and this effect can be particularly severe in those cases where the source plane PSF is extremely stretched (as in the case of strongly lensed sources), possibly introducing systematics in the determination of the gradients.
Therefore, to test the impact of this issue we compute the radial gradients adopting purely circular apertures as well: overall, we find that the choice of the radial binning scheme does not affect the main results and the interpretation presented in this paper, as statistically significant differences are limited to few individual cases.
For the $83\%$ of the sample, the two estimates are consistent within the uncertainties, while for $5$ sources they diverge more than $1\sigma$; however, only in two of the these cases there is also an inversion in the sign of the gradient (from negative to positive). The results are reported in Table~\ref{tab:circ_grads}.

In Fig.~\ref{fig:ellvscircgrad} the metallicity gradients adopting elliptical apertures are plotted against the gradients computed within circular apertures, and the points are colour coded according to the axis ratio of the source plane PSF. 
Some of the points scatter around the 1:1 line and no clear systematic trends are highlighted.
The most deviating points are, not surprisingly, associated  with galaxies whose source plane PSF is extremely elongated. 
For instance, the two different configurations in the case of CSWA164 are compared in Fig.~\ref{fig:ellvscircgrad}: for this source, the circularly-averaged gradient seems to be more representative of the metallicity distribution of individual spaxels across the map, although in both cases the slope remains consistent with zero within its uncertainty.

\begin{table*}
	%\tiny
	\centering
	%\resizebox{\columnwidth}{!}{
	%	\setlength{\extrarowheight}{.3em}
	
	%\extracolsep{\fill}
	\begin{tabular}{@{\extracolsep{\fill}} cccc @{}}
		\hline
		Galaxy & Gradient (Circ. apertures) & Resolved & Diagnostics \\
		&  [dex kpc$^{-1}$]     &               & \\ 
		
		\hline
		\textbf{SINFONI Galaxies} &&& \\
		Horseshoe & 0.011$\pm$0.039 & Yes & R$_{3}$,N$_{2}$ \\
		MACS Arc (North) & -0.021$\pm$0.033 & Yes & R$_{3}$, N${2}$, O$_{3}$O$_{2}$\\
		MACS Arc (South) & 0.002$\pm$0.038 & Yes & R$_{3}$, N${2}$, O$_{3}$O$_{2}$ \\
		CSWA164 & 0.021$\pm$0.025 & Yes & R$_{3}$, N${2}$, O$_{3}$O$_{2}$ \\
		\textbf{KMOS Galaxies} &&& \\
		\textbf{MS2137} &&& \\
		sp1 & 0.023$\pm$0.018 & Yes & R$_{3}$,S$_{2}$ \\
		sp2 & -0.009$\pm$0.022 & Yes & R$_{3}$,O$_{3}$O$_{2}$ \\
		sp3 & 0.023$\pm$0.01 & Yes & R$_{3}$,N$_{2}$ \\
		sp5 & -0.0$\pm$0.022 & Yes &  R$_{3}$,O$_{3}$O$_{2}$ \\
		sp6 & 0.04$\pm$0.023 & Yes &  R$_{3}$,O$_{3}$O$_{2}$ \\
		sp7 & 0.251$\pm$0.008 & Yes & R$_{3}$,N$_{2}$ \\
		sp13 & 0.282$\pm$0.02 & Yes & R$_{3}$,N$_{2}$ \\
		sp15 & 0.014$\pm$0.013 & Yes & R$_{3}$,S$_{2}$ \\
		ph6532 & 0.035$\pm$0.013 & Yes &  R$_{3}$,O$_{3}$O$_{2}$ \\
		ph3729 & 0.007$\pm$0.043 & Yes & R$_{3}$,N$_{2}$ \\
		ph3912 & 0.039$\pm$0.016 & Yes & R$_{3}$,S$_{2}$\\
		ph8073 & 0.005$\pm$0.092 & Marg & R$_{3}$,S$_{2}$ \\
		\textbf{RXJ2248} &&& \\
		GLASS\_00093-99-99 & 0.021$\pm$0.048 & Yes & R$_{3}$,N$_{2}$ \\
		R2248\_LRb\_p1\_M3\_Q4\_58\_\_2 & 0.002$\pm$0.01 & Yes & R$_{3}$,N$_{2}$ \\
		MUSE\_SW\_462-99-99 & -0.0$\pm$0.012 & Yes & R$_{3}$,N$_{2}$ \\
		GLASS\_00333-99-99 & -0.019$\pm$0.055 & Yes &R$_{3}$,S$_{2}$ \\
		R2248\_LRb\_p3\_M4\_Q3\_93\_\_1 & 0.006$\pm$0.007 & Yes & R$_{3}$,N$_{2}$ \\
		R2248\_LRb\_p3\_M4\_Q3\_94\_\_1 & 0.047$\pm$0.046 & Yes & R$_{3}$,O$_{3}$O$_{2}$ \\
		GLASS\_01845-99-99 & 0.016$\pm$0.017 & Yes & R$_{3}$,O$_{3}$O$_{2}$\\
		MUSE\_SW\_45-99-99 & 0.02$\pm$0.032 & Yes & R$_{3}$,S$_{2}$ \\
		MUSE\_SW\_461-99-99 & -0.018$\pm$0.018 & Yes & R$_{3}$,N$_{2}$ \\
		MUSE\_NE\_111-99-99 & -0.005$\pm$0.02 & Marg & R$_{3}$,N$_{2}$ \\
		GLASS\_00800-99-99 & 0.001$\pm$0.021 & Yes & R$_{3}$,N$_{2}$ \\
		MUSE\_NE\_23-99-99 & 0.002$\pm$0.032 & Yes &R$_{3}$,S$_{2}$ \\
		\hline
	\end{tabular}

	\caption{Radial metallicity gradients computed adopting circular apertures in the source plane. For more details see the caption of Table~\ref{tab:grads}.}
	\label{tab:circ_grads} 
\end{table*}

\begin{figure}
	\centering
	\includegraphics[width=0.95\columnwidth]{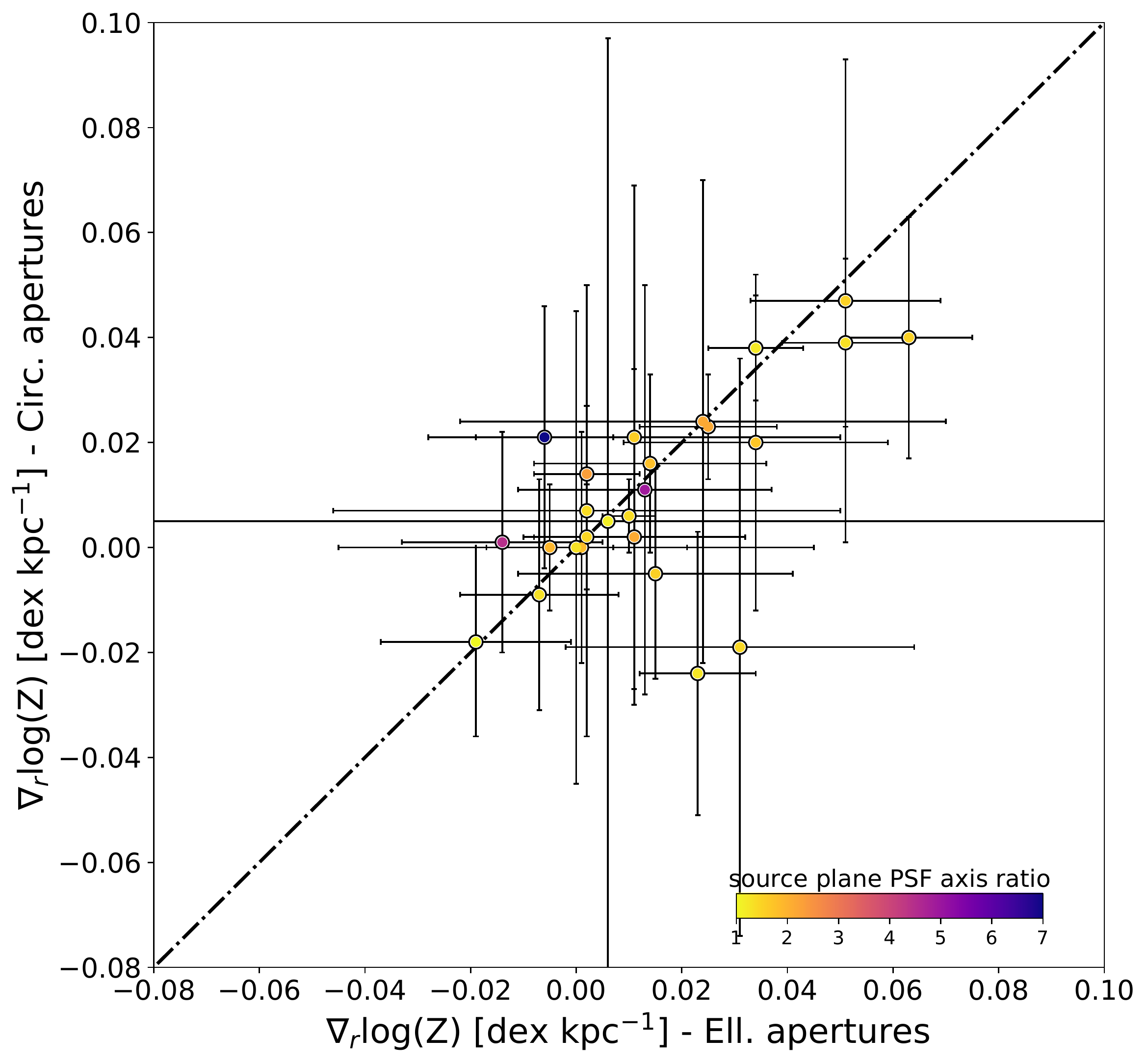} \\
	\vspace{0.3cm}
	\centering
	\includegraphics[width=0.98\columnwidth]{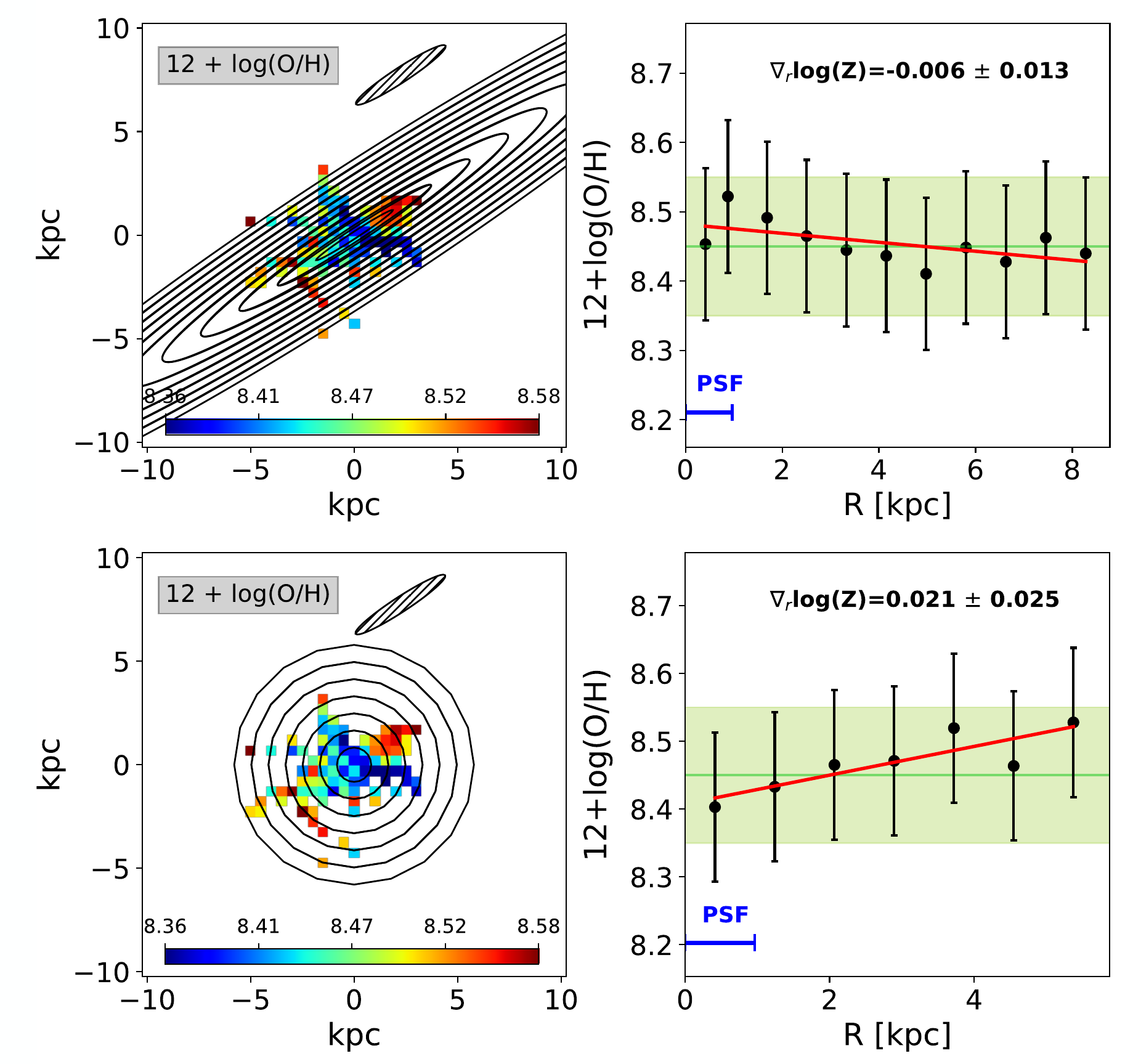}
	
	\caption{\textit{Upper Panel:} Metallicity gradients computed by averaging within elliptical apertures in the source plane are plotted against those derived assuming purely circular apertures. The points are colour coded by the corresponding value of the axis ratio of the PSF in the source plane. The two estimates are consistent for more than $80\%$ of the sources, with only two showing significant  inversion of the slope.
		\textit{Bottom Panels:} The difference between elliptically-averaged (top panel) and circularly-averaged (bottom panel) metallicity gradients is shown for one of the most extreme cases in our sample (i.e. the CSWA164 galaxy, $\mu \approx 13$) .
		Here, the significant elongation of the PSF causes the information from very different radii to be combined within the same annulus, potentially biasing the determination of the gradient.
		In this case, the circularly-averaged gradient seems to be more representative of the metallicity distribution of individual spaxels across the map.
	}
	\label{fig:ellvscircgrad}
\end{figure}

%%%%%%%%%

\renewcommand{\thefigure}{B\arabic{figure}}
\setcounter{figure}{0}
\renewcommand{\thetable}{B\arabic{table}}  
\setcounter{table}{0}
\renewcommand{\thesection}{B\arabic{figure}}
\setcounter{section}{0}

\section*{Appendix B: Reconstructed emission line ratio and metallicity maps}
\label{sect:appenB}

In the upper panels of Figures~\ref{fig:appen_zgrads}-~\ref{fig:append_lastfig} we show, for each galaxy in the \textit{metallicity gradient sample} and from left to right respectively, the source plane HST F160w image, the normalized source plane H$\upalpha$ flux maps, the metallicity map and the metallicities extracted at increasing radii across each galaxy.
In the last panel the linear fit to the metallicity gradient (in red) and the linear size of the PSF-HWHM (in blue) are also shown, together with the global metallicity value and its associated uncertainty (in green).
The shaded black region in the \textit{third} panel reproduce the shape and size of the PSF when mapped back into the source plane.
The elliptical apertures used to derive the radial gradient overlay the 2D metallicity map.
% For the sample of strongly lensed galaxies observed with SINFONI, whose source plane PSF is extremely elongated, we also show the metallicity gradient computed adopting purely circular apertures.

In the bottom panels, we show instead some of the emission line ratios maps in the source plane.
In particular, from left to right, we show [\ion{N}{ii}]/H$\alpha$, [\ion{O}{iii}]/H$\beta$, [\ion{S}{ii}]/H$\alpha$ and [\ion{S}{iii}]/[\ion{S}{ii}]. For z$>2$ galaxies, given the different wavelength range covered by our observations, we show [\ion{O}{iii}]/[\ion{O}{ii}] instead of [\ion{S}{iii}]/[\ion{S}{ii}].

\begin{figure*}
  \centering
      \centering
      \includegraphics[width=0.83\textwidth]{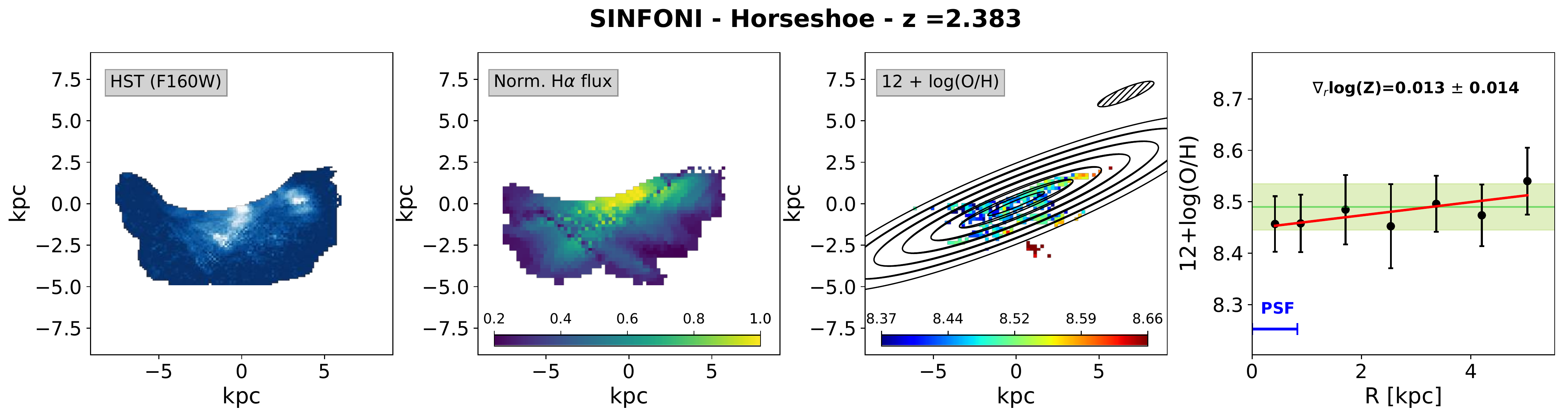}\\
      \includegraphics[width=0.83\textwidth]{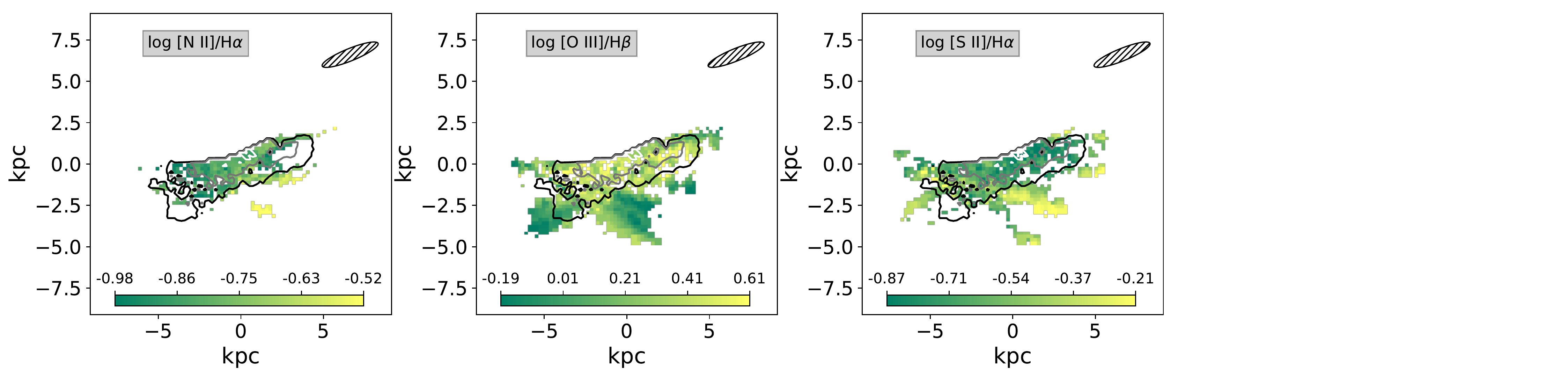}\\
      \centering
      \vspace{0.1cm}
      \includegraphics[width=0.83\textwidth]{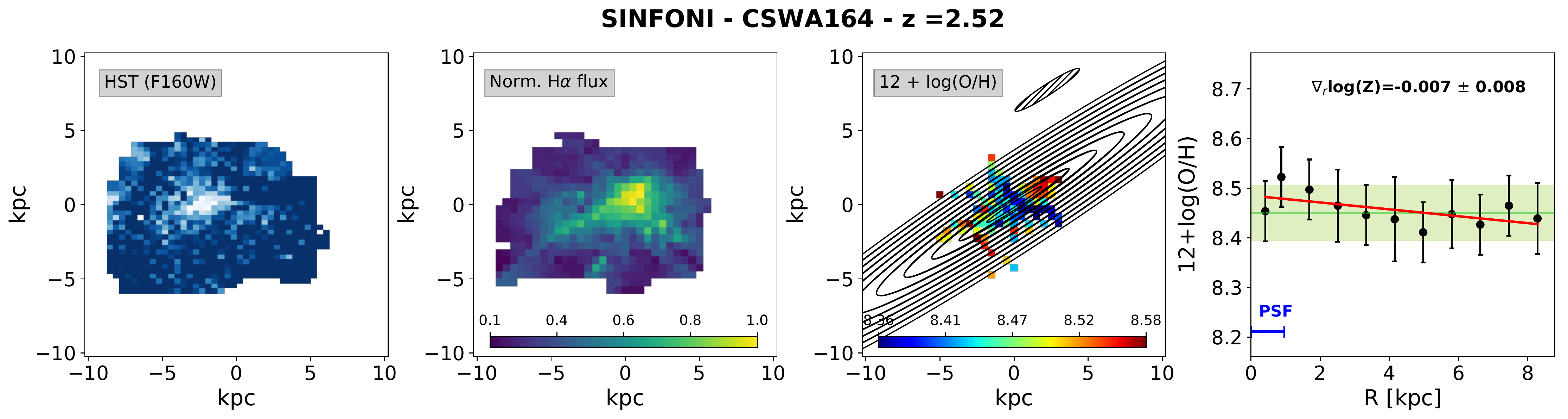}
      \includegraphics[width=0.83\textwidth]{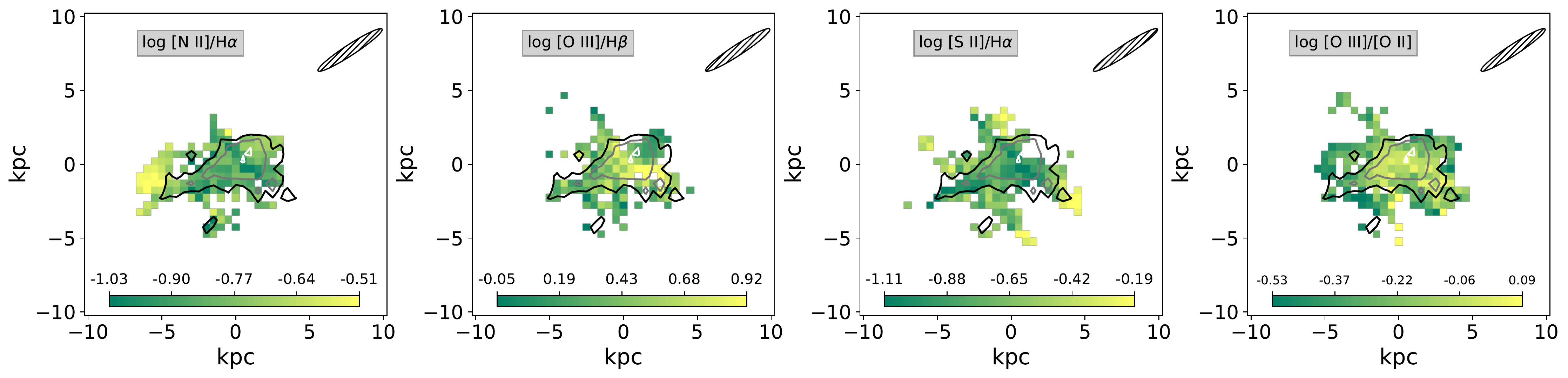}

 \caption{
Source plane HST F160w image, normalised H$\alpha$ flux map, metallicity map, line ratios maps and metallcity gradients for our \textit{metallicity gradient sample}. For details see the text of Appendix B or the caption of Figure~\ref{fig:EGgrad}.
}
   \label{fig:appen_zgrads}
  \end{figure*}

\begin{figure*}
      \centering
      \includegraphics[width=0.83\textwidth]{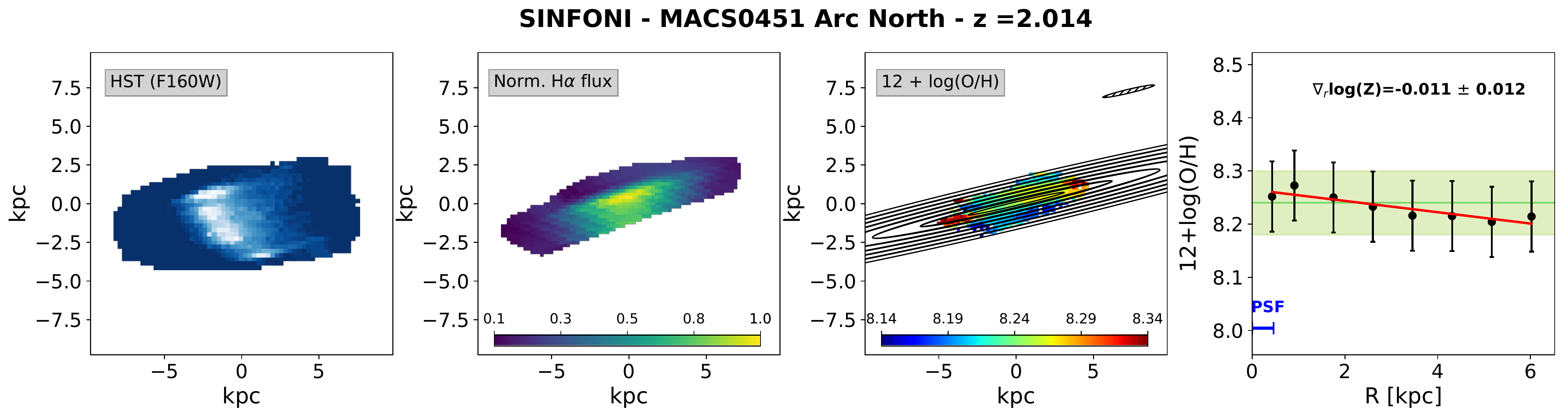}\\
      \includegraphics[width=0.83\textwidth]{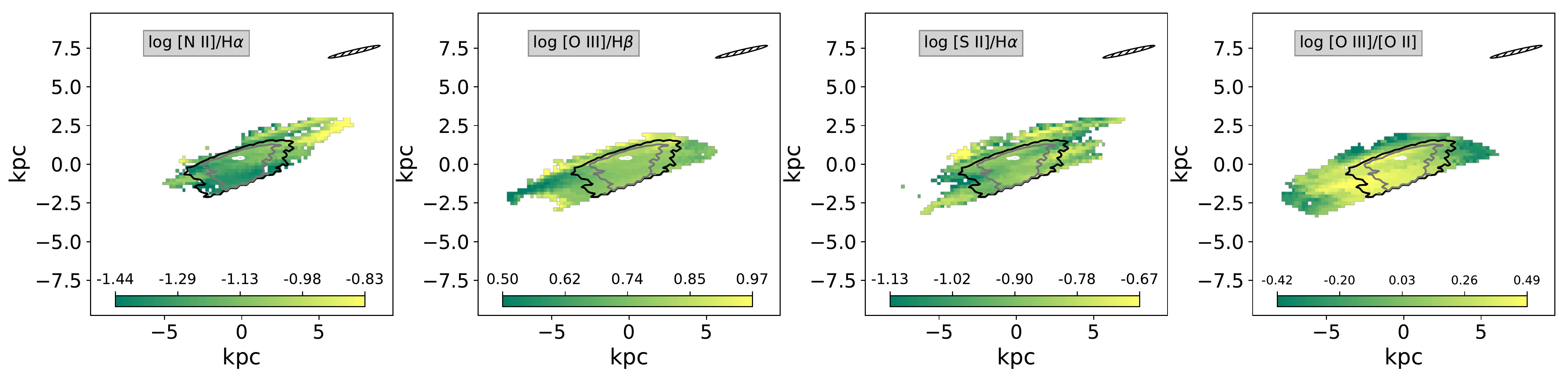}\\
      \vspace{0.1cm}
      \includegraphics[width=0.83\textwidth]{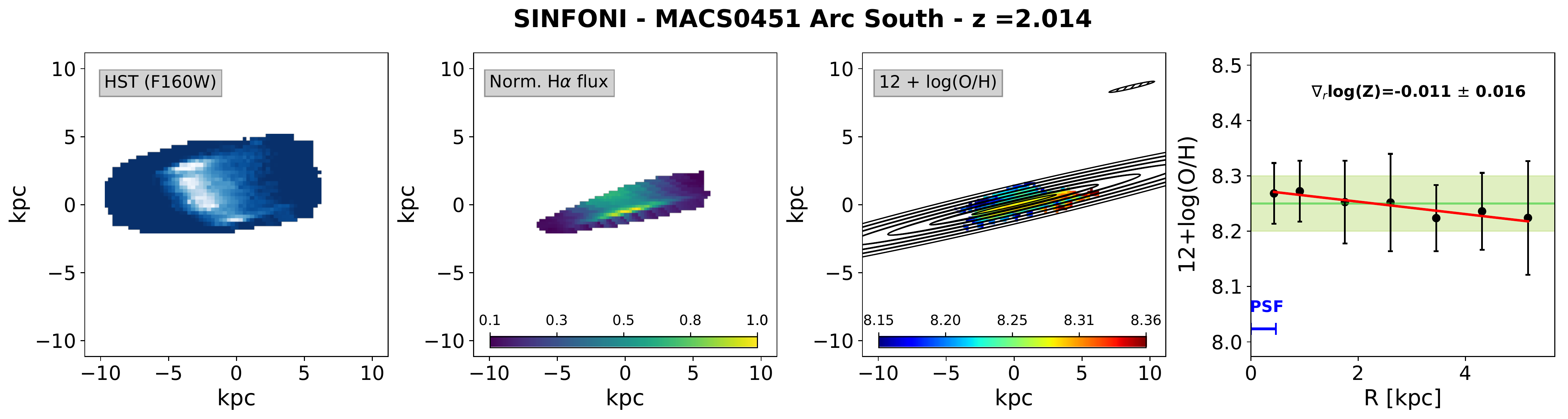}\\
      \includegraphics[width=0.83\textwidth]{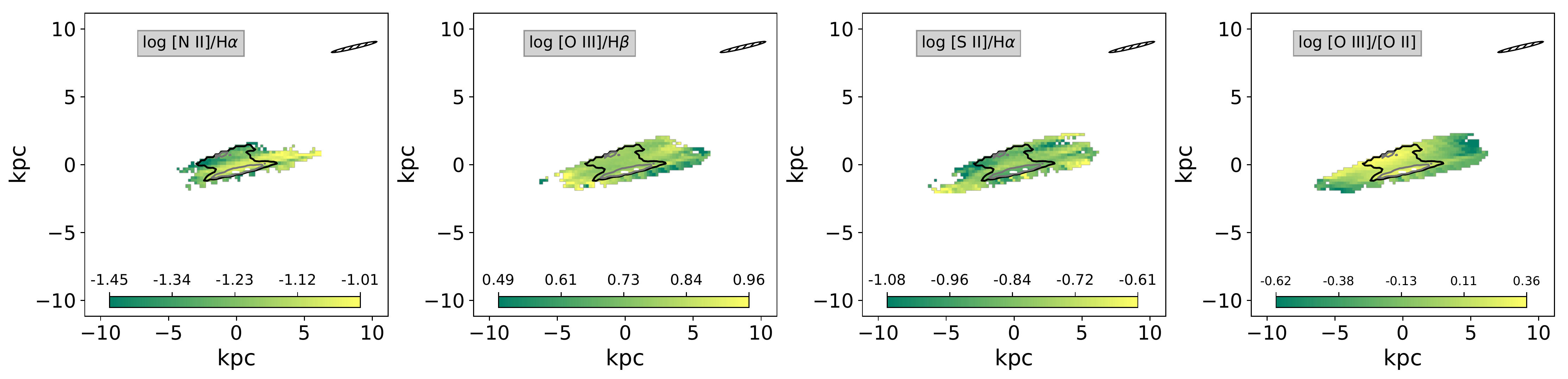}\\
\caption{Same as Fig.~\ref{fig:appen_zgrads}.}
\end{figure*}

  \begin{figure*}
      \centering
      \includegraphics[width=0.83\textwidth]{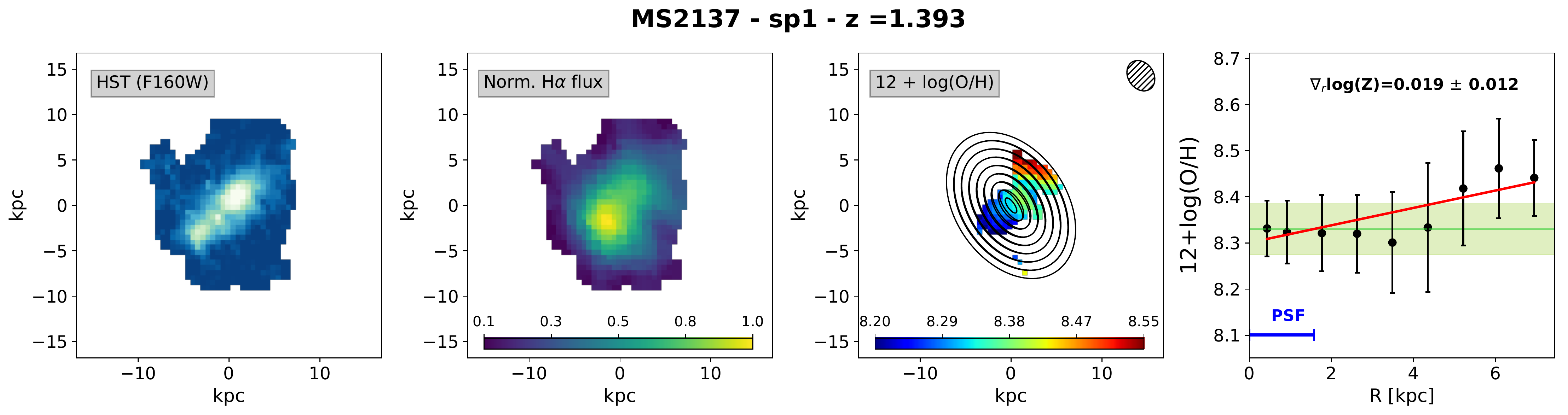}\\
      \includegraphics[width=0.83\textwidth]{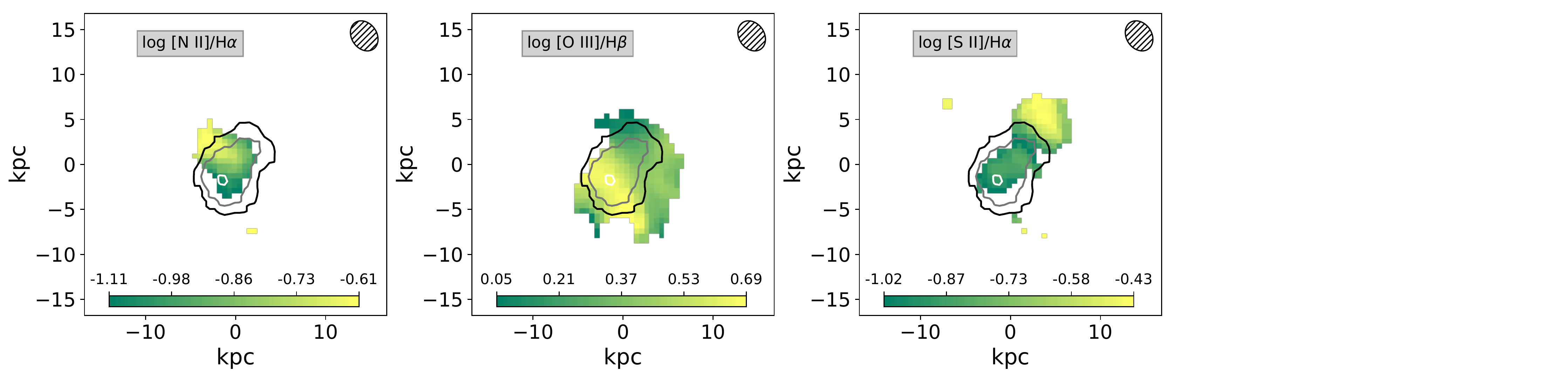}\\
       \vspace{0.1cm}
      \includegraphics[width=0.83\textwidth]{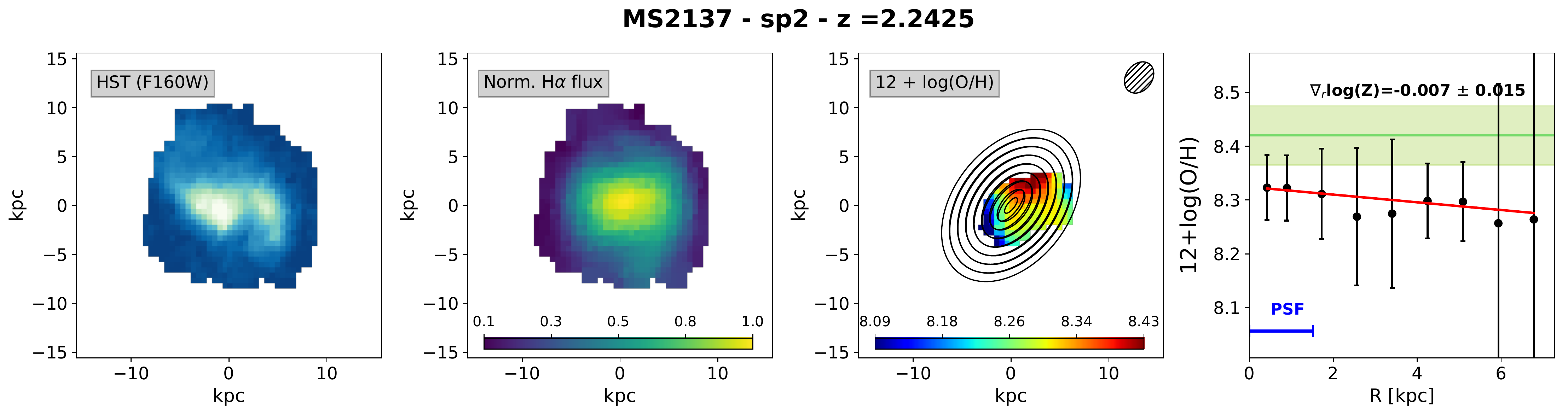}\\
      \includegraphics[width=0.83\textwidth]{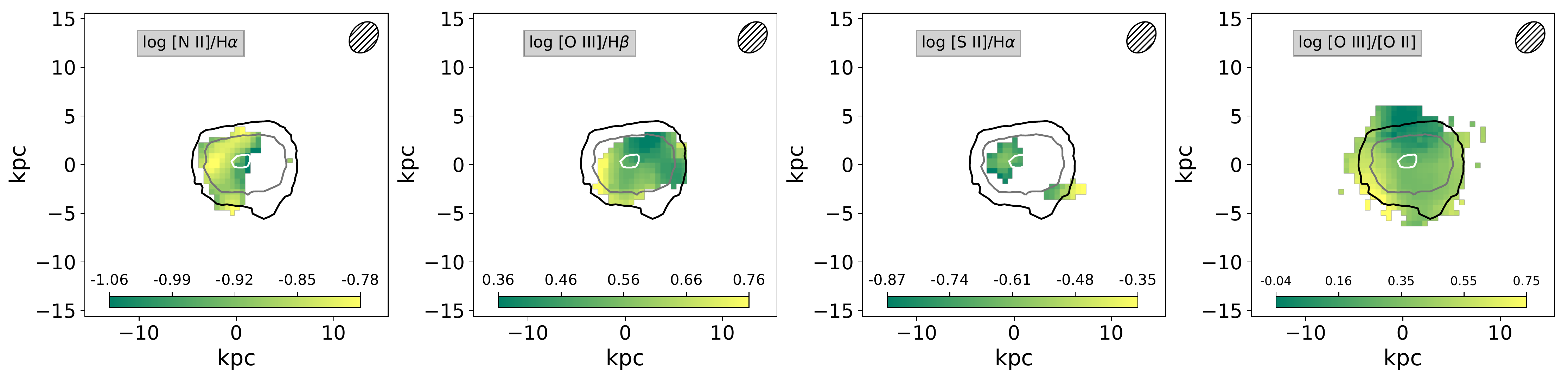}\\
       \vspace{0.1cm}
      \includegraphics[width=0.83\textwidth]{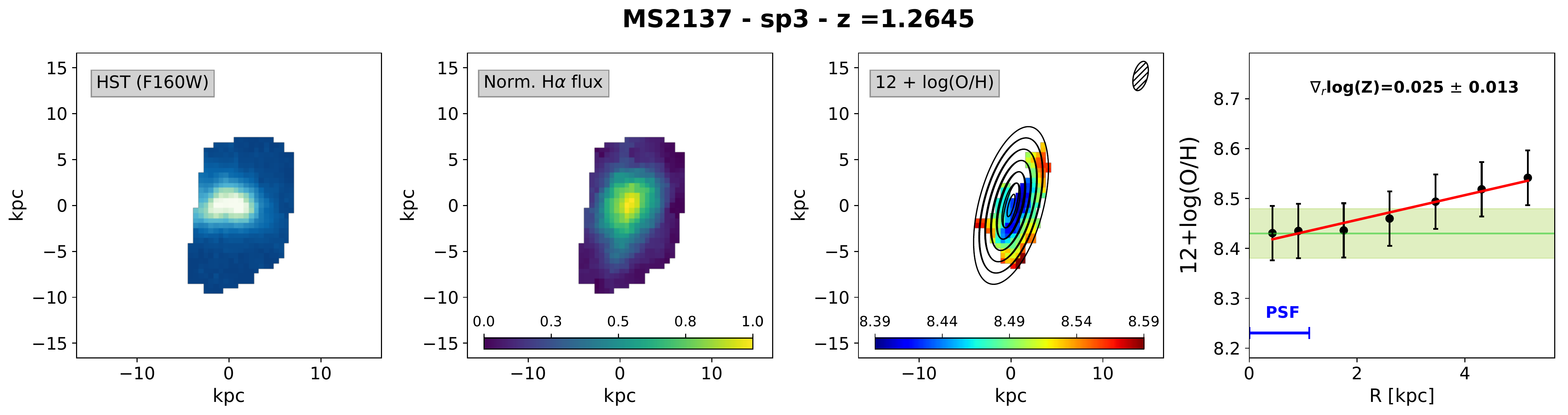}\\
      \includegraphics[width=0.83\textwidth]{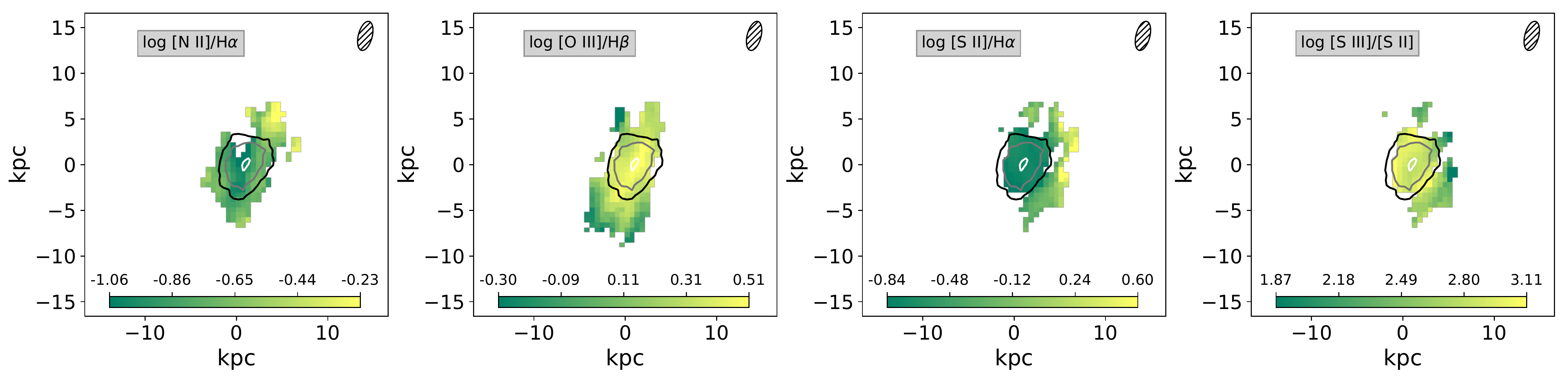}\\
 
 \caption{Same as Fig.~\ref{fig:appen_zgrads}.}

  \end{figure*}

%  \begin{figure*}
  \begin{figure*}
      \centering
      \includegraphics[width=0.83\textwidth]{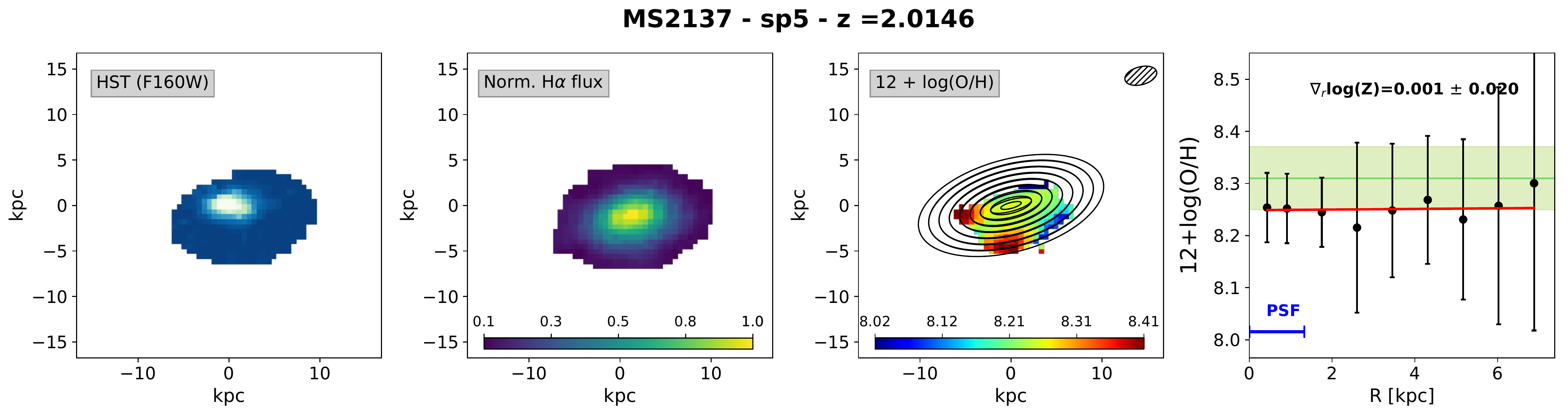}\\
      \includegraphics[width=0.83\textwidth]{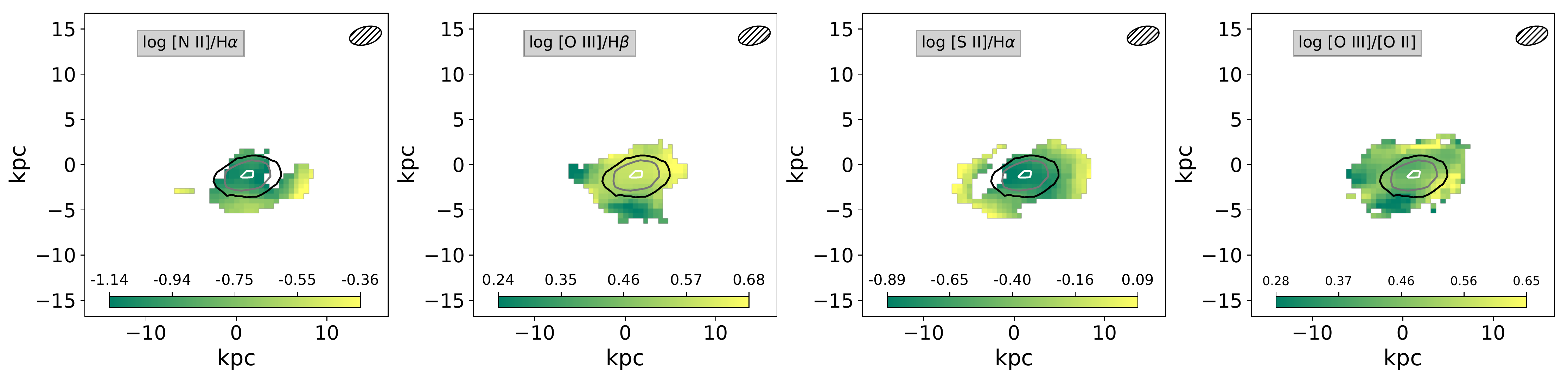}\\
      \vspace{0.1cm}
      \includegraphics[width=0.83\textwidth]{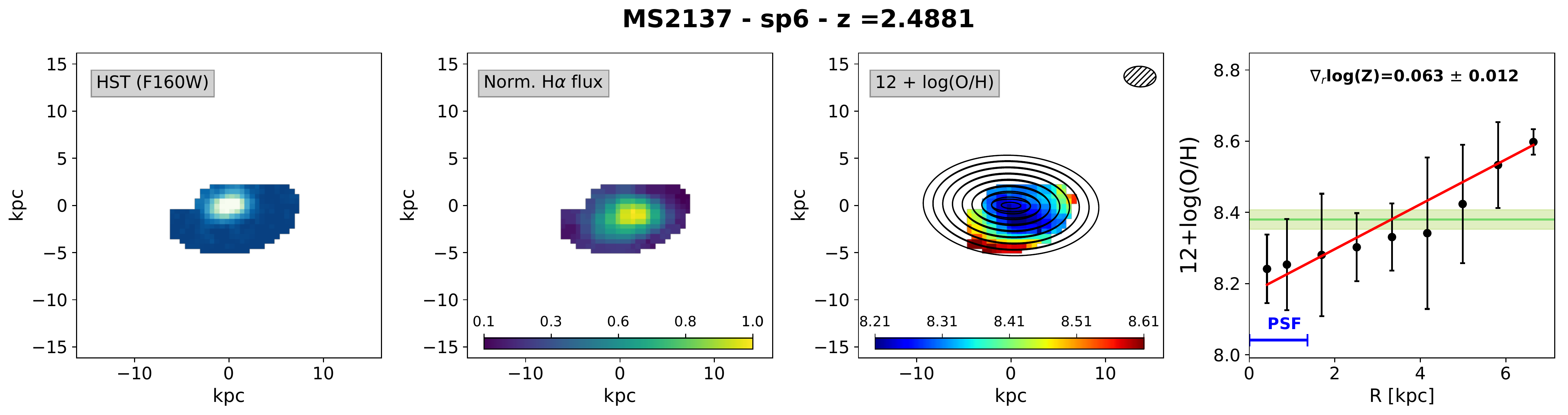}\\
      \includegraphics[width=0.83\textwidth]{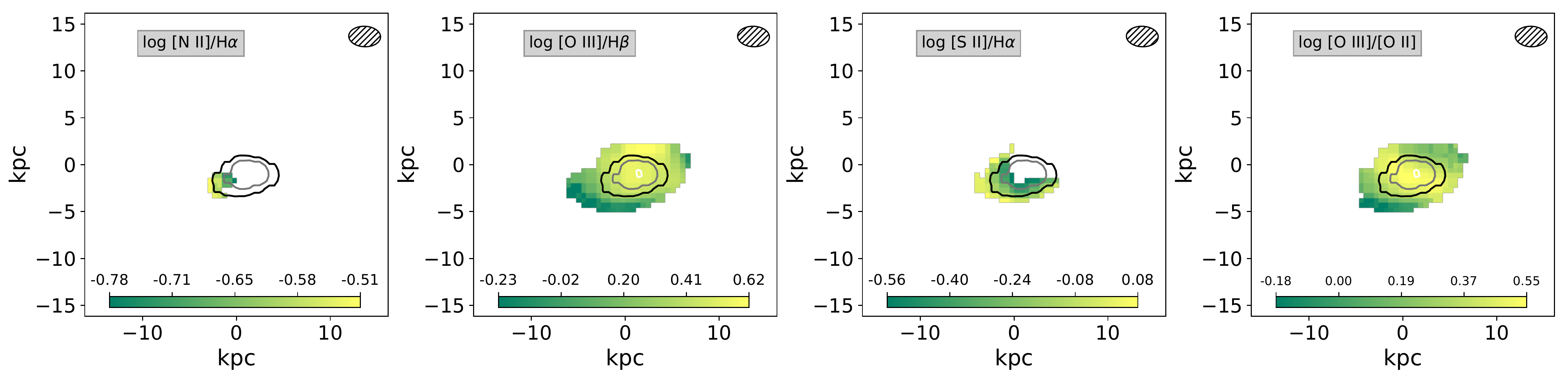}\\
      \vspace{0.1cm}
      \includegraphics[width=0.83\textwidth]{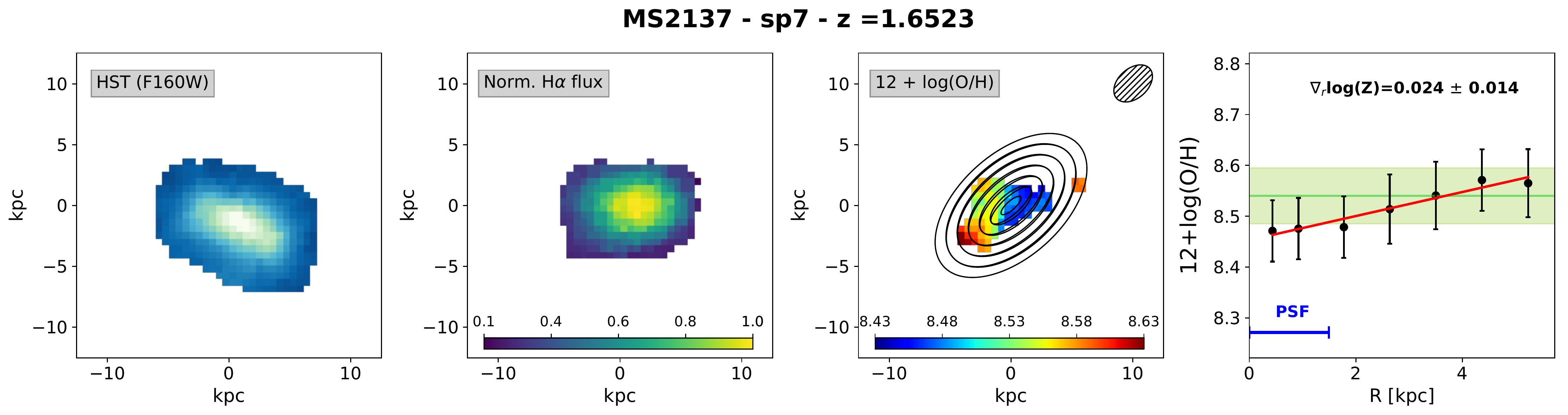}\\
      \includegraphics[width=0.83\textwidth]{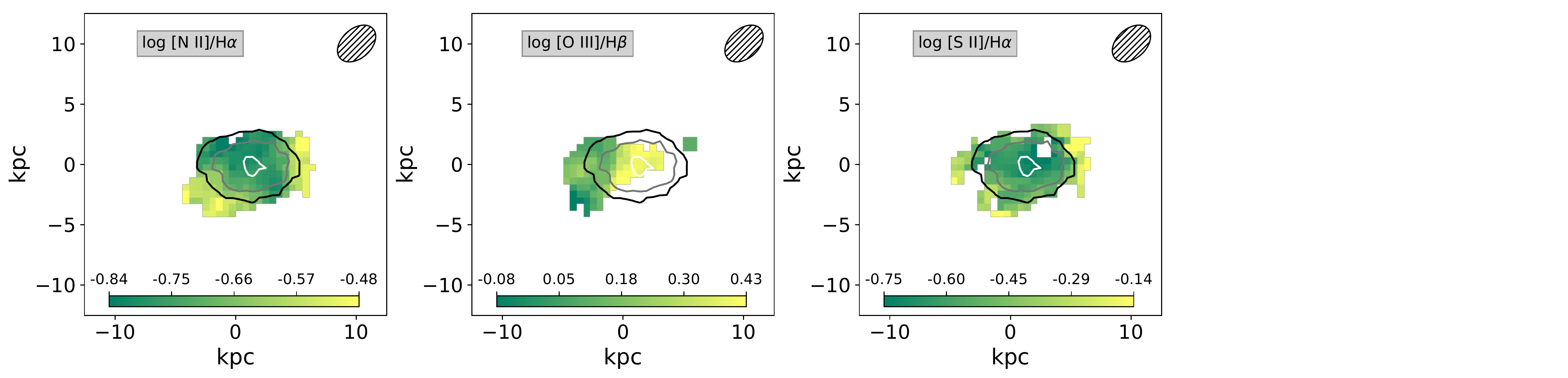}\\

 \caption{Same as Fig.~\ref{fig:appen_zgrads}.}

  \end{figure*}

  \begin{figure*}
      \centering
      \includegraphics[width=0.83\textwidth]{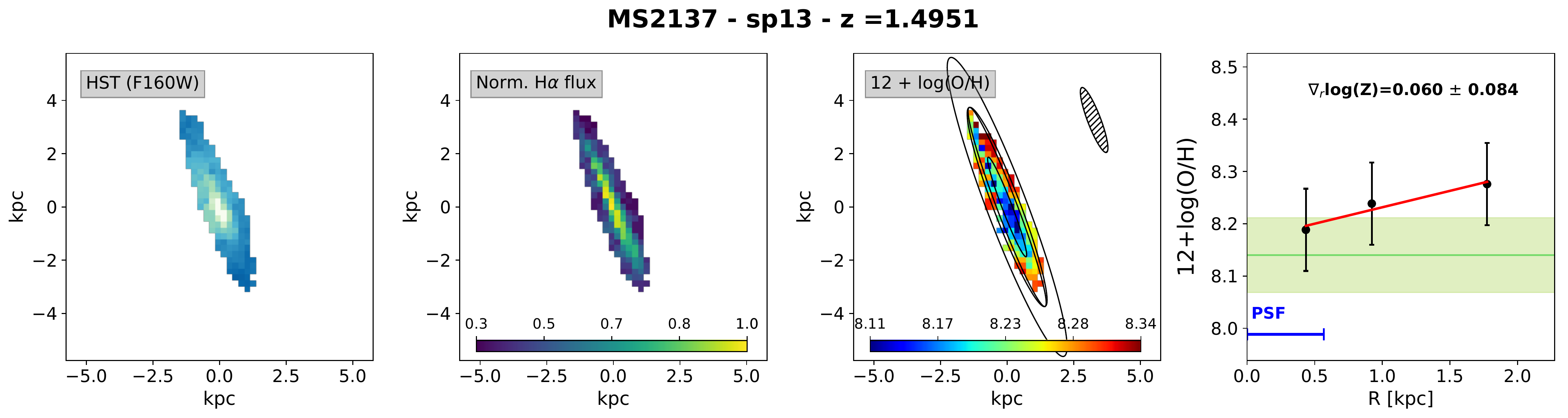}\\
	    \includegraphics[width=0.83\textwidth]{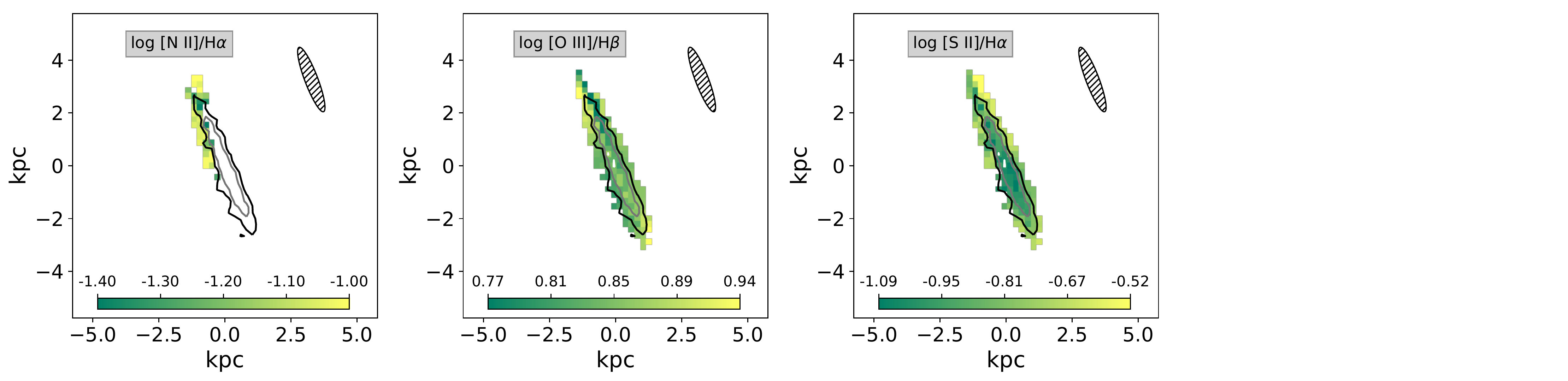}\\
	    \vspace{0.1cm}
      \includegraphics[width=0.83\textwidth]{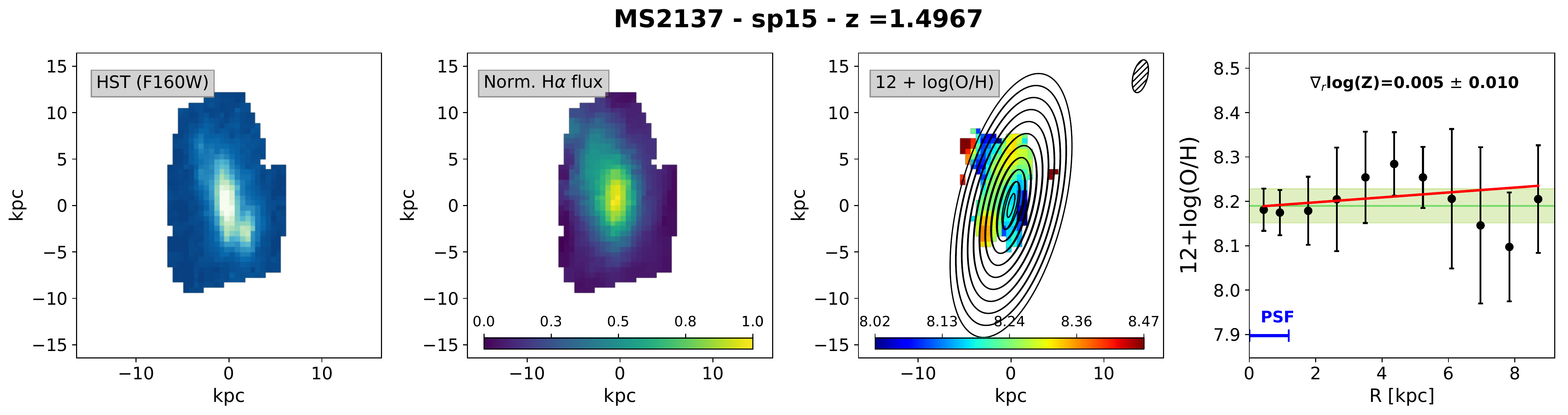}\\
      \includegraphics[width=0.83\textwidth]{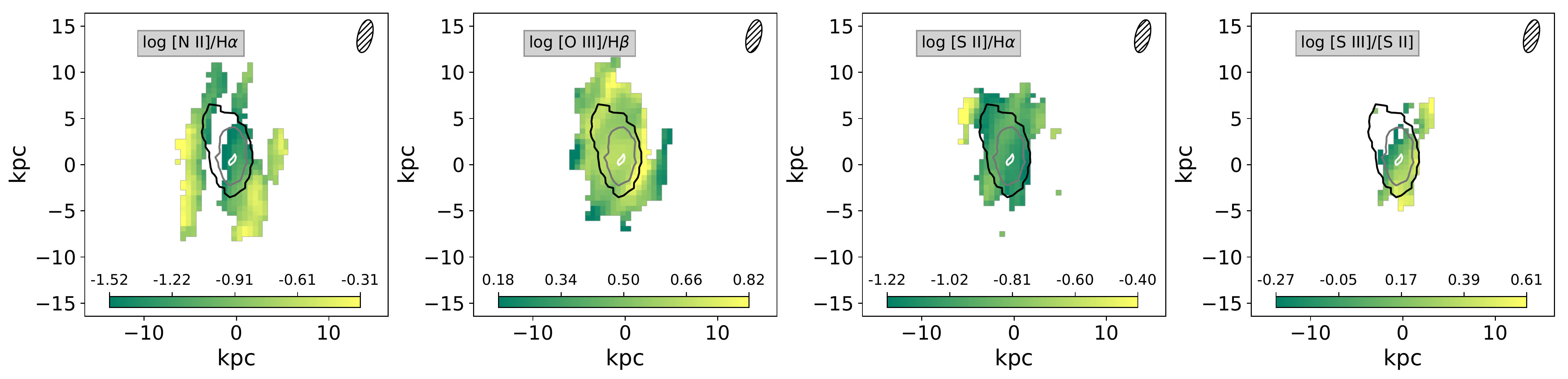}\\
      \vspace{0.1cm}
      \includegraphics[width=0.83\textwidth]{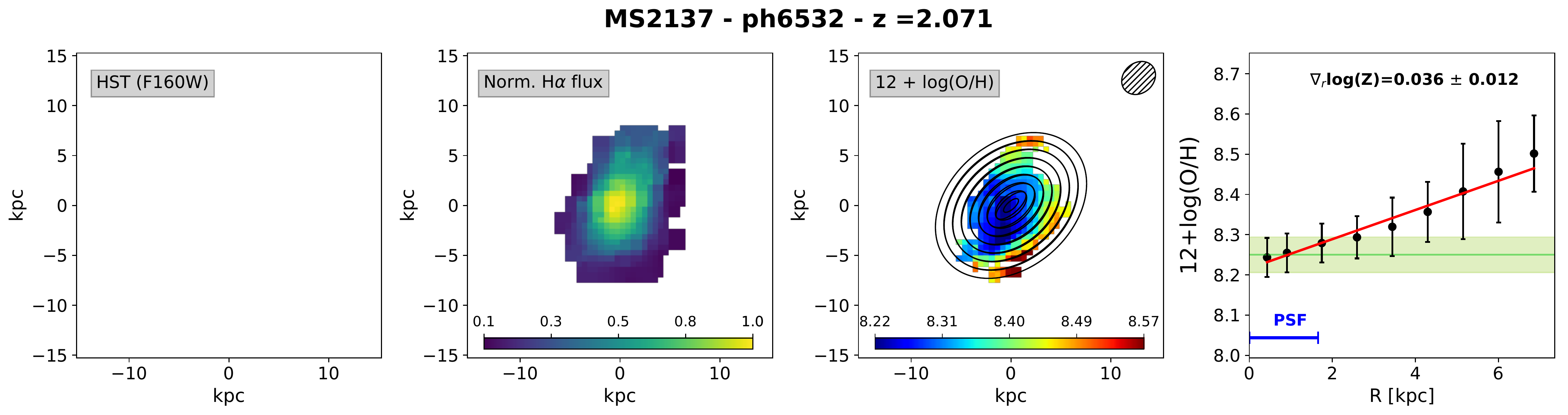}\\
      \includegraphics[width=0.83\textwidth]{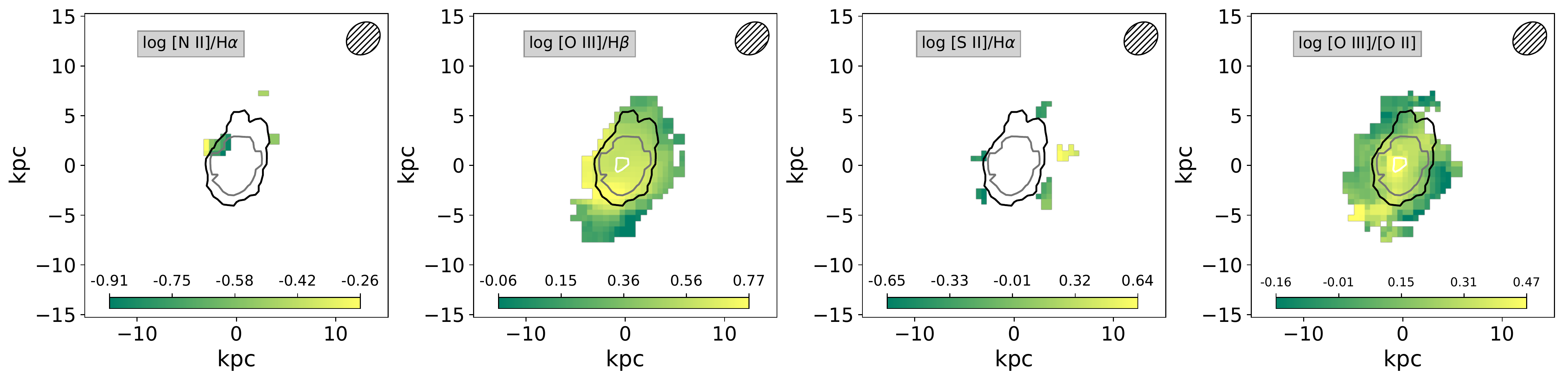}\\

  \caption{Same as Fig.~\ref{fig:appen_zgrads}.}

  \end{figure*}

  \begin{figure*}
      \centering
      \includegraphics[width=0.83\textwidth]{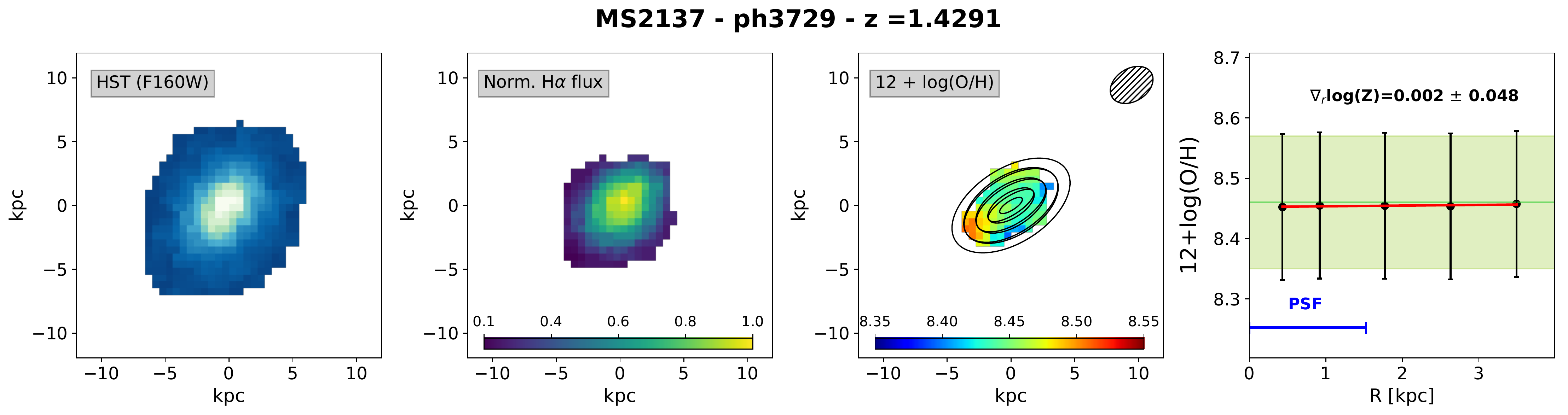}
	  \includegraphics[width=0.83\textwidth]{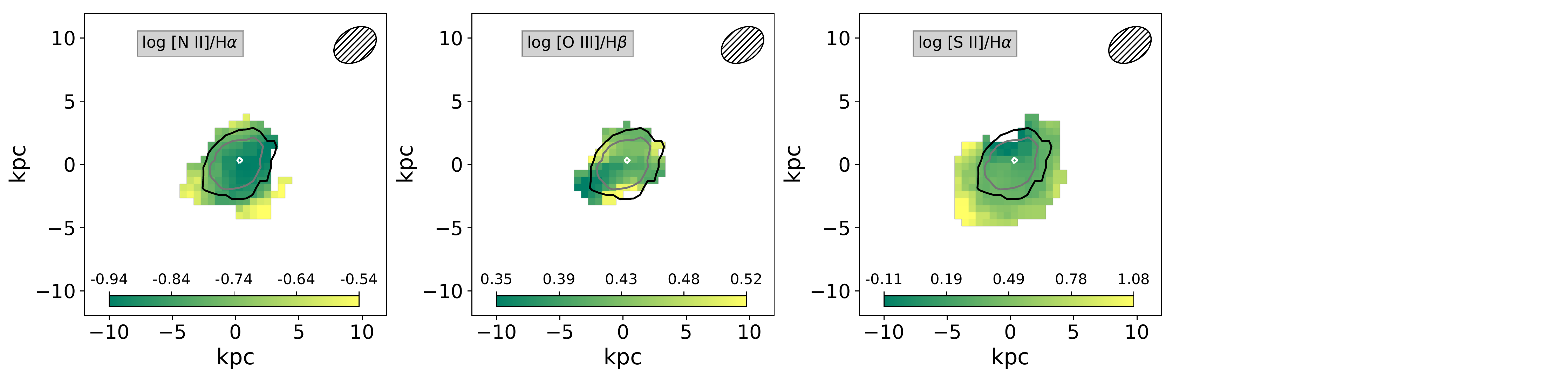}\\
      \vspace{0.1cm}
      \includegraphics[width=0.83\textwidth]{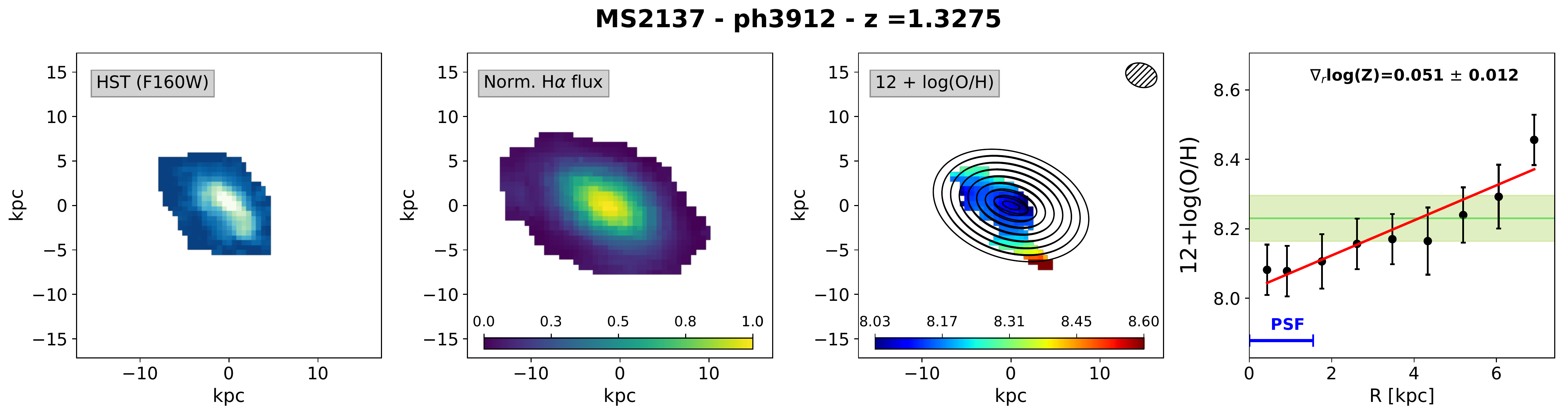}\\
      \includegraphics[width=0.83\textwidth]{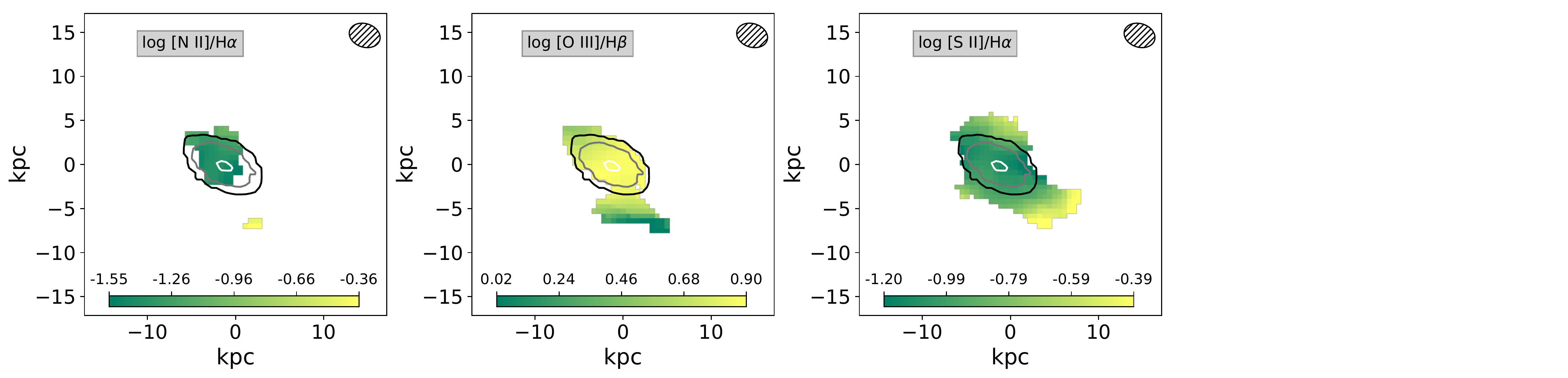}\\
      \vspace{0.1cm}
      \includegraphics[width=0.83\textwidth]{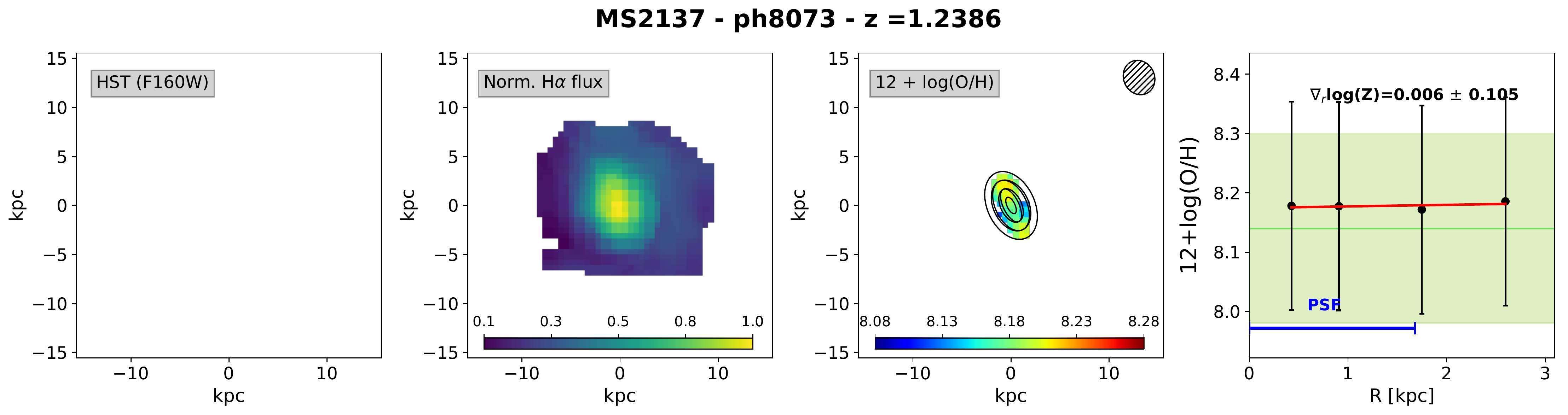}\\
      \includegraphics[width=0.83\textwidth]{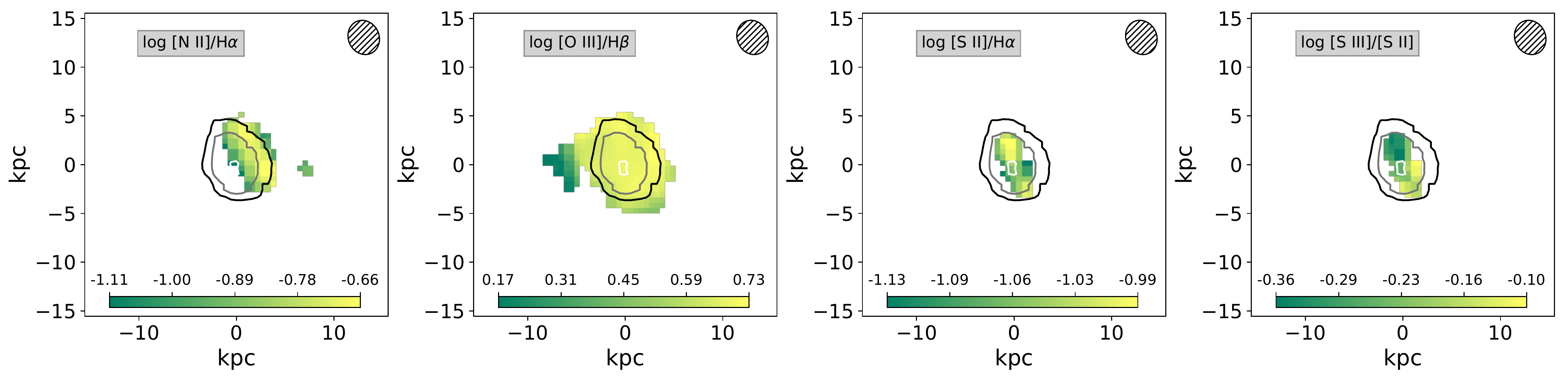}\\
 \caption{Same as Fig.~\ref{fig:appen_zgrads}.}
     
  \end{figure*}

  \begin{figure*}
      \centering
      \includegraphics[width=0.83\textwidth]{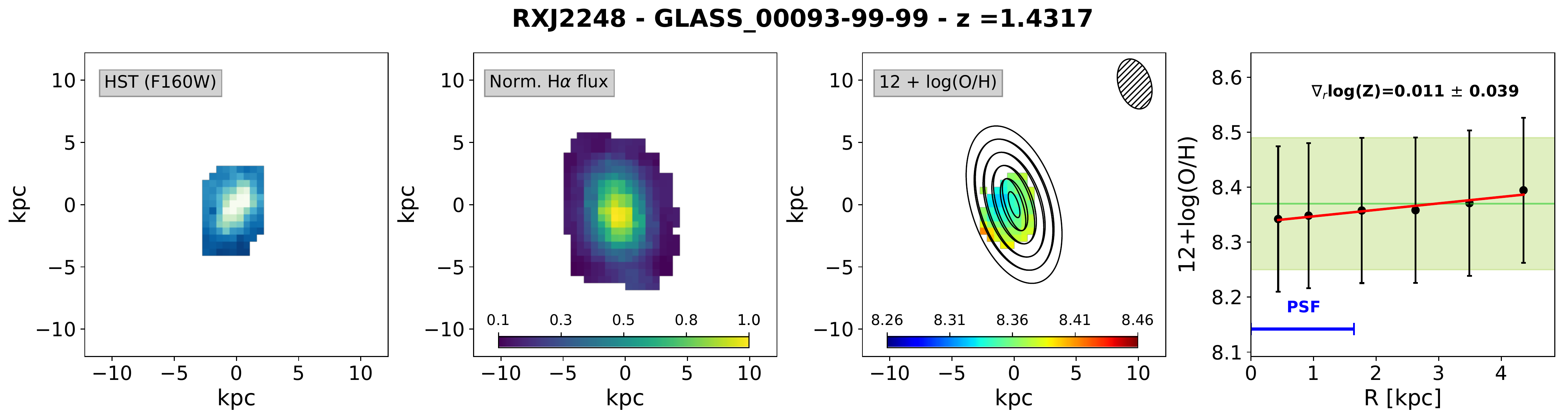}\\
      \includegraphics[width=0.83\textwidth]{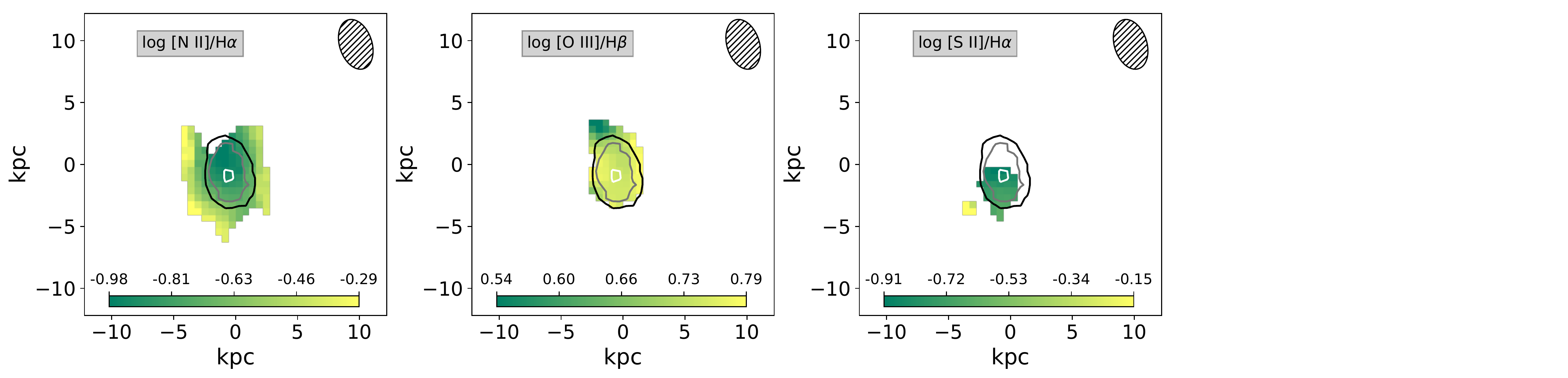}\\
      \vspace{0.1cm}
      \includegraphics[width=0.83\textwidth]{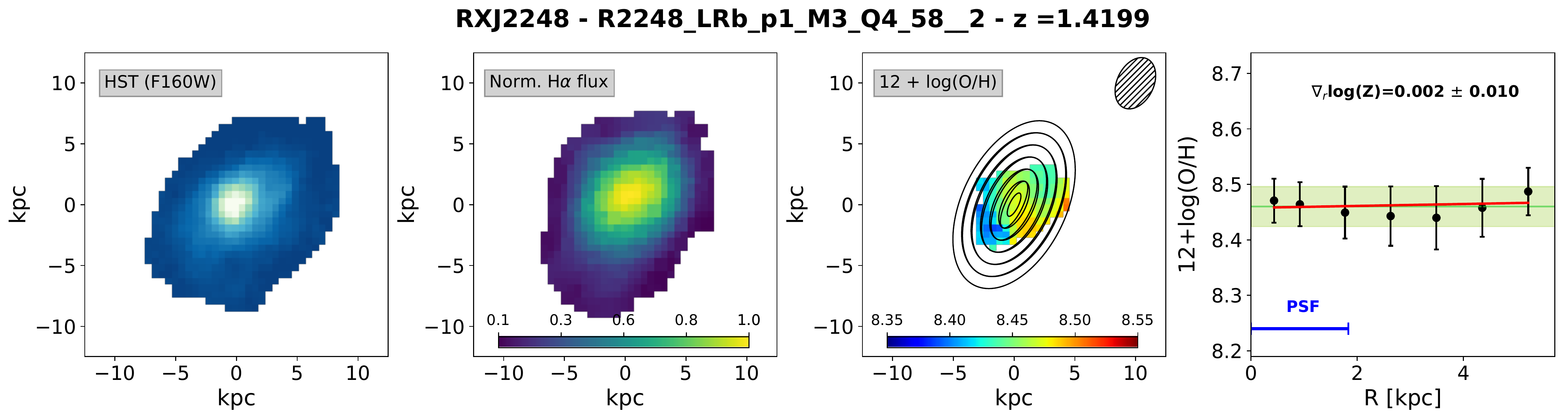}\\
      \includegraphics[width=0.83\textwidth]{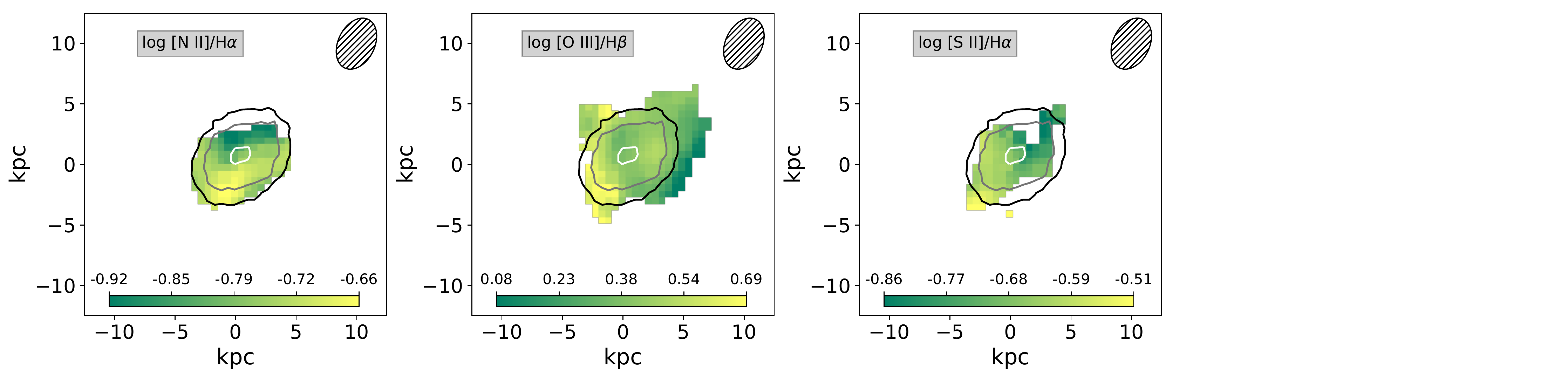}\\
      \vspace{0.1cm}
      \includegraphics[width=0.83\textwidth]{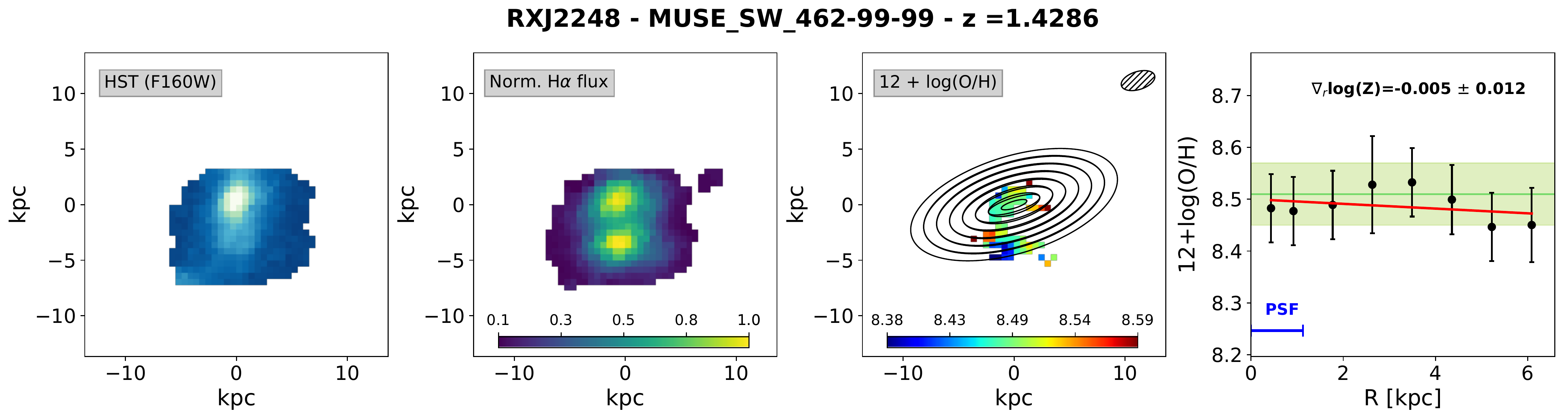}\\
      \includegraphics[width=0.83\textwidth]{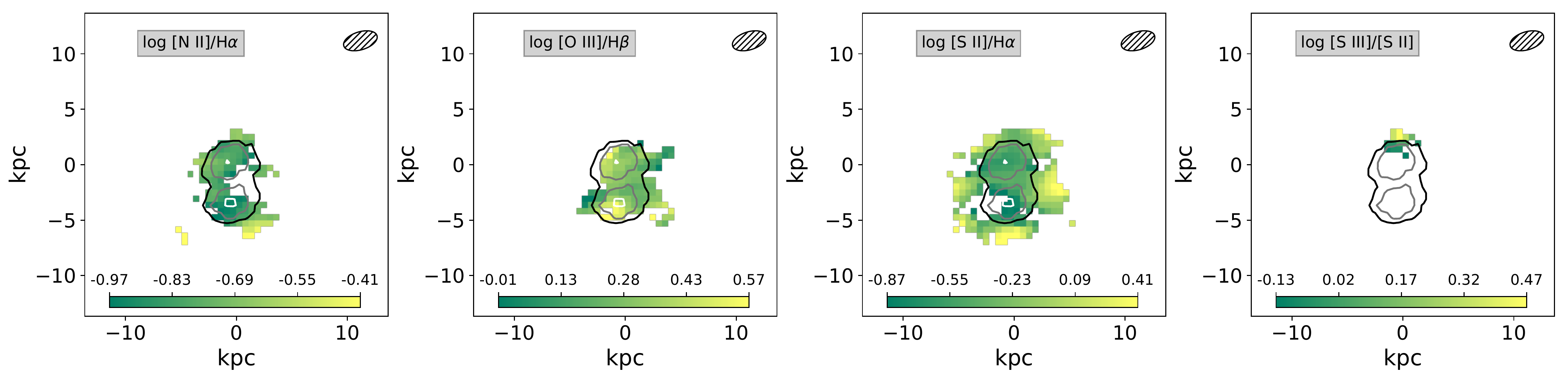}\\
 \caption{Same as Fig.~\ref{fig:appen_zgrads}.}

  \end{figure*}

  \begin{figure*}
      \centering
     \includegraphics[width=0.83\textwidth]{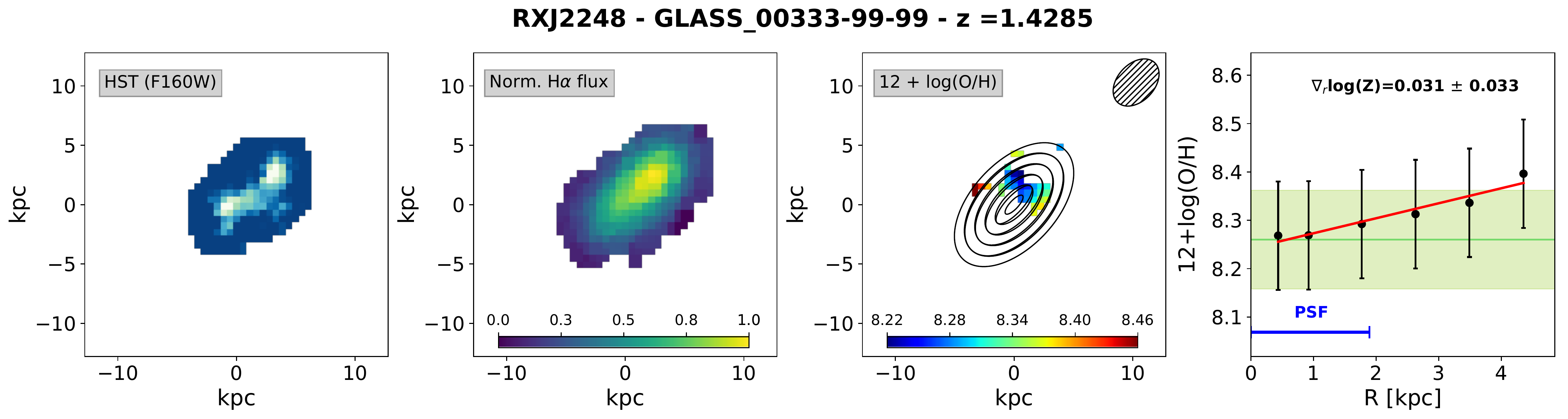}\\
     \includegraphics[width=0.83\textwidth]{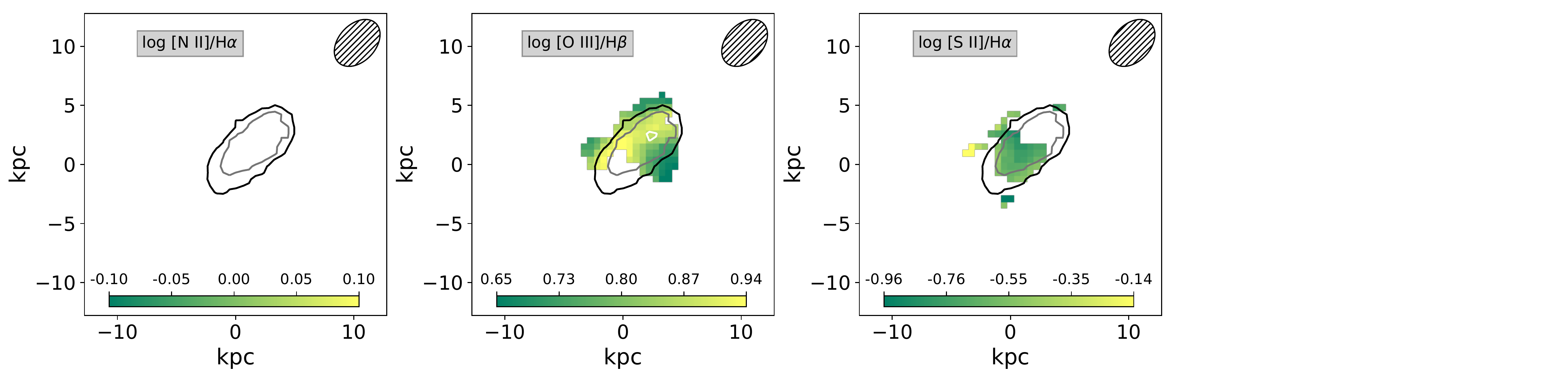}\\
      \vspace{0.1cm}     
     \includegraphics[width=0.83\textwidth]{/grads/R2248_LRb_p3_M4_Q3_93__1_met_grad}\\
     \includegraphics[width=0.83\textwidth]{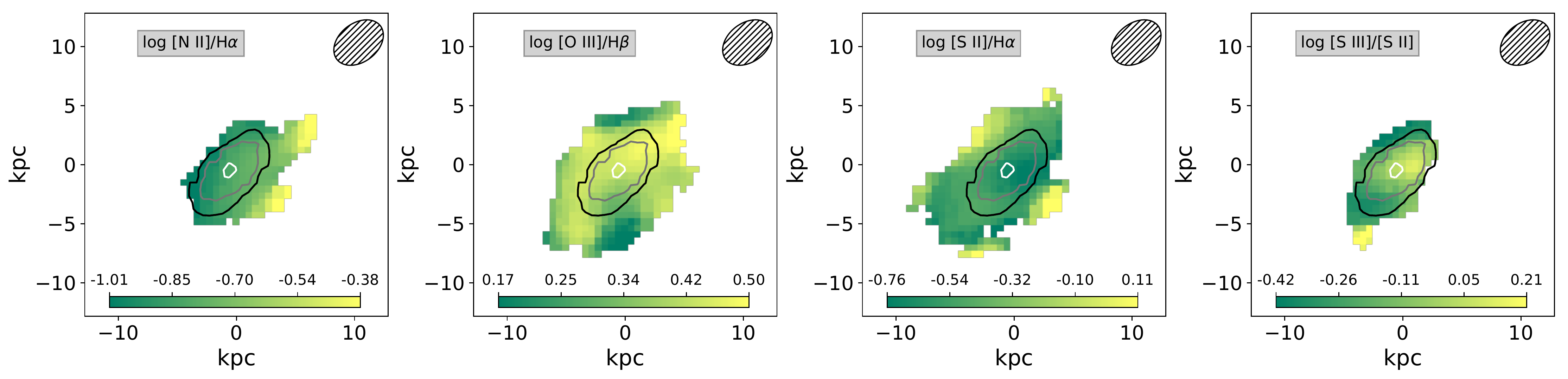}\\
      \vspace{0.1cm}
      \includegraphics[width=0.83\textwidth]{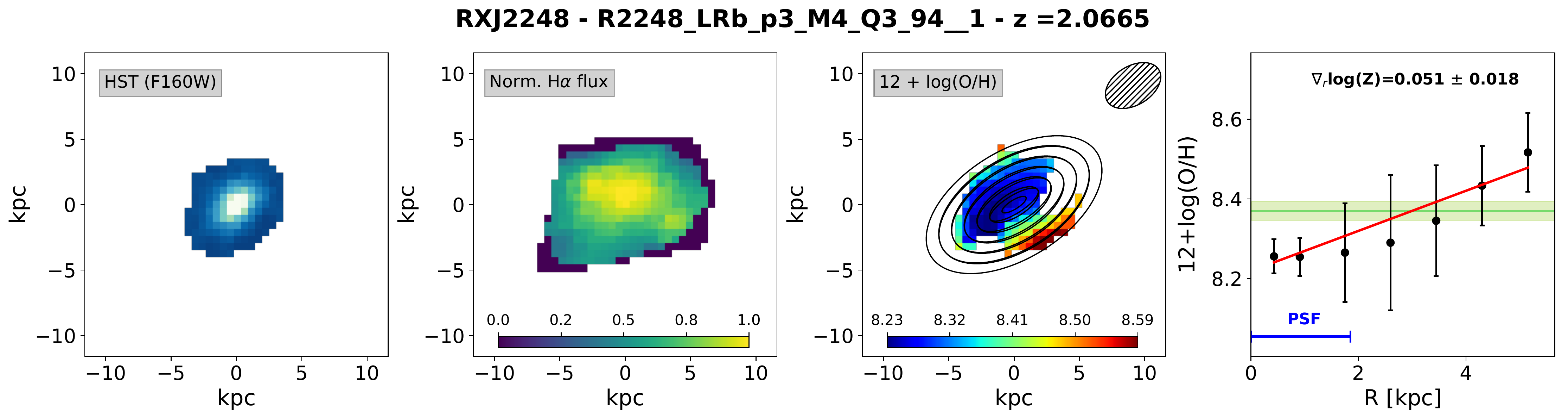}\\
      \includegraphics[width=0.83\textwidth]{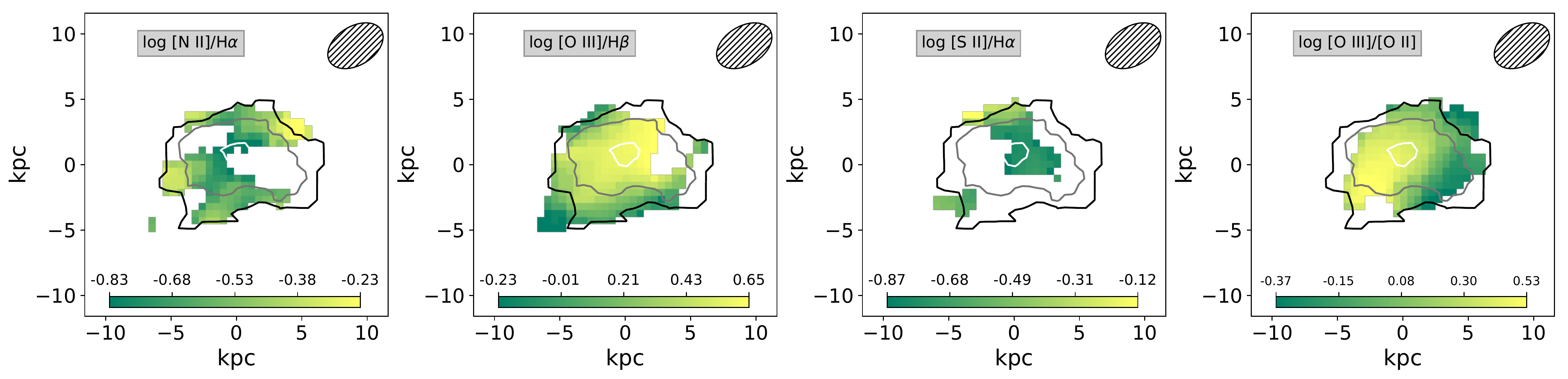}\\
\caption{Same as Fig.~\ref{fig:appen_zgrads}.}

  \end{figure*}

  \begin{figure*}
      \centering
      \includegraphics[width=0.83\textwidth]{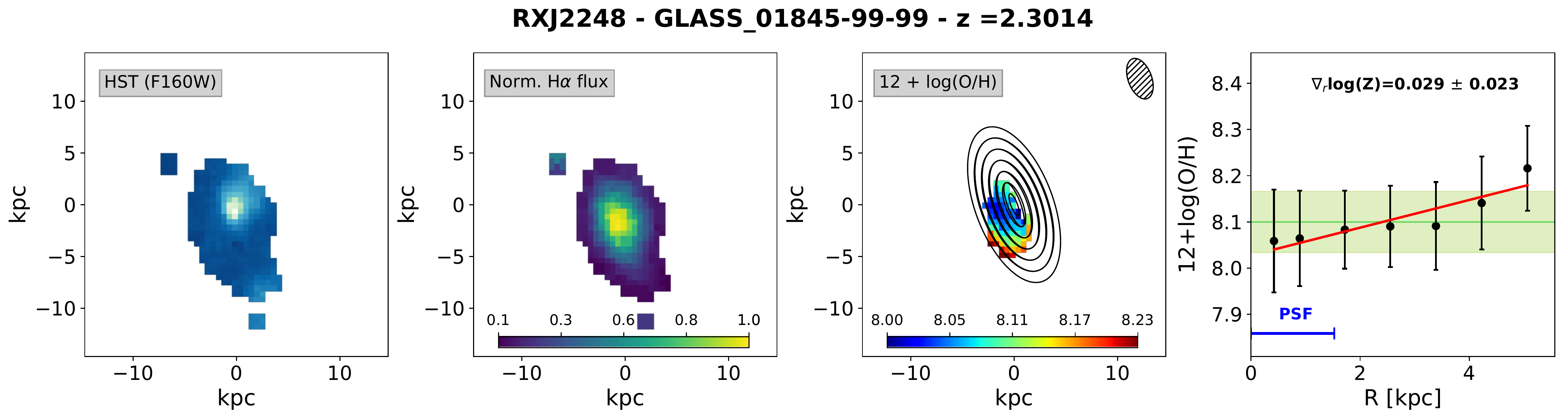}\\
      \includegraphics[width=0.83\textwidth]{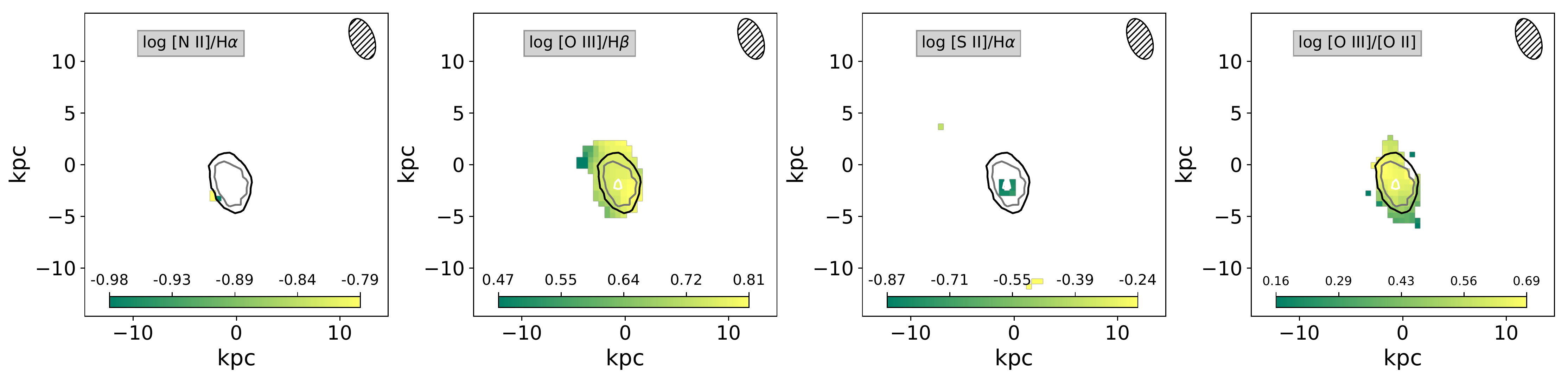}\\
      \vspace{0.1cm}
      \includegraphics[width=0.83\textwidth]{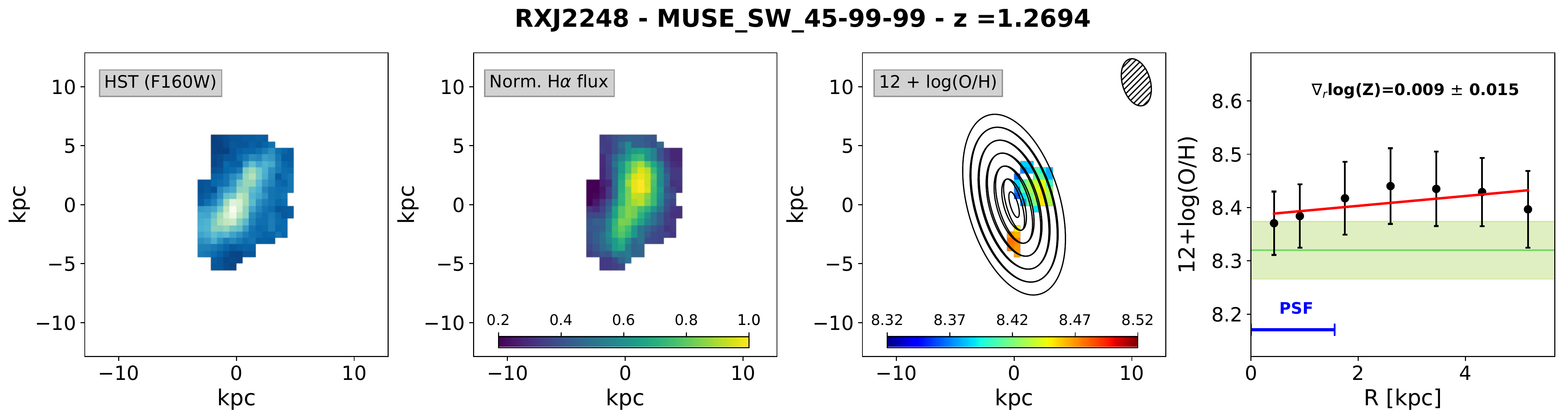}\\
      \includegraphics[width=0.83\textwidth]{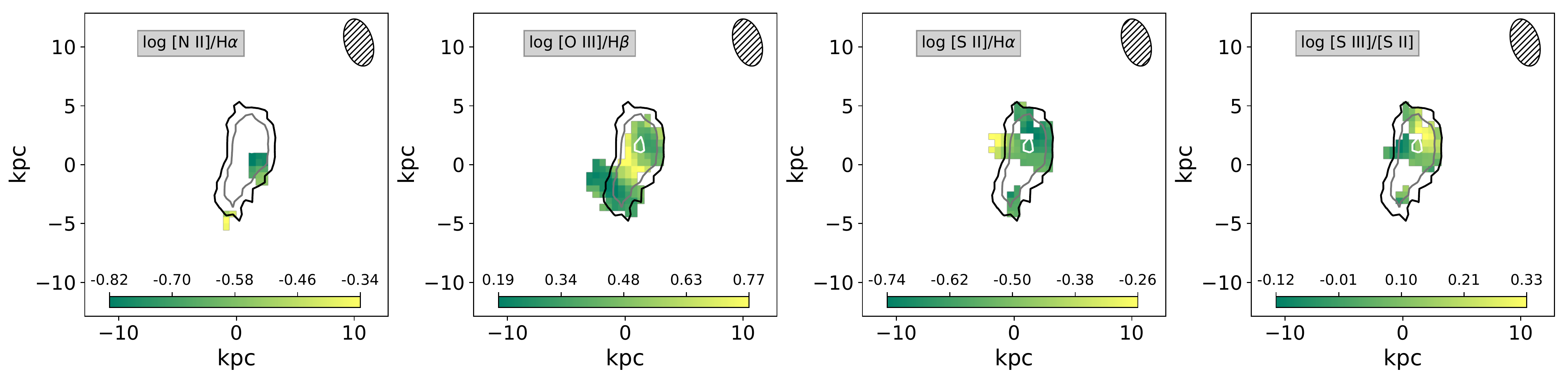}\\
      \vspace{0.1cm}
	\includegraphics[width=0.83\textwidth]{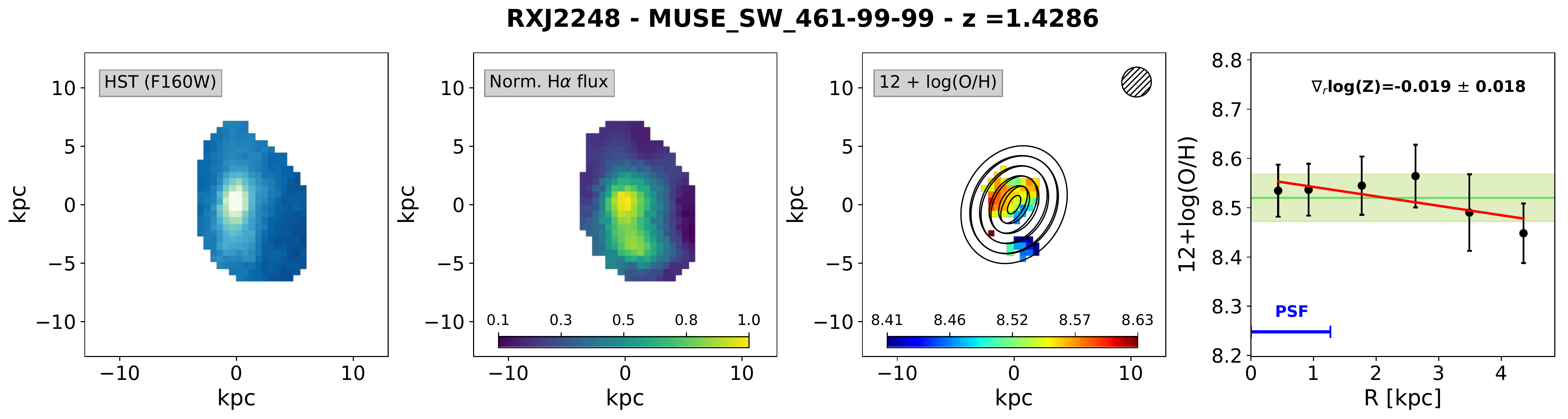}\\
	\includegraphics[width=0.83\textwidth]{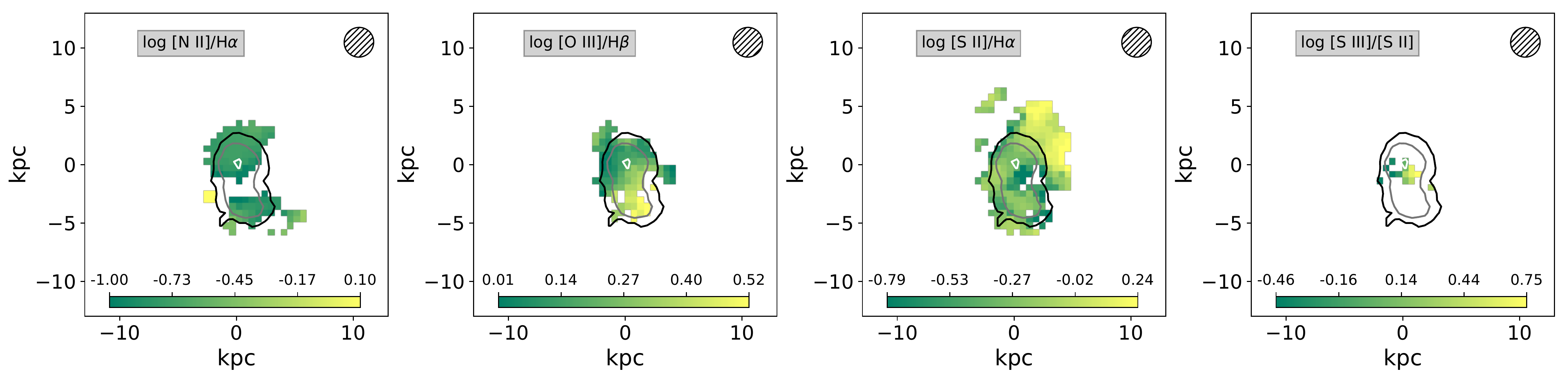}\\

\caption{Same as Fig.~\ref{fig:appen_zgrads}.}

  \end{figure*}

  \begin{figure*}
      \centering
      \includegraphics[width=0.83\textwidth]{/grads/MUSE_NE_111-99-99_met_grad}\\
      \includegraphics[width=0.83\textwidth]{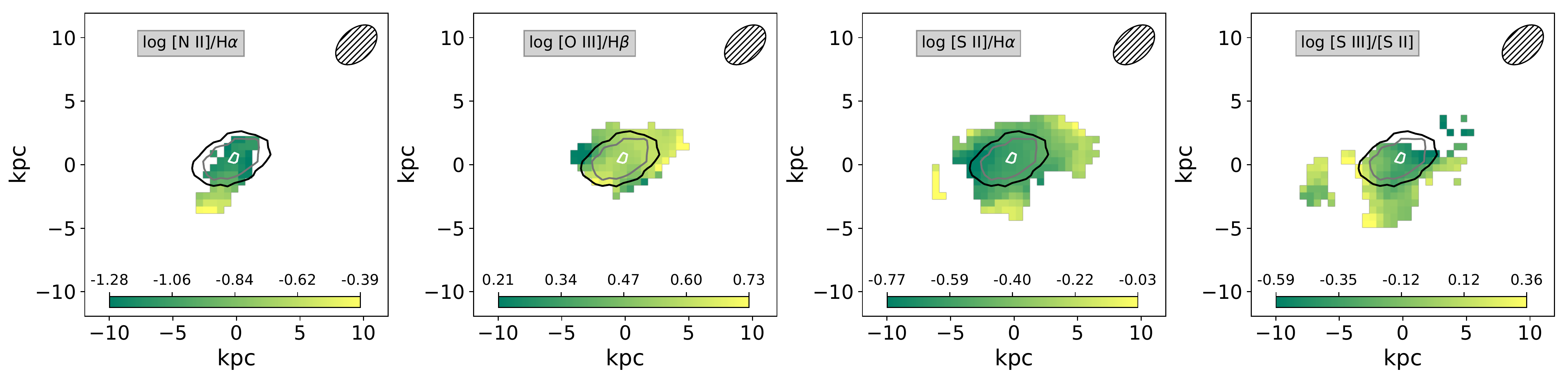}\\
      \vspace{0.1cm}
      \includegraphics[width=0.83\textwidth]{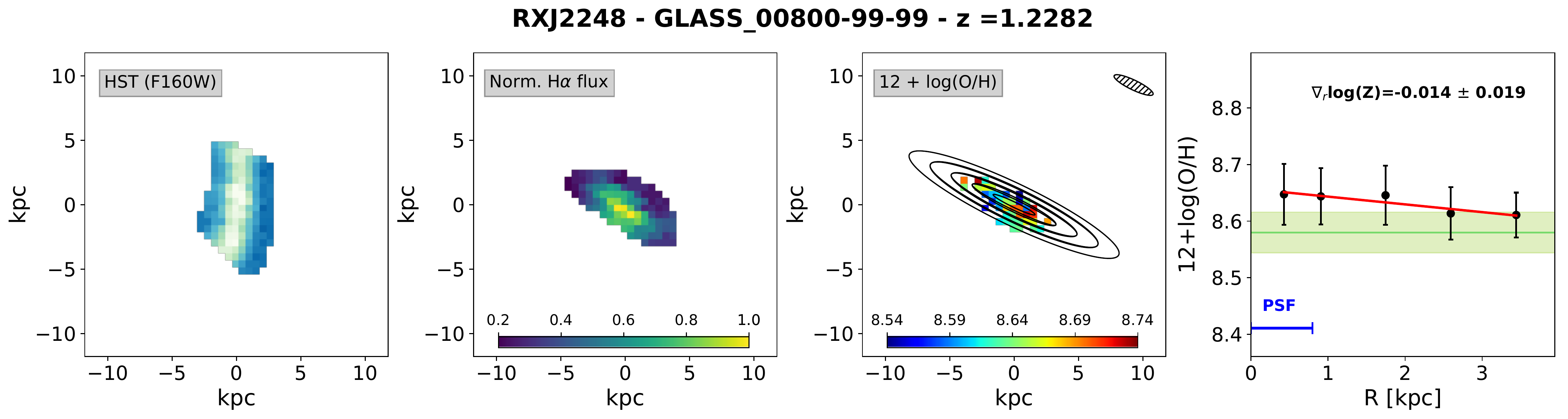}\\
      \includegraphics[width=0.83\textwidth]{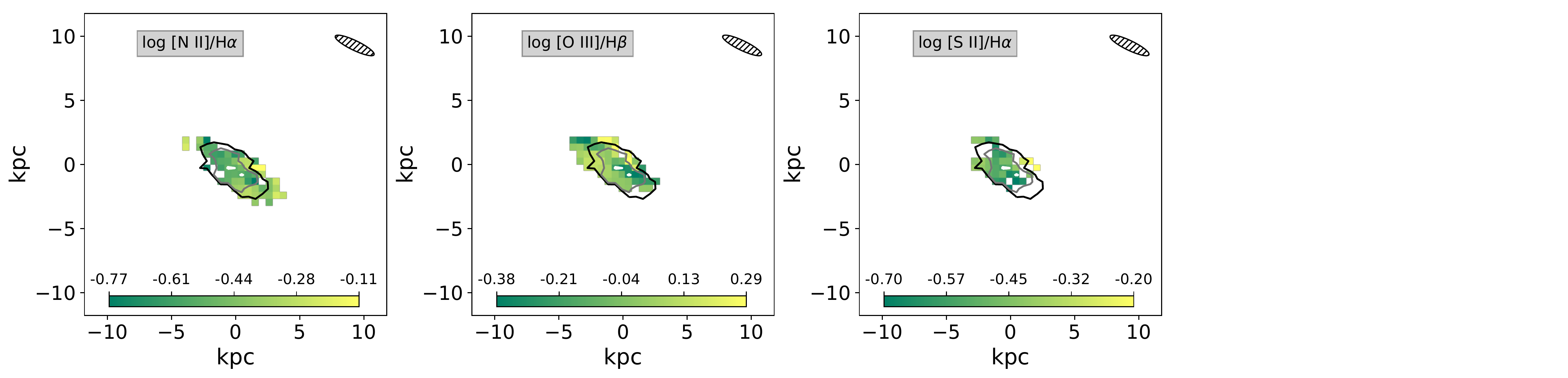}\\
      \vspace{0.1cm}
      \includegraphics[width=0.83\textwidth]{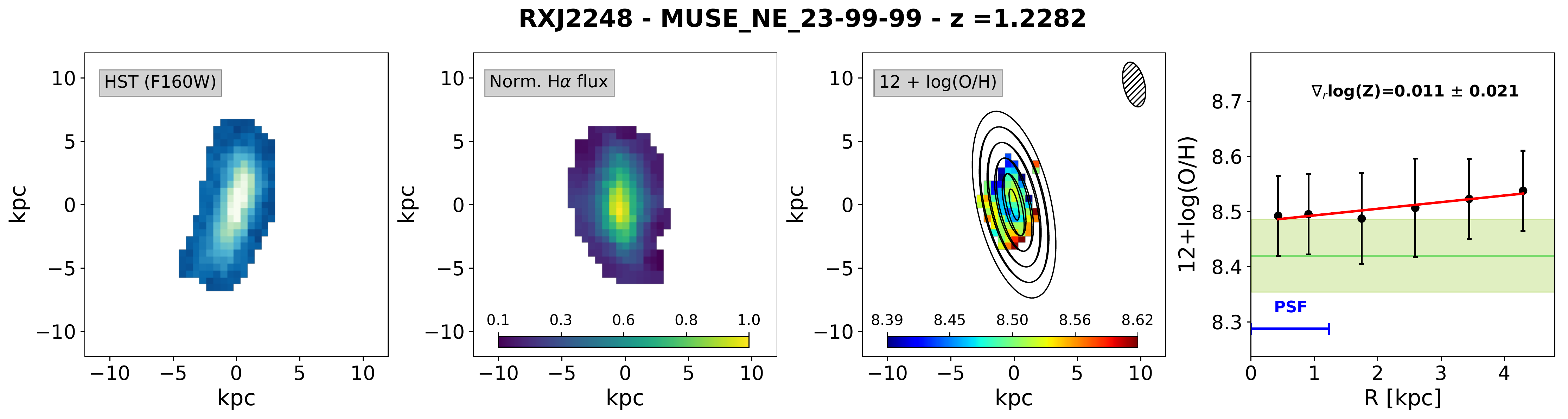}\\
      \includegraphics[width=0.83\textwidth]{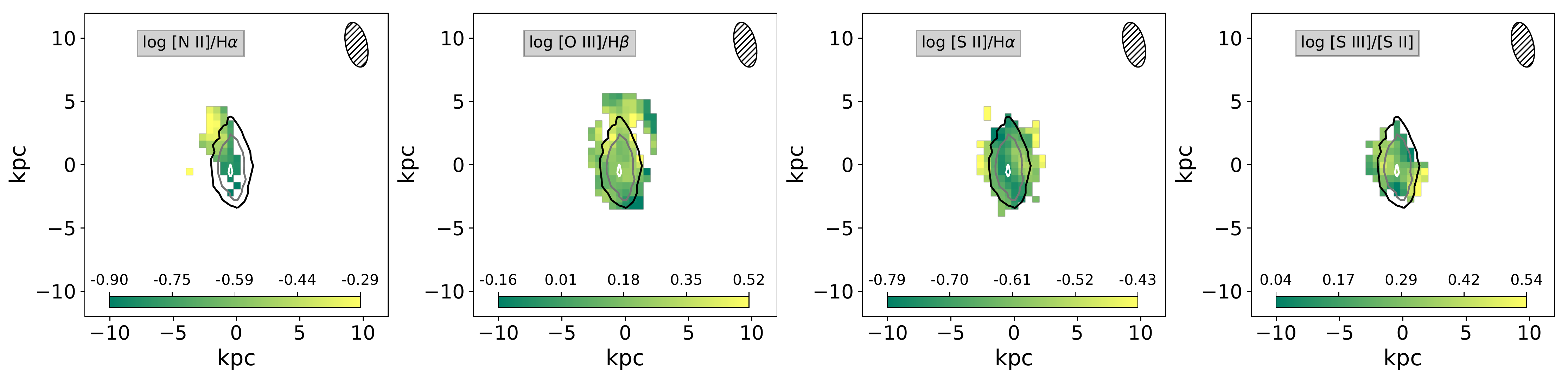}\\

\caption{Same as Fig.~\ref{fig:appen_zgrads}.}
\label{fig:append_lastfig}

  \end{figure*}

\end{document}